# Symmetry breaking and restoration for many-body problems treated on quantum computers

*Brisure et restauration de symétrie pour les problèmes à plusieurs corps traités sur ordinateurs quantiques*

**Doctoral thesis from Paris-Saclay University**

Doctoral school n° 576, Particules, Hadrons, Énergie et Noyau : Instrumentation, Imagerie, Cosmos et Simulation (PHENIICS)
Doctoral specialty: Nuclear structure and reactions
Graduate School: Physics, Referent: Faculty of Sciences of Orsay

Thesis prepared in the research unit UMR9012, IJCLab, France

**Thesis defended at Paris-Saclay, September 21, 2023, by**

**Edgar Andres RUIZ GUZMAN**

**Composition of the jury**

| | |
|---|---|
| **Elias Khan**<br>Professor, Paris-Saclay University, IJCLab, France | President |
| **Mariane Mangin-Brinet**<br>CNRS Research Director, LPSC Grenoble, France | Reviewer & Examiner |
| **José Enrique García-Ramos**<br>Professor, University of Huelva, Spain | Reviewer & Examiner |
| **Thomas Ayral**<br>Research Engineer, Atos, France | Examiner |
| **Francesco Pederiva**<br>Professor, Trento University, and INFN-TIFPA, Italy | Examiner |

**Thesis direction**

**Denis Lacroix**
CNRS Research Director, IJCLab, France

DOCTORAL THESIS

# Contents

# 1 - Introduction

Quantum computers hold the potential to revolutionize computing, with prospective applications across a broad spectrum of scientific fields, from cryptography to drug discovery. Since Richard Feynman first conceptualized this new paradigm of computing in the 1980s [1], it had largely remained theoretical until recent advancements in manipulating physical systems at the quantum level paved the way for the construction of such machines.

The idea of a quantum computer originated as a response to the limitations of classical computers in simulating quantum systems [1]. Despite considerable progress in contemporary science, even the most advanced supercomputers and algorithms have found it challenging to precisely simulate a quantum system with just a few dozen particles. This predicament stems from the fact that the configuration space grows exponentially with each added degree of freedom, rendering the exact simulation of such systems computationally infeasible.

However, the increasing precision in manipulating small physical systems has reframed this challenge as an opportunity. At its core, the quantum computing paradigm involves storing information in quantum states of matter and employing "quantum gates" to manipulate them, precisely performing operations on the stored information. A series of such operations can create a "quantum algorithm", which leverages the principles of superposition and quantum interference to accomplish specific tasks.

The advent of physically realized quantum computers has sparked renewed interest in this technology. However, the current generation of quantum machines is constrained by the inherent noise associated with manipulating the delicate quantum systems they are constructed upon. This has resulted in the term "Noisy Intermediate Scale Quantum" (NISQ) era [2], which describes this transitional phase towards fully fault-tolerant machines. Despite these limitations, the opportunity to experiment with actual quantum devices has stimulated intensified research in the field. Researchers are actively exploring ways to harness the present and future capabilities of quantum computers to address challenges across various domains, including cryptography [3], condensed matter physics [4], nuclear physics [5, 6], quantum chemistry [7, 8, 9], among others. The many-body problem, a common thread across several scientific domains mentioned previously, involves predicting the properties of a multitude of interacting quantum particles. The exponential scaling of its configuration space often leaves the simulation of large systems, those involving many particles, beyond the grasp of traditional ab initio methods [19, 20, 21, 22]. These methods are founded exclusively on quantum mechanics principles and do not rely on empirical parameters adjusted to fit experimental data. However, the increased computational power and a deepened understanding of the underlying physics have rekindled interest in these techniques. Among the ab initio methods, some exploit the symmetries of the system to reduce the size of the Hilbert space under consideration [23, 24, 25, 26, 27]. Additionally, alternative approaches take advantage of symmetry breaking followed by symmetry restoration techniques [28, 29, 30]. These strategies have proven capable of capturing complex correlations among quantum particles at lower computational costs compared to earlier methods. While historically, these methods have been developed within the classical computing paradigm, recent advancements in quantum computing platforms suggest that it is time to consider their potential application in the quantum computing domain. In this context, this thesis concentrates on exploring the feasibility of adapting the symmetry-breaking/restoration strategy to a hybrid quantum-classical framework where quantum computers could demonstrate their utility. Figure 1.1 provides a schematic representation of the thesis's focus, which is the examination of Quantum Many-Body problems. Within this scope, we investigate the potential of quantum computers as an alternative to traditional computing for characterizing certain properties of Many-Body systems, such as the ground-state energy within a spectrum and dynamic attributes like superconductivity.

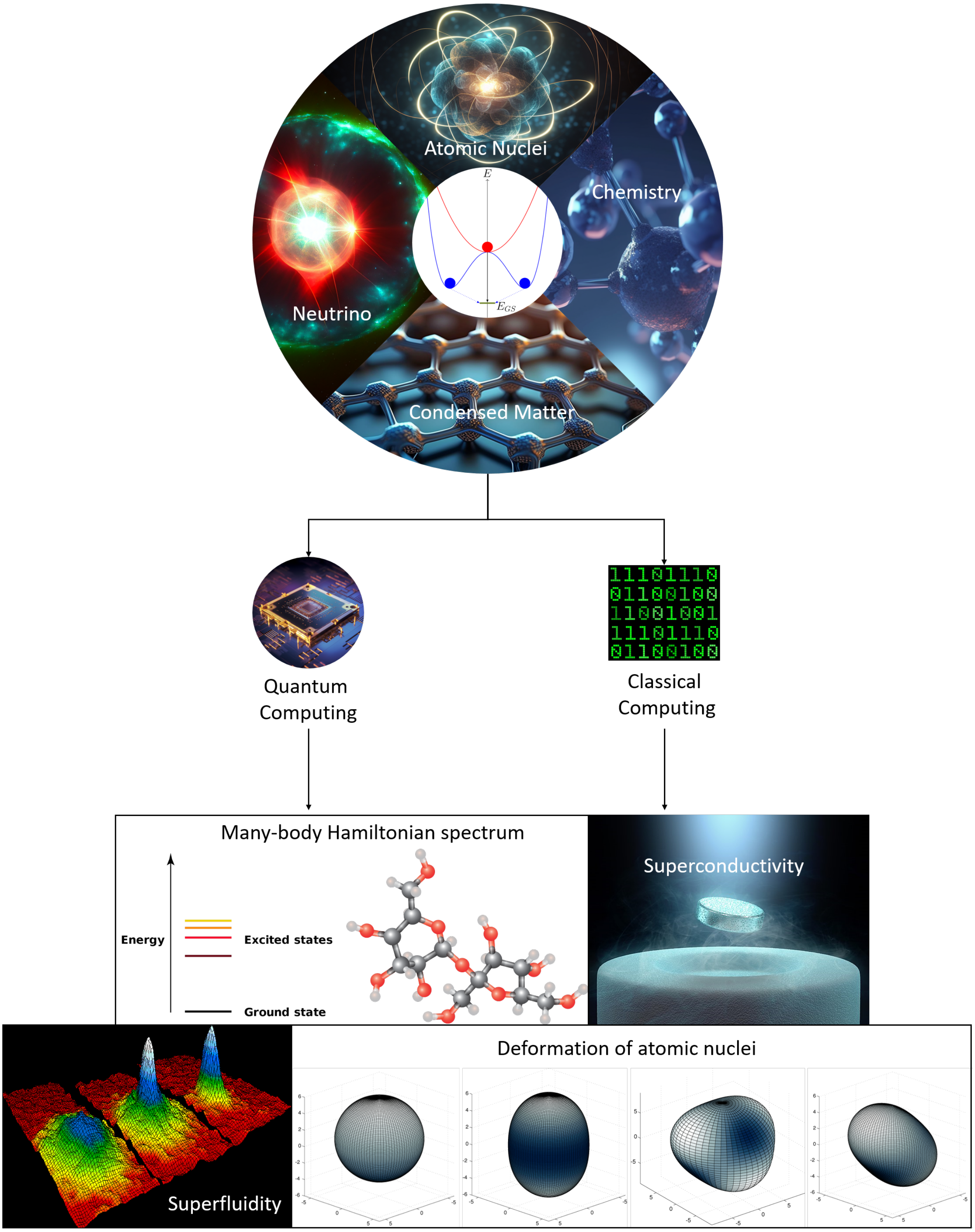

Figure 1.1: Schema showcasing potential applications of the symmetry-breaking/restoring technique, centrally illustrated in the upper circle. The figure elucidates how, by utilizing both classical and quantum computing, we can investigate interesting phenomena in many-body systems, including energy spectrum analysis, superconductivity, superfluidity, and intrinsic deformation within atomic nuclei. This figure is licensed under CC BY-SA 4.0 https://creativecommons.org/licenses/by-sa/4.0/deed.en and is based on the sources [10, 11, 12, 13, 14, 15, 16, 17, 18].

The primary focus throughout this thesis will be on using the symmetry-breaking/symmetry-restoration technique to estimate the ground-state energy of a many-body Hamiltonian. A comprehensive analysis of the classical and quantum computational complexity of this problem reveals that it is unlikely for either classical or quantum computers to find a polynomial time algorithm to resolve it. In such situations, heuristic algorithms–polynomial algorithms without quality guarantees, often taking the form of variational techniques–remain a viable alternative. Quantum heuristic methods, such as quantum variational methods, are expected to enhance the capabilities of classical heuristics, potentially offering solutions in cases previously considered unreachable.

To elucidate the method developed in the thesis, let us start by considering a Hamiltonian with a particular symmetry–meaning when a quantum system evolves using this Hamiltonian, a certain quantity is conserved, such as the number of particles or spin. Each value that this quantity can assume, which is both finite and discrete, delineates a separate block of the total Hamiltonian that can be independently diagonalized. The ground state function corresponding to each block will bear the symmetry value associated with that block. Consequently, one might consider as reasonable to choose an ansatz for a variational procedure to identify the ground state that enforces this symmetry, as it restricts the search to the pertinent segment of the Hilbert space. However, as we will demonstrate in chapter 3, breaking the symmetry of the ansatz may be beneficial, allowing exploration of uncharted parts of the Hilbert space. We can then restore the symmetry using a projector to achieve an approximated ground state that maintains proper symmetries. Aptly, this technique is called the symmetry-breaking/symmetry-restoring method.

Following a broad introduction to quantum computing applied to the many-body problem in the second chapter, the third chapter delves into the details of incorporating the symmetry-breaking/symmetry-restoring method into the quantum computation framework. Specifically, this chapter shows how the integration of this technique with the Variational Quantum Eigensolver (VQE) algorithm allows us to establish two distinct processes, depending on whether the ansatz parameters have been varied before or after performing the projection. The first scenario is known as Quantum Projection After Variation (Q-PAV), while the second is Quantum Variation After Projection (Q-VAP).

Initially, the projective method employed was the Quantum Phase Estimation (QPE) algorithm. We utilized this algorithm because we deal with symmetry operators with known and discrete eigenvalues. Leveraging these properties, the measurement of a certain eigenvalue in the ancillary register of the QPE algorithm serves as a pointer, indicating the system has been projected to a subspace associated with that specific eigenvalue. Throughout this chapter and the remainder of the thesis, we used the Bardeen-Cooper-Schrieffer (BCS) ansatz as a symmetry-breaking ansatz for subsequent projection, along with the Jordan-Wigner transformation to encode the physical problem and associated Hamiltonian operators into quantum circuits.

Considering that the QPE method for projection may incur costly computational resources and might not even converge precisely for certain symmetries such as the spin one, we explored alternative projection methods based on QPE (discussed at the end of chapter 3) and the concept of an "Oracle" in chapter 4. The first type of method, QPE-based, operates on the same principle as the QPE algorithm, specifically by projecting the quantum state into a particular subspace using the associated eigenvalue as a pointer. Meanwhile, Oracle-based methods depend on the "Oracle" operator; these methods hinge on the operator's ability to distinguish the subspace with the correct symmetry from its complementary subspace. Some of the methods presented in both sections are based on well-established algorithms, like Amplitude Amplification, or relatively new ones, such as the Rodeo algorithm. We also introduce some techniques that, to the best of our knowledge, are entirely novel.

Chapter 5 introduces a different approach to symmetry restoration, promising a significant reduction in the use of quantum resources. This advantage is particularly valuable in the NISQ era, where the presence of errors makes each reduction in quantum resource usage beneficial. The method presented here is built upon a recently developed technique for partial state tomography called "Classical Shadows" [31]. This technique optimizes the use of quantum resources to acquire information about any specific set of observables we are interested in. Specifically, it aids in obtaining either the energy of a Hamiltonian or its projected energy. We demonstrated how to achieve the symmetry-restored energy using either particle number symmetry or spin symmetry and discussed potential optimizations of the calculations using the orthogonality of the Pauli matrices.

In the sixth chapter, with Q-PAV and Q-VAP ground state approximations as our points of departure, we present a selection of hybrid quantum-classical methods. These methods serve to either enhance the precision of ground state energy estimation or glean information about the remaining energies in the low-lying Hamiltonian spectrum. Our exploration begins with two methods that are rooted in Hamiltonian moments, denoted as $\langle H^k \rangle$. These moments can be derived from the successive derivatives of the generating function, $F(t) = e^{-itH}$. The first method, known as the $t$-expansion method, combines imaginary time evolution with a Padé approximation to enhance the accuracy of estimating the ground state energy. The second approach, known as the Krylov method, employs the concept of quantum subspace expansion. It constructs a Hamiltonian operator within a smaller subspace, known as the Krylov space, which is nestled within the overall Hilbert space. By diagonalizing a Hamiltonian represented within this Krylov space on a non-orthogonal basis, we can increasingly access the low-lying eigenvalues of the Hamiltonian spectrum as the size of the Krylov space grows. In addition, the Krylov technique aids in accurately estimating the evolution of the survival probability.

Next, we delve into the investigation of the level of precision that is attainable when extracting Hamiltonian moments from quantum computers, utilizing the generating function technique. We realize that as the order of the moment increases (i.e., as $k$ grows larger in $\langle H^k \rangle$), the precision of the Hamiltonian moment estimation decreases, even in the absence of quantum errors. We find that only the methods that are rooted in either a quantum or classical version of the Fourier Transform can yield a satisfactory level of accuracy for further post-processing. However, these methods achieve this level of precision by approximating the Hamiltonian spectrum, which is the main objective of the hybrid quantum-classical methods presented here. In light of this realization, we conclude that the Krylov technique that estimates the Hamiltonian moments from a finite difference method is not a practical solution within the quantum computing framework. In search of alternatives to the Krylov approach within the quantum subspace expansion methods domain, we explored the Quantum Krylov method. This technique directly uses the values of the generating function for the Hamiltonian construction in the subspace. Together with the Q-PAV and Q-VAP methods, we demonstrate how the Quantum Krylov method can enhance the accuracy of ground-state energy estimation and provide information about the low-lying Hamiltonian spectrum.

# 2 - Introduction to Quantum Computing for Many-Body Systems

This chapter provides an overview of the fundamental concepts essential to comprehend the results presented in this thesis. Firstly, we present a brief introduction to the two primary forms of quantum computation, namely, analog and digital computing. Subsequently, we focus on digital quantum computation, which serves as the primary framework used in this study. Here, we discuss why quantum computing has recently brought attention to its application in many-body physics and its potential advantages. Finally, we introduce in the framework of digital quantum computation the two Hamiltonians used throughout the rest of the thesis. For general reviews on this topic, we refer to [19, 32, 33].

## 2.1 . Outline Quantum Computation

Quantum computers are synthetic many-body systems whose state can be controlled using a series of steps, also known as quantum algorithms. Depending on the level of control of their degrees of freedom, they can be differentiated into "analog" and "digital" quantum computers, with the latter having a more precise level of control than the former. As a result, digital quantum computers have the potential for universality, meaning they can theoretically reach any part of the Hilbert space.

### 2.1.1 . Analog Quantum Computation

Analog quantum computers are also called quantum simulators since they are commonly used as proxies of real many-body phenomena. With this goal in mind, their Hamiltonian is tailored as closely as possible to the phenomena to be studied to gain insights into its physics.

Quantum simulators, reviewed in [34, 35], are limited in that each platform can implement only one class of Hamiltonians with partial (often only global) control over their degrees of freedom. These constraints do not allow them to be universal; however, it usually permits them to work with significantly more degrees of freedom than digital quantum computers. Some of the principal platforms for analog quantum computing are (see also Table 2.1):

**Ultracold atoms:** It refers to atoms cooled to extremely low temperatures, usually on the order of a few microKelvins or less. At these temperatures, the atoms move slowly enough to exhibit quantum mechanical properties and can be trapped using electromagnetic fields, which allow precise manipulation and control. Once trapped on a lattice, they can be described using the Fermi- or Bose-Hubbard model conditional on the atomic isotopes used [48]. For example, the one-dimensional Bose-Hubbard model reads:

$$H(t) = \frac{U(t)}{2} \sum_i n_i(n_i - 1) - J(t) \sum_{\langle ij \rangle} \left( b_i^\dagger b_j + b_j^\dagger b_i \right). \tag{2.1}$$

$b_i^\dagger$ and $b_i$ are the bosonic creation and annihilation operators of an atom in the lattice site $i$ so that $n_i = b_i^\dagger b_i$ gives the number of particles on-site $i$. $\langle i, j \rangle$ denotes a summation over all neighboring lattice sites $i$ and $j$. By altering the lasers' amplitude to create the optical lattice, it is possible to temporarily modulate the tunneling rate $J$ between neighboring sites and the on-site interaction energy $U$ between two atoms.

| | Cold atoms | Trapped Ions | Superconducting | Silicon | NV centers | Photons |
|---|---|---|---|---|---|---|
| | 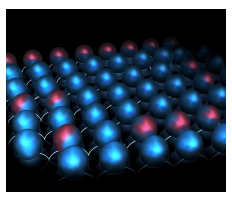 | 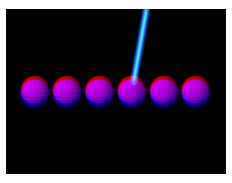 | 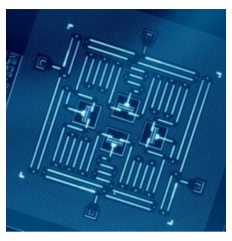 | 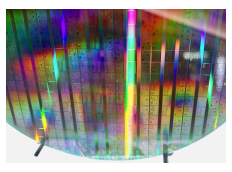 | 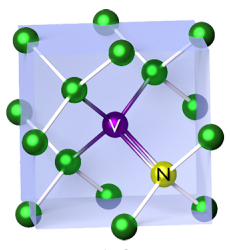 | 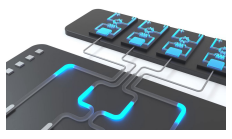 |
| Number of physical qubits | 324 | 32 | 433 | 15 | 10 | 216 |
| Best two-qubit gate fidelity (%) | 99.4 | 99.9 | 99.97 | >99 | 99.2 | 98 |
| Best readout fidelity (%) | 99.1 | 99.9 | 99.4 | 99 | 98 | 50 |
| Best gate time (ns) | 1 | $10^5$ | 20 | $5 * 10^3$ | 10 | <1 |
| Best $T_1$ (s) | >1 | 0.2 | $4 * 10^{-4}$ | $1.2 * 10^{-4}$ | $2.4 * 10^{-3}$ | $\infty$ |
| Temperature (mK) | <1 | <1 | 15 | 100 | $4 * 10^3$ | Room Temperature |
| Scalability (# qubits) | up to $10^4$ | <50 | $\sim 10^3$ | $\sim 10^6$ | $\sim 10^2$ | $\sim 10^6$ |

Table 2.1: Comparison of different kinds of physical implementations of qubits. Table adapted from [36], in this reference, the information was taken from [37, 38, 39, 40, 41, 42, 43]. Gate fidelity is the Fidelity $F = |\langle\psi_N|\psi_I\rangle|^2$ between the two states after the noisy experimental gate $|\psi_N\rangle$ and the state after an ideal gate $|\psi_I\rangle$. Readout fidelity measures the accuracy with which a qubit state can be determined. It is commonly expressed as a real number between 0.5 and 1, where a value of 1 indicates that the qubit state can be determined with certainty, and a value of 0.5 indicates a complete lack of knowledge about the qubit state. The gate time of a quantum gate is the time it takes to implement the gate. $T_1$ is the relaxation time of a qubit, i.e., the decay constant in the probability of measuring the qubit in the state $|1\rangle$ function of time $t$, $p_{|1\rangle} = e^{-t/T_1}$. The temperature value refers to the temperature at which quantum computers based on these qubits must operate. Scalability refers to the estimates of the ability to increase the number of qubits in a quantum computer without compromising the quality of the qubits with the current technologies. The number of physical qubits in the photonic platform 216 corresponds to the number of Gaussian Boson Sampling (GBS) modes. The image related to photon qubits doesn't depict GBS modes; instead, it demonstrates a qubit implementation leveraging photons' Orbital Angular Momentum (OAM). It represents two modes associated with the states $|1\rangle$ and $|-1\rangle$. These states correspond to the conventionally notated states $|0\rangle$ and $|1\rangle$. The intensity profiles of the photon's wave function are displayed in the first row, while the phase profiles are shown in the second row. This figure is licensed under CC BY-SA 4.0 `https://creativecommons.org/licenses/by-sa/4.0/deed.en` and is based on the sources [44, 45, 46, 47, 46].

**Spin qubits:** They are quantum bits based on the intrinsic spin 1/2 degree of freedom of electrons confined in quantum dots. In a spin qubit, the "up" (resp. "down") state corresponds to the electron's spin aligned (resp. anti-aligned) with an external magnetic field. These two states can represent the two classical states of a bit, 0 and 1, and therefore can be used to store and manipulate quantum information. Spin qubits can also be described by the Fermi-Hubbard model or, when neglecting charge fluctuations, by a Heisenberg model [49]:

$$H(t) = J_{\mathrm{ex}}(t)\sum_{\langle ij\rangle}(X_iX_j + Y_iY_j + Z_iZ_j) + \sum_i H_{\mathrm{loc}}^{(i)}, \tag{2.2}$$

with $(X_i, Y_i, Z_i)$ denoting the Pauli matrices acting on the $i^{th}$ spin (see Table 2.2). The exchange constant $J_{\mathrm{ex}} \sim 4J(t)^2/U$ can be turned on and off via the tuning of the tunneling term $J(t)$ between two dots using a gate voltage. The local term $H_{\mathrm{loc}}$ can, for instance, come from a magnetic field with a static ($Z_i$ term) and a rotating ($X_i$ term) component.

**Rydberg atoms:** They are atoms that have been excited to high energy levels. In the Rydberg state, the electron is far from the nucleus, which results in a large spatial extent and strong interactions with other Rydberg atoms. Depending on which atomic levels they target, quantum computing platforms based on Rydberg atoms [50, 51] may implement an Ising Hamiltonian:

$$H(t) = \sum_{ij,i\neq j}\frac{C}{|r_i - r_j|^6}n_in_j + \frac{\Omega(t)}{2}\sum_i X_i - \delta(t)\sum_i Z_i, \tag{2.3}$$

with $n_i = (1 - Z_i)/2$, or a $XY$ Hamiltonian:

$$H(t) = 2\sum_{ij,i\neq j}\frac{C}{|r_i - r_j|^3}(X_iX_j + Y_iY_j) + \Omega(t)\sum_i X_i - \frac{\delta(t)}{2}\sum_i Z_i. \tag{2.4}$$

### 2.1.2 . Digital Quantum Computation

Digital, or gate-based quantum computers, refer to physical setups whose description can be narrowed to an assembly of interacting two-level quantum systems called qubits. While it is possible to work with $d$-level systems called qudits, this text focuses on qubits as they are the most commonly used system in quantum computation. The digital paradigm of quantum computing relies on encoding the information in an interacting set of qubits, and then manipulating them so that the final measurements of their 0 or 1 states result in computation over the information encoded. In the following, each element part of this process, depicted in Fig. 2.1, will be described.

#### Qubit

The quantum bit, or qubit, is any two-level quantum system that can be manipulated to do quantum information processing tasks. Some examples of its physical implementations include transmon qubits in superconducting circuits [53, 54], trapped ions [55, 56, 57], Rydberg atoms [58, 59, 60], photons [61] among others. The state of a qubit $|\psi\rangle$ can be represented by the arbitrary superposition of two computational states $|0\rangle$ and $|1\rangle$, i.e.,

$$|\psi\rangle = c_0|0\rangle + c_1|1\rangle, \tag{2.5}$$

with $c_i \in \mathbb{C}$ and $|c_0|^2 + |c_1|^2 = 1$. Graphically, any qubit's state can be visualized on a Bloch sphere, as shown in Fig. 2.2.

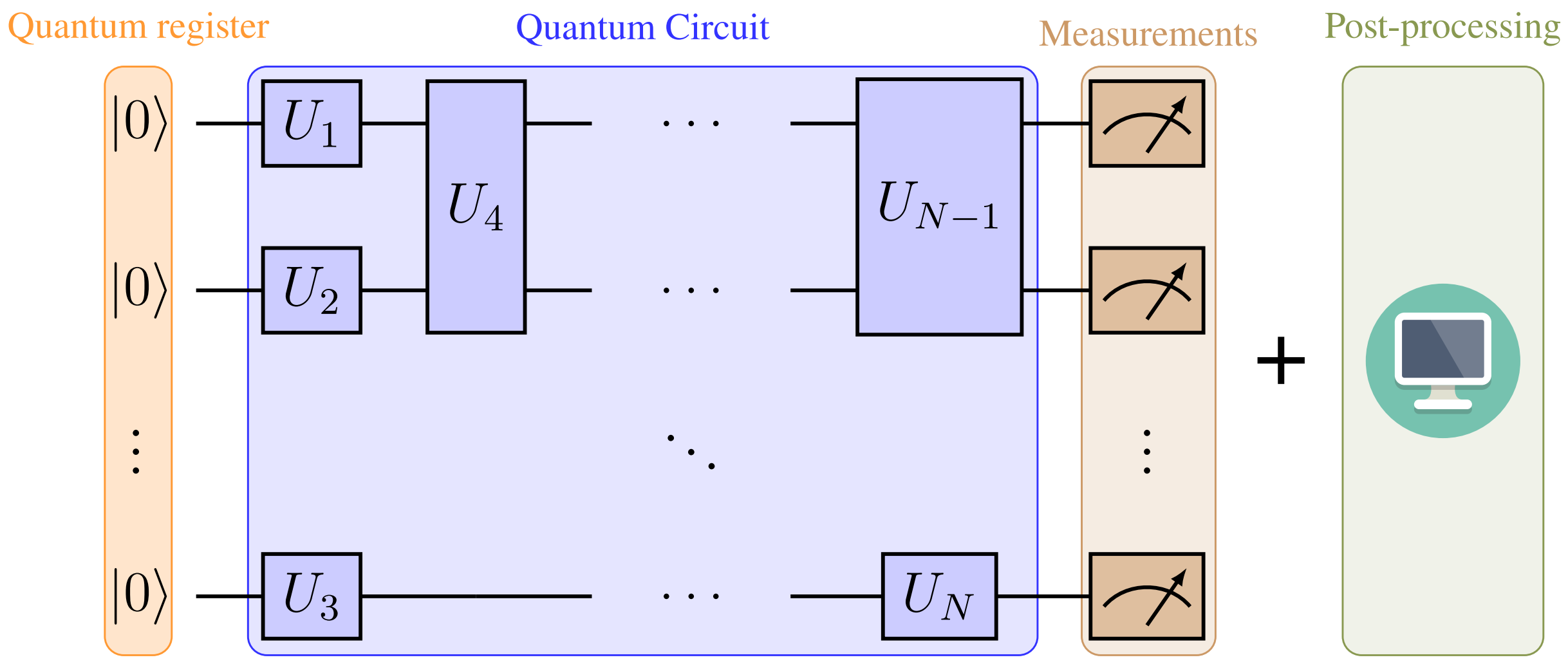

Figure 2.1: Generic process of digital quantum computation. The process begins by initializing a set of qubits, called quantum register, in the $|0\rangle$ state. Then, a set of gates $U_1, U_2, \ldots, U_N$ are applied to the register to perform the computation. In this circuit section, operations are typically expressed using one- and two-qubit quantum gates. This is because it is possible to perform any calculation using these types of gates (see Universal quantum gates in section 2.1). The classical computer records the measurement outcomes of the qubits to produce binary strings to be interpreted. The interpretation is generally made by post-processing the information in the binary strings on a classical computer. We use the Qiskit convention [52] in which the uppermost qubit corresponds to the least significant bit in the binary representation of the states.

A general state of $N$ qubits (also known as quantum register) can be written as the superposition of $2^n$ computational states of the form $\bigotimes_{i=0}^{n-1}|q_i\rangle = |q_{n-1}\ldots q_0\rangle$ with $|q_i\rangle = |0_i\rangle$ or $|1_i\rangle$, e.g., $|0\ldots0\rangle, |0\ldots1\rangle, \ldots, |1\ldots1\rangle$. Each computational state can be replaced by the integer whose binary representation corresponds to the form of the state $|k\rangle = \sum_i q_i 2^i$, i.e.:

$$|\psi\rangle = \sum_{q_{n-1}=0}^{1}\cdots\sum_{q_0=0}^{1}\psi_{q_{n-1}\ldots q_0}|q_{n-1}\ldots q_0\rangle \longrightarrow |\psi\rangle = \sum_{k=0}^{2^n-1}\psi_k|k\rangle. \tag{2.6}$$

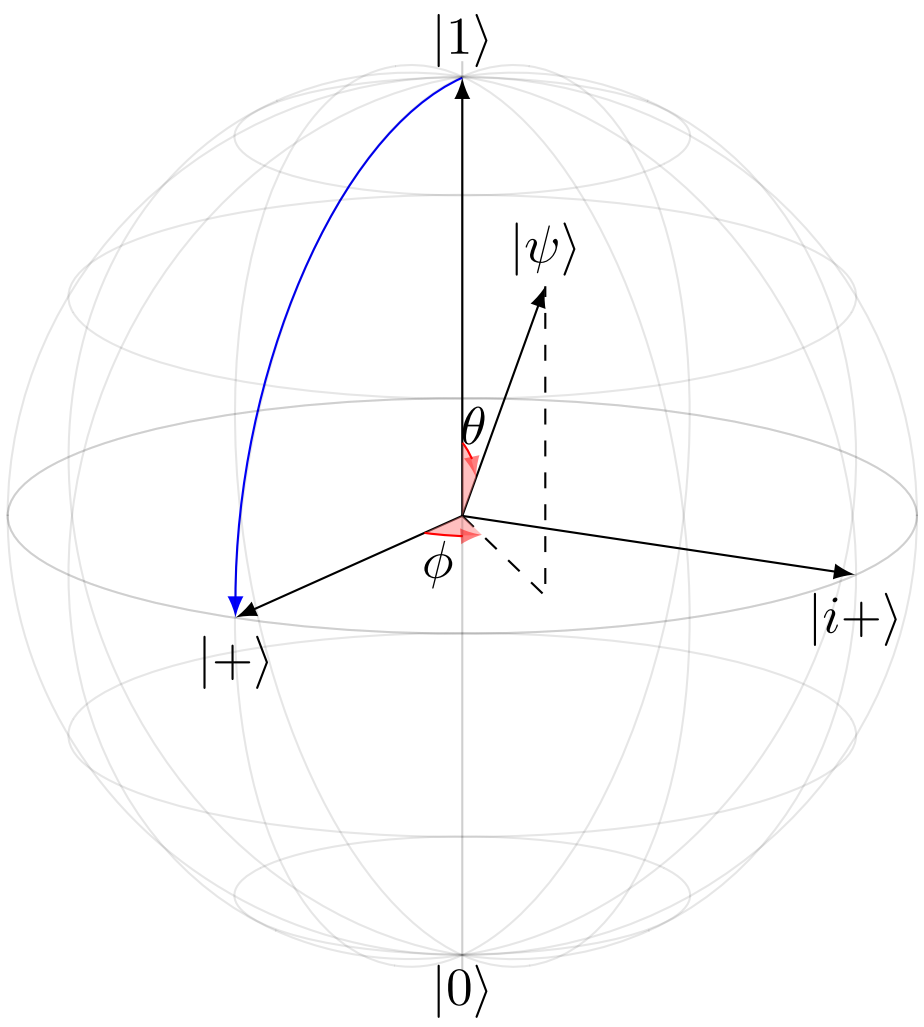

Figure 2.2: Bloch sphere portraying the possible states of a qubit. The state $|+\rangle$ corresponds to $\frac{|0\rangle+|1\rangle}{\sqrt{2}}$ and $|i+\rangle$ to $\frac{|0\rangle+i|1\rangle}{\sqrt{2}}$. In blue is depicted the ideal action of a Hadamard gate $H$ (see Table 2.2) over the state $|0\rangle$, i.e., $H|0\rangle$.

### Quantum Gates

A quantum gate is a unitary operation performed on a set of qubits to change their state. Mathematically, it can be represented as $|\Psi'\rangle = U\,|\Psi\rangle$, where $U$ is the unitary operation, and $|\Psi\rangle$ is the wave function that describes the initial state of the qubits. In physical implementations, quantum gates are achieved by evolving the system under a specific Hamiltonian. Certain gates are standardly used in the quantum circuit model of computation [62]. In Tables 2.2 and 2.3, we present several of the most commonly used single and two-qubit gates.

| Name | Symbol | Matrix | Name | Symbol | Matrix | Name | Symbol | Matrix |
|---|---|---|---|---|---|---|---|---|
| X | $X$ | $\begin{bmatrix}0 & 1\\1 & 0\end{bmatrix}$ | Y | $Y$ | $\begin{bmatrix}0 & -i\\i & 0\end{bmatrix}$ | Z | $Z$ | $\begin{bmatrix}1 & 0\\0 & -1\end{bmatrix}$ |
| Hadamard | $H$ | $\frac{1}{\sqrt{2}}\begin{bmatrix}1 & 1\\1 & -1\end{bmatrix}$ | Phase | $P\left(\varphi\right)$ | $\begin{bmatrix}1 & 0\\0 & e^{i\varphi}\end{bmatrix}$ | Universal | $U\left(\theta,\phi,\lambda\right)$ | $\begin{bmatrix}\cos\left(\frac{\theta}{2}\right) & -e^{i\lambda}\sin\left(\frac{\theta}{2}\right)\\e^{-i\phi}\sin\left(\frac{\theta}{2}\right) & e^{i(\phi+\lambda)}\cos\left(\frac{\theta}{2}\right)\end{bmatrix}$ |
| X-rotation | $R_x\left(\theta\right)$ | $\begin{bmatrix}\cos\left(\frac{\theta}{2}\right) & -i\sin\left(\frac{\theta}{2}\right)\\-i\sin\left(\frac{\theta}{2}\right) & \cos\left(\frac{\theta}{2}\right)\end{bmatrix}$ | Y-rotation | $R_y\left(\theta\right)$ | $\begin{bmatrix}\cos\left(\frac{\theta}{2}\right) & -\sin\left(\frac{\theta}{2}\right)\\\sin\left(\frac{\theta}{2}\right) & \cos\left(\frac{\theta}{2}\right)\end{bmatrix}$ | Z-rotation | $R_z\left(\theta\right)$ | $\begin{bmatrix}e^{-i\frac{\theta}{2}} & 0\\0 & e^{i\frac{\theta}{2}}\end{bmatrix}$ |

Table 2.2: Summary of some standard single-qubit quantum gates. The Clifford group, mentioned in section 2.2.1, can be generated using Pauli gates $X, Y, Z$ in conjunction with the $H$, $S = P\left(\pi/2\right)$) and CNOT gates (see Table 2.3). The phase gate $P\left(\varphi\right)$ is considered a Clifford gate only for $\varphi = \pm\pi/2$. This table has been adapted from [63] and [19].

| Name | Circuit | Matrix | Name | Circuit | Matrix |
|---|---|---|---|---|---|
| CNOT | 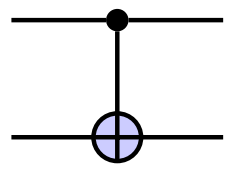 | $\begin{bmatrix}1&0&0&0\\0&0&0&1\\0&0&1&0\\0&1&0&0\end{bmatrix}$ | SWAP | 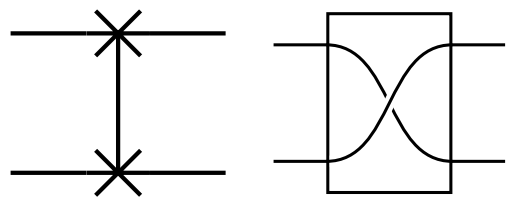 | $\begin{bmatrix}1&0&0&0\\0&0&1&0\\0&1&0&0\\0&0&0&1\end{bmatrix}$ |
| CZ | 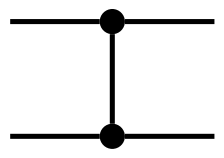 | $\begin{bmatrix}1&0&0&0\\0&1&0&0\\0&0&1&0\\0&0&0&-1\end{bmatrix}$ | fSim | 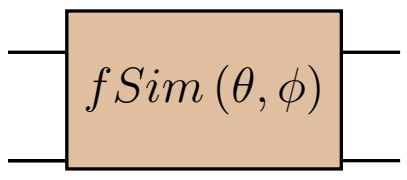  | $\begin{bmatrix}1&0&0&0\\0&\cos(\theta)&-i\sin(\theta)&0\\0&-i\sin(\theta)&\cos(\theta)&0\\0&0&0&e^{i\phi}\end{bmatrix}$ |
| a) | 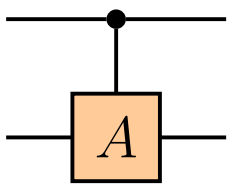  | $\begin{bmatrix}1&0&0&0\\0&a_{00}&0&a_{01}\\0&0&1&0\\0&a_{10}&0&a_{11}\end{bmatrix}$ | b) | 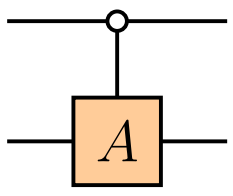  | $\begin{bmatrix}a_{00}&0&a_{01}&0\\0&1&0&0\\a_{10}&0&a_{11}&0\\0&0&0&1\end{bmatrix}$ |
| c) | 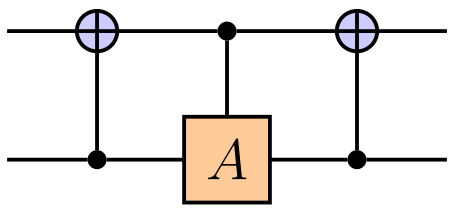  | $\begin{bmatrix}1&0&0&0\\0&a_{00}&a_{01}&0\\0&a_{10}&a_{11}&0\\0&0&0&1\end{bmatrix}$ | d) | 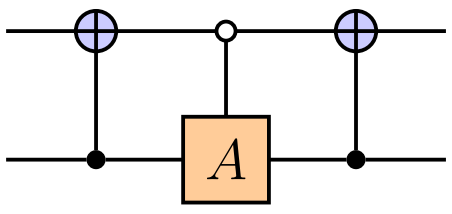  | $\begin{bmatrix}a_{00}&0&0&a_{01}\\0&1&0&0\\0&0&1&0\\a_{10}&0&0&a_{11}\end{bmatrix}$ |

Table 2.3: Summary of some common two-qubit quantum gates. I use the convention where the uppermost qubit line corresponds to the least significant bit in the binary representation of the computational state. The $A$ matrix presented from a) to d) in the table is equal to $\left(\begin{smallmatrix}a_{00}&a_{01}\\a_{10}&a_{11}\end{smallmatrix}\right)$. Table adapted from [64] and [19].

**Universal quantum gates** The Solovay-Kitaev theorem [62, 65] has established that there exist sets of gates known as *universal gate sets*, capable of implementing any unitary operation $U$ on $n$ qubits using a finite sequence of their constituent elements. A commonly used set of universal gates is comprised of one-qubit rotations $R_x(\theta)$, $R_y(\theta)$, and $R_z(\theta)$ (presented in the Table 2.2), and the two-qubit gate “CNOT” (presented in Table 2.3).

### Quantum circuits

A quantum circuit is a quantum computation model that describes a quantum register’s evolution through a sequence of quantum gates (see Fig. 2.1). Typically, the computation begins with the $n$ qubits of the quantum register in the fundamental state, i.e., $|0\rangle^{\otimes n}$. The system is then evolved by applying the gates $U_1U_2\dots U_m$ to the quantum register. This gate sequence is called the *quantum circuit*. Tables 2.2 and 2.3 show some of the standard one- and two-qubit gates applied as part of a quantum circuit, with each line representing the left-to-right time evolution of a single qubit.

### Quantum measurements

After the quantum circuit, the register is in the state $|\psi\rangle = U_mU_{m-1}\dots U_1|0\rangle^{\otimes n}$. We have to perform measurements to extract information about the quantum state. The measurement of any single qubit would collapse its state to either $|0\rangle$ or $|1\rangle$. From Eq. (2.6), a bitstring $q_{n-1}\dots q_0$ (with $q_i = \{0_i, 1_i\}$) is obtained each time a measurement is performed onto all the qubits of the register,

| Measurement | Conversion to measurement in the Z-basis |
|---|---|
| 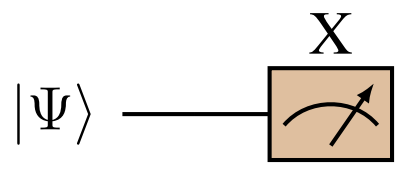 | 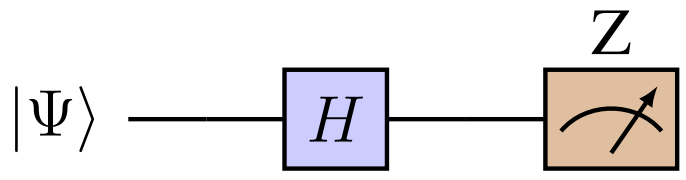 |
| 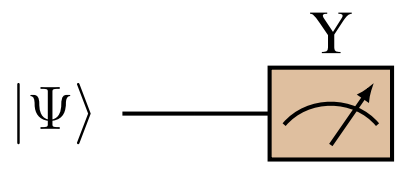 | 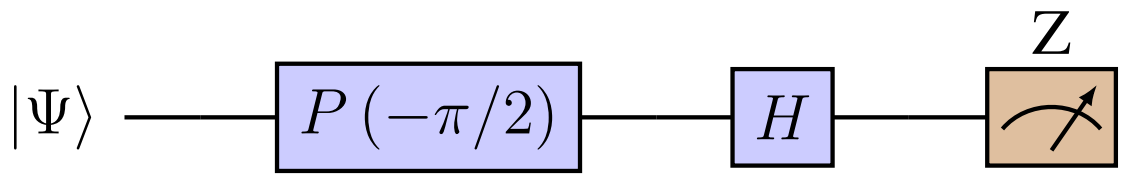 |

Table 2.4: Conversion of measurements in the $X$ or $Y$ basis into measurements in the $Z$ basis. The $P(-\pi/2)$ appears blue because at $\varphi = -\pi/2$, it becomes part of the Clifford group (cf. Table 2.2). Table adapted from [19].

with a probability given by the Born's rule [66]:

$$p_{(|q_{n-1}\ldots q_0\rangle)} = \langle\psi|\hat{P}_{|q_{n-1}\ldots q_0\rangle}|\psi\rangle = |\langle q_{n-1}\ldots q_0|\psi\rangle|^2 = |\psi_{q_{n-1}\ldots q_0}|^2, \tag{2.7}$$

where $\hat{P}_{|q_{n-1}\ldots q_0\rangle} = |q_{n-1}\ldots q_0\rangle\langle q_{n-1}\ldots q_0|$ is the projector over the state $|q_{n-1}\ldots q_0\rangle$. If one decides to perform a partial measurement of the register, i.e., to measure only a set of $m < n$ qubits, the probability of obtaining a bitstring $q_{m-1}\ldots q_0$ would be:

$$p_{(|q_{m-1}\ldots q_0\rangle)} = \langle\psi|\hat{P}_{|q_{m-1}\ldots q_0\rangle}|\psi\rangle = \sum_{q_{n-1}\cdots q_m} |\psi_{q_{n-1}\ldots q_m q_{m-1}\ldots q_0}|^2.$$

The state after the measurement $|\psi'\rangle$ becomes:

$$|\psi'\rangle = \frac{\hat{P}_{|q_{m-1}\ldots q_0\rangle}|\psi\rangle}{\sqrt{p_{(|q_{m-1}\ldots q_0\rangle)}}}$$

**Measurements in a different basis** The computational basis is the set of the eigenstates of the $Z$ Pauli matrix $\{|0\rangle, |1\rangle\}$. These eigenstates are the ones obtained in a quantum measurement. However, if one wants to measure on a different Pauli basis, the qubit's state must be transformed before the measurement. To measure in the $X$ basis, $\{|+\rangle, |-\rangle\}$, or $Y$ basis, $\{|+i\rangle, |-i\rangle\}$, one must use the relations $X = HZH$ and $Y = SHZHS^\dagger$ respectively, where $H$ and $S$ are the ones described in Table 2.2. Table 2.4 shows the circuits used to perform these measurements.

**Measuring an observable** To measure the expected value of an operator, one relates its value with the probabilities of measuring certain bitstrings on the quantum register. For example, let us say we want to measure the expected value of the $Z_0$ operator on a state $|\psi\rangle = \psi_0|0\rangle + \psi_1|1\rangle$ on 1 qubit, i.e., $\langle\psi|Z_0|\psi\rangle$, where the subscript of the Pauli matrices denotes the qubit index. In this case, the expected value of the $Z_0$ operator is:

$$\langle\psi|Z_0|\psi\rangle = (\langle 0|\psi_0^* + \langle 1|\psi_1^*)(\psi_0|0\rangle - \psi_1|1\rangle) = p_0 - p_1,$$

| **Circuit $\mathcal{C}$** | **Expected value** |
|---|---|
| 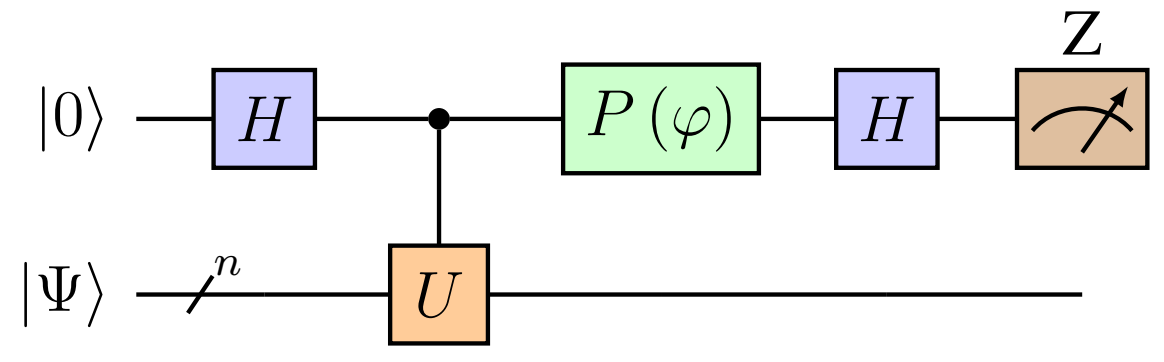 | $\langle\Psi\vert U\vert\Psi\rangle = \langle Z_0\rangle_{\mathcal{C}(\varphi=0)} + i\langle Z_0\rangle_{\mathcal{C}(\varphi=\pi/2)}$ |

Table 2.5: Here, we present a standard interferometry circuit called the Hadamard test [62], which allows for calculating the expected value of an operator $\langle\Psi|U|\Psi\rangle$ by making measurements on an ancillary qubit. On the right, $\langle Z_0\rangle_{\mathcal{C}}$ signifies that the ancillary qubit, labeled as the "$0^{th}$" qubit by convention, must be measured in the $Z$ basis after the execution of the circuit $\mathcal{C}$ on the left. Table adapted from [19].

where $p_i = |\psi_i|^2$. The process with a circuit of $N$ qubits is similar. The probabilities required for computing the expected value are estimated by taking the average of the number of counts $n_c$ a certain bitstring is obtained over the number of times the system was measured, also known as the number of shots $n_s$, i.e., $p_i = n_c/n_s$. In the limit $n_s \to +\infty$, we got $p_i = |\psi_i|^2$. Due to the central limit theorem, $1/\varepsilon^2$ measurements are needed to obtain an accuracy of $\varepsilon$ in the probabilities, or equivalently, in the expected value. A standard circuit used to measure the expected value of a general operator $U$ is given in Table 2.5.

### 2.1.3 . Selected Quantum Algorithms

This section will briefly overview some standard algorithms in quantum computing. The Quantum Phase Estimation and amplitude amplification algorithms are presented to illustrate some cases where quantum computation outperforms classical one. Additionally, the Trotter-Suzuki approximation is discussed as it is an important technique for the implementations presented in this Ph.D. thesis.

#### Quantum Phase Estimation

The Quantum Phase Estimation algorithm (QPE) [62, 67, 68] is a quantum algorithm designed to obtain information about the spectrum of a unitary operator $U$ in a quantum computer. Suppose the eigenvalues of the unitary operator $U$ over $n$ qubits are $\{e^{2\pi i\theta_j}\}_{j=0,\ldots,2^n-1}$ with $\theta_j \in [0,1[$ and have associated eigenstates $\{\phi_j\}$. We can write the system's initial state in this eigenbasis as $|\psi\rangle = \sum_i c_i|\phi_i\rangle$. The QPE algorithm's circuit is shown in Fig. 2.3.

The complete (system + ancillary) register's state just before the measurement is given by:

$$\frac{1}{2^{n_a}}\sum_{m=0}^{2^{n_a}-1}\sum_{k=0}^{2^{n_a}-1} e^{-\frac{2\pi ik}{2^{n_a}}(m-2^{n_a}\theta_i)}|m\rangle\otimes|\psi\rangle, \tag{2.8}$$

where $n_a$ indicates the number of qubits in the ancillary register (cf Fig. 2.3). The prior expression is characterized by prominent peaks corresponding to the values $2^{n_a}\theta_i$ of the ancilla register states.

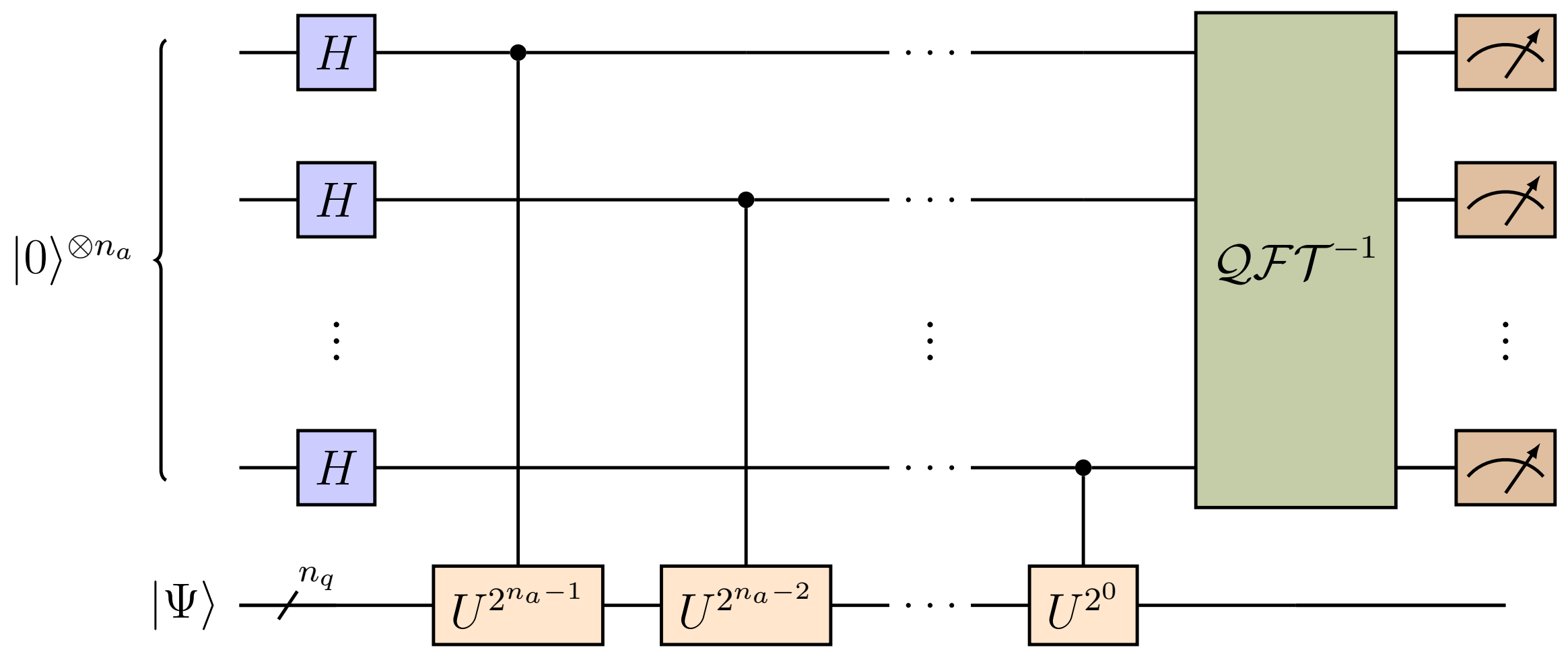

Figure 2.3: Illustration of the QPE circuit used to determine the eigenvalues of an arbitrary unitary operator $U$. The circuit employs the inverse Quantum Fourier Transform (QFT$^{-1}$) [62]. The QPE requires $n_a$ extra ancillary qubits, and the method's precision directly depends on $n_a$.

In other words, these peaks manifest when the ancilla register is in the states $|2^{n_a}\theta_i\rangle$. The values $2^{n_a}\theta_i$ will be approximated by their nearest integers, which leads to measurements of the state $|a_i\rangle$ in the ancilla register. This $a_i$ is deduced from the relationship $2^{n_a}\theta_i = a_i + 2^{n_a}\delta_i$, with $0 \leq |2^{n_a}\delta_i| \leq \frac{1}{2}$. Notably, the probability of the peak at $a_i$, denoted as $p_{|a_i\rangle}$, depends on the initial state and the value of $\delta_i$ [69].

$$p_{|a_i\rangle} = \begin{cases} |c_i|^2 & \text{if } \delta_i = 0, \\ \frac{|c_i|^2}{2^{2n_a}} \left| \frac{1-e^{2\pi i 2^{n_a}\delta_i}}{1-e^{2\pi i\delta_i}} \right|^2 & \text{if } \delta_i \neq 0. \end{cases} \tag{2.9}$$

It can be shown [68] that the factor $\frac{1}{2^{2n_a}} \left| \frac{1-e^{2\pi i 2^{n_a}\delta_i}}{1-e^{2\pi i\delta_i}} \right|^2$ in the case $\delta_i \neq 0$ is greater than $\frac{4}{\pi^2} \approx 0.4$. The probability of the peaks $a_i$ becomes:

$$p_{|a_i\rangle} = \begin{cases} |c_i|^2 & \text{if } \delta_i = 0, \\ \geq \frac{4|c_i|^2}{\pi^2} & \text{if } \delta_i \neq 0. \end{cases} \tag{2.10}$$

The QPE algorithm results in the measurement of the eigenvalues approximations $|a_i\rangle$ in the ancillary register and the projection of the initial state onto the corresponding eigenstate $|\phi_i\rangle$ in the system's register. Figure 2.4 illustrates results obtained using the QPE algorithm with varying $n_a$. This algorithm can be used, in particular, to estimate the eigenvalues $\{E_j\}$ and eigenstates of a Hamiltonian $H$. To do it, it suffices to take the unitary operator as the propagator operator:

$$U = e^{-2\pi i\tau(H-E)}, \tag{2.11}$$

where $\tau$ and $E$ are parameters chosen such that $-\tau\,(E_j - E) \in [0, 1[$. If the eigenvalues are known and can be associated with integers such that $\delta_i = 0$, then, for sufficient $n_a$, the QPE algorithm can be used to project the state $|\Psi\rangle$ into one of the subspaces defined by the eigenvalues $E_j$ [70, 71, 33]. This fact is crucial for the symmetry restoration method described in Chapter 3.

**Shor's algorithm** [3], a landmark in quantum computing, leverages the Quantum Phase Estimation (QPE) algorithm to yield a substantial speed-up over classical methods in tackling the integer factorization problem. This algorithm is devised to identify the prime factors of a composite number $N$–a positive integer that is a product of two or more prime numbers, each of which is greater than 1. Its importance lies in that it runs in poly-logarithmic time, which makes it almost exponentially faster than the most efficient known classical factoring algorithm. A comparison between Shor's algorithm complexity and the best classical algorithm for integer factorization's complexity is shown in Fig. 2.5. The nearly exponential speed-up poses a significant challenge to the most widely-used public-key cryptography schemes because they rely on the hardness of three mathematical problems that Shor's algorithm can efficiently solve: the integer factorization problem, the discrete logarithm problem, and the elliptic-curve discrete logarithm. However, this challenge has also spurred the development of post-quantum cryptography protocols [72]; these are schemes designed to withstand a quantum computer's attack. Two subroutines are necessary to implement Shor's algorithm: the order-finding and the continued fractions algorithm. Next, we give a summary of both of them.

**Order-finding algorithm** [3], is recognized as a specific application of the QPE routine. This quantum algorithm, given positive integers $a$ and $N$ where $a < N$, and that the greatest common divisor (gcd) of $a$ and $N$ is 1 (i.e., $\gcd(a, N) = 1$), is designed to identify the smallest value $r$, commonly referred to as the order, in the equation:

$$a^r \bmod N = 1, \tag{2.12}$$

where mod denotes the modulo operation. To do it, the algorithm uses the QPE algorithm with the unitary operator $U$:

$$U|k\rangle = \begin{cases} |ak \bmod N\rangle & \text{if } 0 \leq k \leq N-1, \\ |k\rangle & \text{if } N \leq k < 2^{n_q} - 1, \end{cases} \tag{2.13}$$

where $|k\rangle$ is the integer representation of a state in the computational basis. The operator $U$ is used in the QPE circuit depicted in Fig. 2.3 together with $|\psi\rangle = |1\rangle$, $n_q = \lceil \log_2(N) \rceil$ and $n_a = 2n_q + 1$. The system input state $|1\rangle$ can be viewed as an equiprobable superposition of the eigenvectors

$$|u_s\rangle = \frac{1}{\sqrt{r}} \sum_{k=0}^{r-1} |a^k \bmod N\rangle, \tag{2.14}$$

of the $U$ operator. This means that $|1\rangle$ can be written as

$$|1\rangle = \frac{1}{\sqrt{r}} \sum_{s=0}^{r-1} |u_s\rangle. \tag{2.15}$$

The eigenvectors $|u_s\rangle$ have eigenvalues $e^{2\pi i \frac{s}{r}}$, this means that the ancillary register's state in the QPE circuit will collapse to $|2^{n_a} \frac{s}{r}\rangle$ with $0 \leq s \leq r-1$. The probability of collapsing to any of the previous states is $1/r$.

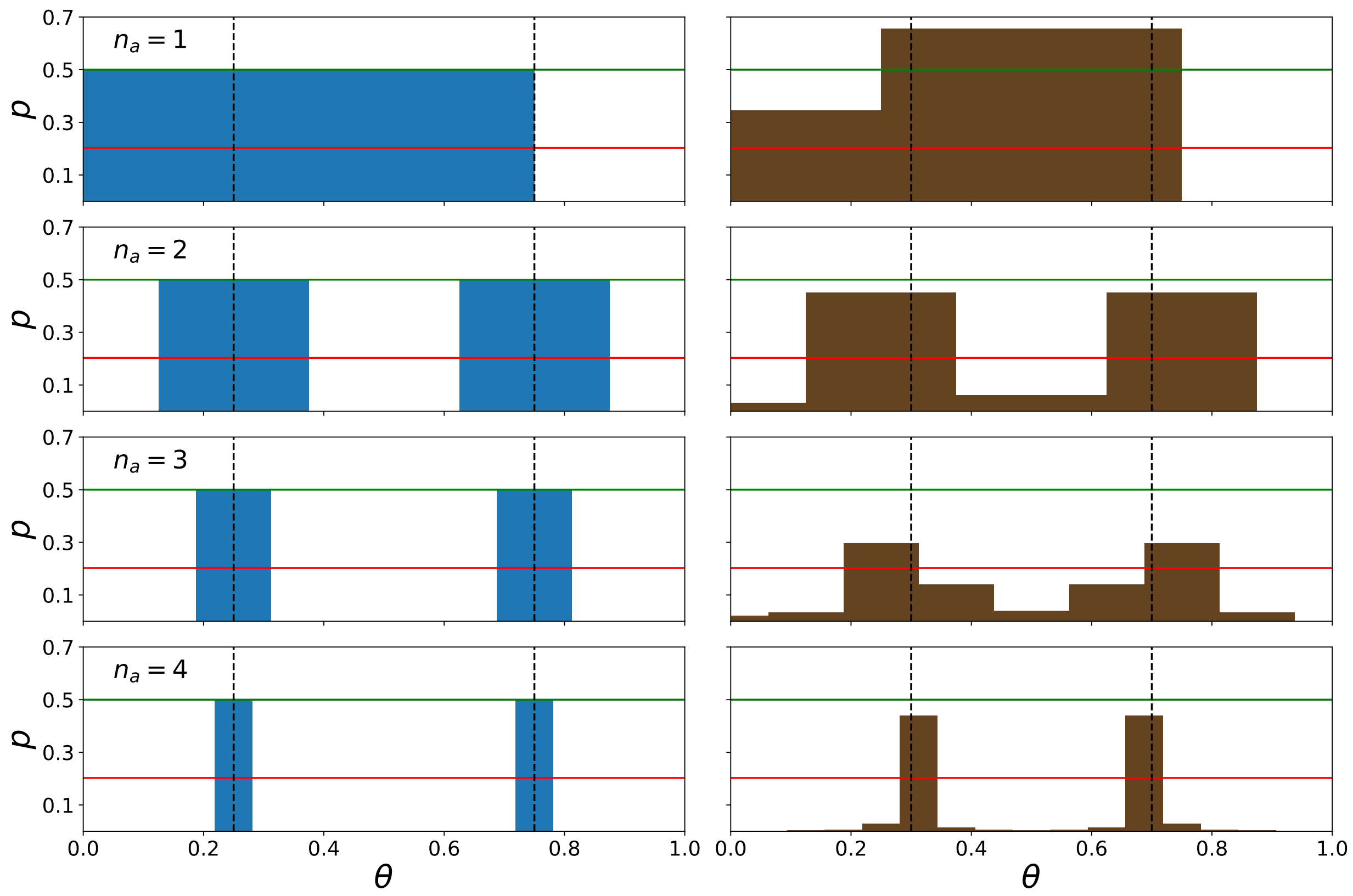

Figure 2.4: Example of the QPE algorithm on a single qubit gate $U$ with eigenvalues $\left\{e^{2\pi i\theta_j}\right\}_{j=0,1}$ and eigenvectors $|j\rangle_{j=0,1}$. $n_a$ is the number of ancillary qubit used in the same row. Two sets of angles $\{\theta_0, \theta_1\}$, represented as vertical black dashed lines, were used: $\left\{\frac{1}{2^2}, \frac{3}{2^2}\right\}$ (Left column) and $\{0.3, 0.7\}$ (Right column). The initial state $|\Psi\rangle$ was set to $\frac{1}{\sqrt{2}}(|0\rangle + |1\rangle)$ so $|c_0|^2 = |c_1|^2 = 0.5$ (Horizontal green line). The red line corresponds to the value $\frac{4|c_i|^2}{\pi^2} = \frac{2}{\pi^2}$. The left column shows the exact convergence of the QPE to the eigenvalues; in this case, we have that $\delta_0 = \delta_1 = 0$. The accuracy of the QPE algorithm increases exponentially with each qubit ancilla; this is represented by the width of the bars, which scales as $\frac{1}{2^{n_a}}$. We can also observe that the peaks have exactly $p = 0.5$. On the other hand, the right column illustrates the case where $\delta_i \neq 0$. Contrary to the left column, the eigenvalues probabilities don't reach 0.5; however, they always remain greater than $\frac{4|c_i|^2}{\pi^2}$. The results depicted in this figure and those in the remainder of the thesis were derived from simulations conducted using Qiskit [52]. Unless otherwise noted, these simulations were performed under noiseless conditions.

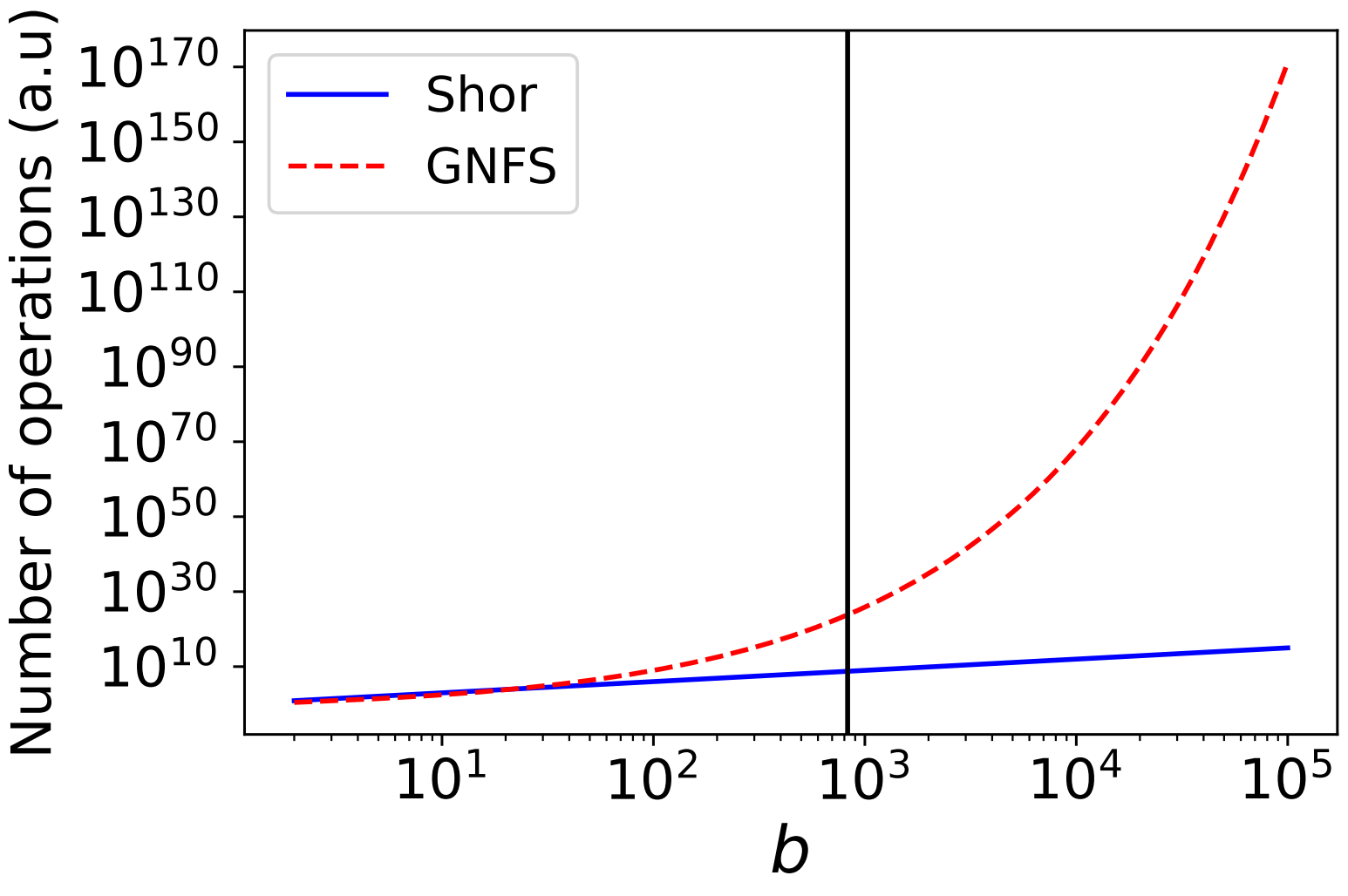

Figure 2.5: Comparison between the Shor's algorithm complexity $\mathcal{O}\left((\log_2 N)^3\right)$[73], and the General number field sieve (GNFS)'s complexity (currently the best classical algorithm to solve integer factorization) $\mathcal{O}\left(e^{\left(\frac{64}{9}\right)^{1/3}(\ln N)^{1/3}(\ln\ln N)^{2/3}}\right)$[74, 75]. $N$ is the integer to be factorized, and $b$ is the number of bits necessary to encode the integer $N$. For large numbers, we can consider $b \approx \log_2 N$. The vertical black line corresponds to the current state of the art of an 829-bit integer factorized on a classical computer [75]. The number of operations is presented in arbitrary units.

**Continued fractions algorithm** Once any of the states $|2^{n_a}\frac{s}{r}\rangle$ has been measured, we will get an estimation over $n_a$ bits of the phase $\varphi = s/r$ by dividing the measured integer by $2^{n_a}$. Now, since $r \leq N \leq 2^{n_q}$ the following inequality holds:

$$\left|\frac{s'}{r'} - \varphi\right| \leq \frac{1}{2r^2}. \tag{2.16}$$

This inequality ensures that $s'/r'$, which will be the candidates for the actual $s/r$, is a convergent of the continued fraction of $\varphi$. The continuous fraction of any real number is the set of positive integers $\{a_i\}_{M+1}$ (including $a_0 = 0$) that allows expressing the number as:

$$[a_0, \ldots, a_M] = a_0 + \cfrac{1}{a_1 + \cfrac{1}{a_2 + \cfrac{1}{\cdots + \cfrac{1}{a_M}}}}. \tag{2.17}$$

The $m^{th}$ convergent for $0 \leq m \leq M$ is defined as $[a_0, \ldots, a_m]$, and the extracted value $r'$ from it is denoted as $r'_m$. To complete the process, the $r'_m$ values obtained from the method are tested in increasing order against Eq. (2.12) until one of them satisfies it.

**Schematic view of the factoring algorithm** The process of a full-fledged factoring algorithm initially involves verifying whether $N$ is even or can be represented in the form $p^q$, where $p \geq 1$ and $q \geq 2$. These verifications can be conducted efficiently using a classical computer. If neither condition is met, the factorization resorts to Shor's algorithm, as detailed in [62]. The procedure is as follows:

1. Randomly choose $x$ such that $1 < x \leq N-1$. If the greatest common denominator (gcd) between $x$ and $N$ is different from 1, i.e., gcd $\neq (x, N) > 1$, one of the factors is $x$. If $\gcd(x, N) = 1$ continue.

2. Use the quantum order-finding and the continued fractions algorithms to determine the order $r$ of $x$ modulo $N$, i.e., to find $r$ in $x^r \bmod N = 1$.

3. If $r$ is odd or $r$ is even but $x^{r/2} \bmod N = -1$ go back to step 1. Otherwise, compute $\gcd\left(x^{r/2} - 1, N\right)$ and $\gcd\left(x^{r/2} + 1, N\right)$, and test to see if one of these values is a non-trivial factor and return the factor if so. Otherwise, the algorithm fails.

It can be shown that $r$ will be even and $x^{r/2} \bmod N \neq -1$ with a probability of at least 1/2 [62].

### Trotter-Suzuki approximation

Some algorithms presented later require using the temporal evolution operator $e^{-iHt}$. For its implementation, we can use the first [76] or higher orders [77, 78] of the Lie-Trotter-Suzuki formula reviewed in [79, 80]. For a time evolution $t$ of a Hamiltonian that can be decomposed as $H = \sum_{i=1}^{m} h_i$, where $h_i$ are non-commuting operators, the first-order Trotter-Suzuki formula is:

$$e^{-iHt} = \left(\prod_{j=1}^{m} e^{-ih_j t/r}\right)^r + \mathcal{O}\left(m^2 t^2 / r\right), \tag{2.18}$$

where $r$ is the number of "Trotter steps". Higher-order formulas can be built by alternating sequences of exponential operators $e^{-ih_j t}$ so that the higher-order error terms cancel. The simplest of these higher-order formulas is the second-order Trotter-Suzuki one, which reads:

$$e^{-iHt} = \left(\prod_{j=1}^{m} e^{-ih_j t/2r} \prod_{j=m}^{1} e^{-ih_j t/2r}\right)^r + \mathcal{O}\left(m^3 t^3 / r^2\right). \tag{2.19}$$

### Amplitude amplification algorithm

The Amplitude amplification algorithm [81, 82] is an algorithm that generalizes the idea behind Grover's algorithm [83] (also known as the quantum search algorithm). This standard algorithm was also used in this thesis. Specifically, it increases the probability of states marked by an oracle. Suppose we have a circuit $U$ that initializes a quantum register on an arbitrary state over $n$ qubits, i.e., $|\psi\rangle = U|0\rangle^{\otimes n} = \sum_{k=0}^{2^n-1} c_k |\phi_k\rangle$ with $\{|\phi_k\rangle\}$ a complete orthonormal base of the Hilbert space. Assume also that we have an operator $O$ (also known as oracle) able to apply a phase $e^{i\pi}$ to the states $|\phi_k\rangle \in \Omega$ whose amplitude we want to amplify, i.e.:

$$O|\phi_k\rangle = f_{|\phi_k\rangle}|\phi_k\rangle \text{ with } f_{|\phi_k\rangle} = \begin{cases} 1 & \text{if } |\phi_k\rangle \notin \Omega, \\ -1 & \text{if } |\phi_k\rangle \in \Omega. \end{cases} \tag{2.20}$$

Here, $\Omega$ can be understood as the subspace of wave functions that possess a specific property. The action of the oracle on the initial state gives:

$$O|\psi\rangle = \underbrace{\sum_{|\phi_k\rangle \notin \Omega} c_k|\phi_k\rangle}_{\equiv|\Psi_B\rangle} - \underbrace{\sum_{|\phi_k\rangle \in \Omega} c_k|\phi_k\rangle}_{\equiv|\Psi_G\rangle}, \tag{2.21}$$

where we have defined the *Good* $|\psi_G\rangle$ and *Bad* $|\psi_B\rangle$ states such that $|\psi\rangle = |\Psi_G\rangle + |\Psi_B\rangle = \sqrt{p_G}|\psi_G\rangle + \sqrt{p_B}|\psi_B\rangle$ with $p_G = |\langle\psi_G|\psi\rangle|^2$ (resp. $p_B = |\langle\psi_B|\psi\rangle|^2$) the probability of measuring a state of the *Good* (resp. *Bad*) states. The amplitude amplification algorithm uses the action of the operator $\mathcal{A} = R_\psi O$ to change the amplitudes of the states in $\Omega$, with $R_\psi$ the reflection operator over the state $|\psi\rangle$ defined as:

$$R_\psi = 2|\psi\rangle\langle\psi| - I = U\left(2|0\rangle^{\otimes n}\langle 0|^{\otimes n} - I\right)U^\dagger = -UO_0U^\dagger, \tag{2.22}$$

where $O_0$ is another specific oracle associated to $\Omega_0 = \{|0\rangle^{\otimes n}\}$. The action of the $\mathcal{A}$ operator is depicted in Fig 2.6. $p_G$ and $\theta$ are related as $p_G = \sin^2(\theta)$. After $m$ applications of the $\mathcal{A}$ operator, the probability of finding the *good* state evolves as:

$$p_G(m) = \sin^2((2m+1)\theta). \tag{2.23}$$

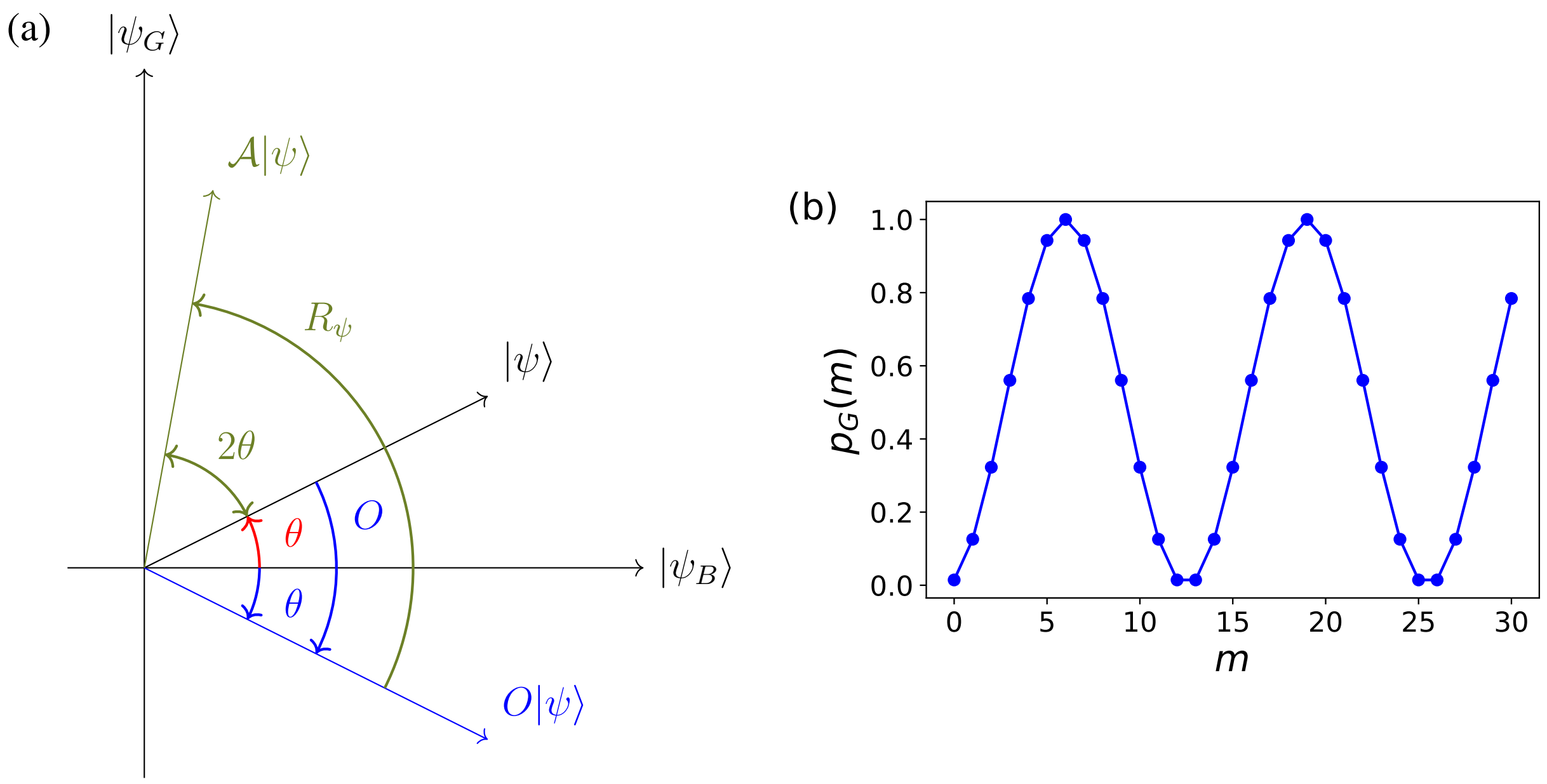

Figure 2.6: (a) Action of the operator $\mathcal{A} = \mathcal{R}_\psi\mathcal{O}$ over the state $|\psi\rangle$ as described in the text. The $O$ operator reflects the *good* component of the $|\psi\rangle$ state; then the $R_\psi$ operator reflects the $O|\psi\rangle$ state with respect to the state $|\psi\rangle$. (b) Example of evolution of $p_G(m)$ for an initial angle $\theta = \pi/26$. As expected, the probability is maximal at $m = \left\lfloor \frac{\pi}{4\theta} \right\rfloor = 6$. If the $\mathcal{A}$ operator continues to be applied, the probability $p_G$ will oscillate, as shown in the Figure.

This probability maximizes when $m = \left\lfloor \frac{\pi}{4\theta} \right\rfloor$. The amplitude amplification algorithm generally increases the probability $p_G$ but does not guarantee its full convergence. In fact, the method often does not converge to $p_G = 1$. To resolve this issue in the case a full projection onto the *good* state is required, it is possible to use the method presented in [84]. In this approach, the general rotation of angle $w = \frac{\pi}{2} - \theta$ needed to fully converge the state $|\psi\rangle$ is decomposed in $n_G$ rotations of angle $\lambda = \frac{w}{n_G}$, where $n_G = \lceil \frac{w}{2\theta} \rceil$. Each $\lambda$ rotation makes use of a generalized $\mathcal{A}$ operator [84], which in turn uses a generalized version of the oracle operator of Eq. (2.20) able to apply an arbitrary phase $e^{i\delta}$ to the states that belong to the "Good" space. This method operates similarly to the original amplitude amplification algorithm but always stops at $p_G = 1$.

## 2.2 . The Many-Body Problem

In this section, I will be delving into the many-body problem that served as the primary focus of my thesis. I will discuss some aspects of its complexity and highlight the potential benefits of treating it using quantum computing.

### 2.2.1 . Introduction

Many-body problem is a term used to describe the study of a physical system with three or more interacting particles, each governed by the laws of quantum mechanics. These physical systems are ubiquitous in many fields of physics, including condensed matter, quantum chemistry, nuclear physics, and quantum chromodynamics, among others. Fig. 2.7 shows an image illustrating some of these fields.

Mathematically, a many-body problem can be described by a set of particles interacting through a Hamiltonian, which in its second-quantized form reads:

$$\begin{aligned} H &= H_{1-\text{body}} + H_{2-\text{body}} + H_{3-\text{body}} + \cdots \\ &= \sum_{\alpha\beta} h_{\alpha\beta} c^\dagger_\alpha c_\beta + \frac{1}{2} \sum_{\alpha\beta\gamma\delta} v_{\alpha\beta\gamma\delta} c^\dagger_\alpha c^\dagger_\beta c_\gamma c_\delta + \ \cdots \ , \end{aligned} \tag{2.24}$$

where $\left(c^\dagger_\alpha, c_\alpha\right)$ denotes a set of (creation, annihilation) operators associated to single particle states and which acts on the particle vacuum $|vac\rangle$. Depending on the system, one may truncate the Hamiltonian up to the $m$-body term. For the Hamiltonians used in this thesis $m = 2$. The main characteristic of many-body problems is that, in many situations, they can not be solved using the mean-field approximation. This method supposes that particles interact through an average one-body potential that depends on the particle density. By reducing the number of degrees of freedom, this technique allows for gaining some insight into the problem at a low computational cost. In most many-body problems, this method proves often insufficient because of the high degree of correlation between the elements of the physical systems — e.g., electrons, nucleons, or spins — and hence their other name: *strongly correlated systems*

#### Difficulties in solving many-body problems

One of the main challenges when solving a many-body problem is the exponential increase in the size of its Hilbert space with the increase of single-particle degrees of freedom. In some cases, this exponential increase can be handled by exploiting the structure of the Hamiltonian to reduce the number of degrees of freedom or by changing the representation of the problem, e.g., some Monte-Carlo methods that trade the exponential difficulty of the size of the Hilbert space to the Monte-Carlo sign problem [88]. Note that some kind of exponentially difficult many-body system can be efficiently simulated in polynomial time; this is the case of "uncorrelated systems" or Clifford circuits. The latter are the circuits that have gates that belong to the Clifford group [89] and which can be simulated with time and space complexity of $\mathcal{O}(n^2)$ with $n$ the number of qubits; this is a result known as the Gottesman-Knill theorem [90, 91]. The gates that belong to the Clifford group are shown in Table 2.2.

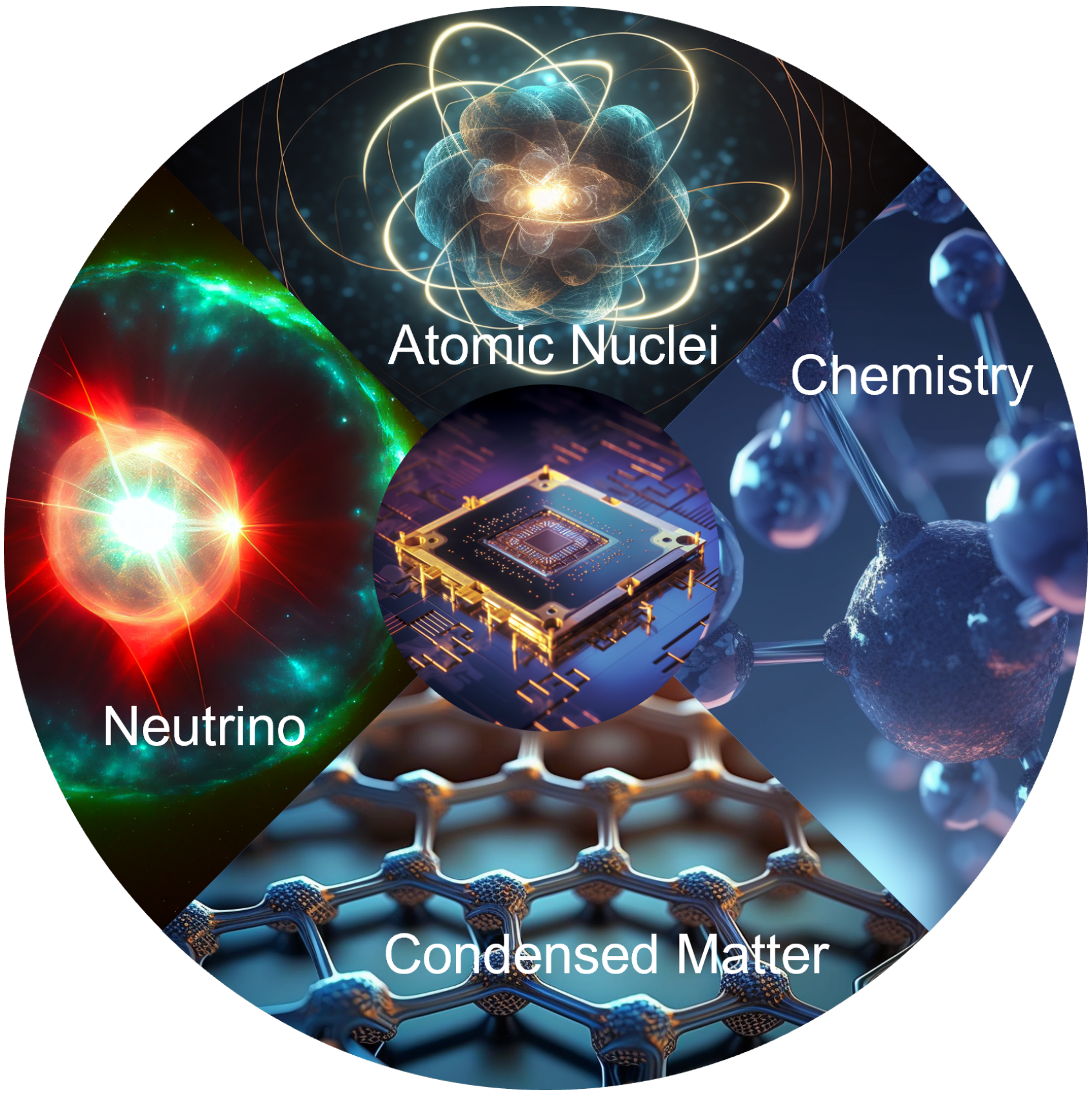

Figure 2.7: Schematic illustration of some many-body systems where quantum computing is now being explored: quantum chemistry [7, 8, 9], condensed matter [4], nuclear [5, 6] and neutrino physics [85, 86, 87].

### Challenges in quantum computing for many-body problems and complexity classes

Quantum computing has gained significant interest in recent decades due to two main developments. Firstly, advancements in experimental control over small devices have brought us closer to the quantum realm, such as the decrease in the number of atoms needed to implement one bit (as predicted by Moore's law) [92]. Secondly, the discovery of quantum algorithms that offer a significant speed advantage over their classical counterparts, such as Shor's algorithm [3] (see Fig. 2.5). Inspired by a similar discussion in [19], this section aims to analyze the expected acceleration that quantum computers could potentially introduce in the resolution of many-body problems. We refer to [93] for an in-depth exploration of this topic. While I must note that I do not claim expertise in this area, I believe the importance of the subject matter justifies its discussion. To address this problem, two quantum complexity classes, Bounded-Error Quantum Polynomial time (BQP) and Quantum Merlin-Arthur (QMA), have been established as quantum analogs to P and NP, respectively [93, 94]. Classical complexity theory categorizes problems into two primary classes: P and NP. Problems in P can be efficiently solved in polynomial time, while those in NP can have their solutions verified in polynomial time. NP problems can be transformed in polynomial time into problems of the class NP-hard. Solutions to problems in this class do not have to be verifiable in polynomial time. On the other hand, NP-complete problems are a subset of NP-hard problems that must belong to the NP class, meaning that their solutions must be verifiable in polynomial time. Unless P=NP, NP-complete problems will likely not have a polynomial-time solution, and because of that, they are often referred to as *exponentially hard* problems. Fig. 2.8 illustrates these classes and their connection.

The current consensus is that BQP does not include NP-complete problems [62], i.e., it is improbable that quantum computers can solve exponentially hard classical problems in polynomial time. In this regard, the factoring problem can be solved using Shor's algorithm (see section 2.1.3) [3] with exponential speedup because factoring is not NP-complete. However, this limitation does not render quantum computers useless against NP-complete problems. It has been shown that in some cases, quantum heuristics, which are algorithms with polynomial time complexity but no guaranteed accuracy, can outperform classical heuristics [96, 97, 98].

The problem of estimating a Hamiltonian's ground state energy $E_0$ can be classified using complexity classes by posing it as a decision problem (problems with yes and no answers). This can be done by choosing numbers $a$ and $b$ such that $b - a > 1/\text{poly}(n)$, and asking whether $E_0 < a$ or $E_0 > b$ [99, 100]. For instance, the ground state estimation problem is QMA-complete for $k$-local Hamiltonians for $k \geq 2$ [99]. These are Hamiltonians whose unitary operators $P_k$ in Eq. (3.13) act on at most $k$ qubits. The QMA-completeness of a problem strongly indicates its computational complexity when approached using quantum computing, i.e., classical and quantum algorithms will likely encounter an exponential wall, e.g., [101, 102]. Restrictions on the value of the coefficients of the terms in the Hamiltonian can lead to a reduction in their computational complexity [103, 104, 105].

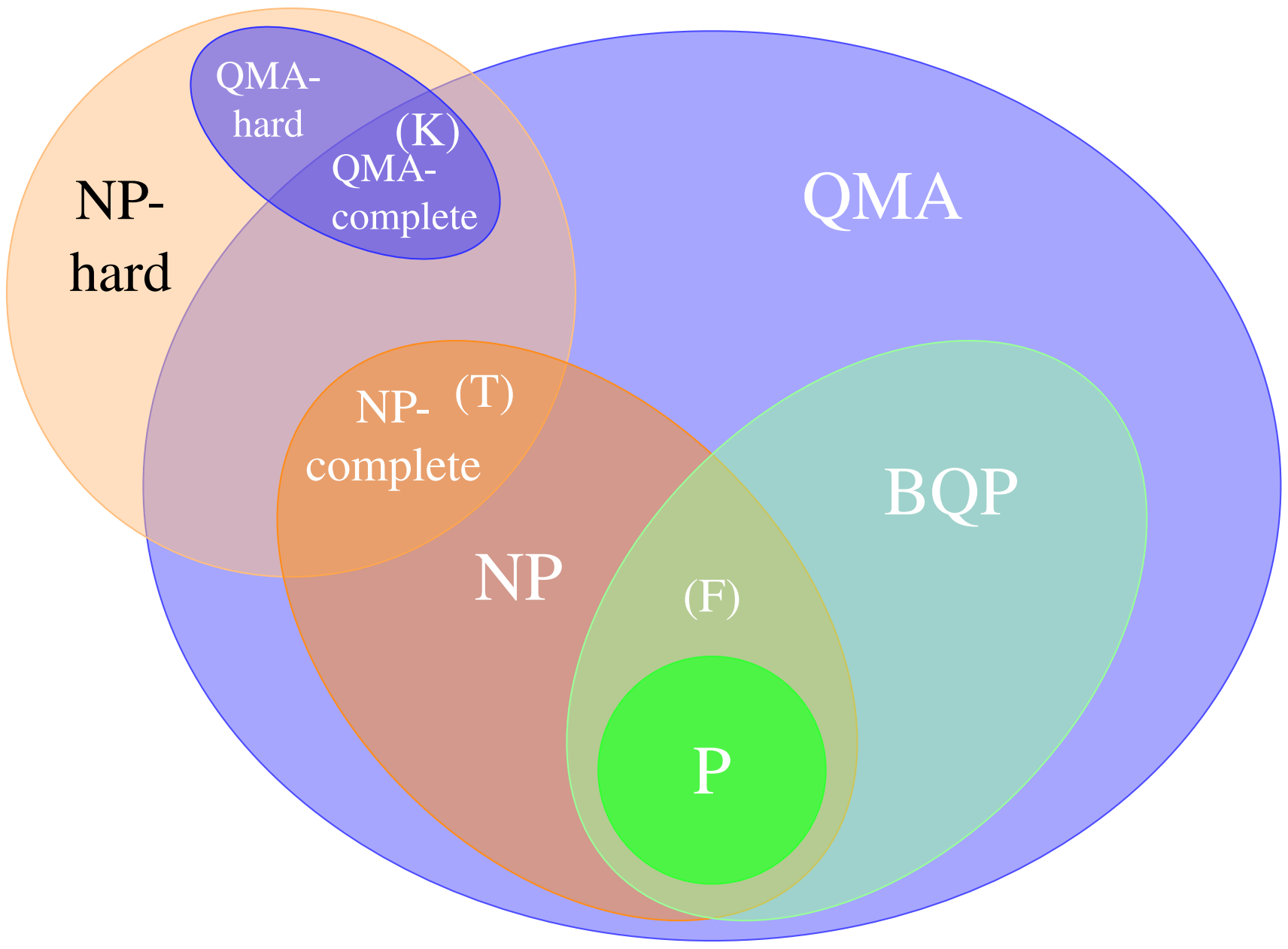

Figure 2.8: Computational complexity classes. The classification presented here is widely accepted; however, many of its relations have not been proven [95]. Three complex problems are symbolized by letters: Factoring (F), Traveling Salesman Problem (T), and k-local Hamiltonian Ground State Estimation (K), which is a commonly encountered challenge in the study of many-body systems. Adapted from [19].

Regarding fermionic and bosonic models, it has been shown that the Bose-Hubbard Hamiltonian is QMA-complete [106]. The Fermi-Hubbard model has been shown to be QMA-complete with an applied external magnetic field [107] and for a specific choice of parameters [108]. It is believed that the generic Fermi-Hubbard Hamiltonian is QMA-complete; however, it has not been yet proven.

It is also important to note that complexity theory can provide a formal evaluation of the feasibility of hybrid quantum-classical algorithms such as the *Variational Quantum Eigensolver* (VQE) (see section 3.3). In particular, the classical optimization procedure for the energy $E(\theta)$ is, in general, NP-hard [109]. However, this does not exclude the possibility of finding heuristics for specific problems. It is important to emphasize that the entire classical variational algorithm faces the same computational challenges as the hybrid quantum-classical algorithm. Ultimately, the most critical question is whether quantum processors can accelerate specific computation tasks *in comparison to* the best classical algorithms.

#### 2.2.2 . Methods to Treat Interacting Fermions on Quantum Computers

It is possible to map a Hamiltonian written in first or second quantization form to a quantum computer [80]. Here, we briefly present the latter approach. Suppose we have a fermionic Hamiltonian written in second-quantized notation as in (2.24) that only contains 1-body and 2-body interactions; this makes it possible to simulate with quantum computing circuits. There are two main steps for representing a fermionic problem in a quantum computer. First, we need a way to map fermionic states in the Fock space of $N$ sites (*aka* orbitals) to the space of $n$ qubits. The fermionic Fock space is spanned by $2^N$ basis vectors $|b_j\rangle^{\otimes N} = |b_{N-1}, \ldots, b_0\rangle$ where $b_j \in \{0, 1\}$ represent the occupation $b_j = 1$ or not $b_j = 0$ of a fermionic particle at orbital $j$. An encoding, denoted by **e**, refers to a mapping that transforms Fock space states to qubit states of the form $|q_{n-1}, \ldots, q_0\rangle$. Mathematically, we can express this as $\mathbf{e} : |b_j\rangle^{\otimes N} \to |q_i\rangle^{\otimes n}$. The simplest encoding is the trivial one, where $n = N$ and each qubit's occupation state corresponds to the occupation of its corresponding orbital state, i.e., $q_j = b_j$. The second step consists of mapping the creation and annihilation operators $\left(c_\alpha^\dagger, c_\alpha\right)$ to quantum gates. As they are fermionic operators, they need to satisfy the anticommutation relations:

$$\{c_\alpha, c_\beta\} = \left\{c_\alpha^\dagger, c_\beta^\dagger\right\} = 0 \qquad \left\{c_\alpha^\dagger, c_\beta\right\} = \delta_{\alpha\beta}, \tag{2.25}$$

with $\{A, B\} = AB + BA$. Following these relations, the action of the creation or annihilation operators on an arbitrary Fock state is:

$$\begin{aligned} c_\alpha^\dagger |b_{N-1}, \ldots, 0_\alpha, \ldots, b_0\rangle &= (-1)^{\sum_{j=0}^{\alpha-1} b_j} |b_{N-1}, \ldots, 1_\alpha, \ldots, b_0\rangle, \\ c_\alpha |b_{N-1}, \ldots, 1_\alpha, \ldots, b_0\rangle &= (-1)^{\sum_{j=0}^{\alpha-1} b_j} |b_{N-1}, \ldots, 0_\alpha, \ldots, b_0\rangle, \\ c_\alpha^\dagger |b_{N-1}, \ldots, 1_\alpha, \ldots, b_0\rangle &= c_\alpha |b_{N-1}, \ldots, 0_\alpha, \ldots, b_0\rangle = 0. \end{aligned} \tag{2.26}$$

There are several ways to perform the encoding and mapping; a comparison of some methods is presented in Table 2.6 and described briefly below.

| Mapping | En-/decoding type | Saved Qubits | $n\,(N,K)$ | Resulting gates | Origin |
|---|---|---|---|---|---|
| Jordan-Wigner\|Parity transform | Linear/linear | None | $N$ | Length-$\mathcal{O}\,(n)$ Pauli strings | [110, 111] |
| Bravyi–Kitaev transform | Linear/linear | None | $N$ | Length-$\mathcal{O}\,(\log_2 n)$ Pauli strings | [112] |
| Checksum codes | Linear/ affine linear | $\mathcal{O}\,(1)$ | $N-1$ | Length-$\mathcal{O}\,(n)$ Pauli strings | [113] |
| Binary addressing codes | Nonlinear/nonlinear | $\mathcal{O}\left(2^{n/K}\right)$ | $\log_2\left(N^K/K!\right)$ | $(\mathcal{O}\,(n))$-controlled gates | [113] |
| Segment codes | Linear/nonlinear | $\mathcal{O}\,(n/K)$ | $N/\left(1+\frac{1}{2K}\right)$ | $(\mathcal{O}\,(K))$-controlled gates | [113] |

Table 2.6: Overview of different encoding and mappings extracted from [113]. The saved qubits are measured relative to the trivial encoding $n = N$. An affine linear function is the composition of a linear function with a translation. The Hamming weight, denoted by $K$, is the number of 1's in the bit string of a given state. A length-$L$ Pauli string is a Pauli operator, which takes the form $\bigotimes_0^{n-1} P_j$, where $P_j = I, X, Y, Z$ are Pauli operators, and $L$ represents the number of terms $P_j$ not equal to the identity operator, denoted by $I$. The methods that use $K$ exploit the particle conservation symmetry. While there are other mapping approaches outlined in [114] and [115], their comparison is not included here because they do not depend on encoding strategies.

### Jordan-Wigner transformation

In this doctoral study, the Jordan-Wigner transformation (JWT) [110] has been adopted as the method for encoding and mapping fermionic problems. This transformation uses the trivial mapping $q_j = b_j$ and the following relations for the transformation of qubit operators:

$$c_\alpha^\dagger = \bigotimes_{k=\alpha+1}^{n-1} I_k \otimes Q_\alpha^+ \bigotimes_{k=0}^{\alpha-1} Z_k,$$

$$c_\alpha = \bigotimes_{k=\alpha+1}^{n-1} I_k \otimes Q_\alpha^- \bigotimes_{k=0}^{\alpha-1} Z_k, \tag{2.27}$$

with $Q_\alpha^\pm = \frac{1}{2}\left(X_\alpha \mp iY_\alpha\right)$ and $X, Y, Z$ the standard Pauli matrices. $Q^+$ (resp. $Q^-$) can be seen as the creation (resp. annihilation) operator in the qubit space. In both operators in (2.27), the $\bigotimes_{k=0}^{p-1} Z_k$ term is added so the qubit operators obey the same anticommutation relations as the fermionic ones in Eq. (2.26).

### Parity encoding

To satisfy the anticommuting relations of Eq. (2.26) using less $Z$ operators, we can use $q_j$ to store the parity of all occupied orbitals up to orbital $j$, i.e., $q_j = \left(\sum_{k=0}^{j} b_k\right) \bmod 2$. Using this encoding, the long tail $\bigotimes_{k=0}^{\alpha-1} Z_k$ is replaced by only $Z_{\alpha-1}$. This technique, parity encoding [112], uses the same amount of qubits as the JWT one, i.e., $n = N$. We can define a transformation matrix $\Pi_n$ that maps states in the Fock spaces, represented as vectors $(b_0, b_1, \ldots, b_{N-1})^T$, to states in the qubit space, represented as vectors $(q_0, q_1, \ldots, q_{N-1})^T$. The matrix $\Pi_n$ and the transformation can be expressed as follows:

$$[\Pi_n]_{ij} = \begin{cases} 1 \text{ if } i \geq j, \\ 0 \text{ if } i < j, \end{cases} \qquad \begin{pmatrix} q_0 \\ q_1 \\ \vdots \\ q_{n-1} \end{pmatrix} = \begin{pmatrix} 1 & 0 & \dots & 0 \\ 1 & 1 & \dots & 0 \\ \vdots & \vdots & \ddots & \vdots \\ 1 & 1 & \dots & 1 \end{pmatrix} \begin{pmatrix} b_0 \\ b_1 \\ \vdots \\ b_{N-1} \end{pmatrix}. \tag{2.28}$$

All the sums are mod 2. As we encode the parity in the qubit states, the qubit creation and annihilation operators of qubit $\alpha$ will depend on the qubit $\alpha - 1$. Specifically, in this case, the creation(+)/annihilation(-) qubit operators $\mathcal{P}^{\pm}$ take the form:

$$\mathcal{P}^{\pm}_{\alpha} = \frac{1}{2}\left(X_{\alpha} \otimes Z_{\alpha-1} \mp iY_{\alpha}\right).$$

When a particle is created or annihilated in orbital $\alpha$, it alters the parity information that all qubits with an index greater than $\alpha$ have. Consequently, we need to apply the $X$ operator to those qubits. The fermionic creation and annihilation operators can be expressed on the parity basis as follows:

$$c^{\dagger}_{\alpha} = \bigotimes_{k=\alpha+1}^{n-1} X_k \otimes \mathcal{P}^{+}_{\alpha},$$
$$c_{\alpha} = \bigotimes_{k=\alpha+1}^{n-1} X_k \otimes \mathcal{P}^{-}_{\alpha}. \tag{2.29}$$

### Bravyi-Kitaev transformation

The parity mapping does not enhance the effectiveness of the JWT, as both approaches use the same number of qubits, and the long tail $\bigotimes_{k=0}^{\alpha-1} Z_k$ must be replaced by $\bigotimes_{k=\alpha+1}^{n-1} X_k$. However, we can improve the encoding efficiency over the two previous approaches by using the Bravyi-Kitaev transformation [112]. This section summarizes the explanations in [111, 116]. We have seen that encoding the fermionic problem in the qubit space requires the occupation of the orbital $j$ and the parity of the set of orbitals with an index less than $j$, i.e., $p_j = \left(\sum_{k=0}^{j-1} b_k\right) \bmod 2$. In the JWT approach, the occupation is encoded locally, while the parity is non-local. On the other hand, the parity encoding encodes the parity locally and the occupation non-locally. The Brayvi-Kitaev transformation (BKT) is a middle ground that balances the locality of the occupation and parity to improve efficiency. More details on the BKT is given in appendix A.

#### 2.2.3 . Application to Some Model Hamiltonians

This section introduces two model Hamiltonians used in this thesis to benchmark the different methods developed. Specifically, the JWT, Trotter-Suzuki, and QPE methods are applied to treat these many-body systems. It should be noted that all results presented here were obtained using classical emulators of quantum computers without noise.

### Pairing Hamiltonian

The pairing Hamiltonian [117, 118, 119, 120], standardly used in nuclear physics or small superconducting systems, consists of fermions distributed in a set of doubly degenerated single-particle levels $p = 0, \ldots, N-1$. Its second-quantized form reads:

$$\hat{H} = \sum_p \varepsilon_p \hat{N}_p - g \sum_{pq} \hat{P}_p^\dagger \hat{P}_q. \tag{2.30}$$

The operators $\hat{N}_p$, $\hat{P}_p$ are, respectively, the pair occupation and pair creation operators defined as:

$$\hat{N}_p = a_p^\dagger a_p + a_{\bar{p}}^\dagger a_{\bar{p}}, \qquad \hat{P}_p^\dagger = a_p^\dagger a_{\bar{p}}^\dagger, \tag{2.31}$$

where $\left(a_p^\dagger, a_{\bar{p}}^\dagger\right)$ are the creation operators of time-reversed states or spin-up/spin-down single-particle states (as in [121]) associated with the energies $\varepsilon_p$ and $g$ is the coupling strength. We will consider a seniority-zero scheme with no broken pairs, as in [121, 122, 71, 123]. Using this scheme, we will directly encode the occupation on pairs in the qubits as in the JWT introduced in section 2.2.2. As the qubit pair does not have the anticommutation relations of individual fermions, we can dispense of the $Z$ operators necessary to maintain them. The operators in the pairing Hamiltonian expressed in terms of qubit operators take the form:

$$\hat{P}_p^\dagger = \frac{1}{2}\left(X_p - iY_p\right) = \begin{bmatrix} 0 & 0 \\ 1 & 0 \end{bmatrix}_p, \qquad \hat{N}_p = I_p - Z_p = \begin{bmatrix} 0 & 0 \\ 0 & 2 \end{bmatrix}_p. \tag{2.32}$$

Here, we perform the replacement $\varepsilon_p \to \varepsilon_p + g/2$ where the $g/2$ term is added to compensate for the shift induced by the scattering of each pair by itself. With this replacement, the complete pairing Hamiltonian becomes:

$$\hat{H} = \sum_p \varepsilon_p \left(I_p - Z_p\right) - \frac{g}{2} \sum_{p>q} \left(X_p X_q + Y_p Y_q\right). \tag{2.33}$$

**Pairing evolution operator** is necessary for the implementation of some algorithms as the QPE one (see section 2.1.3). To implement it, we initially separate the pairing Hamiltonian into the non-commutative one- and two-body terms:

$$\hat{H} = \underbrace{\sum_p \varepsilon_p \left(I_p - Z_p\right)}_{H_\varepsilon} + \underbrace{-\frac{g}{2} \sum_{p>q} \left(X_p X_q + Y_p Y_q\right)}_{H_g}.$$

As all the $I_p - Z_p$ operators commute between each other, we can apply the first-order Trotter-Suzuki approximation from section 2.1.3 as follows:

$$e^{-iHt} = \left(e^{-iH_\varepsilon t/r} \prod_{p>q=0}^{N-2} e^{ig\frac{t}{2r}\left(X_p X_q + Y_p Y_q\right)}\right)^r + \mathcal{O}\left(\left(\frac{N\left(N-1\right)}{2} + 1\right)^2 t^2/r\right),$$

with:

$$e^{-iH_\varepsilon t} = \bigotimes_{p=1}^{N-1} P\left(-2t\varepsilon_p\right).$$

$P(\varphi)$ is the phase gate (see Fig. 2.2). We mention that it is possible to reduce the factor $\frac{N(N-1)}{2}+1$ further by considering that the operators $XX+YY$ commute when they act over two sets of different qubits, i.e.,$[X_pX_q+Y_pY_q, X_{p'}X_{q'}+Y_{p'}Y_{q'}]=0$ if $p,q\neq p',q'$. The quantum circuit capable of implementing the unitary operator $e^{-i\frac{\theta}{2}(XX+YY)}$ is shown in Fig. 2.9. An example of using the QPE algorithm and the temporal evolution operator using the pairing Hamiltonian with the JWT is shown in Fig. 2.10. We can see in the figure the gradual convergence towards the exact eigenvalues (vertical solid black lines) as the number of qubit ancilla used in the QPE increases.

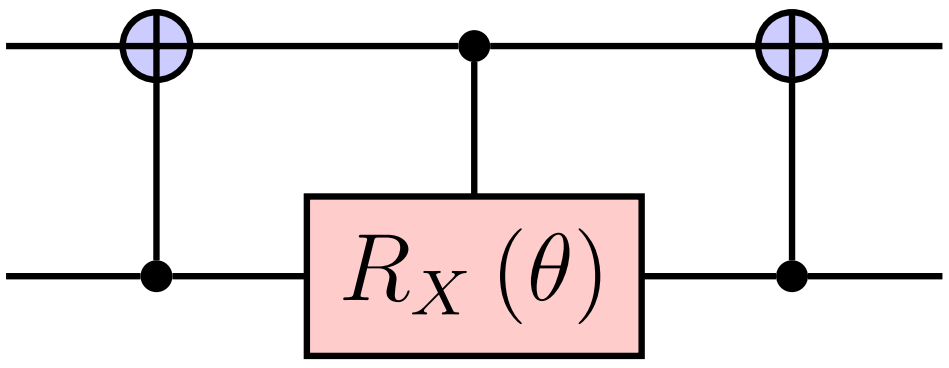

Figure 2.9: Circuit able to implement the unitary operator $e^{-i\frac{\theta}{2}(X\otimes X+Y\otimes Y)}$.

### Hubbard model

The one-dimensional Fermi-Hubbard model [124, 125] describes a set of $N$ interacting fermions with spin on a set of $M$ lattice sites labeled by $i=0,1,\ldots,M-1$. It can be written as a sum of a hopping $H_J$ and interaction $H_U$ terms, i.e., $H=H_J+H_U$. These terms can be written as:

$$H_J=-J\sum_{i,\sigma}\left(a^\dagger_{i+1,\sigma}a_{i,\sigma}+a^\dagger_{i,\sigma}a_{i+1,\sigma}\right), \qquad H_U=U\sum_i n_{i,\uparrow}n_{i,\downarrow}, \tag{2.34}$$

with $n_{i,\sigma}=a^\dagger_{i,\sigma}a_{i,\sigma}$ and $\sigma=\{\uparrow,\downarrow\}$. To apply the JWT, we will assign the spin-up single-particle states $i=0,\ldots,M-1$ to the qubits $\alpha=0,\ldots,M-1$ and the spin-down ones to the qubits $\alpha=M,\ldots,2M-1$. The advantage of this mapping is that the qubits operators become local. The Hamiltonian terms, written using the qubits operators of Eq. (2.27), become:

$$\begin{aligned} H_J &= J\sum_{\alpha=0,\alpha\neq M-1}^{2M-2}\left(Q^+_{\alpha+1}Q^-_\alpha+Q^+_\alpha Q^-_{\alpha+1}\right), \\ H_U &= \frac{U}{4}\sum_{\alpha=0}^{M-1}\left(I_\alpha-Z_\alpha\right)\left(I_{\alpha+M}-Z_{\alpha+M}\right). \end{aligned} \tag{2.35}$$

**Hubbard Evolution operator** Following a procedure similar to the pairing Hamiltonian and considering that:

$$Q^+_{\alpha+1}Q^-_\alpha+Q^+_\alpha Q^-_{\alpha+1}=\frac{1}{2}\left(X_{\alpha+1}X_\alpha+Y_{\alpha+1}Y_\alpha\right),$$

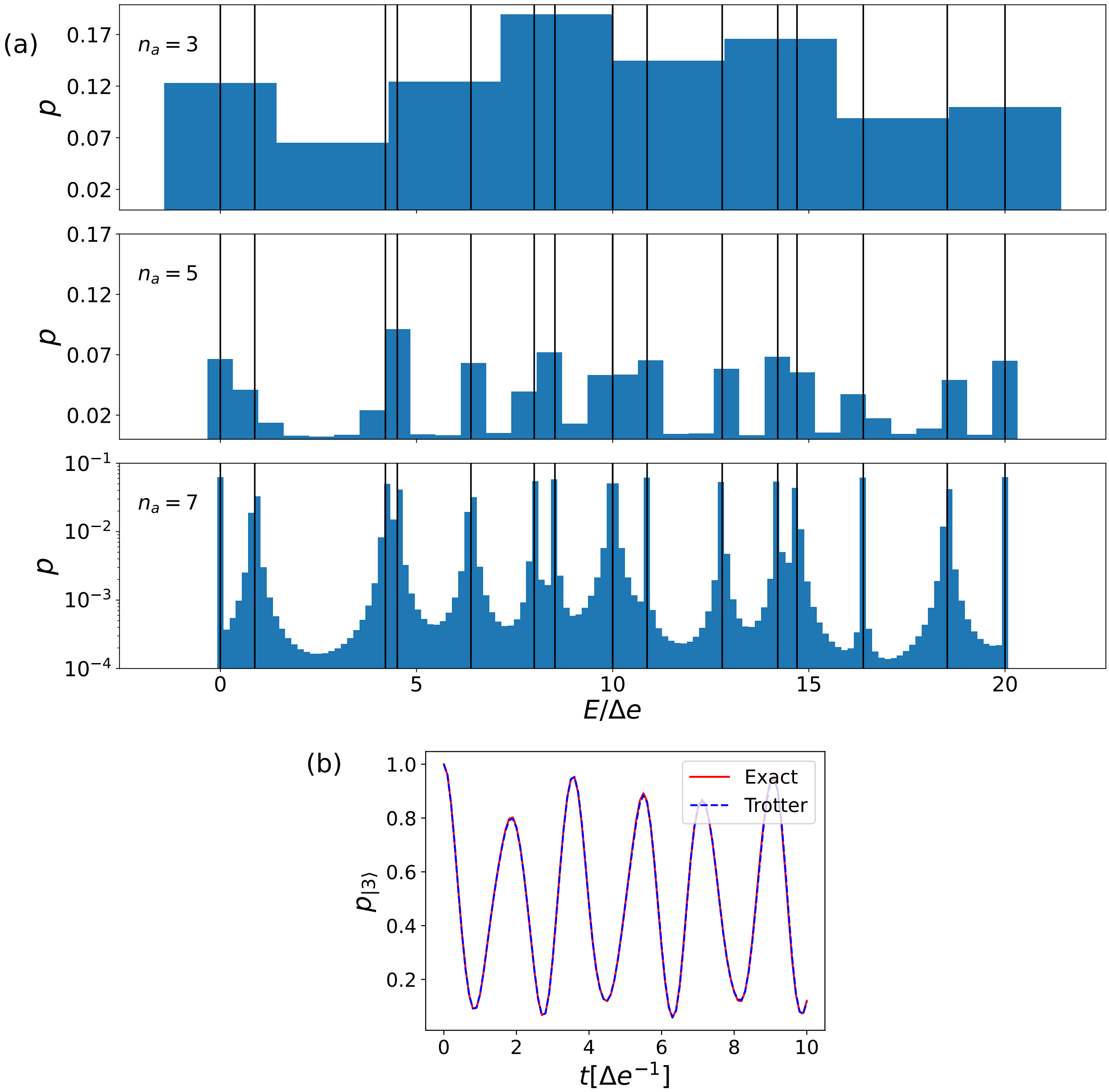

Figure 2.10: (a) QPE algorithm applied to the pairing Hamiltonian shown in Eq. (2.33) of 4 levels (i.e. 4 qubits) with one-body energies $\varepsilon_{p=1,\dots,4} = p\Delta e$ ($\Delta e = 1$) and coupling strength $g = 1$. If the eigenenergies are unknown, we can choose $E = -|H|$ and $\tau = -1/2\,|H|$ in Eq. (2.11) with $|H| = \sum_k |c_k|$ where $H = \sum_k c_k U_k$ and $U_k$ are unitary operators. The values of $\tau$ and $E$ are chosen to ensure that $-\tau\,(H - E) \in [0, 1[$. However, in this case, the eigenvalues are known. Thus, to display the spectra with maximum resolution, we set $E = E_{min}$, $\tau = -1/\left(\mathcal{E}_{max} - E_{min}\right)$ and $\mathcal{E}_{max} = E_{max}\left(\frac{2^{n_a}}{2^{n_a}-1}\right)$ where $E_{min}$ and $E_{max}$ correspond to the minimum and maximum energies, respectively. (b) Evolution of the probability of occupation of the state $|3\rangle$ depicted for the pairing Hamiltonian with the same settings as in (a). Here, the $|3\rangle$ state corresponds to the two pairs occupying the two lowest levels, i.e., $|0011\rangle$. The blue dashed line represents the Trotter evolution, wherein each Trotter step involves a temporal evolution of $\Delta t = 0.1$. The exact evolution is obtained through direct diagonalization of the Hamiltonian on a classical computer.

the first order Trotter-Suzuki of the Hubbard Hamiltonian reads:

$$e^{-iHt} = \left( e^{-iH_U t/r} \prod_{\alpha_{\text{even}},\alpha\neq M-1}^{2M-2} e^{iJ\frac{t}{2r}(X_{\alpha+1}X_\alpha + Y_{\alpha+1}Y_\alpha)} \prod_{\alpha_{\text{odd}}=1,\alpha\neq M-1}^{2M-3} e^{iJ\frac{t}{2r}(X_{\alpha+1}X_\alpha + Y_{\alpha+1}Y_\alpha)} \right)^r + \mathcal{O}\left(3^2 t^2/r\right),$$

with:

$$e^{-iH_U t} = \bigotimes_{p=1}^{N-1} CP\left(-tU\right),$$

where the controlled-phase gate $CP\left(\theta\right)$ is shown in Fig.2.11. Contrary to the pairing evolution operator, the Trotter error does not scale with the number of sites considered. An example of using the QPE algorithm and the temporal evolution operator using the Hubbard Hamiltonian with the JWT is shown in Fig. 2.12.

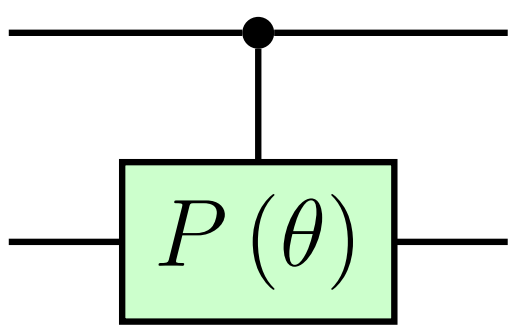

Figure 2.11: Gate able to implement the unitary operator $e^{i\theta(I-Z)\otimes(I-Z)}$.

## 2.3 . Conclusion

In this chapter, we have introduced the framework of the problem, beginning with an overview of the various types of quantum computing and the platforms used for their implementation. We then focus on the digital quantum computing paradigm, describing its basic building blocks: qubits, quantum register, gates, and other relevant elements. Additionally, we provide a brief overview of important algorithms commonly used in quantum computing, including some relevant to this thesis. We explore how to address many-body problems on a quantum computer, offering detailed explanations of different encoding types. Finally, we introduce two schematic Hamiltonians used in the following chapters as a test bench.

In the upcoming sections, we will introduce the problem of breaking and restoring symmetries in many-body systems. This will be useful to enhance the accuracy of ground state estimation in the presence of symmetries by improving upon the standard ansatzes used in the Variational Quantum Eigensolver (VQE) [126]. We will then explore various methods of accessing the energy in the symmetry-conserving state, including projection or other techniques. Finally, some post-processing methods that improve the accuracy of the ground-state energy estimation will be explained, some of which offer approximations of excited state energies.

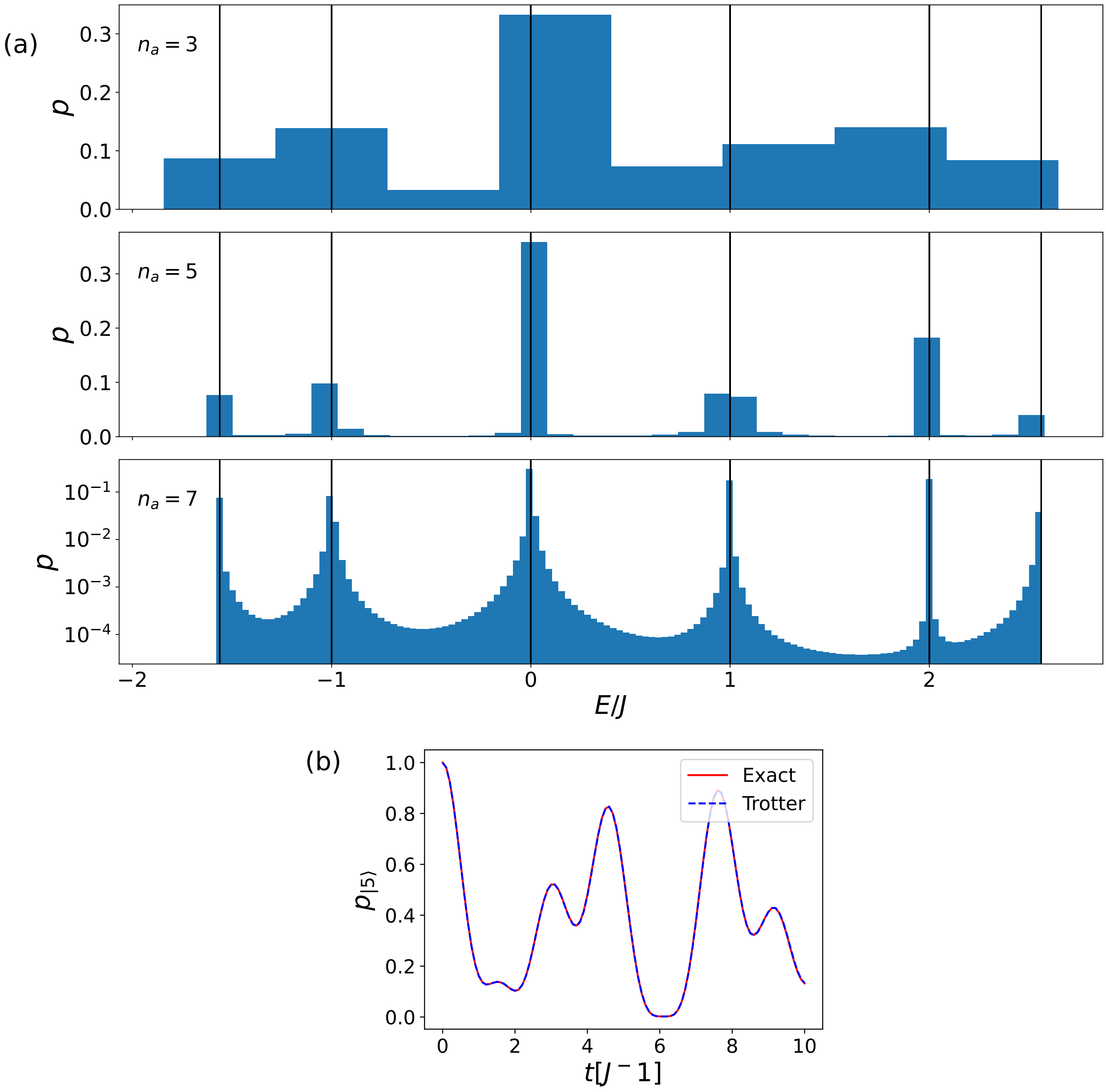

Figure 2.12: Same as in Fig. 2.10 but using the Hubbard Hamiltonian in Eq. (2.35) with $M = 2$ sites, i.e., 4 qubits, and constants $U = J = 1$. Contrary to the pairing case, here we used the state $|5\rangle$ in panel (b) since an initial state of $|3\rangle$ would not allow changes in the system. The Trotter-step for the evolution was again $0.1$.

# 3 - Restoration of Symmetries

## 3.1 . Introduction

The use of symmetries is widespread in treating problems in physics [127]. In particular, when considering many-body physics problems, symmetries are essential to facilitate their resolution by restricting the relevant Hilbert space [128, 129, 130]. However, in certain situations, breaking symmetries can be helpful in describing complex correlations between particles. In variational principles, the symmetry-breaking (SB) technique [28, 29] is commonly employed to capture specific particle correlations. This method also enables reducing the complexity of the ansatz since an ansatz that enforces symmetry and thus can account for these correlations is often more complicated than one that does not. This technique applied in quantum computers has been explored in [70, 121, 71]. An illustration of the ability of a symmetry-breaking ansatz to capture superfluid correlations is shown in Fig. 3.1. The figure illustrates the ground state (GS) energy obtained for a small superfluid system using standard perturbation theory (STP) with no SB and using the Bardeen-Cooper-Schrieffer (BCS) approximation where the particle number symmetry is broken. The results demonstrate that SPT falls short in capturing specific correlations, but the SB ansatz effectively describes them.

Although breaking symmetries can be useful, the exact solution to the problem lies in the section of the Hilbert space with the proper symmetry. This means that at some point, the symmetry-breaking wave function, ultimately, has to be projected into the correct subspace. This process is known as symmetry restoration (SR). The schematics of the SB/SR process are shown in Fig. 3.2. This technique is useful in systems that present spontaneous symmetry-breaking (SSB). An example of a ferromagnetic material that undergoes SSB is the 2-dimensional Ising model. For temperatures exceeding the Curie temperature $T_c$, thermal fluctuations cause the system's ground state to have a magnetization $M = 0$. In this context, magnetization serves as the order parameter, represented as $v = M$ in Fig. 3.2, correlating with the depicted red curve. Conversely, beneath the Curie temperature $T_c$, the system undergoes a quantum phase transition. Even though the ground state energy maintains a magnetization $M = 0$, it is now composed of a combination of symmetry-breaking states. The blue curve is associated with this case. Another example of a system that encounters several SSB is the atomic nuclei. The symmetries broken in this system are, for instance, translation invariance, rotational invariance, and particle number [132]. A detailed discussion of the SB/SR technique and its connection to a quantum phase transition of the system can be found in [132, 133]. In quantum systems, one might think SSB implies the creation of entanglement, but Ref [134] suggests this is not necessarily the case.

One can differentiate two possible ways of using the Symmetry-breaking/Symmetry restoration process (SB/SR) in conjunction with a variational method: either one varies the parameters of a symmetry-breaking ansatz before or after the symmetry restoration. The former case is called projection after variation (PAV), and the latter variation after projection (VAP) [121, 135, 136, 30]. To explain them in detail, let us consider a certain symmetry $S$ that an ansatz $|\Psi(\boldsymbol{\theta})\rangle$ breaks, where $\boldsymbol{\theta}$ is a set of parameters. For a given Hamiltonian $H$, the expectation value of the SB state is given by

$$E_{SB}(\boldsymbol{\theta}) = \langle \Psi(\boldsymbol{\theta}) | H | \Psi(\boldsymbol{\theta}) \rangle. \tag{3.1}$$

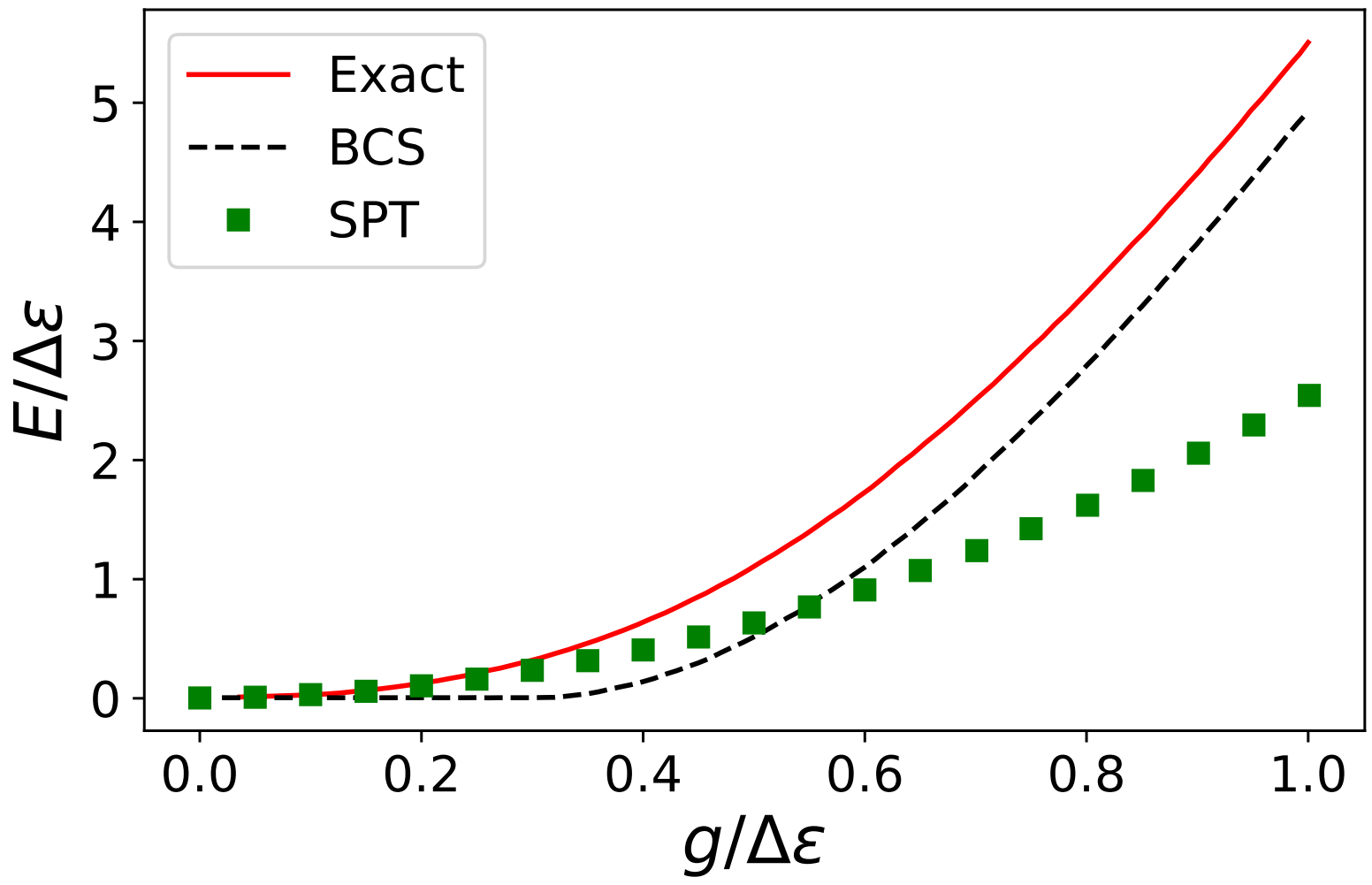

Figure 3.1: Data extracted from [131]. The figure shows the correlation energy of the ground state as a function of the coupling strength for the pairing Hamiltonian from Eq. (2.30) with 8 levels, i.e., $N = 8$, and 8 particles, the ground state has no broken pairs. The correlation energy is defined as the difference between the Hartree-Fock energy $E_{HF} = 2\sum_{p=0,N_{\text{pairs}}-1} \varepsilon_p$ (with $N_{\text{pairs}} = N/2 = 4$) and the ground-state energy obtained with each different approach. Here, the exact ground state energies are obtained via direct diagonalization, while the Bardeen–Cooper–Schrieffer (BCS) ones are obtained using the method described in section 3.2 of [120]. The BCS approach allows the wave function to mix different particle numbers ($U(1)$ symmetry breaking). The standard perturbation theory (SPT) ground state energies are obtained by adding $E_{HF}$ with the standard second-order correction $E^{(2)}$, as described in [131]. It is worth noting that the SPT method does not incorporate the effects of symmetry breaking (SB). Standard perturbation theory matches the exact result below the threshold $g/\Delta\varepsilon \approx 0.3$, but significantly underestimates the correlation for larger values of $g$. This shows that the system's nature becomes highly nonperturbative for strong couplings. In contrast, the BCS ansatz allows for the incorporation of nonperturbative physics, even in cases of strong interactions, by breaking the symmetry associated with particle number conservation.

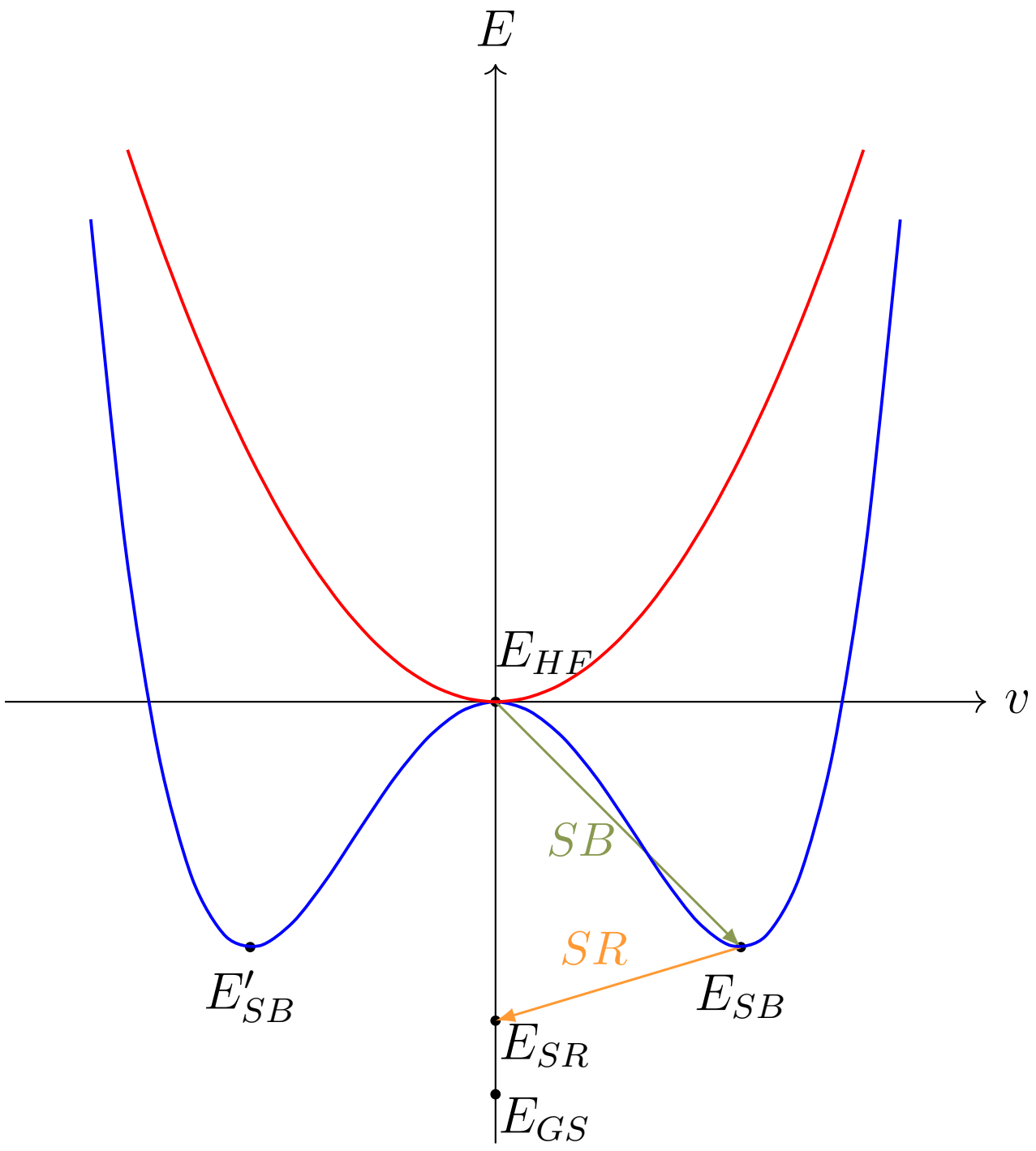

Figure 3.2: Schematic representation of the Symmetry-Breaking/Symmetry-Restoration (SB/SR) method applied to find a Hamiltonian's ground state using a variational Hartree-Fock (HF) technique. Here, a quantity $v$ indicates either symmetry-breaking ($v \neq 0$) or the absence of it ($v = 0$). Suppose we consider an HF ansatz $|\Psi_{HF}\rangle$ that is allowed not to respect a certain symmetry. We then perform a variational minimization constraining the value of $v$. We show in the figure two types of behaviors for the energy. In red, the energy landscape has a minimum at $v = 0$; therefore, the HF solution leads to a symmetry-respecting solution. In blue is shown a case where it is advantageous to break the symmetry, i.e., the energy at $v = 0$ is higher than the HF solution at $v \neq 0$. Notably, the symmetry-breaking solution $E_{SB}$ is closer to the exact ground state energy $E_{GS}$. This demonstrates that one can achieve an improved solution by permitting the wave function to break the Hamiltonian's symmetries. However, for realistic solutions, symmetry must be restored by projecting the SB solution onto the Hilbert space with the correct symmetry. This typically results in an energy $E_{SR}$ close to the exact solution. Usually, SR is carried out by averaging over the distinct system configurations that possess the same SB energy–in this case, the configurations corresponding to $E_{SB}$ and $E'_{SB}$. The red curve (or blue curve) could correspond to a physical system wherein a quantum phase transition has not (or has) occurred.

The projection of the ansatz wave function after the minimization leads to the PAV energy

$$E_{PAV}(\boldsymbol{\theta}) = \frac{\langle\Psi(\boldsymbol{\theta})|H\mathcal{P}_S|\Psi(\boldsymbol{\theta})\rangle}{\langle\Psi(\boldsymbol{\theta})|\mathcal{P}_S|\Psi(\boldsymbol{\theta})\rangle}, \tag{3.2}$$

where, $\mathcal{P}_S$ denotes a generic projector associated with the restoration of the $S$ symmetry. To obtain Eq. (3.2) we utilized the relations $\mathcal{P}_S^2 = \mathcal{P}_S$ and $[H, \mathcal{P}_S] = 0$. In the VAP method, we directly minimize the energy in Eq. (3.2), which is referred to as $E_{VAP}(\boldsymbol{\theta})$. By employing the variational principle, and since both PAV and VAP states belong to the same Hilbert subspace that preserves the symmetry, we naturally obtain the property $E_{GS} \leq E_{VAP} \leq E_{PAV}$ at the minimum of the VAP technique [71]. Here, $E_{GS}$ represents the exact ground-state energy. In the quantum computing context, the use of symmetries can be useful for correcting the errors of noisy platforms that spontaneously break the symmetry of the wave function even out of an SB/SR process [33, 137, 138, 139, 140]. The following section will present some common symmetries in many-body problems and their corresponding treatment in quantum computers. The next two sections will introduce the ingredients to perform SB, SR, PAV, and VAP on quantum computers, which is one of the main objectives of this thesis. Finally, I will illustrate the different procedures using the pairing Hamiltonian.

## 3.2 . Common Symmetries

If a physical system possesses symmetries, then the coefficients in Eq. (2.6) exhibit specific properties associated with the symmetry. Below, we explain the consequences of respecting some symmetries on the wave function. For the particle number and parity symmetries, I will consider a standard JWT over $n$ fermions (see section 2.2.2) where the occupation (resp. absence) of a particle in orbital $j$ is associated with the state $|1\rangle$ (resp. $|0\rangle$) of qubit $j$. On the other hand, for the total spin azimuthal projection and the total spin symmetries, I will consider $n$ particles with spins $1/2$, which are mapped onto a quantum computer by assuming that each particle is described by one qubit. The states $|0\rangle_j$ and $|1\rangle_j$ correspond to the spin-up and spin-down state of particle $j$, respectively. The spin components of each particle $j$ are given by $\mathbf{S}_j = \frac{1}{2}(X_j, Y_j, Z_j)$, and for the total set of particles, we define the total spin operator as $\mathbf{S} = \sum_j \mathbf{S}_j$.

### 3.2.1 . Particle Number

The particle number operator $N$ is given by

$$N = \frac{1}{2}\sum_{i=0}^{n-1}(I_i - Z_i). \tag{3.3}$$

States within the computational basis serve as eigenstates of this operator, possessing eigenvalues $p = 0, \cdots, n$. These eigenvalues correspond to the number of 1s present in the binary representation of the states. Let us assume we're considering the subspace associated with a specific value of $p$. This subspace is characterized by a significant degree of degeneracy, with the number of states in the computational basis satisfying $N|k\rangle = p|k\rangle$ equalling $C_n^p$. When considering the imposed symmetry, it follows that the pertinent Hilbert space for the problem is considerably smaller than $2^n$. Moreover, any state represented by Eq. (2.6) has, at most, $C_n^p$ non-zero components. In classical computing, these symmetries can be utilized by solely considering the subspace of states exhibiting the appropriate symmetry. The unitary operator associated with this symmetry can be written in terms of phase gates (see table 2.2), i.e.,

$$e^{i\gamma N} = \bigotimes_{k=0}^{n-1} P_k(\gamma), \tag{3.4}$$

with $\gamma \in \mathbb{R}$. The index $k$ in Eq. (3.4) indicates that the gate acts on the qubit $k$.

#### 3.2.2 . Parity

The parity operator, denoted by $\pi_p$, has two eigenvalues that partition the states of the computational basis into two subgroups: those with odd and those with even numbers of 1s. A possible choice for this operator is to take $\pi_p = \bigotimes_j Z_j$. When acting on a state of the computational basis, we have

$$\pi_p|q_{n-1}\dots q_0\rangle = (-1)^{\sum_j q_j}|q_{n-1}\dots q_0\rangle, \tag{3.5}$$

which shows that these states are eigenstates of the operator $\pi_p$ with eigenvalues $+1$ or $-1$ if the number of 1s is even or odd, respectively. If, for instance, the state in Eq. (2.6) is an eigenstate of $\pi_p$ with parity $+1$, this automatically implies that the components on the computational states with parity $-1$ are zero. In [141], this symmetry was used to mitigate the error in quantum computers. To construct the unitary operator associated with the parity symmetry, it is enough to take $\gamma = \pi$ in Eq. (3.4).

#### 3.2.3 . Total Spin Azimuthal Projection

The total spin azimuthal operator $S_z = (\sum_j Z_j)/2$ has for eigenstates the states in the computational basis. Any state $|k\rangle$ of this basis has an eigenvalue $m(k) = [n_0(k) - n_1(k)]/2$ where $n_0(k)$ (resp. $n_1(k)$) is the number of 0s (resp. 1s) in the binary representation of $k$. Since the total number of qubits is fixed, we also have the constraint $n_0(k) + n_1(k) = n$. Using Eq. (3.3) in addition to the previous considerations, we have that

$$m(k) = \frac{n}{2} - \langle k|N|k\rangle. \tag{3.6}$$

We can see that the particle number symmetry and the total spin azimuthal symmetry have a linear relation, which means that the considerations on the number of relevant states for each eigenvalue in both symmetries are the same. Using the relation in Eq. (3.6), the unitary operator associated with the symmetry can be constructed as,

$$e^{i\gamma S_z} = e^{i\gamma n/2}e^{-i\gamma N}, \tag{3.7}$$

with $\gamma \in \mathbb{R}$.

#### 3.2.4 . Total Spin

The total spin operator $\mathbf{S}^2$ can be written as [128]:

$$\mathbf{S}^2 = \frac{1}{4}\sum_{jl}\left(X_jX_l + Y_jY_l + Z_jZ_l\right). \tag{3.8}$$

This operator can be rewritten by using its connection with the permutations group [128, 142]:

$$\mathbf{S}^2 = \frac{n(4-n)}{4} I + \sum_{j<l,l=0}^{n-1} P_{jl}, \tag{3.9}$$

where $P_{jl}$ is the transposition operator $P_{jl} = \frac{1}{2}(I + X_j X_l + Y_j Y_l + Z_j Z_l)$ which permutes the state of two qubits, i.e. $P_{jl}|s_j s_l\rangle = |s_l s_j\rangle$. Because of these transposition operators, the eigenstates of $\mathbf{S}^2$ have specific symmetries with respect to the permutation of qubits; we can then make use of Young tableaux [128] to construct the aforementioned eigenstates as shown in Fig. 3.3. It has been shown that all eigenstates of $(\mathbf{S}^2, S_z)$ can be constructed using the iterative procedure of Fig. 3.3. These states form a complete basis of the Hilbert space, called the total spin basis (TSB). It is worth noting that the subspace corresponding to the $(S, M)$ eigenvalues is degenerate. The degeneracy is associated with the number of paths leading to a given Young tableau as depicted in Fig. 3.3. Nonetheless, each eigenstate in a particular block corresponds to a single path, and this fact was recently exploited in Ref.[143] to construct specific eigenstates on a quantum computer. Furthermore, the TSB can serve as an alternative basis to the computational basis for quantum computing. This idea underlies permutational quantum computing (PQC) introduced in Refs.[144, 145] and further discussed in Ref.[146].

### Unitary operator associated with the total spin

Given that the total spin operator in Eq. (3.8) has no-commuting terms, we should use Trotter-Suzuki expansion (see section 2.1.3) to obtain its associated unitary operator. First, we rewrite $\mathbf{S}^2$ given by Eq. (3.8) as:

$$\mathbf{S}^2 = \frac{3n}{4} + \frac{1}{2} \sum_{i<j,j=1}^{n-1} \left(X_i X_j + Y_i Y_j + Z_i Z_j\right). \tag{3.10}$$

Then, we develop as

$$e^{i\gamma \mathbf{S}^2} = e^{i\gamma 3n/4} e^{i\frac{\gamma}{2}\sum_{i<j,j=1}^{n-1}(X_i X_j + Y_i Y_j)} e^{i\frac{\gamma}{2}\sum_{i<j,j=1}^{n-1}(Z_i Z_j)} = e^{i\frac{\gamma}{2}\sum_{i<j,j=1}^{n-1}(X_i X_j + Y_i Y_j)} \prod_{i<j,j=1}^{n-1} e^{i\frac{\gamma}{2} Z_i Z_j}.$$

We recognize the $R_{zz}(\theta) = e^{-i\frac{\theta}{2} Z\otimes Z}$ gate; a decomposition of this gate in terms of the universal standard set of gates ($R_x$, $R_y$, $R_z$ and CNOT) is shown in Fig 3.4. Since the terms $X_i X_j + Y_i Y_j$ do not commute; we use the Trotter-Suzuki approximation as:

$$e^{i\frac{\gamma}{2}\sum_{i<j,j=1}^{n-1}(X_i X_j + Y_i Y_j)} \approx \left( \prod_{i<j,j=1}^{n-1} e^{i\frac{\gamma}{2n_t}(X_i X_j + Y_i Y_j)} \right)^{n_t}, \tag{3.11}$$

where $n_t$ is the number of Trotter-Suzuki steps. A circuit to implement an operator $e^{-i\frac{\theta}{4}(X\otimes X + Y\otimes Y)}$ is shown in Fig. 2.9. $\theta$ should be equal to $-\frac{2\gamma}{n_t}$ in order to implement the gates in Eq (3.11). An alternative integral form of the $S^2$ projector will be introduced in Chapter 5.

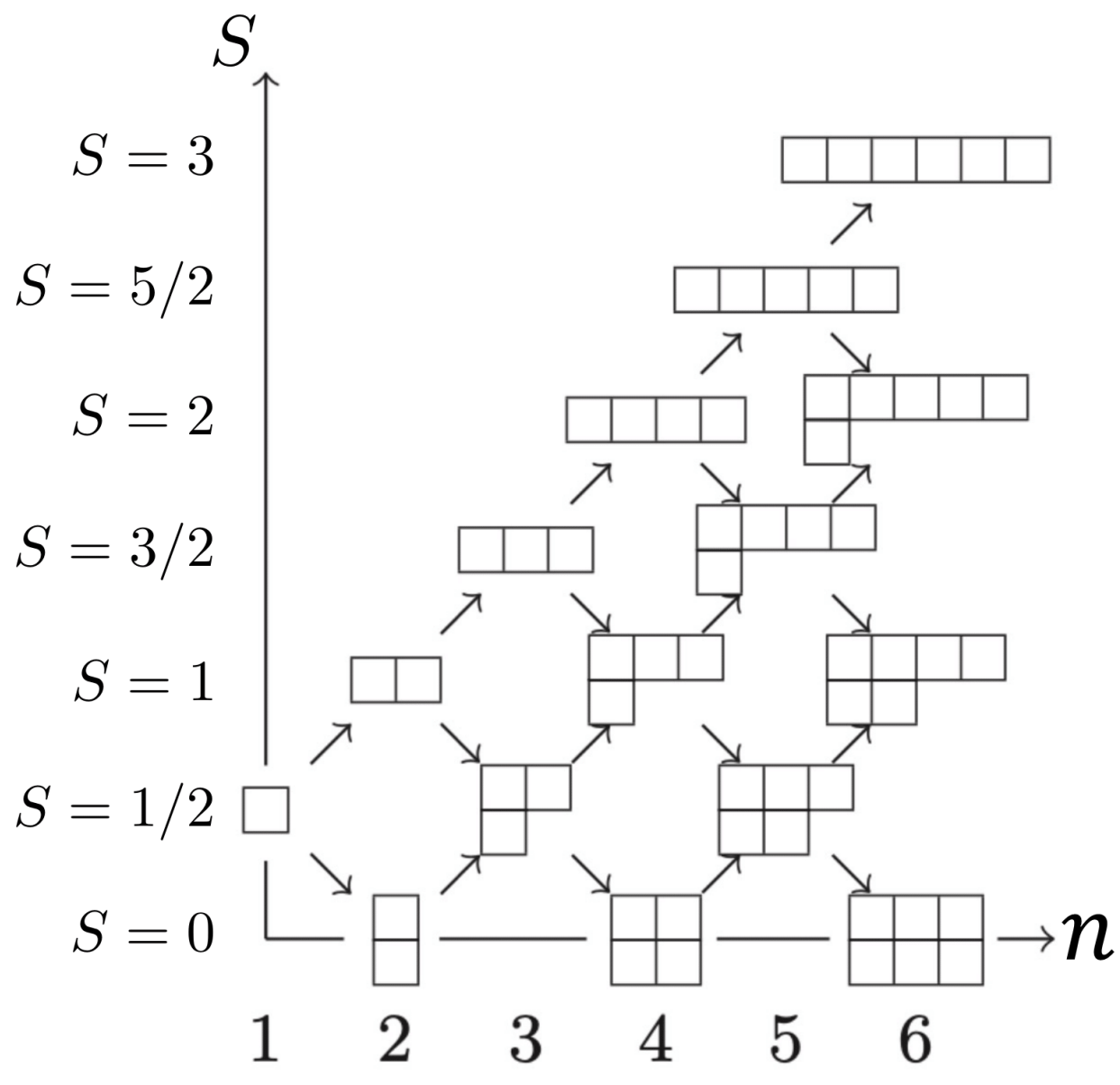

Figure 3.3: Illustration of the iterative procedure for constructing the eigenstates $|S, M\rangle$ of $(\mathbf{S}^2, S_z)$ using Young tableaux [128] with different numbers of qubits $n$. The tableaux are restricted to having only two horizontal rows with lengths $l_u$ (upper line) and $l_d$ (lower line), and they must satisfy the condition $S = l_u - l_d \geq 0$, where $S$ is the total spin. Each row is composed of blocks (squares). Symmetry-preserving states can be constructed by assigning a qubit to each block, with possible values of $0$ and $1$. The resulting eigenstate is symmetric (resp. antisymmetric) with respect to the exchange of indices within the same row (resp. column). In particular, this means that having two $0$s or two $1$s in the same vertical line is not possible. The state with only one horizontal line is, therefore, fully symmetric.

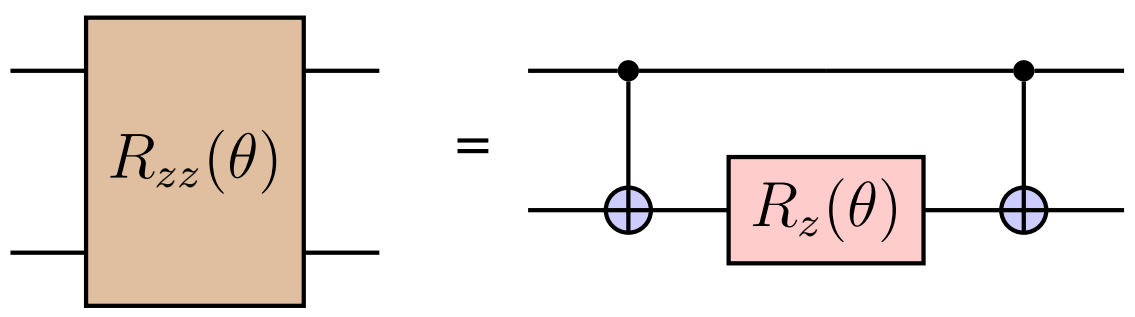

Figure 3.4: Decomposition of the $R_{zz}(\theta) = e^{-i\frac{\theta}{2}Z\otimes Z}$ gate.

### 3.3 . Variational Quantum Eigensolver

The Variational Quantum Eigensolver (VQE) algorithm is the method that allows performing the variational minimization of the energies $E_{SB}$, $E_{PAV}$, and $E_{VAP}$ from Eq. (3.1) and Eq. (3.2) on a quantum computer. Introduced in [126] and reviewed in [147], the VQE algorithm is a widely used hybrid quantum-classical algorithm that aims at finding an approximation of the ground state of a Hamiltonian by minimizing an energy $E_{VQE}(\boldsymbol{\theta})$ over a parametrized state called *ansatz*, $|\Psi(\boldsymbol{\theta})\rangle$. This ansatz is generated by a parametrized trial circuit which describes the unitary operation $U(\boldsymbol{\theta})$, i.e., $|\Psi(\boldsymbol{\theta})\rangle = U(\boldsymbol{\theta})|\mathbf{0}\rangle$ with $|\mathbf{0}\rangle$ corresponding to the fundamental state of the quantum register $|0\rangle^{\otimes n}$:

$$E_{\text{VQE}} = \min_{\boldsymbol{\theta}} \left[\langle \mathbf{0}|U^{\dagger}(\boldsymbol{\theta})|H|U(\boldsymbol{\theta})|\mathbf{0}\rangle\right] \equiv \min_{\boldsymbol{\theta}}\left[E(\boldsymbol{\theta})\right], \tag{3.12}$$

where $\boldsymbol{\theta} \equiv \{\theta_p\}$ is a set of parameters that defines the trial state. By taking the Hamiltonian decomposition:

$$H = \sum_k \alpha_k P_k, \tag{3.13}$$

we can obtain the parameterized observable $E(\boldsymbol{\theta})$:

$$E(\boldsymbol{\theta}) = \sum_k \alpha_k \langle P_k \rangle_{\boldsymbol{\theta}}, \tag{3.14}$$

where we use the shorthand notation $\langle . \rangle_{\boldsymbol{\theta}} \equiv \langle \Psi(\boldsymbol{\theta})|.|\Psi(\boldsymbol{\theta})\rangle$ and the operators $\{P_k\}$ as a set of Pauli chains, i.e., $P_k = \{I, X, Y, Z\}^{\otimes n}$. Fig. 3.5 gives a schematic view of the VQE algorithm.

Each $\langle P_k \rangle_{\boldsymbol{\theta}}$ is computed in the quantum processor by preparing the trial wave function $|\Psi(\boldsymbol{\theta})\rangle$ and measuring in the Pauli basis determined by $P_k$ as indicated in subsection "Measurements in a different basis" of section 2.1.2. An example of this process is shown in Fig. 3.6. Generally, to achieve a precision of $\varepsilon$ on the operator's expectation value, we need to perform $\mathcal{O}(1/\varepsilon^2)$ measurements, also called shots, of the circuit [147].

### 3.4 . Different QPE-Based Projections

Implementing the projection $\mathcal{P}_S$ in Eq. (3.2) directly in quantum computers is not straightforward because it is not unitary, whereas quantum computers can only perform unitary operations. The main workaround for this limitation is to use the non-unitary nature of measurements to implement the projectors. This section will explore some of these approaches, particularly those based on the QPE algorithm that were developed during this thesis. For a more in-depth review of projecting methods, refer to [33]. It is worth noting that it is possible to perform the VAP procedure without explicitly implementing the projector [123]. We first present several methods inspired by the QPE technique below and compare them in an example.

#### 3.4.1 . Symmetry Restoration with QPE

Initially proposed in [70], using the QPE algorithm (see section 2.1.3) to restore the symmetry of a state is natural since symmetries have eigenvalues linked to integer numbers. To project a wave function using the QPE, consider a Hermitian operator $S$ acting on $n$ qubits whose eigenvalues, written in ascending order, are $\{\lambda_0 \leq \cdots \leq \lambda_\Omega\}$ ($\Omega$ depends on $n$), i.e., $S|s_i\rangle = \lambda_i |s_i\rangle$, where $|s_i\rangle$ are the associated eigenstates. The connection of the eigenvalues with the integer numbers $m_i$ can

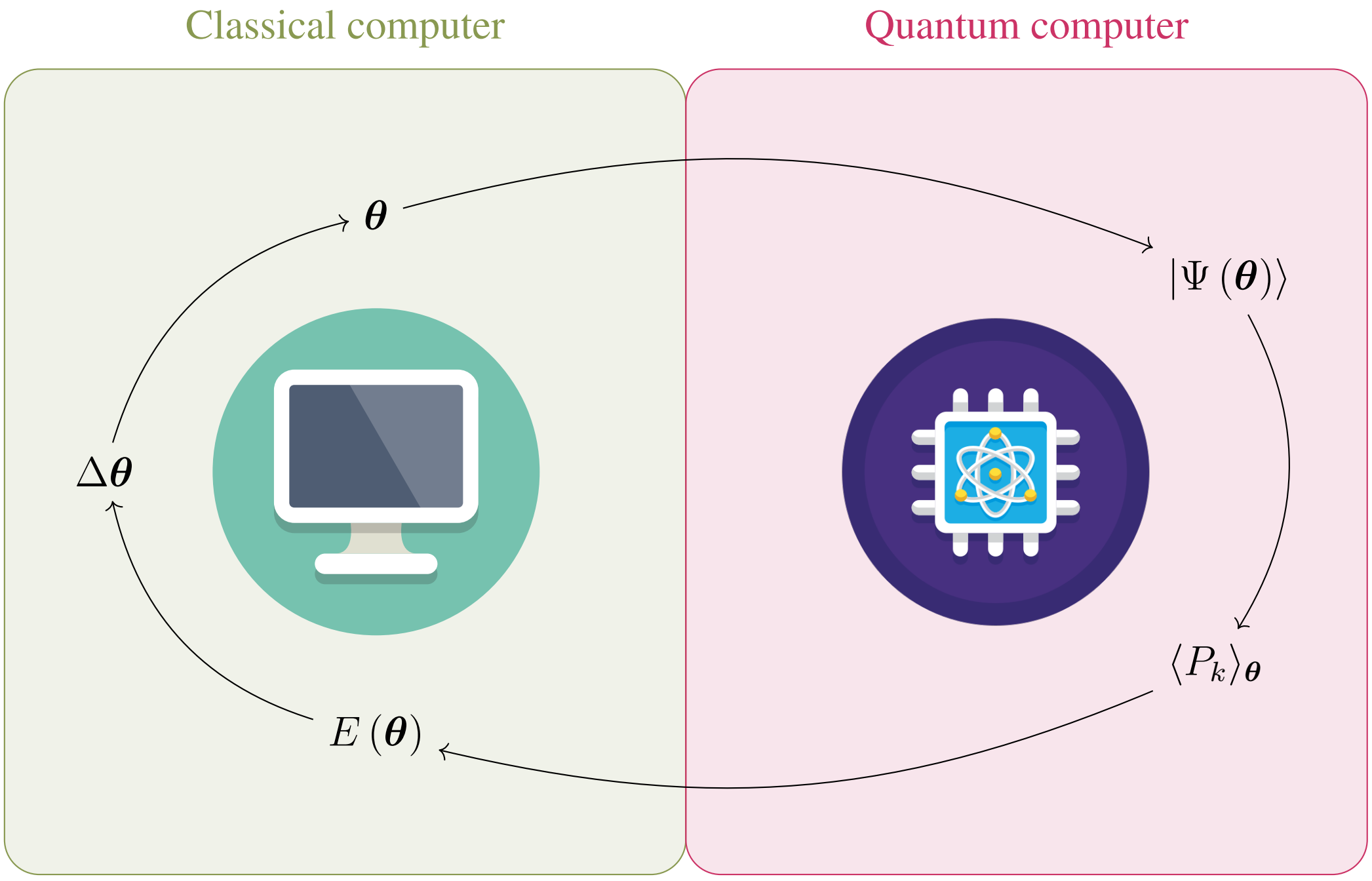

Figure 3.5: Schematic representation of the VQE algorithm using a classical computer [148] *(Left)* and a quantum computer [149] *(Right)*. A set of parameters $\boldsymbol{\theta}$ is passed onto the quantum computer to prepare the parametric wave function $|\Psi(\boldsymbol{\theta})\rangle$. Then, the quantum processor computes all the expectation values $\langle P_k \rangle_{\boldsymbol{\theta}}$ which are passed onto the classical computer to calculate $E(\boldsymbol{\theta})$ and the change in the parameters $\Delta\boldsymbol{\theta}$ to lower the energy. In this case, $E(\boldsymbol{\theta})$ plays the role of a cost function. The loop is repeated until (ideally) convergence to the ground state energy.

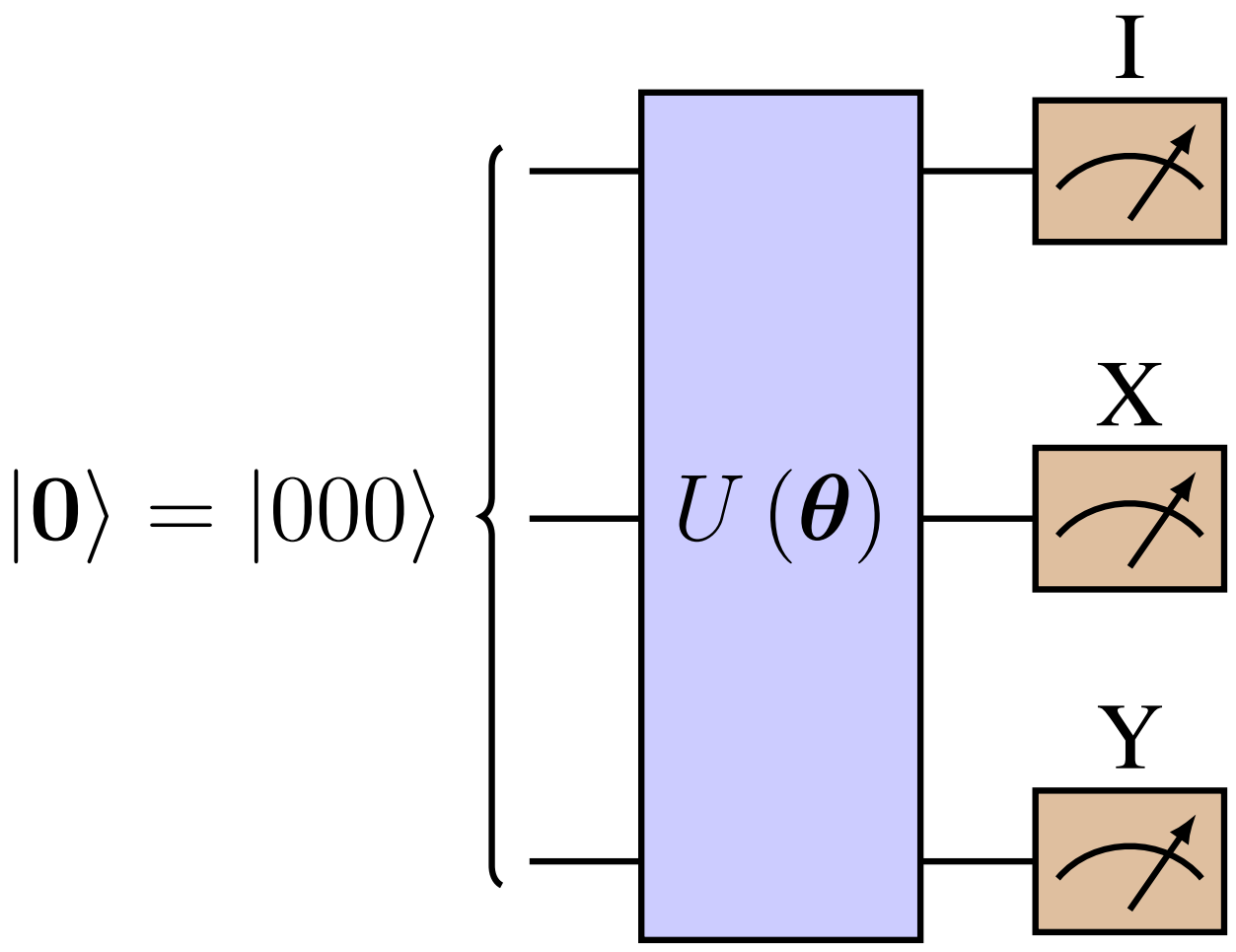

Figure 3.6: Example of a circuit used for estimating the expectation value $\langle \Psi(\boldsymbol{\theta}) | YXI | \Psi(\boldsymbol{\theta}) \rangle$, where $\Psi(\boldsymbol{\theta}) = U(\boldsymbol{\theta}) |\mathbf{0}\rangle$. The measurements in the different Pauli basis are performed as indicated in Table 2.4 of section 2.1.2. Given that $I$ and $Z$ are measured in the same eigenbasis, by measuring $\langle YXI \rangle_{\boldsymbol{\theta}}$ we have immediate access to $\langle YXZ \rangle_{\boldsymbol{\theta}}$. Further optimization methods to reduce the number of repetitions can be found in [147].

be made by subtracting the lowest eigenvalue $\lambda_0$ and dividing by a constant $c$, i.e., $m_i = \frac{\lambda_i - \lambda_0}{c}$. As the ancilla register in Fig. 2.3 collapses to $|2^{n_a}\theta_i\rangle$, with $n_a$ the number of qubits ancilla, we can take $\theta_i = m_i/2^{n_a}$ so the integer $m_i$ is directly read in the ancillary register once the measurement is made. To ensure that all the integers $\{m_i\}$ can be encoded in the ancillary register, $n_a$ has to satisfy the condition $n_a \geq \log_2(m_\Omega + 1)$ and thus it can be taken as $n_a = \lceil \log_2(m_\Omega + 1) \rceil$. Using the previous considerations, the unitary operator $U$ in Fig. 2.3, which can be used to restore the symmetry $S$ can be written as

$$U_S = e^{2\pi i \left[\frac{S - \lambda_0}{c 2^{n_a}}\right]}. \tag{3.15}$$

Since $n_a$ has been selected in such a way that the ancillary register collapses to $|m_i\rangle$ upon measurement, we can compare this situation to the left column of Figure 2.4. In both cases, we have that for all $|2^{n_a}\theta_i\rangle$ measured, $\delta_i = 0$. This means that whenever we measure $|m_i\rangle$ in the ancillary register, the system register will collapse to the associated eigenstate $|s_i\rangle$, effectively restoring the symmetry.

### 3.4.2 . Iterative QPE-Like Technique

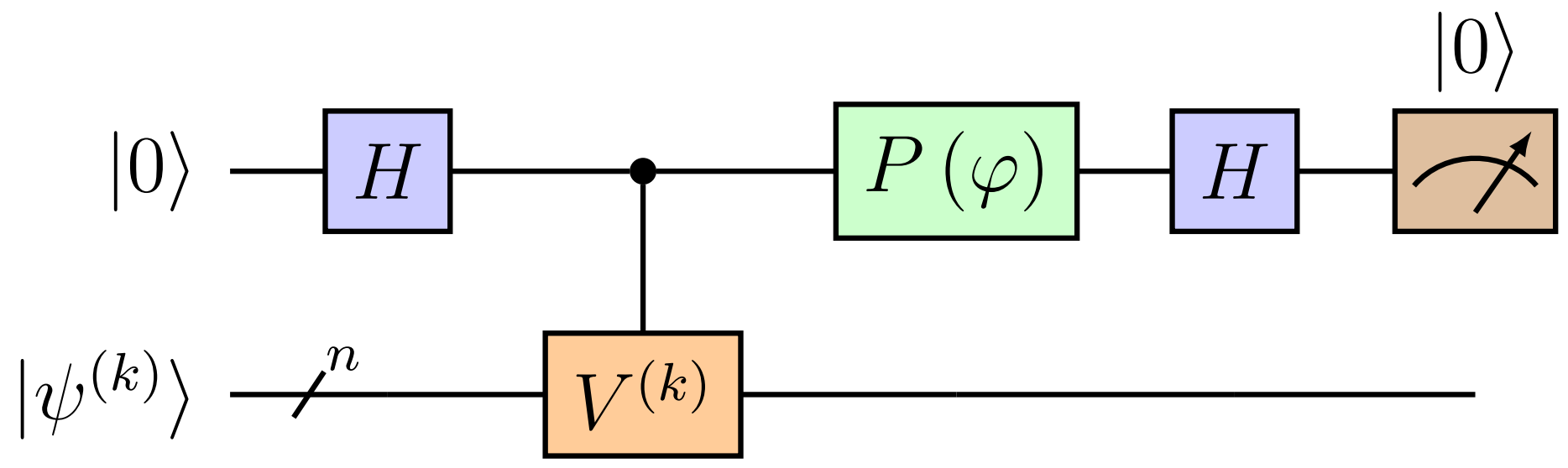

Figure 3.7: Illustration of the circuit used at step $(k)$ in the "IQPE" method. This circuit corresponds to a Hadamard test (shown in table 2.5) where $V^{(k)} = e^{i\phi_k S/c}$, $\phi_k = \pi/2^k$ and $\varphi = -\phi_k \lambda'/c$ with $\lambda'$ the eigenvalue we want to project onto. $|\psi^{(k)}\rangle$ is the state after $k$ applications of the circuit.

The QPE-based method for projection presented in the previous section can require a significant number of gates due to the evolution gates and the inverse QFT (see Fig. 2.3). As a less-costly alternative to it, we present a method called Iterative QPE (IQPE) projection introduced in [33]. This technique applies $n_{IQPE}$ times the circuit shown in Fig. 3.7 to project the wave function successively into a target space; the schematic view of this process is illustrated in Fig. 3.8. To explain how it works, suppose we have a set of eigenvalues $\{\lambda_i\}$ of a symmetry $S$ with their associated integers $\{m_i\}$ as described in the previous section. If we retain only those measurements of the ancilla qubit in the state $|0\rangle$, after the $n_{IPQE}$ applications, we obtain the state

$$\frac{1}{\mathcal{N}} \prod_{k=0}^{n_{IQPE}-1} \left[ I + e^{i\phi_k \left(\frac{S - \lambda'}{c}\right)} \right] |\psi\rangle, \tag{3.16}$$

where $\phi_k = \pi/2^k$, $\mathcal{N}$ is a normalization constant, and $\lambda'$ the eigenvalue we want to project onto. It can be shown that the operator applied to the state $|\psi\rangle$ in Eq. (3.16) is the projector on the eigenvalue $\lambda'$ [150]. We should emphasize that the selection of the operator set $\hat{V}^{(k)}$ in Fig. 3.7 is not necessarily unique. Consequently, making a suitable choice among them could help reduce $n_{IQPE}$, the complexity of $\left\{\hat{V}^{(k)}\right\}$, or potentially both. If we assume that the set of integers $\{m_i\}$ is continuous between $0$ and $m_\Omega$, $n_{IPQE}$ can be obtained via the relation presented in [150]:

$$2^{n_{IQPE}-1} \leq \max\left(m', m_\Omega - m'\right) < 2^{n_{IPQE}}, \tag{3.17}$$

or, equivalently,

$$n_{IQPE} = \lfloor \log_2 \left[\max\left(m', m_\Omega - m'\right)\right] \rfloor + 1, \tag{3.18}$$

where $m'$ is the integer associated with the eigenvalue $\lambda'$ we want to project onto.

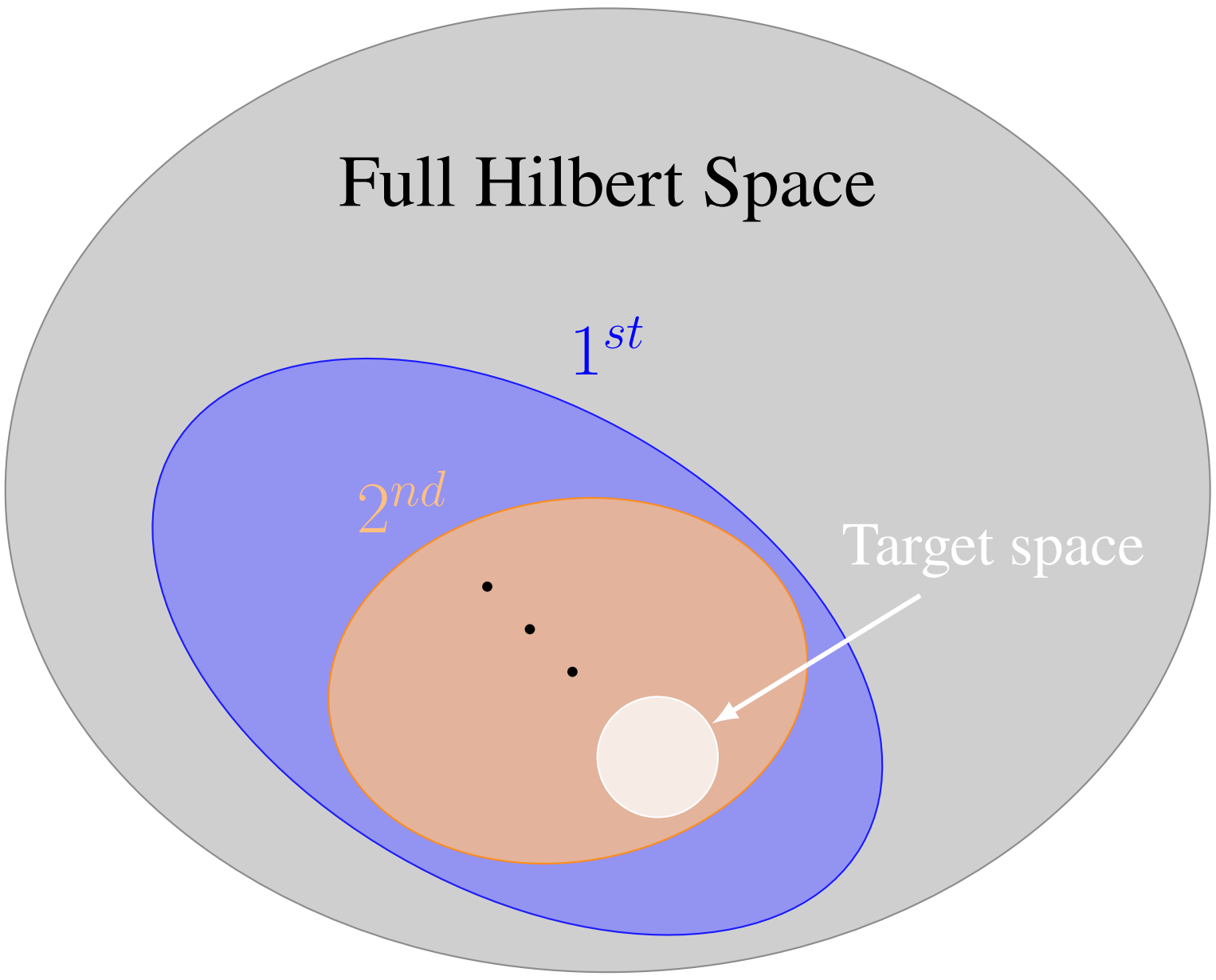

Figure 3.8: Schematic representation of the Iterative QPE-like method that utilizes a series of IQPE-like circuits to project a quantum state onto a targeted sector of the Hilbert space. The application of the IQPE-like circuits–$1^{st}$,$2^{nd}$,...– progressively projects the wave function into more and more restricted regions of the Hilbert space until it finally reaches the target space.

#### 3.4.3 . Rodeo Algorithm

An alternative method, similar to the IQPE-like algorithm, to project a wave function is the *Rodeo* algorithm [151, 152, 153]. Like the QPE, this technique approximates the eigenvalues and eigenstates of a general Hamiltonian $H$. However, we can again replace $H$ with a symmetry operator $S$ and utilize this method to project the state into one of the eigenstates of $S$. In the following, we outline the general implementation of this technique and mention how it can be applied to the symmetry restoration problem.

Consider an initial state $|\psi\rangle$ that decomposes into eigenstates $\{|\phi_j\rangle\}$ of a Hamiltonian $H$, with associated eigenvalues $\{E_j\}$; that is, $|\psi\rangle = \sum_j c_j |\phi_j\rangle$. We can again use the circuit in Fig. 3.7 to estimate the eigenvalues $E_j$ and project onto the eigenstate $|\phi_j\rangle$ if we choose $V^{(k)} = e^{-iH\tau_k}$ and $\varphi(k) = E\tau_k$, where $E$ is a freely-varying parameter, $\{\tau_k\}_{k=1,\ldots,n_R}$ is a set of evolution times and $n_R$ is the number of times the circuit is applied. It can be shown [151, 152] that the probability of obtaining only the $|0\rangle$ state in all of the consecutive $n_R$ measurements of the ancilla qubit is:

$$p_{0^{n_R}}(E,\{\tau_k\}) = \sum_j |c_j|^2 \prod_{k=1}^{n_R} \cos^2\left((E_j - E)\frac{\tau_k}{2}\right). \tag{3.19}$$

We can choose the $\{\tau_k\}$ values to make the function $P_{0^{n_R}}(E)$ peak around the eigenvalues $E_j$. Below, we discuss two main options:

- *Fixed times prescription:* Let us consider $\tau_k = \tau/2^{k-1}$ and $\tau = \frac{\pi 2^{n_R-2}}{|E_{up}-E_{low}|}$, where $E_{up}$ (resp. $E_{low}$) is an upper bound (resp. lower bound) on the spectrum of the Hamiltonian. We can use Eq. (3.19) to obtain the probability distribution $p_{0^{n_R}}$ for different values of $E$, as illustrated in Fig. 3.9. From the position of the peaks and their relative amplitudes in the distribution, we can approximate eigenenergies and the corresponding weights of the eigenstates in the decomposition of $|\psi\rangle$, denoted by $|c_j|^2$.

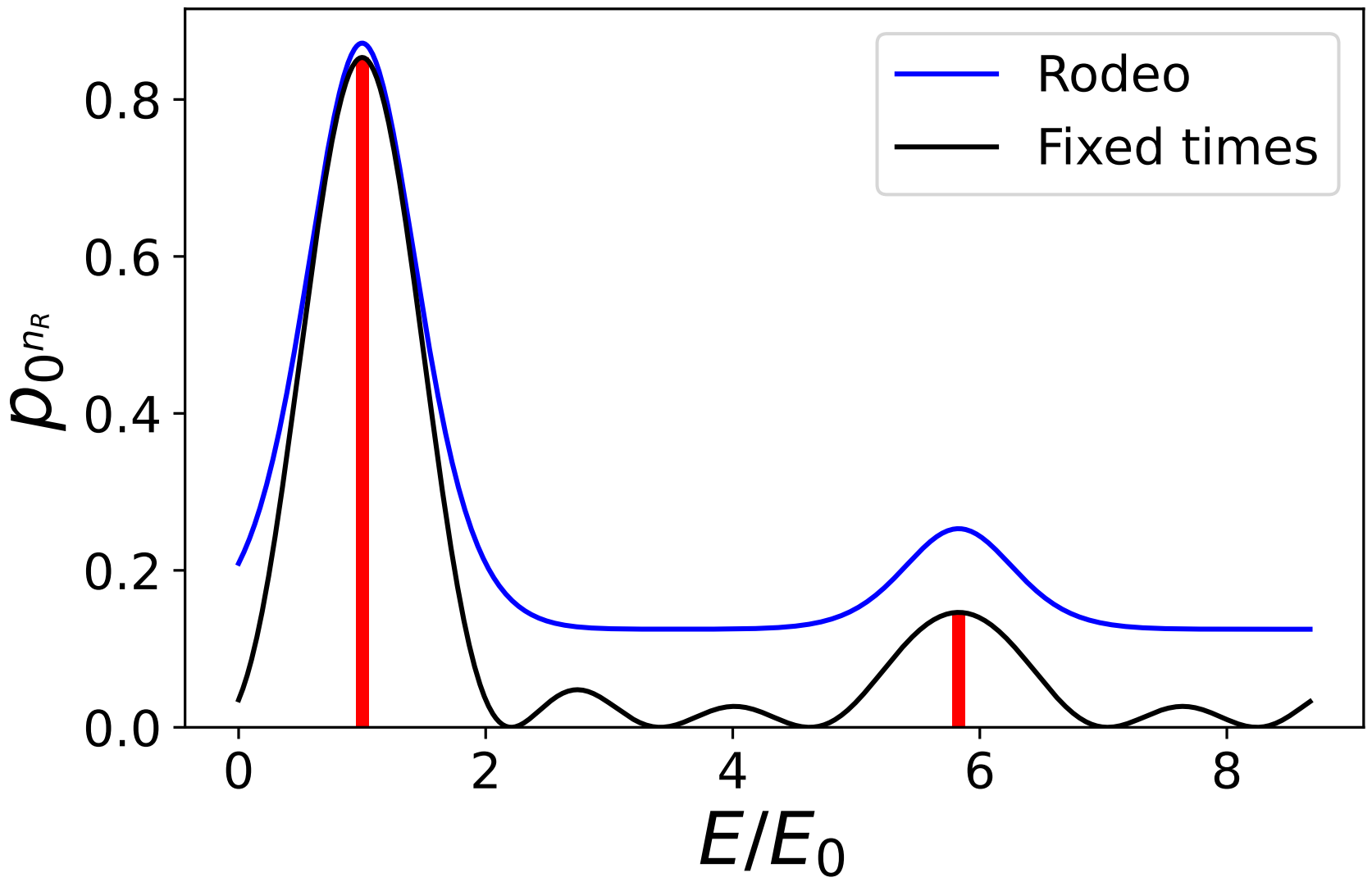

Figure 3.9: Illustration of the use of the fixed times (black line) and Rodeo (blue line) prescriptions. The red bars indicate the (exact) decomposition of the initial wave function in the eigenbasis of the Hamiltonian $\{|E_j\rangle\}$, i.e., the points $(E_j, |c_j|^2)$ from $|\psi\rangle = \sum_j c_j|\phi_j\rangle$. Here the system has two eigenvalues and $n_R$ is taken as 3. In the Rodeo case, each eigenenergie $E_j$ contributes to a flat background proportional to $|c_j|^2/2^{n_R}$ (see Eq. (3.20)).

- *Rodeo prescription:* As proposed in [151], the Rodeo method assumes that each time $\tau_k$ is generated by a standard Gaussian distribution of probability with mean 0, i.e., $\frac{1}{\sigma\sqrt{2\pi}}e^{-\tau_k^2/2\sigma^2}$ where $\sigma$ is an adjustable parameter. By marginalizing over the set of $\tau_k$ using their joint

probability, we obtain

$$p_{0^{n_R}}(E) = \sum_j |c_j|^2 \left[\frac{1 + e^{-(E_j - E)^2\sigma^2/2}}{2}\right]^{n_R}. \tag{3.20}$$

This function peaks strongly around the eigenenergies. An illustration of the Rodeo prescription is also given in Fig. 3.9.

There are two main advantages to using the Rodeo method instead of the fixed times approach. Firstly, it can flatten the probabilities away from the energies, which helps identify peaks. Secondly, the additional parameter $\sigma$ can be utilized as a resolution to scan a particular energy range efficiently. After finding an approximation of the eigenvalues $E_j$, we fix $E = E_j$ and apply the circuit $n_R$ times again. This suppresses the contributions of all other eigenstates outside the interval $[E_j - \epsilon, E_j + \epsilon]$ by a factor $\delta$, where $\delta$ obeys the relation $n_R\sigma / |c_j|^2 = \mathcal{O}\left[\log \delta / \left(|c_j|^2 \epsilon\right)\right]$ [152]. In particular, we can observe in Eq. (3.20) that the pollution in preparing a given eigenstate is exponentially suppressed with respect to $n_R$ by the factor $1/2^{n_R}$. To restore the symmetry of a wave function, we can use the last procedure together with $H = S/c$ and $E = \lambda'/c$ with $c$ as described in previous sections and $\lambda'$ the eigenvalue we want to project onto.

### 3.4.4 . Comparison of the QPE-Based Methods for the Particle Number Projection

To analyze the three methods introduced previously, let us take the case of the particle number projection. Suppose again that we have a set of fermions mapped on a quantum computer using the JWT; the QPE method will project a wave function $|\psi\rangle$ onto a state with a given number of particles $p'$ with a probability equal to the overlap of this state with the original wave function, i.e., if we assume the decomposition of $|\psi\rangle = \sum_{p=0}^{n} c_p |p\rangle$, the projection onto $p'$ will succeed with a probability of $|c_{p'}|^2$. This suggests that the QPE technique is capable of procuring the decomposition $\left(p, |c_p|^2\right)$ for $p = 0, \ldots, n$. With this decomposition in hand, we can investigate how quickly the IPQE and Rodeo algorithms project the state into the correct subspace. Fig. 3.10 shows how these two methods project an equiprobable wave function $|\psi_e\rangle$ defined on $n = 8$ qubits onto a state of 4 particles. $|\psi_e\rangle$ can be written in the computational as $|\psi_e\rangle = \frac{1}{\sqrt{2^n}} \sum_k |k\rangle$ and in the particle number basis as $|\psi_e\rangle = \frac{1}{2^n} \sum_p C_n^p |p\rangle$. We can observe that the IPQE method has completely converged on $n_I = 3$ where $n_I$ is the number of applications of the IPQE circuit in Fig. 3.7. On the other hand, for Rodeo's algorithm, the projection accuracy increases with $\sigma$ but still generally leaves in some contributions from other states. Table 3.1 compares several metrics for the three QPE-based methods. Both the IQPE and Rodeo methods significantly reduce the number of ancillary qubits used in the QPE to just one. Moreover, neither method requires the execution of an inverse QFT, making them less resource-intensive to implement. Comparing the Rodeo and IQPE methods, the IQPE technique offers the distinct advantage of converging precisely to the projected state. In contrast, the Rodeo method may necessitate more iterations to achieve similar accuracy.

## 3.5 . Symmetry-Breaking/Symmetry-Restoring Using the Pairing Hamiltonian

Having discussed the process of performing projections on quantum computers associated with symmetry restoration, I will now demonstrate how this technique can be applied to approximate the ground state of a Hamiltonian. Specifically, I will focus on the pairing Hamiltonian of Eq. (2.33).

| Method | # ancilla | # measurements | Gate resources |
|---|---|---|---|
| QPE | $n_{QPE} = \lceil log_2 (n+1) \rceil$ | $\sim p_G$ | $n_{QPE}$ Controlled-evolution gates.<br>$n_{QPE}$ Hadamard gates.<br>An inverse QFT over $n_{QPE}$ qubits. |
| IQPE | 1 | $\sim p_G$ | $n_{IQPE}$ IQPE circuits, each:<br>- 2 Hadamard gates.<br>- 1 Controlled evolution gates.<br>- 1 Phase gate. |
| Rodeo | 1 | $\sim p_G, \sigma^{-1}$ | $n_R$ IQPE circuits |

Table 3.1: Comparative analysis of various QPE-based projection methods specifically for the case of particle number projection. The columns, from left to right, present the method, the number of ancillary qubits required, the probability of retaining an event upon measurement, and a concise overview of the quantum subroutines needed for the method's implementation. For the particle number projection $m_\Omega = n$ with $n$, the number of qubits in the system wave function. $p_G = \langle \Psi_G | \Psi \rangle$ denotes the amplitude of the desired symmetry (Good component). In the case of the IQPE method, $N$ is the particle number we want to project onto, and $n_{IQPE}$ is determined by Eq. 3.18. For the rodeo algorithm, $n_R$ depends on the suppression factor $\delta$ desired for the unwanted eigenstates; as mentioned in the text, this factor also depends on $\sigma$ and $p_G$. All of the methods mentioned give access to the projected state. Table adapted from [33].

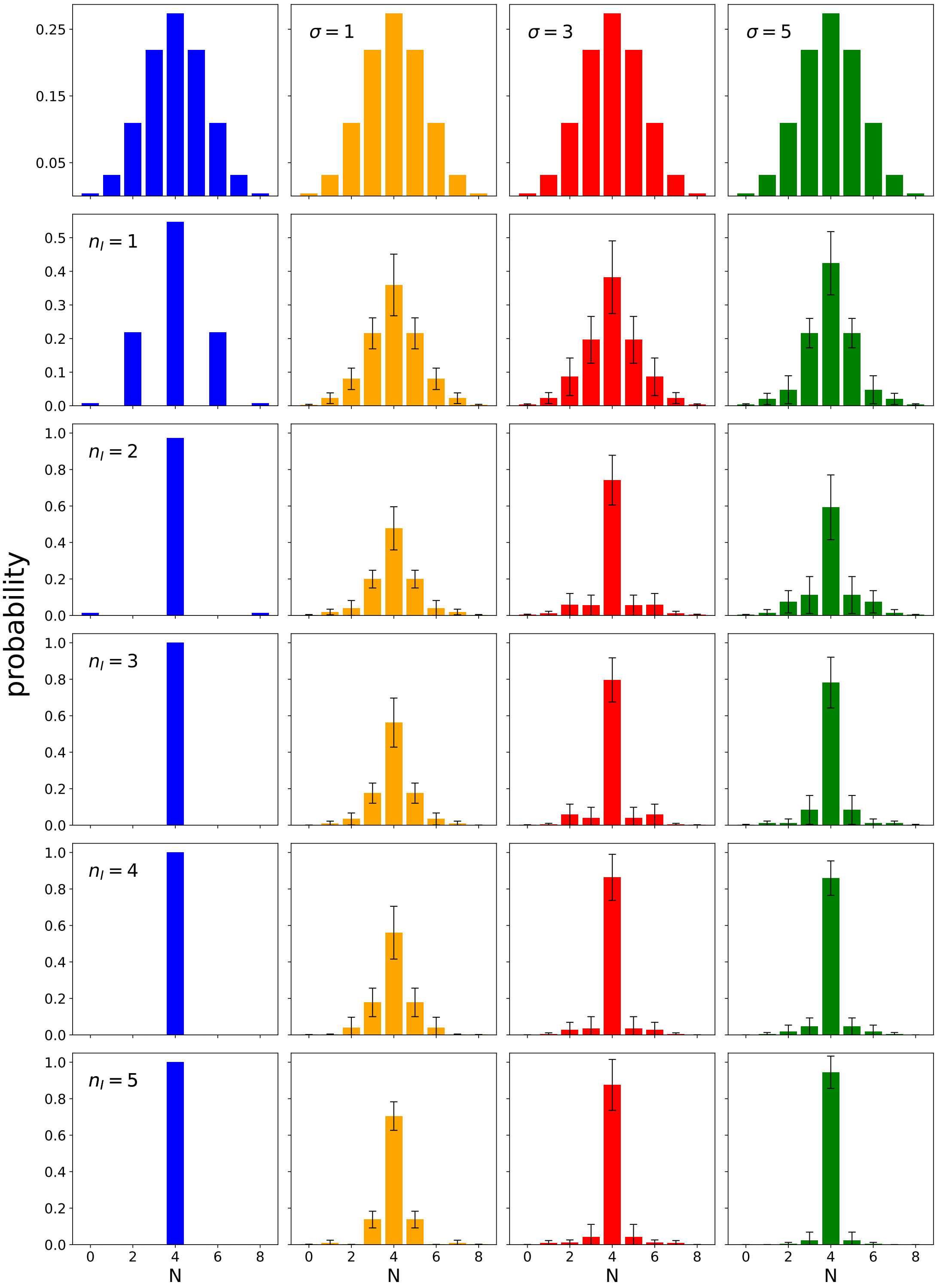

Figure 3.10: Comparison between the IPQE-like method (blue) and the Rodeo algorithm (orange for $\sigma = 1$, red for $\sigma = 3$, and green for $\sigma = 5$) for projecting an equiprobable state $|\psi_e\rangle = \frac{1}{2^n}\sum_p C_n^p |p\rangle$ over $n = 8$ qubits onto a state with $N = 4$ particles. The values $\sigma = 1, 3, 5$ were chosen so that the overlap between 2 consecutive terms in Eq. (3.20) was proportional to 1, 3, and 5 standard deviations $\gamma = 1/\sigma$ of the Gaussian in Eq. (3.20). The QPE decomposition of the successive iterations of the IQPE-like and Rodeo circuits are shown in different rows and numbered by the number of iterations $n_I$. The first row depicts the initial QPE decomposition of the equiprobable state $|\psi_e\rangle$. For the Rodeo algorithm, the height of the bars represents the mean of 10 runs, and the error bars indicate the corresponding standard deviation.

To begin, let us define the symmetry-breaking Bardeen–Cooper–Schrieffer (BCS) ansatz over $n$ qubits, following the convention of [120]:

$$|\psi(\boldsymbol{\theta})\rangle = \bigotimes_{k=0}^{n-1} \left[\sin(\theta_k)|0\rangle_k + \cos(\theta_k)|1\rangle_k\right], \tag{3.21}$$

$$= \bigotimes_{k=0}^{n-1} R_y^k(\pi - 2\theta_k)|0\rangle_k, \tag{3.22}$$

with $\boldsymbol{\theta} \equiv \{\theta_k\}$ and $R_y^k$ the y-rotation gate from Table 2.2 applied to the qubit $k$. Here, we have again used the JWT to encode a pair's absence (resp. occupation) in a certain level $k$ in the qubit state $|0\rangle_k$ (resp. $|1\rangle_k$). To minimize the energy while constraining the number of pairs in the ground state to $A_p$, we minimize the following cost function:

$$\mathcal{C}(\boldsymbol{\theta}) = \langle H - \lambda_F (N - A_p)\rangle_{\boldsymbol{\theta}}, \tag{3.23}$$

where $\lambda_F$ is the Fermi energy, $H$ is the pairing Hamiltonian, $N$ is the pairs number operator, and $A_p$ is the number of pairs we want to project onto. Here $\langle . \rangle_{\boldsymbol{\theta}}$ is a short-hand notation meaning that the expectation value is taken on the state $|\psi(\boldsymbol{\theta})\rangle$. By adapting the classical variational procedure for this problem presented in [120] (see also Fig. 3.5), the minimization of Eq. (3.23) is performed by the following iterative scheme [71]:

1. Initial values for the the parameters $\{\theta_k\}$ and the Fermi energy $\lambda_F$ are chosen.

2. While the condition $|\langle N\rangle - A_p| \leq \varepsilon_{tol}$, where $\varepsilon_{tol}$ is a tolerance fixed *a priori*, the following steps are performed:

   2.1 The cost function $\mathcal{C}(\boldsymbol{\theta})$ of Eq. (3.23) is minimized with respect to the set of parameters $\boldsymbol{\theta_k}$. This minimization is performed using the VQE algorithm (see section 3.3) and the relation:

   $$\langle N\rangle = \sum_{k=0}^{n-1} \cos^2(\theta_k). \tag{3.24}$$

   2.2 Using the set of optimized parameters $\{\theta'_k\}$, we compute a new Fermi energy $\lambda'$.

   2.3 We then use the new set of $\{\lambda', \theta'_k\}$ to restart the process at 2.1.

A comparison of the results of this algorithm with the classical method described in [120] is shown in Fig. 3.11. Once the BCS ground state approximation has been obtained, we projected it to get the Q-PAV state using any of the projecting methods described previously. $Q$ here stands for quantum, given that a quantum computer is involved in the procedure [71]. Finally, the Q-VAP approximation is obtained by projecting each time the ansatz $|\psi(\boldsymbol{\theta})\rangle$ in the VQE minimization. A comparison of different ground state energy approximations obtained with the 3 methods –BCS, Q-PAV, and Q-VAP– is shown in Fig. 3.12. In this figure, we use a pairing Hamiltonian over 8 levels (i.e., 8 qubits) and set $A_p = 4$. The figure shows the error $\Delta E/E\,(\%)$ defined as:

$$\frac{\Delta E}{E}(\%) = \left|\frac{E_c^{approx} - E_c^{exact}}{E_c^{approx}}\right| \times 100 \tag{3.25}$$

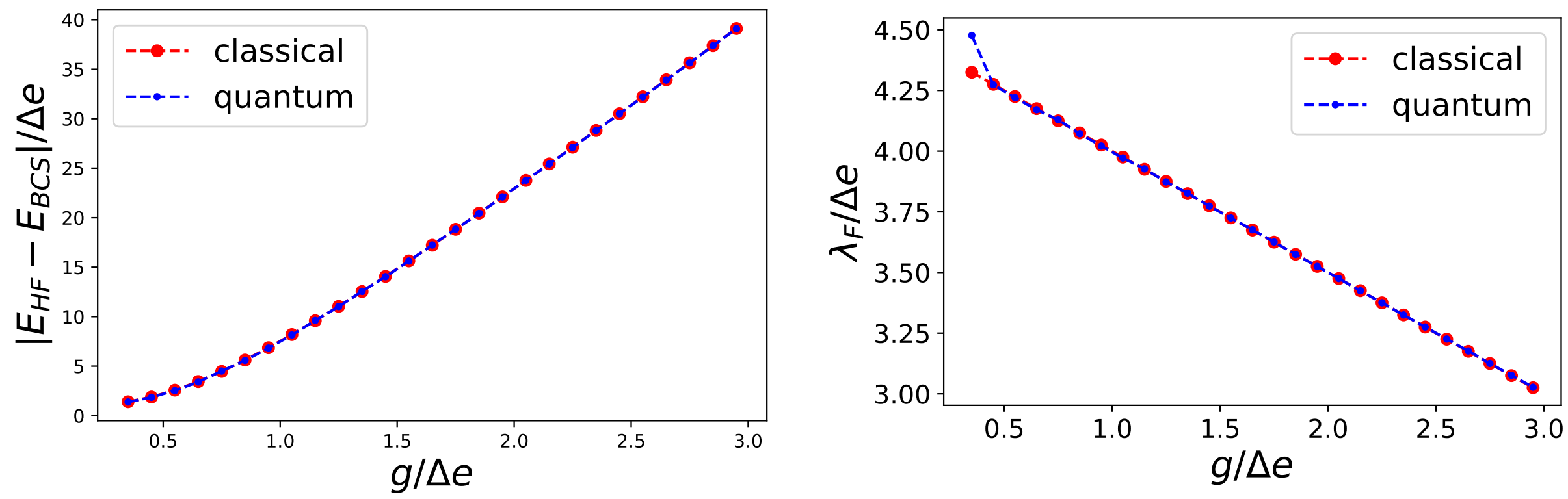

Figure 3.11: Comparison between the classical method explained in [120] and the quantum method described in section 3.5 for finding the ground state with a BCS ansatz for the pairing Hamiltonian. Displayed are the results for the absolute value of the correlation energy defined in section 3.5 (left) and the Fermi energy (right). These are functions of the unitless coupling constant $g/\Delta e$ with $\Delta e$ being the spacing between the energies of two sequential energy levels $\varepsilon_p$ in the pairing Hamiltonian of Eq. (2.33), i.e., $\varepsilon_{p=1,\ldots,n} = p\Delta e$. In this case, $\Delta e = 1$. Using the JWT, the number of levels (or qubits) $n$ was set to 8. The number of pairs to project onto was $A_p = 4$ (8 particles).

where $E_c$ is the correlation energy defined as the total energy minus the reference Hartree-Fock (HF) energy. This latter equals the system's energy when filling the $A_p$ least energetic doubly degenerated levels. To further compare the ground state approximation these three methods are reaching, we can use the QPE algorithm (see section 2.1.3) to observe the different eigenstate decomposition of each method's state. This comparison is shown in Fig. 3.13. It can be observed that the Q-VAP method yields a significantly improved purification/approximation of the ground state by effectively removing the excited state components.

## 3.6 . Conclusion

In this chapter, I have introduced the symmetry-breaking/symmetry-restoring framework to study many-body systems in quantum computing. This framework uses the VQE algorithm and a projection to execute the Q-PAV and Q-VAP methods. I presented three QPE-based projecting methods and the specific requirements needed to use them in the case of symmetry restoration of some common symmetries. Afterward, I compared the QPE-based projection methods in the context of particle number projection, concluding that the IQPE method is the most efficient. Finally, we applied the previously introduced BCS, Q-PAV, and Q-VAP techniques to estimate the ground state of the pairing Hamiltonian. Our analysis revealed that the Q-VAP technique significantly outperforms the other methods. The upcoming chapter will present various methods for implementing the symmetry-restoring projection using a generalized oracle (see section 2.1.3).

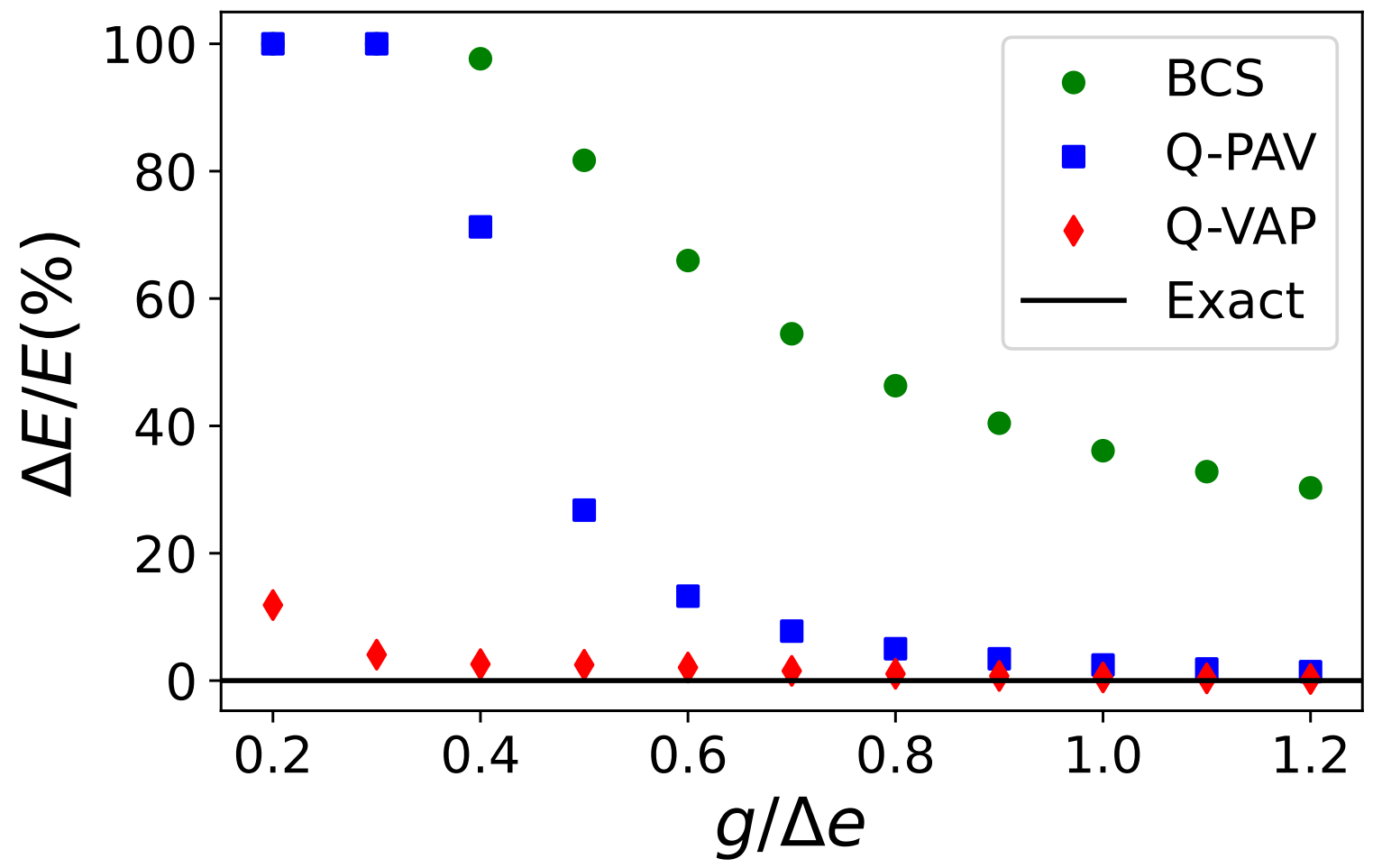

Figure 3.12: Illustration of the precision in energy, as defined by Eq. (3.25), for the pairing problem with four pairs on eight equidistant energy levels ($n = 8$) using the BCS (green circles), Q-PAV (blue squares), and Q-VAP (red diamonds) methods. The values of $g/\Delta e$ range from 0.2 to 1.2. The solid black line represents the exact result ($\Delta E/E = 0$). The results were obtained using the hybrid quantum-classical minimization and projection procedures outlined in section 3.5. For this case, the transition from normal to superfluid occurs at approximately $g_c \approx 0.29\Delta e$. Below this critical value, the Q-PAV and Hartree-Fock (HF) states are identical. We can observe how the relation $E_{GS} \leq E_{Q-VAP} \leq E_{Q-PAV} \leq E_{BCS}$ is consistently maintained with the Q-VAP method reaching the best approximation for the ground state (GS) energy each time.

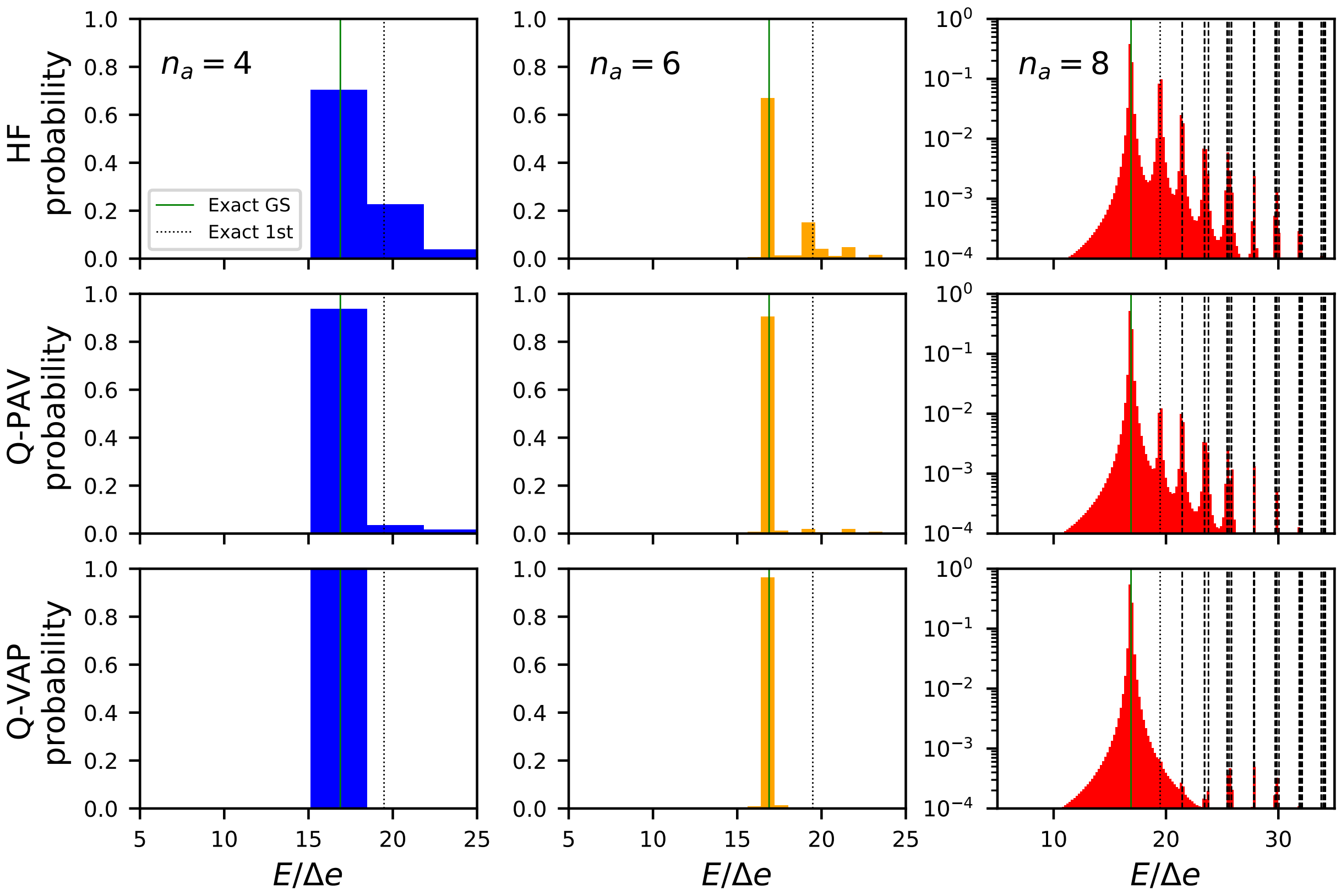

Figure 3.13: Comparison of the results obtained with the QPE method applied to the HF (top), the Q-PAV (middle), and Q-VAP (bottom) wave function. The different methods were applied to the pairing Hamiltonian with the same characteristics as Fig. 3.12 but fixing $g/\Delta e$ to $0.5$. The QPE results (histograms) are shown for $n_a = 4$ (blue), $n_a = 6$ (orange), and $n_a = 8$ (red) ancillary qubits. The solid green and black dotted lines indicate the ground-state and first-excited-state energies, respectively. In the rightmost column, the probabilities are shown in a logarithmic scale to enable the detection of small components in the QPE. The vertical dashed lines indicate the exact eigenvalues of the Hamiltonian, starting from the second excited eigenvalue upwards. The histogram's width corresponds to the QPE method's resolution for a given number of ancillary qubits $n_a$.

# 4 - Restoring Symmetry Using Oracles

In the previous chapter, we introduced the symmetry-breaking/symmetry-restoring technique and its implementation on a quantum computer. In particular, we presented three distinct approaches based on quantum phase estimation to perform the symmetry restoration part of the method, demonstrating how these typically non-unitary operations can be carried out through indirect measurements leading to projections. In this chapter, we will explore alternative projection techniques based on the concept of *oracle*. The idea of an oracle was initially introduced in the amplitude amplification algorithm of Section 2.1.3. Generally, the circuit associated with an oracle demands significant quantum resources, which we will discuss further. We will demonstrate how the Linear Combination of Unitaries (LCU) technique can be employed to reduce the computational cost. Notably, the Oracle method will showcase how to carry out the Q-VAP without explicitly performing any projection. We will provide some illustrations using the pairing model.

## 4.1 . General Oracle

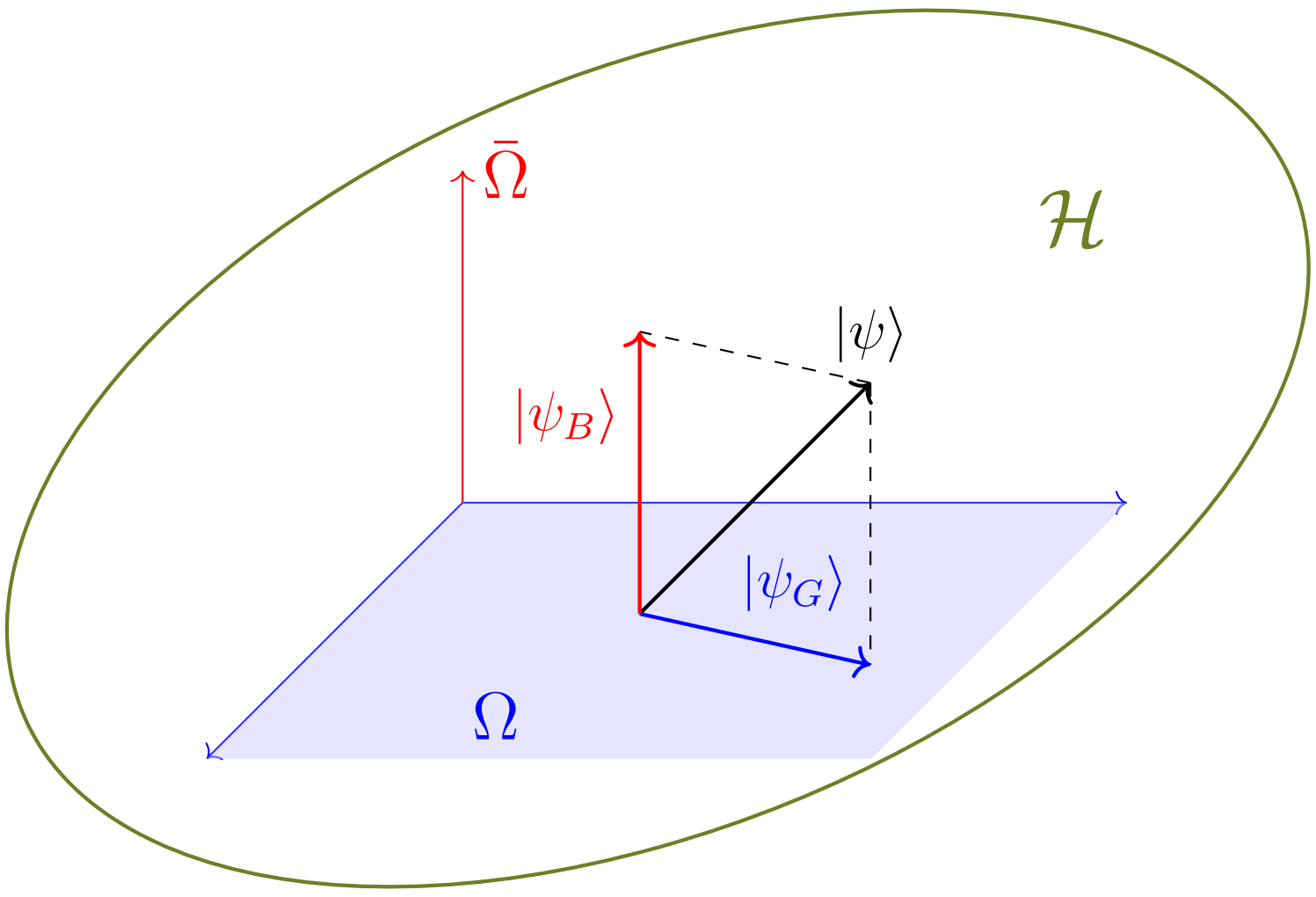

Figure 4.1: Illustration of the two distinct subspaces, $\Omega$ and $\bar{\Omega}$, used in the definition of the general oracle in Eq. (4.1) and section 2.1.3. The blue plane and red axis represent schematically the subspace $\Omega$ and $\bar{\Omega}$ respectively. Any state $|\psi\rangle$ can be decomposed into its projections over the subspaces $\Omega$ and $\bar{\Omega}$, which are denoted by $|\psi_G\rangle$ and $|\psi_B\rangle$, respectively. $G$ (resp. $B$) stands for *Good* (resp. *Bad*). The total Hilbert space $\mathcal{H}$ is spanned by these subspaces, $\Omega$ and $\bar{\Omega}$.

In the following sections, we will explore various methods that necessitate the generalization of the notion of oracle introduced in Section 2.1.3. The fundamental assumption of an oracle is that the total Hilbert space can be partitioned into two complementary subspaces, denoted as $\Omega$ and $\bar{\Omega}$. These subspaces encompass all states possessing or lacking a specific property. States with the property are referred to as “Good states”, while those without are called “Bad states”. Generally, a system’s state is a superposition of both good and bad components. An illustration of the partition between the $\Omega$ and the $\bar{\Omega}$ subspaces is shown in Fig. 4.1. The amplitude amplification algorithm (see section 2.1.3) employs the oracle defined in Eq. (2.20) to enhance the amplitudes of the good states at the expense of the amplitudes of the bad ones. We consider here a generalization of the oracle operator commonly used in the amplitude amplification algorithm. Following Ref. [123], we can define a more general version of this operator that applies different phases to the “Good” and “Bad” states, i.e.,

$$O_\Omega|\phi\rangle = f_{|\phi\rangle}|\phi\rangle \text{ with } f_{|\phi\rangle} = \begin{cases} e^{i\varphi} & \text{if } |\phi\rangle \in \Omega, \\ e^{i\mu} & \text{if } |\phi\rangle \notin \Omega. \end{cases} \tag{4.1}$$

If we have access to the projector onto the $\Omega$ subspace, denoted as $P_\Omega$, the general oracle operator in Eq. (4.1) can then be expressed as:

$$O_\Omega = e^{i(\varphi P_\Omega + \mu(I-P_\Omega))} = e^{i\varphi P_\Omega} e^{i\mu(I-P_\Omega)}, \tag{4.2}$$

where we used the fact that $[P_\Omega, I - P_\Omega] = 0$. Given that $I$ and $P_\Omega$ commute, this operator can be implemented in the general case using the Trotter-Suzuki approximation (see section 2.1.3) on the operator $e^{i(\varphi-\mu)P_\Omega}$, the oracle then becomes

$$O_\Omega = e^{i\mu} e^{i(\varphi-\mu)P_\Omega},$$

with $e^{i\mu}$ a global phase. By exploiting the property for projectors $P_\Omega^n = P_\Omega$ for $n \geq 1$ and thus $e^{ixP_\Omega} = I + (e^{ix} - 1) P_\Omega$ for $x \in \mathbb{R}$, the previous equation can be further simplified as

$$O_\Omega(\varphi, \mu) = e^{i\varphi} P_\Omega + e^{i\mu}(I - P_\Omega). \tag{4.3}$$

Note that the oracle standardly used in the Grover algorithm (see section 2.1.3: amplitude amplification) can be retrieved by taking $(\varphi, \mu) = (\pi, 0)$.

### 4.2 . Grover-Hoyer Method

Although the amplitude amplification method from section 2.1.3 can increase the probabilities of the “Good” state, it does not always fully converge to the subspace $\Omega$. In fact, in order to achieve full convergence to $\Omega$, the expression $\frac{1}{2}\left[\frac{\pi}{2\theta} - 1\right]$ must be an integer, where $\theta$ represents the initial angle between the initial state $|\psi\rangle$ and its projection onto the “Bad” subspace $|\psi_B\rangle$ (see Fig. 4.1). Fig. 4.2 provides an example of a situation in which the amplitude amplification method is unable to project the state fully.

To perform the full projection to the $\Omega$ subspace, we can use the amplitude amplification method supplemented by the Hoyer method briefly described in section 2.1.3, taken from [84]. The Grover-Hoyer technique was applied for the symmetry-restoration case in [33]. This method, which I discuss in more detail here, aims to create an operator able to rotate an angle $\pi/2 - \theta$ no matter which initial $\theta$ (see Fig. 4.3).

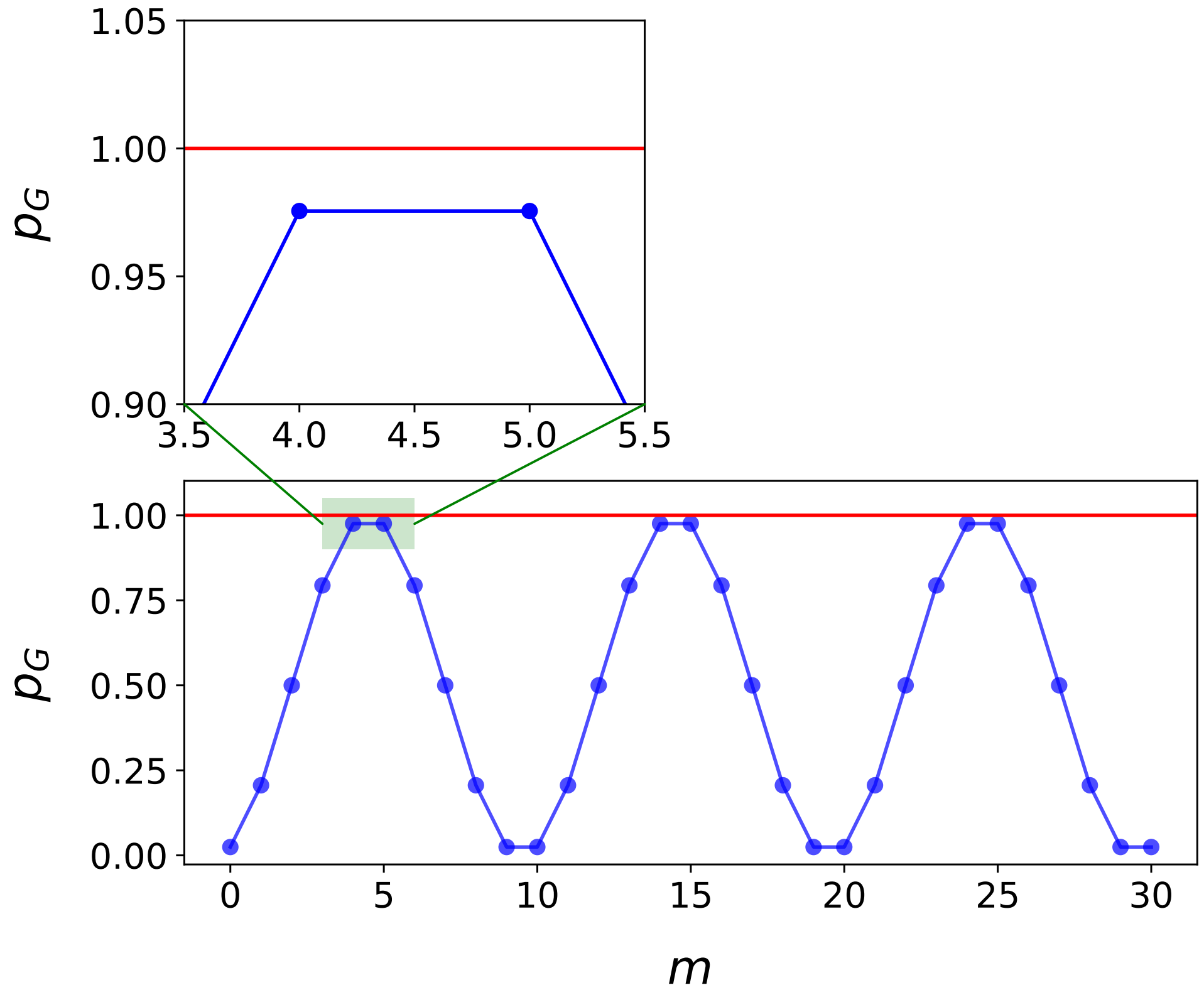

Figure 4.2: Example of applying the amplitude amplification algorithm to project state $|\psi\rangle$ onto state $|\psi_G\rangle$. The initial angle $\theta$ in Eq. (2.23) is set to $\pi/20$. The zoomed-in area illustrates that this method does not fully converge for this value $\theta$.

Let us define our problem again. We begin with a state $|\Psi\rangle$ that we would like to project onto a subspace $\Omega$. As in section 2.1.3, the initial state can be seen as a superposition of "Good" and "Bad" states, i.e., $|\Psi\rangle = |\Psi_G\rangle + |\Psi_B\rangle = \sqrt{g}|\psi_G\rangle + \sqrt{b}|\psi_B\rangle$ with $\{|\psi_G\rangle, |\psi_B\rangle\}$ the orthonormal basis and $g = |\langle\psi_G|\Psi\rangle|^2$, $b = |\langle\psi_B|\Psi\rangle|^2$. To improve clarity, we have changed the notation $p_{G/B}$ to $g/b$. We need to use the general oracle operator from Eq. (4.1) with $\mu = 0$, which we refer to as $O_\varphi$ in the following. Additionally, we utilize a generalized version of the reflection operator, denoted as $U_\psi$ [84].

$$O_\varphi = \begin{pmatrix} 1 & 0 \\ 0 & e^{i\varphi} \end{pmatrix} \begin{vmatrix} |\psi_B\rangle \\ |\psi_G\rangle \end{vmatrix}, \qquad \begin{aligned} U_\psi &= \left(1 - e^{i\phi}\right) |\psi\rangle\langle\psi| - I, \\ &= \left(1 - e^{i\phi}\right) \begin{pmatrix} b & \sqrt{bg} \\ \sqrt{bg} & g \end{pmatrix} \begin{vmatrix} |\psi_B\rangle \\ |\psi_G\rangle \end{vmatrix} - I. \end{aligned} \tag{4.4}$$

Both operators $O_\varphi$ and $U_\psi$ implement unitary operations on the subspace spanned by $|\Psi_B\rangle$ and $|\Psi_G\rangle$. The notation $\begin{vmatrix} |\psi_B\rangle \\ |\psi_G\rangle \end{vmatrix}$ represents the ordering of the terms in the matrix with respect to the basis. For example, the operator $O_\varphi$ does not apply any phase to the Bad state $|\psi_B\rangle$ while it applies a phase of $e^{i\varphi}$ to the Good state $|\psi_G\rangle$. Given that $g + b = 1$, the general Grover operator can be

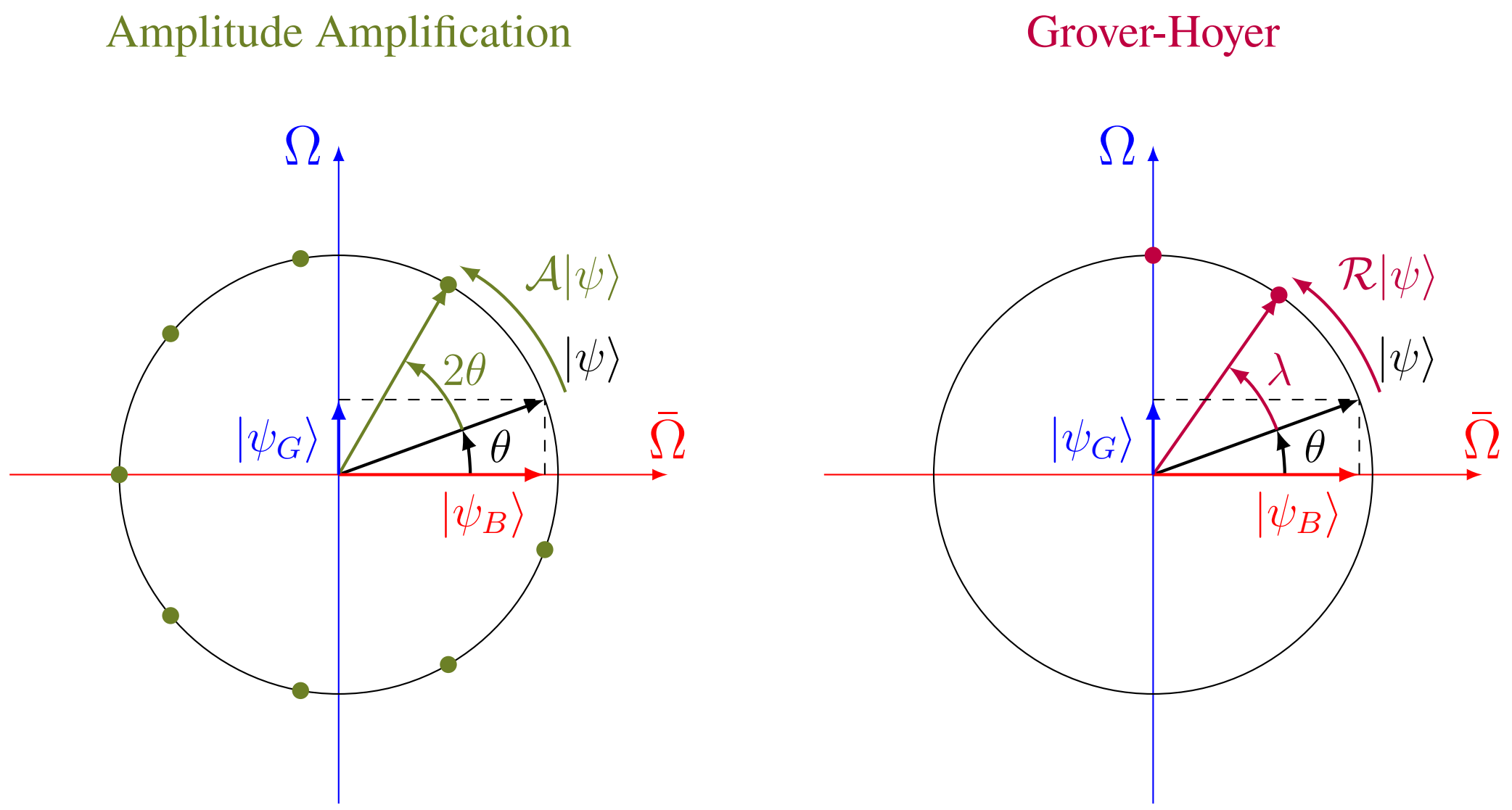

Figure 4.3: Comparison between the amplitude amplification (AA) algorithm (left) and the Grover-Hoyer method described in Section 4.2 (right) for an initial angle of $\theta = \pi/9$. The AA method fails to converge since $\frac{1}{2}\left[\frac{\pi}{2\theta} - 1\right]$ is not an integer. In contrast, the Grover-Hoyer method is capable of performing arbitrary rotations $R$ with an angle $\lambda \leq 2\theta$. In this case, two applications of the operator $R$ are sufficient to fully project the state onto the "Good" subspace.

written as:

$$G = U_\psi O_\varphi = \begin{pmatrix} -\left[\left(1-e^{i\phi}\right)g + e^{i\phi}\right] & \left(1-e^{i\phi}\right)\sqrt{g(1-g)}e^{i\varphi} \\ \left(1-e^{i\phi}\right)\sqrt{g(1-g)} & \left[\left(1-e^{i\phi}\right)g - 1\right]e^{i\varphi} \end{pmatrix}. \tag{4.5}$$

### 4.2.1 . Recovering Grover Algorithm

We can notice that if we set $\phi = \varphi = \pi$ as in the original Grover algorithm (see section 2.1.3), we have:

$$G(\phi = \varphi = \pi) = \begin{pmatrix} 1-2g & -2\sqrt{g(1-g)} \\ 2\sqrt{g(1-g)} & 1-2g \end{pmatrix}.$$

As $g = \sin^2(\theta)$ where $\theta$ is the angle between the state $|\Psi\rangle$ and $|\psi_B\rangle$ in Fig. 4.3 :

$$G(\phi = \varphi = \pi) = \begin{pmatrix} 1-2\sin^2(\theta) & -2\sin(\theta)\cos(\theta) \\ 2\sin(\theta)\cos(\theta) & 1-2\sin^2(\theta) \end{pmatrix} = \begin{pmatrix} \cos(2\theta) & -\sin(2\theta) \\ \sin(2\theta) & \cos(2\theta) \end{pmatrix}$$

The Grover operator results in a $2\theta$ rotation as in the original Grover's algorithm.

### 4.2.2 . F Matrix Decomposition

Here, we show how Eq. (4.5) can be transformed into a general rotation. Let us start by defining the matrix $F$ as:

$$F = e^{iv}\begin{pmatrix} 1 & 0 \\ 0 & e^{iu} \end{pmatrix}\begin{pmatrix} \cos(\lambda) & -\sin(\lambda) \\ \sin(\lambda) & \cos(\lambda) \end{pmatrix}\begin{pmatrix} 1 & 0 \\ 0 & e^{-iu} \end{pmatrix}. \tag{4.6}$$

If we impose $F = G$, where $G$ is given by Eq. (4.5), we obtain the following relations:

$$\begin{cases} \tan\left(\frac{\varphi}{2}\right) = \tan\left(\frac{\phi}{2}\right)(1-2g), \\ \cos(\phi) = 1 - \frac{\sin^2(\lambda)}{2g(1-g)}, \qquad 0 \le \lambda \le 2\theta, \\ u = \frac{\pi - \varphi}{2}, \\ v = \mathrm{Arg}\left(-\frac{1}{\cos(\lambda)}\left[\left(1-e^{i\phi}\right)g + e^{i\phi}\right]\right). \end{cases} \tag{4.7}$$

From these equations, we can derive the formula for an arbitrary rotation $R(\lambda)$:

$$R(\lambda) = \begin{pmatrix} \cos(\lambda) & -\sin(\lambda) \\ \sin(\lambda) & \cos(\lambda) \end{pmatrix} = e^{-iv} \begin{pmatrix} 1 & 0 \\ 0 & e^{-iu} \end{pmatrix} G(\phi, \varphi) \begin{pmatrix} 1 & 0 \\ 0 & e^{iu} \end{pmatrix}. \tag{4.8}$$

valid for an angle $\lambda \le 2\theta$. The full details of the derivation are shown in Appendix B.

#### 4.2.3 . Complete Searching Algorithm

Suppose we want to implement an arbitrary rotation of angle $w = \frac{\pi}{2} - \theta$ so we converge exactly to the good state $|\psi_G\rangle$. First, we check whether $k = \frac{1}{2}\left[\frac{\pi}{2\theta} - 1\right]$ is a whole number. If that's the case, we apply the standard amplitude amplification operator $\mathcal{A}$ from section 2.1.3 $k$ times. Otherwise, i.e., if *k* is not a whole number, we compute $\lambda$ and then *m* to apply *m* rotations of angle $\lambda$. The number of rotations *m* and the angle $\lambda$ should satisfy $m\lambda = w$ together with $\lambda \le 2\theta$. This means that:

$$m \ge \frac{w}{2\theta}.$$

The lowest value of *m* that satisfy the previous inequality is:

$$m = \left\lceil \frac{w}{2\theta} \right\rceil, \tag{4.9}$$

$\lambda$ is then selected to be:

$$\lambda = \frac{w}{m}. \tag{4.10}$$

Using Eq (4.8), we can implement *m* rotations of angle $\lambda$ using the following operator:

$$R^m(\lambda) = \left(e^{-iv} A^\dagger G(\phi, \varphi) A\right)^m,$$

where $A = \begin{pmatrix} 1 & 0 \\ 0 & e^{iu} \end{pmatrix} = O_{(\varphi = u)}$. This can be written as:

$$R^m(\lambda) = e^{-imv} \underbrace{A^\dagger G(\phi, \varphi) A \ldots A^\dagger G(\phi, \varphi) A}_{=\left(A^\dagger G(\phi,\varphi) A\right)^m}.$$

Finally, as $AA^\dagger = I$, we get:

$$R^m(\lambda) = e^{-imv} A^\dagger G^m(\phi, \varphi) A. \tag{4.11}$$

This operator exactly projects the state $|\Psi\rangle$ to the "Good" space (see schematic representation in Fig. 4.3).

### 4.3 . Combining Oracle with Hadamard Test

Eq. (4.3) suggest a way to restore a state's symmetry. We can define the projector from Eq. (4.3) for any pair of phases $(\varphi, \mu)$ as

$$P_\Omega = \frac{O_\Omega(\varphi, \mu) - e^{i\mu} I}{e^{i\varphi} - e^{i\mu}}. \tag{4.12}$$

In particular, if we take the angles of the oracle in Eq. (2.20) used for the amplitude amplification algorithm, i.e., $(\varphi = \pi, \mu = 0)$, the projector operator becomes

$$P_\Omega = \frac{1}{2}\left(I - O_\Omega(\pi, 0)\right). \tag{4.13}$$

This operator can be implemented using the Hadamard test circuit shown in Table 2.5 without the phase gate. Indeed, by taking $U = O_\Omega(\pi, 0)$, the state just before the measurement is

$$|0\rangle \otimes \frac{1}{2}\left(I + O_\Omega(\pi, 0)\right)|\Psi\rangle + |1\rangle \otimes \frac{1}{2}\left(I - O_\Omega(\pi, 0)\right)|\Psi\rangle, \tag{4.14}$$

or, in terms of the projector:

$$|0\rangle \otimes \left(I - P_\Omega\right)|\Psi\rangle + |1\rangle \otimes P_\Omega|\Psi\rangle. \tag{4.15}$$

The previous equation means that the state $|\Psi\rangle$ will be projected into the subspace $\Omega$ (resp. $\bar{\Omega}$) when the ancilla qubit is measured in the state $|1\rangle$ (resp. $|0\rangle$). Hence, given that the oracle can be efficiently implemented using a circuit, its combination with the Hadamard test provides a highly efficient method for performing projection. It is also worth noting that only a single ancillary qubit is required in this scenario.

### 4.4 . Practical Implementation of Oracles

One of the main difficulties in the previously discussed methods is the actual implementation of an oracle. As a first approach, given the decomposition of the projector operator as a weighted sum of Pauli chains (as in Eq. (3.13)), we can use a Trotter-Suzuki approximation of Eq. (4.3) to implement the oracle as mention in section 4.1. Another way, particularly well-suited for symmetry restoration, is the implementation of a projector which arises from the relation:

$$\delta_{nm} = \frac{1}{M+1}\sum_{k=0}^{M} e^{2i\pi\frac{k(m-n)}{M+1}} \tag{4.16}$$

where $m, n \in [0, M]$. Similar to section 3.4, we consider a set of eigenvalues $\{\lambda_0, \lambda_1, \ldots, \lambda_\Omega\}$ linked with integers $\{0, m_1, \ldots, m_\Omega\}$ via the relation $m_i = \frac{\lambda_i - \lambda_0}{c}$ and $c \in \mathbb{R}$. These eigenvalues are associated to a set of eigenstates $\{|s_i\rangle\}$ of a Hermitian symmetry operator $S$, i.e., $S|s_i\rangle = \lambda_i|s_i\rangle$. The projector over a certain eigenvalue $\lambda'$ takes the form:

$$P_S = \sum_{k=0}^{M} \alpha_k e^{i\phi_k S}, \tag{4.17}$$

with

$$\alpha_k = \frac{e^{-i\phi_k \lambda'}}{M+1} \qquad \phi_k = \frac{2\pi k}{c(M+1)}, \tag{4.18}$$

and $M = m_\Omega$. Using Eq. (4.3), we can obtain a similar form for the general oracle operator:

$$O_S = \sum_{k=0}^{M} \beta_k e^{i\phi_k S}, \tag{4.19}$$

with

$$\beta_k = e^{i\mu}\delta_{0k} + \frac{\left(e^{i\varphi} - e^{i\mu}\right) e^{-i\phi_k \lambda'}}{M+1} \qquad \phi_k = \frac{2\pi k}{c(M+1)}. \tag{4.20}$$

Both equations (4.17) and (4.19) have the form of a linear combination of unitaries operators. The following sections offer a detailed discussion on how to utilize effectively the decomposition described above for practical implementations.

### 4.4.1 . Implicit Projection

We are usually interested in computing the expectation value of an operator $A$ over the projected state $P_S|\Psi\rangle$, i.e.,

$$\frac{\langle \Psi | P_S A P_S | \Psi\rangle}{\langle \Psi | P_S P_S | \Psi\rangle} = \frac{\langle A P_S\rangle}{\langle P_S\rangle}, \tag{4.21}$$

where we utilized the property that $P_S^2 = P_S$ and assumed $[A, P_S] = 0$. Additionally, on the right-hand side of the equation, we used the shorthand notation $\langle . \rangle = \langle \Psi | \cdot | \Psi \rangle$. The expectation value can be calculated without explicitly projecting the wave function. Taking advantage of Eq. (4.17), one can compute separately the different expectation values $\langle A e^{i\phi_k S}\rangle$ and $\langle e^{i\phi_k S}\rangle$:

$$\langle A P_S\rangle = \sum_k \alpha_k \langle A e^{i\phi_k S}\rangle, \qquad \langle P_S\rangle = \sum_k \alpha_k \langle e^{i\phi_k S}\rangle. \tag{4.22}$$

Similarly, the expectation values using the oracle $O_S$ can be retrieved by simply changing $\alpha_k$ for $\beta_k$ from Eq. (4.20) in Eq. (4.22). In contrast to other methods that utilize a single circuit for the projection, the primary disadvantage of this approach is the necessity to compute $2M$ expectation values instead of just one. However, the circuit depth required for calculating each expectation value is significantly reduced. We have that the expectation values using either the projector or the oracle hold the relation:

$$\frac{\langle A P_S\rangle}{\langle P_S\rangle} = \frac{\langle A O_S\rangle - e^{i\mu}\langle A\rangle}{\langle O_S\rangle - e^{i\mu}}. \tag{4.23}$$

This method usually reduces the resources for the symmetry-restoring process. However, it has the drawback of not giving access to the projected state $P_S|\Psi\rangle$ itself. Below, we present an alternative method allowing us to construct the projected state.

### 4.4.2 . Linear Combination of Unitaries Algorithm

Considering that both equations (4.17) and (4.19) are expressed as a linear combination of unitary (LCU) operators, the LCU algorithm [154] can be employed to implement either the projector or the oracle. The LCU algorithm is designed to implement operators regardless of whether they are unitary. Let us assume that we have the decomposition of an operator $A$ in terms of the unitary operators $V_k$, as follows:

$$A = \sum_{k=0}^{m-1} \alpha_k V_k. \tag{4.24}$$

Fig. 4.4 shows the circuit capable of implementing the $A$ operator on a wavefunction $|\psi\rangle$. It uses two operators called $B$ and $E$ that we discuss below. A set of $n_{\mathrm{LCU}}$ ancillary qubits with $2^{n_{\mathrm{LCU}}} \geq m$ is used. After applying operation $B$ to the ancillary register, a set of states $V_k|\Psi\rangle$ are sequentially associated with the different computational states of the ancillary register using controlled operations. Finally, operation $E$ is applied to the ancillary register. By measuring the qubits in the ancillary register and selecting the events where the state $|0\rangle^{\otimes n_{LCU}}$ is measured, we obtain that the system state, after each of these measurements, collapses to the desired $A|\Psi\rangle$ state. The choice of $B$ and $E$ is not unique. For Hamiltonian simulation [155, 156, 157], the $B$ proposed to be used is:

$$B|0\rangle^{\otimes n_{LCU}} = \frac{1}{\mathcal{N}_{\mathcal{B}}} \sum_{k=0}^{2^{n_{LCU}}-1} \sqrt{\alpha_k}|k\rangle_{\mathrm{LCU}}, \tag{4.25}$$

with $\alpha_{k\geq m} = 0$, $\alpha_k \in \mathbb{R}^+$, and with the normalization constant $\mathcal{N}_{\mathcal{B}} = \sqrt{\sum_k \alpha_k}$. In this case, $E = B^\dagger$. Another proposal that reduces the number of operations in the ancilla register is presented in Ref. [158]. Here, the $B$ operator is taken as:

$$B|0\rangle^{\otimes n_{LCU}} = \frac{1}{\mathcal{N}_{\mathcal{H}}} \sum_{k=0}^{2^{n_{LCU}}-1} \alpha_k|k\rangle_{\mathrm{LCU}}, \tag{4.26}$$

with $\alpha_{k\geq m} = 0$, $\alpha_k \in \mathbb{C}$ and $\mathcal{N}_{\mathcal{H}} = \sqrt{\sum_k |\alpha_k|^2}$. For this method, we use $E = H^{\otimes n_{LCU}}$. A slightly better choice for $E$ is to use $E = L^\dagger$, where $L$ is the operator that initializes an equiprobable distribution in the ancillary register up to the state $m$, i.e.:

$$L|0\rangle^{\otimes n_{LCU}} = \frac{1}{\sqrt{m}} \sum_{k=0}^{m-1} |k\rangle_{\mathrm{LCU}}. \tag{4.27}$$

This replacement can be done because the only requirement for the operator $E$ to work in the LCU algorithm is to have equal components in its $E_{0,j\leq m}$ entries. Using $E = L^\dagger$ defined in Eq. (4.27), the probability of measuring $|0\rangle^{\otimes n_{\mathrm{LCU}}}$, denotes by $p_0$, scales as $\sim 1/m$ and leads to a reduction of the number of rejected events compared to taking $E = H^{\otimes n_{\mathrm{LCU}}}$. We mention that the construction of such an operator is polynomial in the number of qubits $\mathcal{O}\left(Ln_{LCU}^2\right)$[159].

The success probability $p_0$ associated with the LCU approach is discussed in detail in Appendix C. For the projector, the success probability is similar or equal to that of the QPE, i.e., $\sim g$. The Oracle implementation is different, since the success probability does not depend on the quantity $g$ but solely on the number of system qubits $n$. Unlike the QPE case, this success probability is independent of the initial state, which may be advantageous when the initial state is not optimized to have significant overlap with the good state. The Oracle will be implemented with the same probability without regard to the initial state's overlap with the good state. It should be noted that, in this case, the LCU success probability decreases linearly as $1/n$.

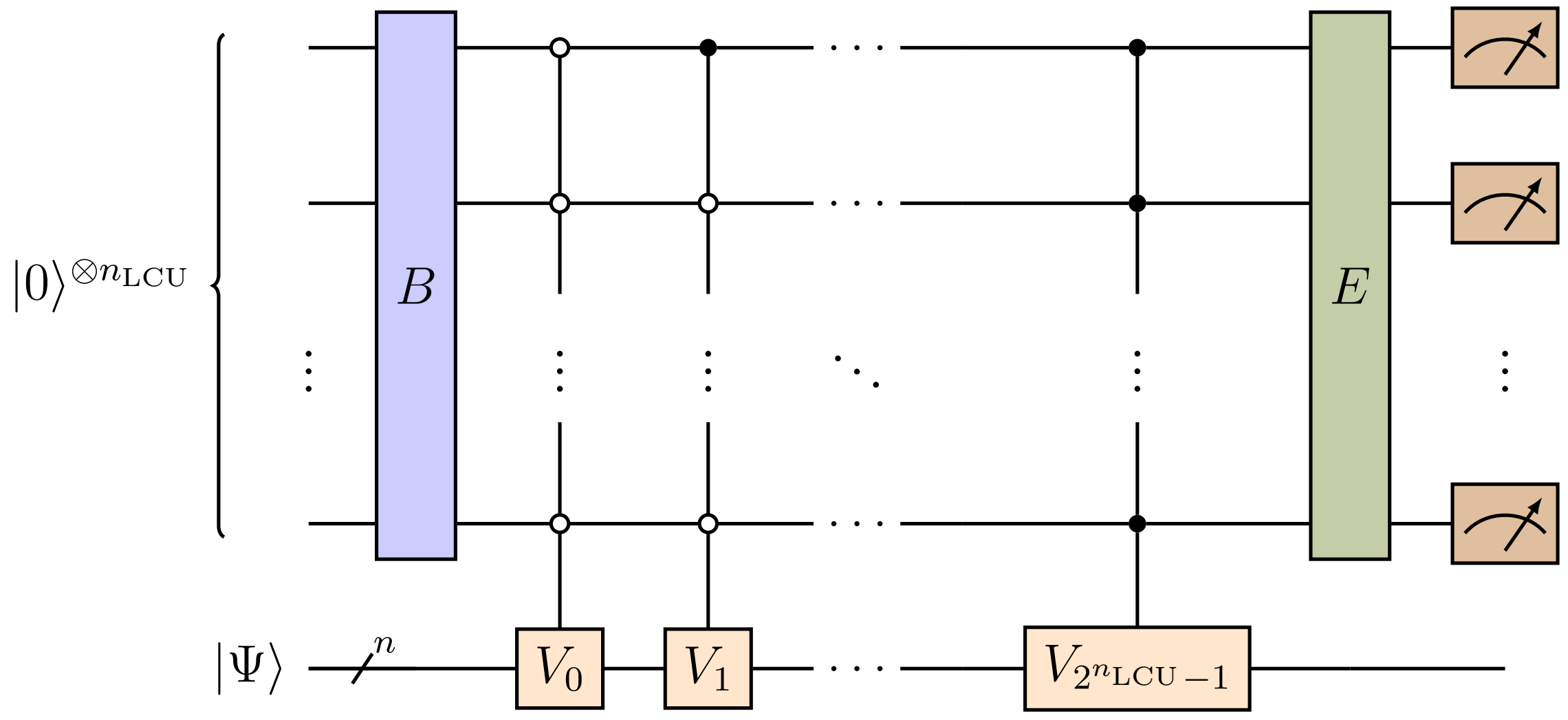

Figure 4.4: Generic LCU circuit with $n_{\text{LCU}}$ ancilla qubits. This circuit can implement a linear combination of up to $2^{n_{\text{LCU}}}$ unitary operators. The filled circles represent controlled operations triggered by the state $|1\rangle$, while open circles represent operations triggered by the state $|0\rangle$. Figure taken from [123].

### 4.5 . Comparison of Different Oracle-Based Projection Methods

In the various sections above, I have presented multiple methods that either provide access to the expectation values of observables exclusively (implicit projection) or to both observables and projected states (Oracle+Hadamard, LCU, Grover/Hoyer). These methods exhibit differences in their underlying principles and in the quantum resources required for their implementation. This section examines this aspect, akin to the comparison presented in Table 3.1 for the QPE-based projections. Table 4.1 presents a comparison in terms of the number of ancilla qubits, the number of measurements, and the gate resources required to implement each method. The Grover-Hoyer method is the most resource-intensive due to the need for implementing two quantum state preparation (QSP) subroutines and $n_G + 2$ generalized oracles. QSP is a quantum routine that enables the preparation of any state from an initialized register in the state $|0\rangle^{\otimes n}$. Generally, without ancilla qubits, the number of gates (single qubit plus CNOT) required to implement this subroutine scales exponentially with the number of qubits [160]. Implementing an oracle operator can also be quite costly; for instance, an oracle operator for the number of particles has exponential complexity when implemented without ancillary qubits [161, 162]. Considering the complexity of these two quantum subroutines, the Implicit projection method emerges as the least resource-intensive option. When comparing more broadly with the QPE-based methods, two approaches stand out as the best choices for symmetry restoration, depending on whether the projected state is needed or not. If the projected state is required, the IQPE method is preferable. Otherwise, the implicit projection method serves as the optimal choice.

| Method | # ancilla | # measurements | Gate resources |
|---|---|---|---|
| Oracle+Hadamard | 1 | $\sim p_G$ | 1 controlled oracle<br>2 Hadamard gates |
| Grover-Hoyer | 0 | - | 2 generalized oracles<br>$n_G$ generalized Grover operators, each:<br>- 1 generalized oracle.<br>- 2 QSP over $n$ qubits.<br>- 1 phase gate controlled by $n-1$ qubits. |
| Implicit projection | 1 | - | $2M$ Hadamard tests, each:<br>- 2 Hadamard gates.<br>- 1 control operator.<br>- 1 controlled evolution operator. |
| LCU | $n_{LCU} = \lceil log_2 (M+1) \rceil$ | $\sim p_G$ | $M$ Controlled-evolution gates.<br>2 QSP circuit over $n_{LCU}$ qubits. |

Table 4.1: Comparative analysis of various oracle-based projection methods for the case of particle number projection, similar to that presented in Table 3.1. In this table, $M$ represents the number of terms in the LCU form of the projector, as provided by Eq. (4.17). For the number of particles, $M = n$, with $n$ being the number of qubits of the wave function register. In the Grover-Hoyer method, we have $n_G = \lceil \frac{w}{2\theta} \rceil$ with $w = \frac{\pi}{2} - \theta$ and $\sin^2(\theta) = p_G$. The generalized oracle mentioned in the Grover-Hoyer method corresponds to the oracle in Eq. (4.4). We assume that the expectation values in the implicit projection method are obtained using a standard Hadamard test (see Table 2.5). All methods, except the implicit projection one, provide access to the projected state. Table adapted from [33].

### 4.6 . Implicit Projection Using an Oracle in the Q-VAP Technique

As outlined in [123], we can leverage an oracle for symmetry restoration. This section presents a method to calculate the energy associated with the symmetry-restored subspace using an oracle to yield the Q-VAP results for the pairing Hamiltonian. To begin, let us consider the expectation value $\langle U \rangle$ computed by the Hadamard test circuit in Table 2.5 when $U$ is taken as $U = VO_S$ ($V$ is a generic operator):

$$\begin{aligned} \langle VO_S \rangle &= (\langle \Psi_G| + \langle \Psi_B|) \, V \left( e^{i\varphi} |\Psi_G\rangle + e^{i\mu} |\Psi_B\rangle \right), \\ &= e^{i\varphi} \left( \langle V \rangle_G + \langle \Psi_B | V | \Psi_G \rangle \right) + e^{i\mu} \left( \langle \Psi_G | V | \Psi_B \rangle + \langle V \rangle_B \right), \end{aligned} \tag{4.28}$$

where we have used the shorthand notation $\langle \Psi_{G/B} | . | \Psi_{G/B} \rangle = \langle . \rangle_{G/B}$ and the action of the oracle over an arbitrary wave function $|\Psi\rangle$, i.e., $O_S|\Psi\rangle = e^{i\varphi}|\Psi_G\rangle + e^{i\mu}|\Psi_B\rangle$. If we consider that $V$ also preserves the symmetry $S$, we have that $\langle \Psi_B|V|\Psi_G\rangle = \langle \Psi_G|V|\Psi_B\rangle = 0$ and thus:

$$\langle VO_S \rangle = e^{i\varphi} \langle V \rangle_G + e^{i\mu} \langle V \rangle_B. \tag{4.29}$$

If we take $(\varphi = 0, \mu = \pi/2)$, we can directly have access to the expectation value over the good state $\langle V \rangle_G$ by computing the real part of the expectation value in Eq. (4.29) using the Hadamard test in Table 2.5, i.e.,

$$p_0 - p_1 = \langle \Psi_G | V | \Psi_G \rangle, \tag{4.30}$$

where $p_0$ (resp. $p_1$) denotes the probability to measure 0 or 1 in the ancillary qubit. An illustration of the use of Eq. (4.30) using the probability $p_0$ and $p_1$ with a limited set of measurements $N_e$ is shown in Fig. 4.5.

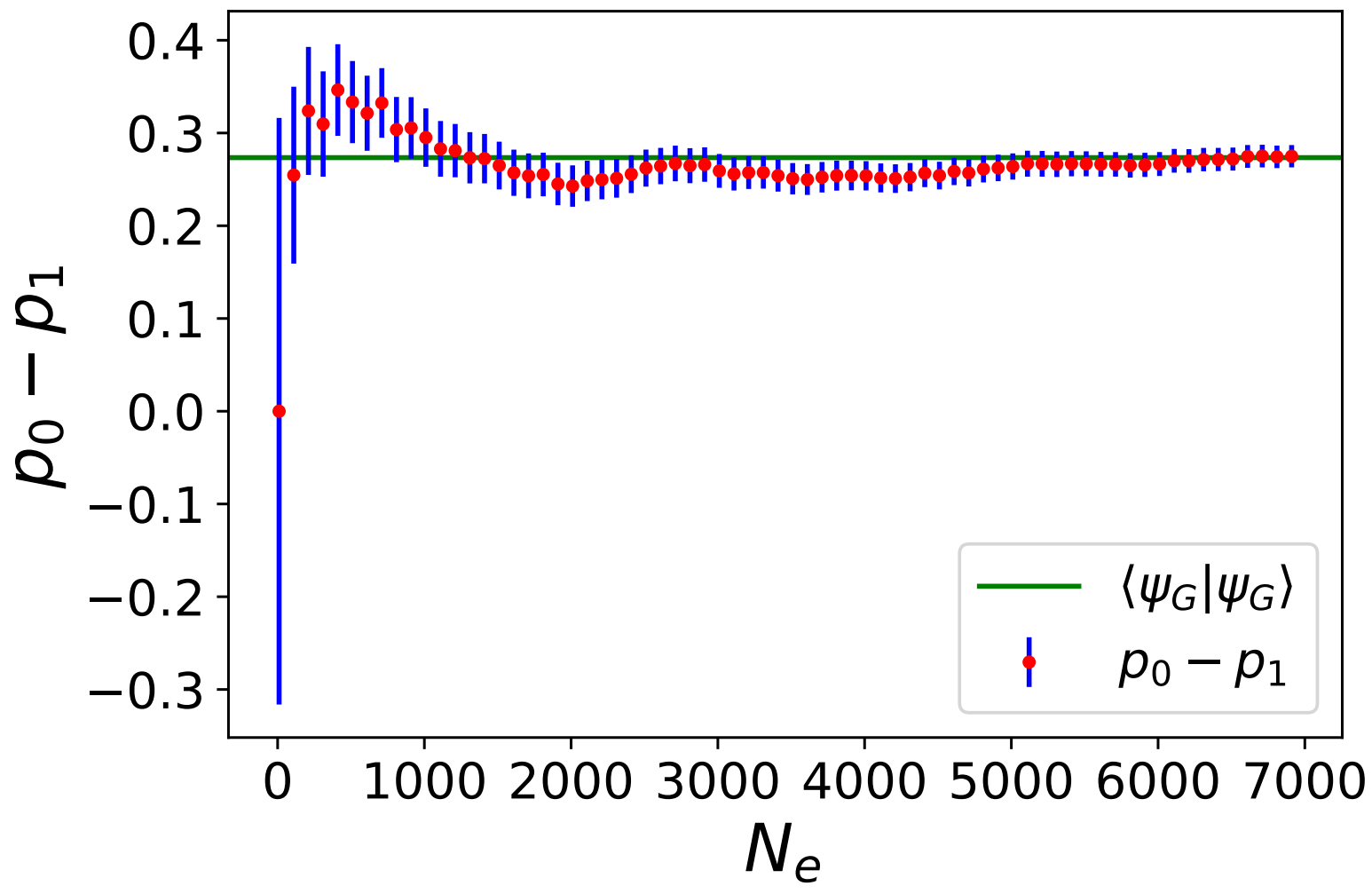

Figure 4.5: Illustration of the estimation of the difference of probabilities $p_0 - p_1$ in Eq. (4.30) as a function of the number of measurements (*aka* events) $N_e$ when using an oracle to obtain $\langle \Psi_G | V | \Psi_G \rangle$. The green horizontal line corresponds to its exact value. The blue error bars shown in the figure are calculated as $1/\sqrt{N_e}$. In this specific example, we consider $V = I$ and $(\varphi = 0, \mu = \pi/2)$, for which the oracle provides the amplitude of the initial state belonging to $\Omega$. The initial state is encoded on 8 qubits and corresponds to the equiprobable state $|\Psi\rangle = \frac{1}{2^4} \sum_k |k\rangle$. The "Good" states are defined as the states containing four 1s in their binary representation. If the initial state represents a many-body system encoded on the quantum register using the Jordan-Wigner fermion-to-qubit mapping from section 2.2.2, the projected state corresponds to a many-body system with exactly 4 particles, while the initial state mixes all possible particle numbers. The oracle is constructed using the method discussed in section 4.4.2.

We should also mention that Eq. (4.29) is a relatively simple trigonometric function of the angles $(\varphi, \mu)$. It depends solely on the two parameters $\langle \Psi_G | V | \Psi_G \rangle$ and $\langle \Psi_B | V | \Psi_B \rangle$. As a result, knowing the value of $\langle \Psi | V O_S | \Psi \rangle$ as given by Eq. (4.29) for a few selected values of $(\varphi, \pi)$ allows us to obtain the same expectation values for all values of these angles. For the case presented in Fig. 4.5 with $V = I$, a single pair of angles is sufficient, since:

$$\langle \Psi_G | \Psi_G \rangle + \langle \Psi_B | \Psi_B \rangle = 1.$$

The corresponding function $\mathrm{Re}[\langle \Psi | O_S(\varphi, \mu) | \Psi \rangle]$ is illustrated in Fig. 4.6, along with a few important sets of angles. To obtain the expectation value over the good state normalized, it would be enough to compute the expectation value of $\mathrm{Re}\{\langle O_S(0, \pi/2) \rangle\} = \langle P_S \rangle = p_G$ with the

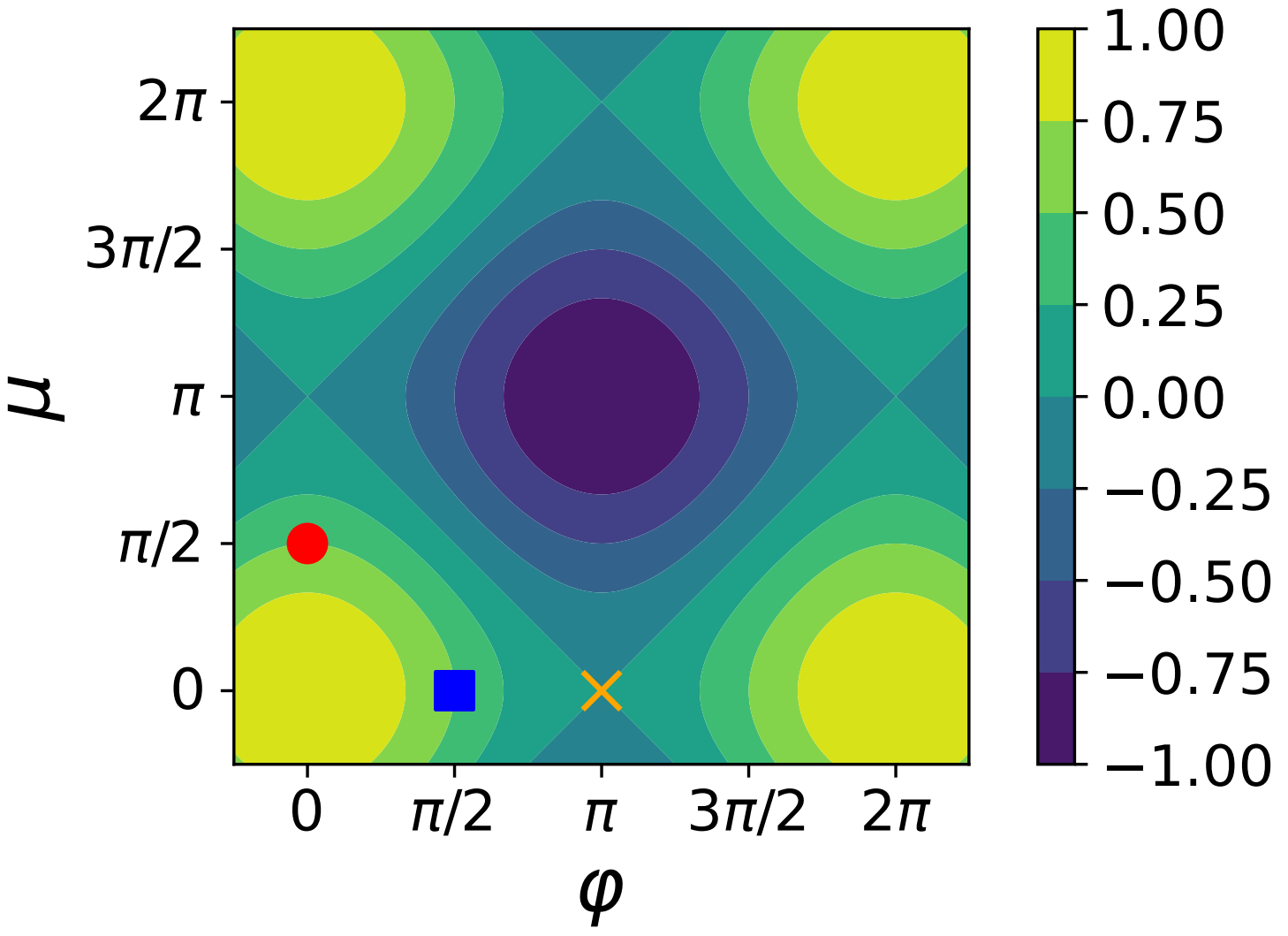

Figure 4.6: Contour plot of the $\mathrm{Re}[\langle\Psi|O_S(\varphi,\mu)|\Psi\rangle]$ values, derived from Eq. (4.29), as a function of the angles $\varphi$ and $\mu \in [-\pi/4, 2\pi + \pi/4]$ for the same initial state and oracle considered in Fig. 4.5. The symbols highlight specific angles: the angles used in Fig. 4.5 [red filled circles], the case $(\varphi = \pi/2, \mu = 0)$ that leads to $p_0 - p_1 = \langle\Phi_B|\Phi_B\rangle$ [blue filled square], and the standard prescription for the oracle typically used in the Grover search algorithm, i.e., $(\varphi = \pi, \mu = 0)$ [orange cross].

Hadamard test, i.e.,

$$\frac{\langle VP_S \rangle}{\langle P_S \rangle} = \frac{\mathrm{Re}\left\{\langle VO_S \rangle\right\}}{\mathrm{Re}\left\{\langle P_S \rangle\right\}}. \tag{4.31}$$

This method can be used to directly perform the Q-VAP without executing any explicit projection using the variational approach described in section 3.1. A comparison between the results obtained with the Q-VAP using the QPE projection and the Q-VAP using the oracle described in this section is presented in Fig. 4.7. The two methods strictly give the same results.

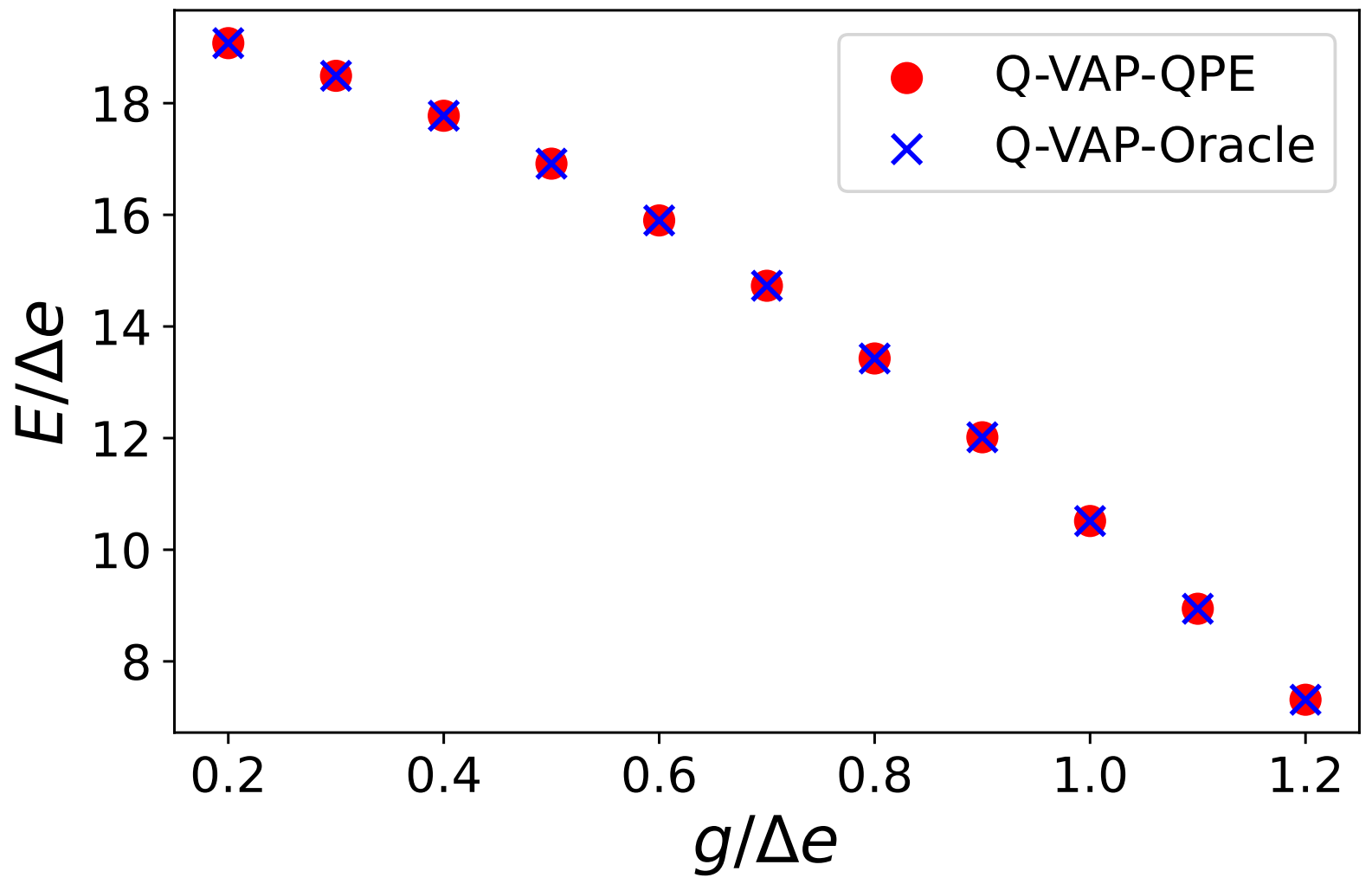

Figure 4.7: Illustration of the approximate ground state energy obtained using the Q-VAP approach with two different projections: One based on the QPE algorithm (red filled circles) described in Section 3.4.1, and the other based on the use of the oracle (blue crosses) described in Section 4.6. The model Hamiltonian employed was the pairing Hamiltonian of Eq. (2.33) with $4$ pairs on 8 energy levels of energies $\varepsilon_{j=1\dots,8} = j\Delta e$ ($\Delta e = 1$). I display here the ground state energies found as a function of the coupling strength $g$.

### 4.7 . Conclusion

In this chapter, I have introduced four distinct oracle-based projection methods. The specificity of each method is systematically detailed and compared will each other. While these methods are generally more resource-intensive than the QPE-based ones, they can offer certain advantages in some cases, such as not requiring ancillary qubits. Moreover, I have presented a method for implementing either the projector or a general oracle operator associated with symmetries using the LCU algorithm. I then compared the various QPE-based projection methods in Table 3.1 and Oracle-based methods in Table 4.1 in terms of the resources needed for their implementation. Our analysis concludes that the IQPE method is preferable if access to the projected state is required, while the implicit projection approach is more suitable if it is not needed. Lastly, we introduced a technique called Implicit Q-VAP, which leverages the oracle operator to directly obtain the expectation value of an operator within the projected subspace. This technique has been validated using the Q-VAP methodology with the QPE projection, and its equivalence has been demonstrated. The methodologies outlined here remain too resource-intensive for implementation on existing quantum devices. The subsequent chapter will focus on a more feasible approach rooted in the principles of quantum tomography.

# 5 - Classical Shadows

Current Noisy Intermediate-Scale Quantum (NISQ) platforms struggle to implement Quantum Phase Estimation (QPE)-based and Oracle-based methods due to their demanding depth and ancilla qubits requirements. Significant advancements in quantum computing platforms are needed for these techniques to be practical, potentially awaiting a fault-tolerant era. This chapter delves into the promising potential of newly advanced tomography techniques [62, 163, 164, 31], capable of recovering expectation values with significantly reduced resources compared to traditional methods. This even extends to the complex task of deriving expectation values of a projected state. In line with the primary objective of this thesis, this chapter aims to develop a method for symmetry restoration. To reach this goal, we first lay the groundwork with a comprehensive introduction to the general tomography principle, followed by exploring the classical shadow paradigm [31]. We then demonstrate how this paradigm can be specifically applied to address the symmetry restoration problem. This approach paves the way for a more resource-efficient methodology in quantum computations, potentially facilitating practical applications on existing quantum platforms.

## 5.1 . Quantum State Tomography

Quantum state tomography refers to the procedure of reconstructing a quantum state via repeated projective measurements [62]. This process necessitates the preparation of numerous copies of the quantum state $\rho$ because the collection of projective measurements must be performed on a tomographically complete basis $\mathcal{U}$ to characterize the state entirely. A basis is deemed tomographically complete if, for any pair of distinct density matrices $\rho$ and $\sigma$, there exists an operator $U \in \mathcal{U}$ and a state $|b\rangle$ such that $\langle b|U\sigma U^\dagger|b\rangle \neq \langle b|U\rho U^\dagger|b\rangle$, where $b = \{0,1\}^{\otimes n}$, and $|b\rangle = |b_{n-1}, \ldots, b_1, b_0\rangle$. The Pauli matrices $X, Y, Z$ constitute such a basis for a single-qubit state. As previously mentioned, measurements in the set of tomographically complete operators $\mathcal{U}$ are essential because non-orthogonal states are indistinguishable. For instance, the single-qubit states $|0\rangle$ and $|+\rangle = \frac{1}{\sqrt{2}}\left(|0\rangle + |1\rangle\right)$ cannot be distinguished using a single measurement in either the $Z$ or $X$ basis (the measurement in the $X$ basis is shown in Tab. 2.2).

### 5.1.1 . Single Qubit Tomography Using the Pauli Operators

Let us examine the practical implementation of one-qubit tomography. The set $\left\{I/\sqrt{2}, X/\sqrt{2}, Y/\sqrt{2}, Z/\sqrt{2}\right\}$ forms an orthonormal set of matrices under the Hilbert-Schmidt inner product for any $2 \times 2$ matrix, denoted generically as $M$. The Hilbert-Schmidt inner product is defined by $\langle A, B\rangle_{HS} = \mathrm{Tr}\left(A^\dagger B\right)$. Specifically, when $M = \rho$, we obtain:

$$\rho = \frac{\mathrm{Tr}\left(\rho\right) I + \mathrm{Tr}\left(X\rho\right) X + \mathrm{Tr}\left(Y\rho\right) Y + \mathrm{Tr}\left(Z\rho\right) Z}{2}. \tag{5.1}$$

The quantities $\mathrm{Tr}\left(A\rho\right)$ denote the average values of the observable $A$. To introduce the concept of sampling, it is essential to comprehend that the estimation of $\mathrm{Tr}\left(A\rho\right)$ can be viewed as a probabilistic process solely on the basis that diagonalizes $A$. Assume:

$$A = \sum_k \lambda_k |k\rangle\langle k|, \tag{5.2}$$

which leads to:

$$\mathrm{Tr}\left(A\rho\right) = \sum_k \lambda_k \langle k|\rho|k\rangle. \tag{5.3}$$

In terms of measurement, given that the measure is made on the $|k\rangle$ basis, the measurement of a particular eigenvalue $\lambda_k$ will be achieved with a probability $p_k = \langle k|\rho|k\rangle$. By repeating the measurement with the same initial state $\rho$, we derive a set of measured eigenvalues $\{\lambda_k\}$. The quantity $\mathrm{Tr}\left(A\rho\right)$ can then be estimated using:

$$\mathrm{Tr}\left(A\rho\right) = \frac{1}{m}\sum_k \lambda_k, \tag{5.4}$$

where $m$ represents the total number of measurements. For instance, to estimate $\mathrm{Tr}\left(Z\rho\right)$, we perform measurements on the quantum state in the $Z$ basis, yielding a series of outcomes $\{z_i\}_m$. Here, $z_i = 1$ or $-1$ depends on whether the measurement results in the $|0\rangle$ or $|1\rangle$ state. The empirical average of these outcomes estimates $\mathrm{Tr}\left(A\rho\right)$, thus $\mathrm{Tr}\left(Z\rho\right) \approx \frac{1}{m}\sum_i z_i$. Using the central limit theorem, we infer that the distribution of estimates approximates a Gaussian for large $m$, with a mean equal to $\mathrm{Tr}\left(Z\rho\right)$ and standard deviation $\sigma = \Delta\left(Z\right)/\sqrt{m}$, with $\Delta\left(Z\right)$ as the standard deviation for a single measurement. Since $\Delta\left(Z\right) \leq 1$, we obtain that $\sigma$ is at most $1/\sqrt{m}$. To acquire all the coefficients in Eq. (5.1), one needs to estimate several traces utilizing different Pauli matrices. Given their non-commuting nature, these matrices do not share a common eigenbasis; thus, each trace must be estimated independently by applying the appropriate unitary transformation before the measurement. Considering the $Z$ basis as the computational basis $\{|0\rangle, |1\rangle\}$, the unitary transformations required for estimating $\mathrm{Tr}\left(X\rho\right)$ and $\mathrm{Tr}\left(Y\rho\right)$ are $H$ and $HS^\dagger$, respectively (see Table 2.4). As we will explore shortly, this methodology is precisely what is implemented in the classical shadow technique. Extending this procedure to a general density matrix of $n$ qubits, we obtain:

$$\rho = \sum_{\vec{v}} \frac{\left[\mathrm{Tr}\left(\left[\bigotimes_{k=1}^n P_{v_k}\right]\rho\right)\bigotimes_{k=1}^n P_{v_k}\right]}{2^n}, \tag{5.5}$$

where the sum is taken over all possible vectors $\vec{v} = (v_1, \ldots, v_n)$, with each entry $v_k$ being chosen from the set of operators $\{I, X, Y, Z\}$. We observe that a full tomography of $\rho$ requires the estimation of $4^n$ traces through measurement.

### 5.2 . Classical Shadows

The primary drawback of quantum state tomography, as discussed in the previous section, is the exponential scaling of the number of parameters $\mathrm{Tr}\left(\left[\bigotimes P_{i_k}\right]\rho\right)$ required to recreate the quantum state $\rho$. Additionally, to limit the overall error to $4^n/\sqrt{m}$, it is necessary to estimate each parameter using a minimum of $m$ measurements per parameter. Furthermore, collecting, storing, and processing these measurement outcomes necessitates an exponential amount of memory. Therefore, the reconstruction of the comprehensive density matrix $\rho$ not only entails an exponential number of measurements in relation to $n$, but also demands an exponential amount of classical memory and computational resources. Several methods, such as matrix product state (MPS) tomography [165] and neural network tomography [166, 167] have been proposed to mitigate these constraints under certain conditions. However, for general quantum systems, these methods still require an exponential quantity of samples [31].

In contrast, “Shadow Tomography”, introduced in [168], posits that fully reconstructing the density matrix of a quantum system may be unnecessary for certain tasks. This approach can predict properties like expectation values of a set of observables without requiring complete state characterization. Utilizing a polynomial number of state copies, it is capable of predicting an exponential number of target functions, including fidelity, expectation values, two-point correlators, and entanglement witnesses. The “Classical Shadow” technique [31] extended this concept to develop an efficient protocol for obtaining a minimal classical sketch $S_\rho$ (the classical shadow) of an unknown quantum state $\rho$, which can be used to predict arbitrary linear function values $\{o_i\}$ associated with the operators $\{O_i\}$:

$$o_i = \mathrm{Tr}\left(O_i \rho\right), \qquad 1 \leq i \leq M. \tag{5.6}$$

The creation of the classical shadow is based on the repetition of the next procedure: Apply a unitary transformation $U$ to the quantum state $\rho$, i.e.

$$\rho \rightarrow U \rho U^\dagger, \tag{5.7}$$

and then measure all the qubits in the computational basis. This action results in the collapse of the wavefunction into one of the states of the computational basis, which takes the form of a bitstring of length $n$. In other words, the state $|b\rangle$ is expressed as a sequence of $n$ elements, each of which is either 0 or 1, i.e., $|b\rangle = \{0, 1\}^n$. The unitary $U$ is selected from a predefined random ensemble $\mathcal{U}$. Considering that the unitaries $U$ can be efficiently implemented on a classical computer over the bitstring, a classical description $U^\dagger|b\rangle\langle b|U$ is stored in classical memory.

Two different ensembles are usually considered for selecting the random unitaries $U$, tensor products of random single-qubit Clifford circuits, or random $n$-qubit Clifford circuits. The former case is the one discussed above in section 5.1. The average $\mathbb{E}$ over both the choice of $U$ and the outcomes can be viewed as a quantum channel $\mathcal{M}$ that maps the original $\rho$ to a set of classical representations $U^\dagger|b\rangle\langle b|U$:

$$\mathbb{E}\left[U^\dagger|b\rangle\langle b|U\right] = \mathcal{M}\left(\rho\right). \tag{5.8}$$

The form of the quantum channel depends on the ensemble of random unitary transformations $\mathcal{U}$. If this ensemble is tomographically complete, we can invert the quantum channel to reconstruct the state:

$$\rho = \mathbb{E}\left[\mathcal{M}^{-1}\left(U^\dagger|b\rangle\langle b|U\right)\right]. \tag{5.9}$$

An illustration of the state reconstruction of 1 qubit using classical shadows in Eq. (5.9) and quantum state tomography Eq. (5.1) is shown in Fig. 5.1. Each term in the average $\mathbb{E}$ in Eq. (5.9), i.e., point in the Bloch sphere, can be viewed as a single classical snapshot $\hat{\rho}$:

$$\hat{\rho} = \mathcal{M}^{-1}\left(U^\dagger|b\rangle\langle b|U\right). \tag{5.10}$$

Here, we adopt the notation from [31] where $\rho$ refers to the exact density matrix of the quantum state, while $\hat{\rho}$ represents a snapshot of this density obtained using the Classical Shadows technique. A collection of $N$ snapshots is termed a “classical shadow” $S\left(\rho, N\right)$ of size $N$:

$$S\left(\rho, N\right) = \left\{\hat{\rho}_1 = \mathcal{M}^{-1}\left(U_1^\dagger|b_1\rangle\langle b_1|U_1\right), \ldots, \hat{\rho}_N = \mathcal{M}^{-1}\left(U_N^\dagger|b_N\rangle\langle b_N|U_N\right)\right\}. \tag{5.11}$$

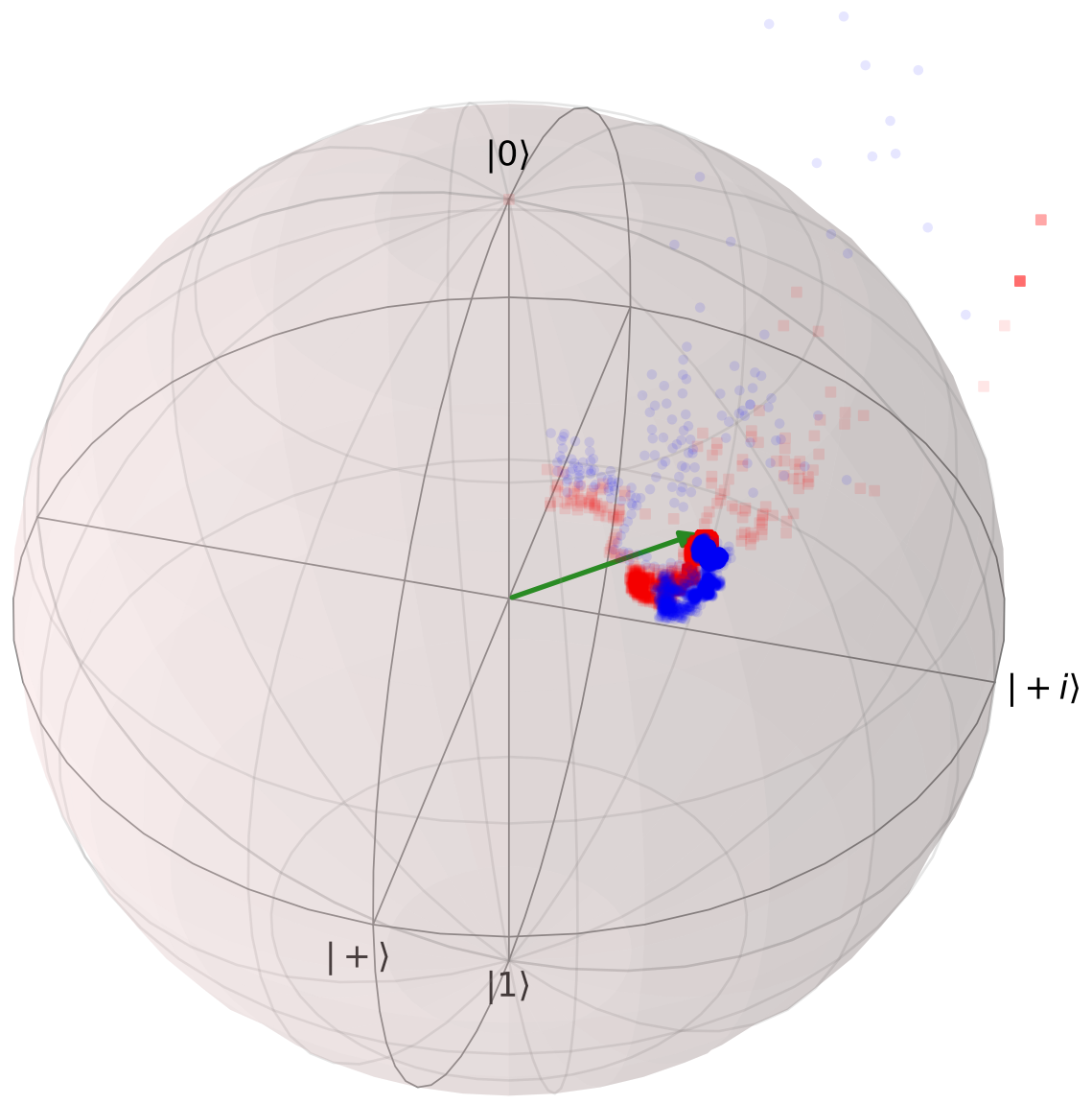

Figure 5.1: Illustration of the process of reconstructing the state $\rho = |\psi\rangle\langle\psi|$ described on 1 qubit, where $|\psi\rangle \approx \frac{70}{76}|0\rangle + \frac{30}{76}e^{i\pi/4}|1\rangle$. The blue dots represent the progression of the state reconstruction using the classical shadow method, as described in Eq. (5.9), while the red squares denote the analogous process for quantum state tomography as discussed in section 5.1.1. Each method employs a total of $10^4$ data points. The green arrow is a marker for the exact state $\rho$. Although it is difficult to see, both methods converge toward the exact state.

Given that the average of the classical shadow approximates the quantum state $\rho$ (see Eq. (5.9)), it becomes feasible to estimate the expectation value of any observable $O$ using an empirical mean over the individual snapshots:

$$\langle O \rangle = \frac{1}{N} \sum_{i=1}^{N} \mathrm{Tr}\left(O\hat{\rho}_i\right). \tag{5.12}$$

It has been proven that using a classical shadow of size $N$, it is possible to predict $L$ arbitrary linear functions $\mathrm{Tr}\left(O_1\rho\right), \ldots, \mathrm{Tr}\left(O_L\rho\right)$ up to an additive error $\epsilon$ if:

$$N \geq \mathcal{O}\left((\log L)\left(\max\left\{||O_i||^2_{\text{shadow}}\right\}_{i=1,\ldots,L}\right)/\epsilon^2\right), \tag{5.13}$$

[31]. The shadow norm $||O_i||^2_{\text{shadow}}$ depends on the unitary ensemble that is chosen. To make the approximation of any $\langle O \rangle$ more robust to outlier corruption, Ref. [31] uses a median of means estimator of the form:

$$\langle O \rangle (N, K) = \text{median}\left\{\langle O \rangle^{(1)}, \ldots, \langle O \rangle^{(K)}\right\}, \tag{5.14}$$

where

$$\langle O \rangle^{(k)} = \mathrm{Tr}\left(O\hat{\rho}_k\right) \quad \text{and} \quad \hat{\rho}_k = \frac{1}{\lfloor N/K \rfloor} \sum_{j=(k-1)\lfloor N/K \rfloor + 1}^{k\lfloor N/K \rfloor} \hat{\rho}_j, \tag{5.15}$$

for $1 \leq k \leq K$. This estimator partitions the shadow into $K$ equally-sized segments and calculates each segment's mean $\langle O \rangle^{(k)}$. The approximated value of $\langle O \rangle$ is then derived from the median of these computed means. Using this estimator, the authors of Ref. [31] demonstrate that, in order to attain an error of $\epsilon$ in the estimation of all expectation values for the set of Hermitian operators, $\{O_{i=1,\ldots,L}\}$, i.e.,

$$|\langle O_i \rangle_{\text{shadow}} - \langle O_i \rangle_{\text{exact}}| \leq \epsilon, \tag{5.16}$$

one has to extract $N$ and $K$ from the following equations:

$$\begin{cases} K = 2\ln\left(2L/\delta\right) \\ M = \dfrac{34}{\epsilon^2}\max_{1\leq i\leq L}\left|\left|O_i - \dfrac{\mathrm{Tr}\left(O_i\right)}{2^n}\mathbb{I}\right|\right|^2_{\text{shadow}} \\ N = \lceil MK \rceil \end{cases}, \tag{5.17}$$

with $\delta$ the probability that the bound in Eq. (5.16) fails. In the following sections, we illustrate how the classical shadow approach operates to estimate observables utilized to calculate the energy of a Hamiltonian. We then outline the procedure for obtaining this energy in the symmetry-restoration case.

### 5.3 . Estimation of Pauli Observables Using Classical Shadows

Suppose we have an observable $A$, like a Hamiltonian, expressed as the weighted sum of Pauli chains, where each Pauli chain $O_i$ is a tensor product of Pauli operators $P_j$ with $P_j \in \{I, X, Y, Z\}$ for $j = 1, \ldots, n$:

$$A = \sum_i \alpha_i O_i. \tag{5.18}$$

Given the form of the observables $O_i$, to estimate the expectation values $\langle O_i \rangle$, it is advantageous to measure in the Pauli basis as shown in Table 2.4. Thus, we should consider the tomographically complete ensemble associated with the Pauli matrices $X$, $Y$, and $Z$:

$$\mathcal{U} = \left\{ \bigotimes_{j=0}^{n-1} U_j \right\} \quad \text{with} \quad U_j = \left[ H, HS^\dagger, I \right], \tag{5.19}$$

given that we have the following relations:

$$X = HZH \qquad Y = SHZHS^\dagger \quad \text{and} \quad Z = IZI, \tag{5.20}$$

where $H$ and $S$ correspond to the Hadamard and $S$ gates, respectively (see Table 2.2). When considering this ensemble $\mathcal{U}$, the inverted quantum channel can be procured from the inverse of every single qubit's quantum channel, i.e., $\mathcal{M}^{-1} = \bigotimes_{j=0}^{n-1} \mathcal{M}_j^{-1}$. Every single snapshot in Eq. (5.10) can be shown (supplementary material of Ref. [31]) to have the form:

$$\hat{\rho} = \bigotimes_{j=1}^{n-1} \left( 3U_j^\dagger |b_j\rangle\langle b_j| U_j - \mathbb{I} \right). \tag{5.21}$$

The computation of the expectation value of each Pauli chain $O$ in the Hamiltonian over every snapshot $\hat{\rho}$ takes the form:

$$\mathrm{Tr}\left( O\hat{\rho} \right) = \mathrm{Tr}\left( \bigotimes_{j=0}^{n-1} P_j \left( 3U_j^\dagger |b_j\rangle\langle b_j| U_j - \mathbb{I} \right) \right) = \prod_{j=0}^{n-1} \mathrm{Tr}\left( P_j \left( 3U_j^\dagger |b_j\rangle\langle b_j| U_j - \mathbb{I} \right) \right). \tag{5.22}$$

First, we observe that irrespective of the selected measurement base $U_j$ and the state $|b_j\rangle$ measured, each term in the product of Eq. (5.22) when $P_j = I_j$, $\mathrm{Tr}\left( 3U_j^\dagger |b_j\rangle\langle b_j| U_j - \mathbb{I} \right)$, will yield 1. For the alternative combinations of $P_j$, $U_j$, and $|b_j\rangle$, we can leverage the properties of the Pauli matrices, specifically their tracelessness and their orthogonality under the Hilbert-Schmidt inner product $\mathrm{Tr}\left( A^\dagger B \right)$. We first note that for $P_j = \{X_j, Y_j, Z_j\}$, each term in the product reduces to:

$$3\,\mathrm{Tr}\left( P_j \left( U_j^\dagger |b_j\rangle\langle b_j| U_j \right) \right), \tag{5.23}$$

given that $\mathrm{Tr}\left( P_j \right) = 0$. Considering that we can express any Pauli matrix using the measurement basis operators as $P_j = U_j^\dagger Z_j U_j$, the trace in Eq. (5.23) will be either $1$ or $-1$, depending on whether the measurement resulted in $|0_j\rangle$ or $|1_j\rangle$, respectively. From this, we conclude that $\mathrm{Tr}\left( P_j \left( 3U_j^\dagger |b_j\rangle\langle b_j| U_j - \mathbb{I} \right) \right)$ can have only four potential values: it becomes $1$ if $P_j = \mathbb{I}$, $\pm 3$ if $U_j$ corresponds to the unitary associated with the Pauli operator $P_j$, and 0 otherwise. As previously noted, the $+$ (respectively, $-$) sign in $\pm 3$ corresponds to the outcome $|b_j\rangle = |0\rangle$ (respectively, $|b_j\rangle = |1\rangle$). Table 5.1 showcases the possible values of the trace for different Pauli/measurement basis combinations. Consequently, if a single $U_j$ does not correspond to the operator $P_j$, the entire product evaluates to 0. This distinctive property simplifies the estimation of an observable's expectation value into merely counting the number of compatible measurements in the classical shadow and applying the suitable sign to each result. A measurement protocol $\bigotimes_{j=0}^{n-1} U_j$ is defined as compatible with an observable $O=\bigotimes_{j=0}^{n-1} P_j$ if the non-identity parts of $O$ match the measurement protocol. Such property is particularly attractive for applications since it can significantly reduce the number of measurements required to estimate an observable. This process is known as "derandomization", where we select measurement protocols explicitly to glean information about the expectation value of a specific set of observables [169]. While this procedure is assured of performing at least as well as the randomized one, it has been observed to perform substantially better when the observables have few identity parts.

| Measurement $U$ | Operator $P$ | | | |
|---|---|---|---|---|
| | $\mathbb{I}$ | $X$ | $Y$ | $Z$ |
| $H$ | $(1,1)$ | $(+3,-3)$ | $(0,0)$ | $(0,0)$ |
| $HS^\dagger$ | $(1,1)$ | $(0,0)$ | $(+3,-3)$ | $(0,0)$ |
| $\mathbb{I}$ | $(1,1)$ | $(0,0)$ | $(0,0)$ | $(+3,-3)$ |

Table 5.1: Potential outcomes of the function $\mathrm{Tr}\left(P\left(3U^\dagger|b\rangle\langle b|U-\mathbb{I}\right)\right)$, given a specific measurement basis $U$ and operator $P$. The results correspond to the measurement states $(|0\rangle,|1\rangle)$ of the resultant qubit.

## 5.4 . Projected Observables

I show here that it is possible to perform symmetry restoration using the classical shadow paradigm. This section will illustrate how to restore the particle number and total spin symmetries in a post-processing step. Both projections take advantage of the fact that their respective projectors can be written as a weighted sum of local one-body operators, i.e., operators composed of the tensorial product of single-qubit gates.

### 5.4.1 . Particle Number Projection

Let us consider the general projector of Eq. (4.17) for the particle number case:

$$P_N=\frac{1}{n+1}\sum_{k=0}^{n} e^{2\pi ik\left(\hat{N}-N\right)/(n+1)}=\sum_k \alpha_k e^{i\phi_k\hat{N}}, \tag{5.24}$$

with $\alpha_k=\frac{1}{n+1}e^{-i\phi_k N}$, $\phi_k=\frac{2\pi k}{n+1}$, where $\hat{N}$ is the particle number operator, $N$ is the number of particles to project onto, and $n$ is the maximum number of particles that can be in the system; in our case, $n$ is equal to the number of qubits. Using Eq. (3.4), this operator can be written in terms of the tensorial product of single-qubit phase gates (see Table 2.2):

$$P_N=\sum_k \alpha_k \bigotimes_{i=0}^{n-1} R_i\left(\phi_k\right), \tag{5.25}$$

where we have used the notation $R_j\left(\phi_k\right)$ for the phase gates in order to avoid confusion with the Pauli gates $P_j$. We can directly apply this operator to each snapshot in a post-processing step. The projected expectation value of an observable $O$ from Eq. (5.22) over a single snapshot $\hat{\rho}$ becomes:

$$\mathrm{Tr}\left(OP_N\hat{\rho}\right) = \sum_k \alpha_k \prod_{j=0}^{n-1} \mathrm{Tr}\left(P_j R_j\left(\phi_k\right)\left(3U_j^\dagger|b_j\rangle\langle b_j|U_j - \mathbb{I}\right)\right). \tag{5.26}$$

Expressing the phase gates in terms of Pauli matrices, i.e.:

$$R\left(\lambda\right) = e^{\lambda/2}\left[\cos\left(\lambda/2\right)\mathbb{I} - iZ\sin\left(\lambda/2\right)\right], \tag{5.27}$$

we get:

$$\begin{aligned}\mathrm{Tr}\left(OP_N\hat{\rho}\right) = \sum_{k=0}^{n} \alpha_k e^{in\phi_k/2} \prod_{j=0}^{n-1} \Big\{ & \cos\left(\phi_k/2\right)\mathrm{Tr}\left[P_j\left(3U_j^\dagger|b_j\rangle\langle b_j|U_j - \mathbb{I}\right)\right] \\ & -i\sin\left(\phi_k/2\right)\mathrm{Tr}\left[P_jZ_j\left(3U_j^\dagger|b_j\rangle\langle b_j|U_j - \mathbb{I}\right)\right]\Big\}. \end{aligned} \tag{5.28}$$

Again, leveraging the orthogonality of the Pauli operators, the trace values are constrained to 1, $\pm 3$, or 0 (see Table 5.1). We can establish a derandomization process akin to what was detailed in Section 5.3. This is accomplished by acknowledging that, for any given observable $\bigotimes_{j=0}^{n-1} P_j$ with $X_j$ and/or $Y_j$ operators, any measurement protocols including $Z_j$ in the qubits corresponding to the $X_j, Y_j$ operators will result in the associated product equaling zero. For instance, for an observable of the form $O = I_3X_2Y_1Z_0$, all the measurement protocols, such as $P_3P_2Z_1P_0$ and $P_3Z_2P_1P_0$ will yield a product of zero and therefore, could be disregarded. An important aspect to point out is that having an estimate for the set of values $\mathrm{Tr}\left(P_jR_j\left(\phi_k\right)\left(3U_j^\dagger|b_j\rangle\langle b_j|U_j - \mathbb{I}\right)\right)$, as depicted in Eq. 5.26, provides us with the means to get projections onto all particle numbers $N$. This is feasible as the information about $N$ is purely encoded in the coefficients $\alpha_k$. Hence, it is only necessary to compute all products once to achieve the projected expectation value of the observable $O$ for any given particle number $N$. As a demonstration, Fig. 5.2 presents the probabilities of the projection over different particle numbers for a Gaussian state spanning four qubits. Fig. 5.3 further showcases the convergence of the expectation value of the projector $P_{N=2}$ as a function of the size of the classical shadow, i.e., the number of snapshots.

### 5.4.2 . Spin Projection

Projection onto total spin symmetry introduces greater complexity. Our initial consideration was to utilize the $S^2$ operator in a projector analogous to Eq. (5.24). This strategy, however, was found to be unviable due to the presence of two-body operators in the $S^2$ operator. Such inclusion prevents the operator $e^{i\phi_k S^2}$ from being decomposed into a tensor product of single-qubit operators, thereby hindering our application of the orthogonality of the Pauli operators to streamline computations. To circumvent this issue, we adopt the integral form of the projector for the spin states $|s,m\rangle$, which possesses total spin $s\left(s+1\right)$ and projection on the z-axis $m$ (where $\hbar = 1$). This form was proposed in [170]:

$$P_{s,m} = |s,m\rangle\langle s,m| = \frac{2s+1}{8\pi^2}\int_0^{2\pi}\int_0^{\pi}\int_0^{2\pi} \sin(\beta)\left(D^s_{m,m}(\alpha,\beta,\gamma)\right)^* e^{-i\alpha S_z}e^{-i\beta S_y}e^{-i\gamma S_z}d\alpha d\beta d\gamma, \tag{5.29}$$

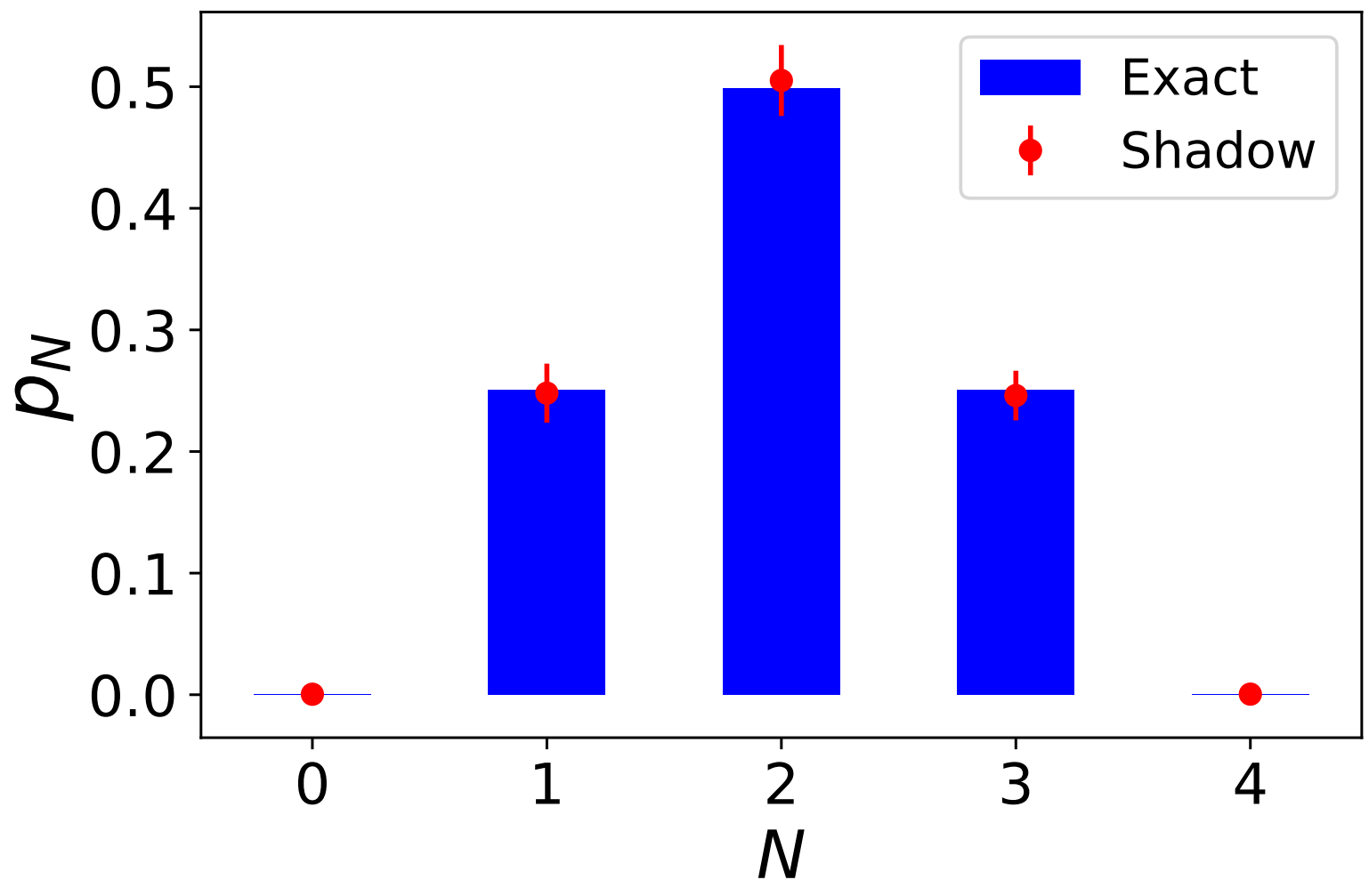

Figure 5.2: Illustration of the projection over the particle number $N = 0, 1, 2, 3, 4$ employing the classical shadow procedure delineated in Section 5.4.1. We consider a Gaussian state encompassing the register $k = 0, 1, \ldots, 2^n - 1$, where $n = 4$. This state takes the form $|\psi_G(k)\rangle = \frac{1}{\mathcal{N}} e^{-\frac{1}{2}\left(\frac{k-\mu}{\sigma}\right)}$, with a median $\mu = (2^n - 1)/2$ and a standard deviation $\sigma = \mu/3$. $\mathcal{N}$ is a normalization constant. The results derived from the classical shadows were acquired using $10^4$ shots, repeated 50 times. The red data points and their corresponding error bars denote the mean value and standard deviation derived from these 50 trials.

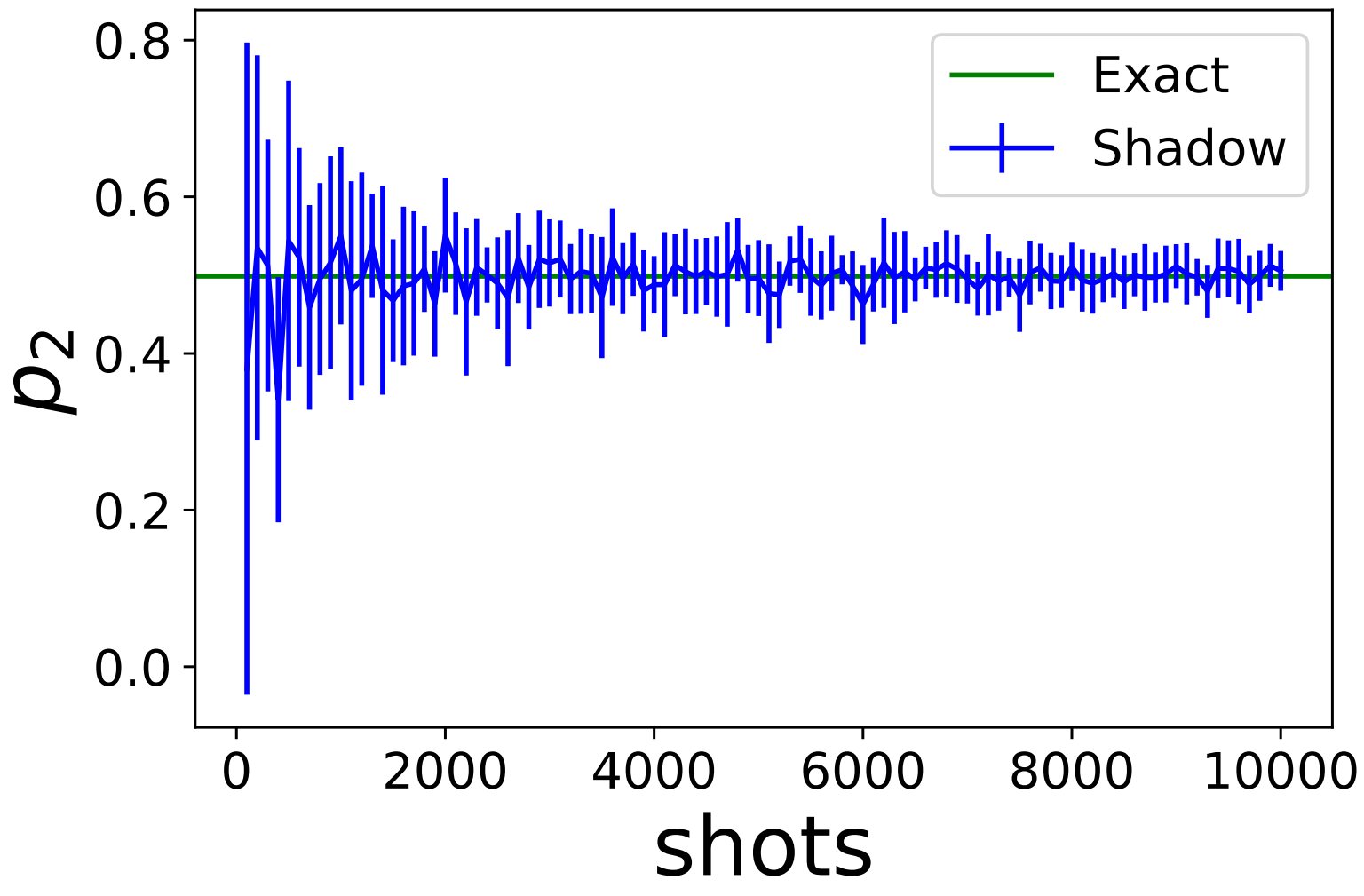

Figure 5.3: Convergence of the expectation value of the projector in particle number $p_2 = \langle P_{N=2} \rangle$ against the number of "shots" within the classical shadow for the state $|\psi\rangle$ featured in Fig. 5.2. The results derived from the classical shadows are based on ten repetitions of the number of shots. The blue data points and their corresponding error bars represent the average and standard deviation of these ten repetitions, respectively. The solid green line is the expected result.

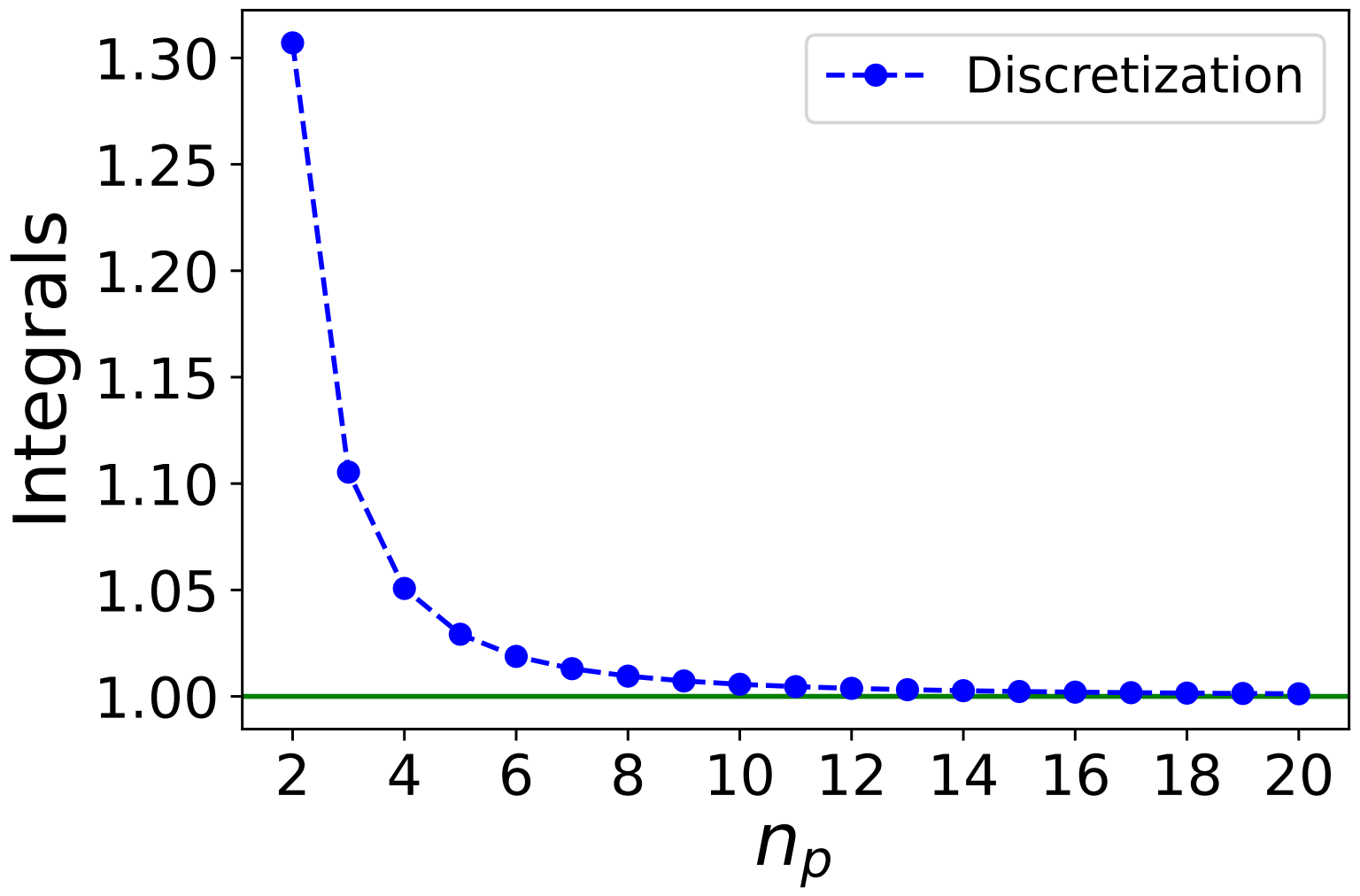

Figure 5.4: Comparison of the numerical values of the discretized integrals, as given in Eq. (5.29), without incorporating the function – i.e., $\frac{1}{8\pi^2}\int_0^{2\pi}\int_0^{\pi}\int_0^{2\pi}\sin(\beta)d\alpha d\beta d\gamma$ – against the number of points $n_p$ per integral. The green line corresponds to the exact value.

where $D^s_{m,m}(\alpha,\beta,\gamma)$ is the $s,m$ component of the Wigner D matrix $D(\alpha,\beta,\gamma)$. Discretizing the integrals and using the relations:

$$S_Y = \frac{1}{2}\sum_{j=0}^{n-1} Y_j \qquad S_Z = \frac{1}{2}\sum_{j=0}^{n-1} Z_j, \tag{5.30}$$

we obtain the following form of the spin projector:

$$P_{s,m} = \frac{2s+1}{8\pi^2}\Delta\alpha\Delta\beta\Delta\gamma\sum_{p,q,k}\sin(\beta_q)\left(D^s_{m,m}(\alpha_p,\beta_q,\gamma_k)\right)^*\bigotimes_{j=0}^{n-1} R_z(\alpha_p)\,R_y(\beta_q)\,R_z(\gamma_k) \tag{5.31}$$

where, $\Delta\alpha$, $\Delta\beta$, and $\Delta\gamma$ denote the step sizes for the respective angles $\alpha$, $\beta$, and $\gamma$, i.e., $\alpha_p = p\Delta\alpha$, $\beta_q = q\Delta\beta$, and $\gamma_k = k\Delta\gamma$ with $p,q,k = 0,\dots,n_p$, and $n_p$ indicates the number of points considered for each integral. The quantity of terms in the $p,q,k$ sum – i.e., the number of $R_z(\alpha_p)\,R_y(\beta_q)\,R_z(\gamma_k)$ operators factored into the calculation – is directly proportional to the precision of the discretization. Fig. 5.4 illustrates the impact of discretization on the integrals expressed in Eq. (5.29). The projected expectation value of an observable $O = \bigotimes_{j=0}^{n-1} P_j$ over a single snapshot $\hat{\rho}$ becomes:

$$\begin{aligned}\mathrm{Tr}\left(OP_{s,m}\hat{\rho}\right) = \frac{2s+1}{8\pi^2}\Delta\alpha\Delta\beta\Delta\gamma\sum_{p,q,k}\sin(\beta_q)\left(D^s_{m,m}(\alpha_p,\beta_q,\gamma_k)\right)^* \\ \prod_{j=0}^{n-1}\left[c_I(\alpha_p,\beta_q,\gamma_k)\,\mathrm{Tr}\left(P_j\hat{\rho}_j\right) + c_X(\alpha_p,\beta_q,\gamma_k)\,\mathrm{Tr}\left(P_jX_j\hat{\rho}_j\right)\right. \\ \left.+c_Y(\alpha_p,\beta_q,\gamma_k)\,\mathrm{Tr}\left(P_jY_j\hat{\rho}_j\right) + c_Z(\alpha_p,\beta_q,\gamma_k)\,\mathrm{Tr}\left(P_jZ_j\hat{\rho}_j\right)\right],\end{aligned} \tag{5.32}$$

with:

$$
\begin{cases}
c_I(\alpha_p,\beta_q,\gamma_k) = \cos(\alpha_p)\cos(\beta_q)\cos(\gamma_k) - \dfrac{1}{4}\sin(\alpha_p)\cos(\beta_q)\cos(\gamma_k) \\
c_X(\alpha_p,\beta_q,\gamma_k) = \dfrac{i}{4}\left[\sin(\alpha_p)\sin(\beta_q)\cos(\gamma_k) - \cos(\alpha_p)\sin(\beta_q)\sin(\gamma_k)\right] \\
c_Y(\alpha_p,\beta_q,\gamma_k) = -\dfrac{i}{2}\left[\cos(\alpha_p)\sin(\beta_q)\cos(\gamma_k) + \dfrac{1}{4}\sin(\alpha_p)\sin(\beta_q)\sin(\gamma_k)\right] \\
c_Z(\alpha_p,\beta_q,\gamma_k) = -\dfrac{i}{2}\left[\cos(\alpha_p)\cos(\beta_q)\cos(\gamma_k) + \sin(\alpha_p)\cos(\beta_q)\cos(\gamma_k)\right]
\end{cases}, \quad (5.33)
$$

and $\hat{\rho}_j = 3U_j^\dagger|b_j\rangle\langle b_j|U_j - \mathbb{I}$. In this case, no derandomization process can be applied since no term in the productorial will be zero for any selection of $P_j$ and measurement basis $U_j$. We can once again observe that all traces possess only the four possible values listed in Table 5.1. Moreover, computing all the productorials once provides access to the projected expectation values for all combinations of the $s, m$ parameters. Like in the previous section, we present the amplitude decomposition for a specific quantum state, using the $(S^2, S_z)$ eigenbasis denoted by $|s, m\rangle$. This is illustrated in Figure 5.5. Additionally, Fig. 5.6 demonstrates the convergence of the expectation value of the projector $P_{|1,0\rangle}$ in relation to the size of the classical shadow.

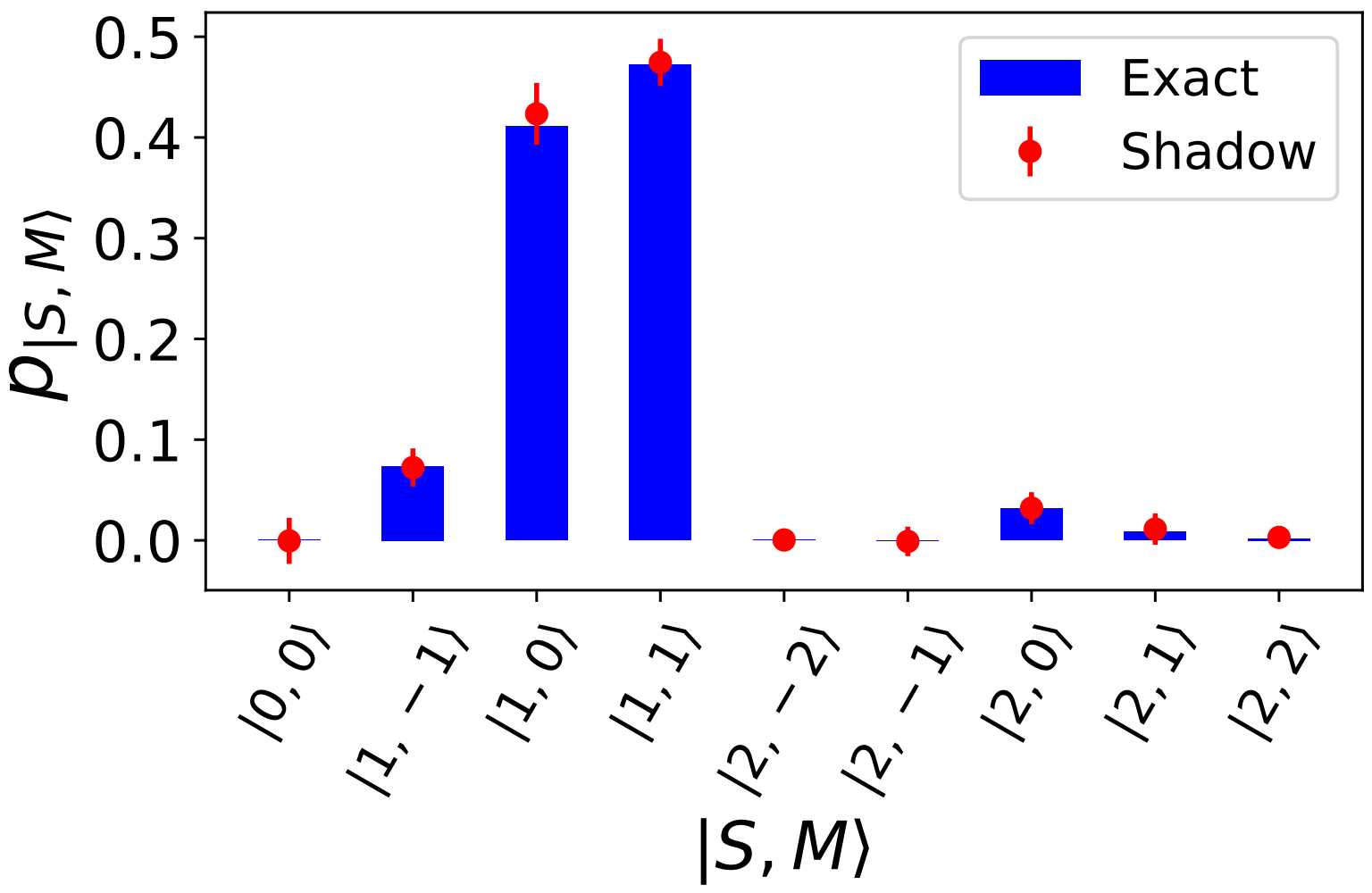

Figure 5.5: Illustration of the projection onto different basis states, denoted as $|s, m\rangle$, for a system composed of 4 qubits. The red dots are obtained using the classical shadow procedure, as detailed in Section 5.4.2. We used a Gaussian state situated within the eigenspace of the $S^2$ operator. To be specific, given the $S^2$ operator's decomposition as $S^2 = UDU^\dagger$, we can express $|\psi\rangle = U|\psi_G(k)\rangle$, where $|\psi_G(k)\rangle$ is the state used in Fig. 5.2. The results derived from the classical shadows were acquired using $10^4$ shots, repeated 50 times. The red data points and their respective error bars symbolize the mean and standard deviation calculated from these 50 trials, respectively. Results displayed here are obtained with $n_p = 10$ points to discretize each integral.

### 5.4.3 . Projected Energy

We can also derive the projected energy by calculating the expectation value of all the operators $O_k$ in the decomposition $HP = \sum_k \alpha_k O_k$, where $H$ represents the Hamiltonian as given in Eq.

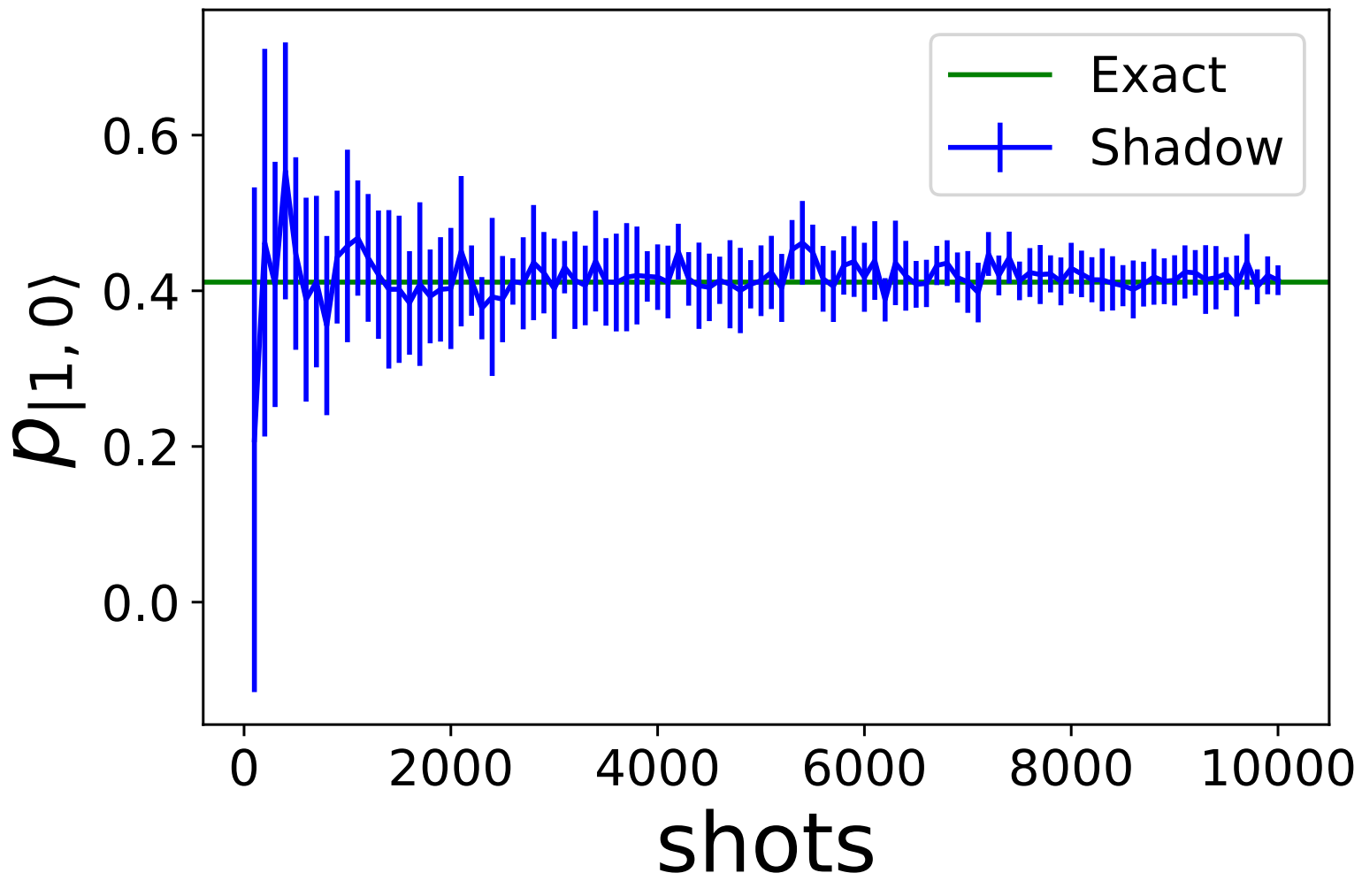

Figure 5.6: Convergence of the expectation value of the projector $P_{|1,0\rangle}$ as the number of "shots" in the classical shadow increases, specifically for the state $|\psi\rangle$ depicted in Fig. 5.5. The experiment was replicated with the same number of "shots" ten times to derive these results from the classical shadows. The blue data points and their corresponding error bars represent the average outcome and standard deviation across these ten repetitions, respectively.

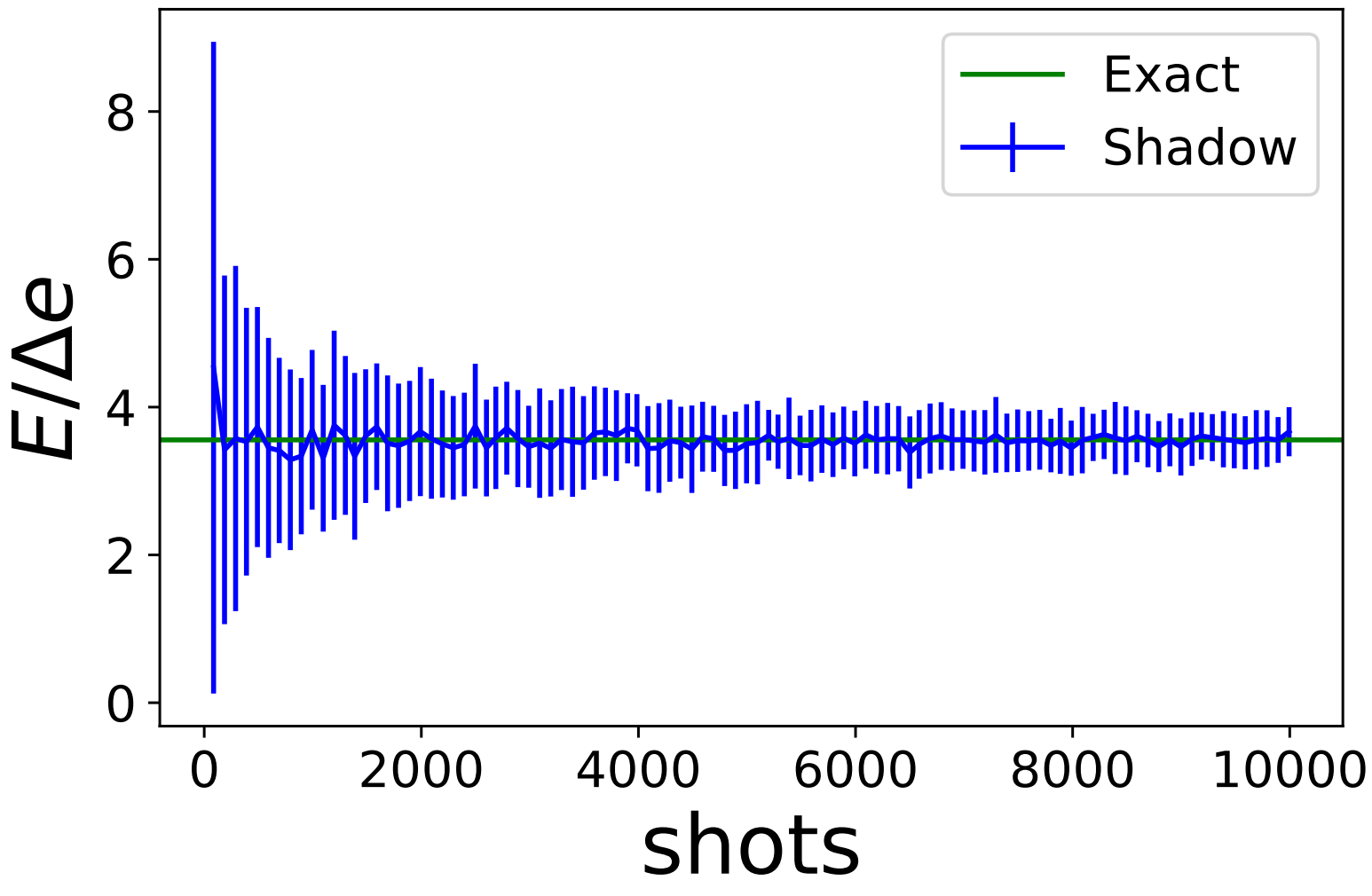

Figure 5.7: Convergence of the projected energy, employing the same wave function as in Fig. 5.2, with the pairing Hamiltonian and $P_{N=2}$. The depicted results were obtained from the classical shadows by repeating the process with the same number of 'shots' 50 times. The blue data points and their corresponding error bars represent the mean outcome and the standard deviation derived from these 50 iterations, respectively. Despite significant fluctuations, we can observe a rapid convergence of the average energy value to the exact projected energy.

(5.18), and $P$ stands for either the particle number or spin projector. The projected energy can be computed utilizing either Eq. (5.28) for the particle number projection or Eq. (5.32) for the spin projection. Fig. 5.7 illustrates the projected energy of the wave function employed in Fig. 5.2, in conjunction with the pairing Hamiltonian within the subspace with $N = 2$ pairs. This figure shows that the correct energy is indeed obtained as the number of shots increases.

## 5.5 . Conclusion

In our pursuit to discover alternative methods for minimizing the resources required for the symmetry restoration process, we've delved into the classical shadows paradigm. By harnessing suitably structured forms of the particle number and spin projectors–represented as weighted sums of tensor products of single qubit operators–we showed how the orthogonality of the Pauli operators under the Hilbert-Schmidt inner product facilitates efficient computation of projected expectation values. On this basis, we examined the potential extension of the original derandomization principle to such quantities, concluding that its extension is feasible solely for the particle number projection. Additionally, we observed that computing the various productorials only once grants access to all different projected expectation values. Lastly, we presented an example of extracting the projected energy using the pairing Hamiltonian and the particle number projector. This chapter presents an exploratory study into the use of the classical shadow formalism for symmetry restoration. The results provide a constructive starting point for further research in this field.

# 6 - Hybrid Quantum-Classical Computations and Post-Processing Techniques

In previous sections, we have presented several methods for restoring symmetries. By combining them with variational approaches, we were able to implement the equivalent of the VAP approach on a quantum computer. However, these variational techniques have an important drawback: they primarily focus on the ground state and do not provide insight into the excited state spectra. To address this limitation, in this final chapter, we will explore various hybrid quantum-classical techniques that could potentially enhance the accuracy of ground state energy estimation, reveal the spectrum of excited states, or accomplish both goals simultaneously. We will present examples of these techniques throughout the chapter, using the model Hamiltonians introduced in section 2.2.3 as outlined in Ref. [122].

Our discussion will begin with the application of the generating function concept to extract information about Hamiltonian spectra by calculating Hamiltonian moments, denoted as $\langle H^k \rangle$. These moments will be employed to approximate an imaginary time evolution and to implement a Krylov method for estimating the lowest eigenenergies of a Hamiltonian. In this context, we will emphasize the challenges of achieving high precision when obtaining these moments. Finally, we will investigate an alternative approach for performing a quantum Krylov approximation that directly employs the generating function values.

## 6.1 . Generating Function $\langle F(t) \rangle$ for $\langle H^k \rangle$

The generating function (GF) $F(t)$ of Hamiltonian moments $\langle \psi | H^k | \psi \rangle$ over a state $|\psi\rangle$ is the expectation value of the propagator associated to the Hamiltonian $H$, i.e.:

$$F(t) = \langle \psi | e^{-itH} | \psi \rangle. \tag{6.1}$$

The generating function in Eq. (6.1) can be decomposed in terms of the eigenenergies $\{E_k\}$ associated with the eigenstates $\{|\phi_k\rangle\}$ of the Hamiltonian $H$:

$$F(t) = \sum_k |c_k|^2 e^{-itE_k}, \tag{6.2}$$

where we have assumed the decomposition $|\psi\rangle = \sum_k c_k |\phi_k\rangle$. The GF is already employed, explicitly or implicitly, within the context of quantum computing. For example, the QPE approach (see section 2.1.3) applied to an operator $U_A = e^{2\pi i A}$ essentially computes an approximation of the generating function associated with the operator $A$ on a set of ancillary qubits before the application of a quantum inverse Fourier transforms used to obtain the probability distribution of $A$'s eigenvalues. Moreover, the GF serves as a crucial component of the time-series method, as discussed in Ref. [171]. On a quantum computer, the values $F(t)$ can be retrieved using the standard Hadamard test in Table 2.5 together with the propagator $e^{-itH}$ [122]. Figure 6.1 displays the real and imaginary parts of the generating function obtained for certain instances of the pairing and Hubbard Hamiltonians (see section 2.2.3). The lines represent the generating function (GF) calculated classically through direct diagonalization of the Hamiltonian. The symbols indicate the results obtained with the quantum computer (QC) simulator using the Hadamard test circuit depicted in Table 2.5. Each data point in the figure is derived from the average of $10^4$ events using a QC emulator, specifically the IBM Qiskit toolkit [52]. As anticipated, given the emulator's simulation of a noiseless QC, results garnered from the quantum and classical computers show perfect alignment. The only prerequisites for this convergence are a sufficient number of measurements and a suitably small numerical time-step, $\Delta t$, which safeguards the validity of the Trotter-Suzuki approximation. We employed $\Delta t.J = 0.02$ and $\Delta t.\Delta e = 0.002$ for the Fermi-Hubbard and pairing models, respectively.

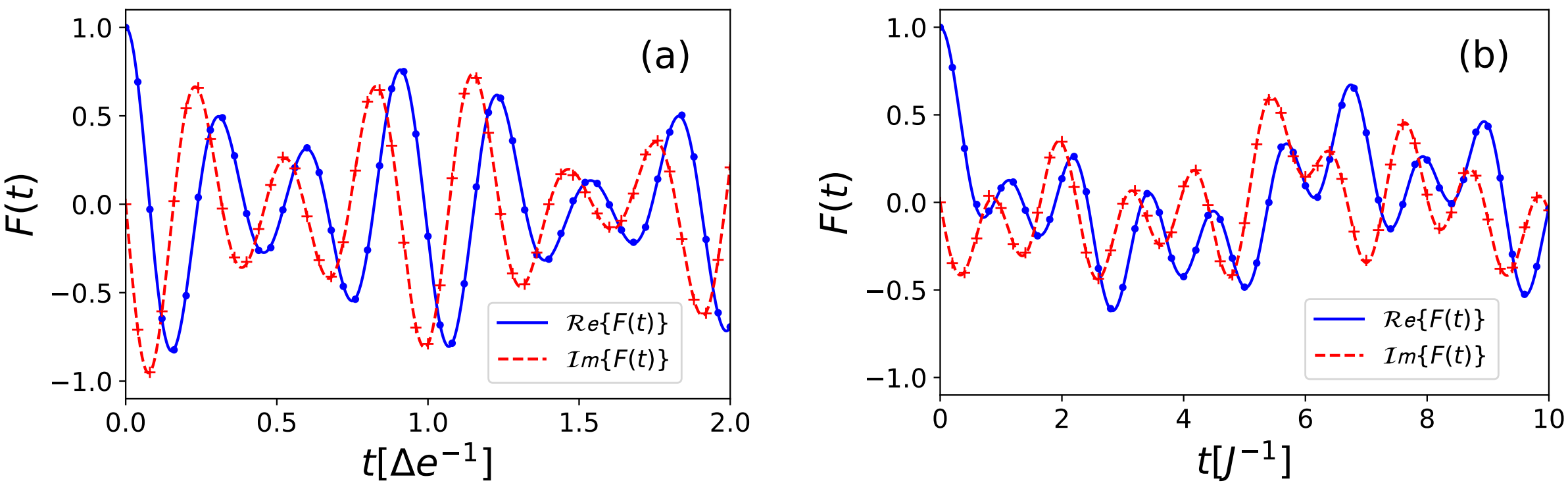

Figure 6.1: Panel (a): Illustration of the real (solid blue line) and imaginary (red dashed line) parts of the generating function obtained classically for the pairing model in Eq. (2.33) with four pairs on eight doubly degenerated single-particle levels with energy $\varepsilon_p = p\Delta e$ (with $p = 1, \ldots, 8$) and $g_{pq} = \text{cte} = g/\Delta e = 1$. The initial condition corresponds to the Slater determinant, where the four lowest single-particle energies are occupied. In panel (b), we show the same quantities obtained for the Fermi-Hubbard model with four particles on four sites for the parameters $U/J = 1$. The initial condition corresponds to a spin-saturated case where the initial state is an average over the $C_4^2 = 6$ Slater determinants, with pairs of particles with opposite spins randomly occupying two sites among the four possibilities. Results displayed with points in both panels correspond to calculations obtained with a noiseless quantum computer emulator, Qiskit [52], and the Hadamard test circuit of Table 2.5. Each point in these figures is obtained by averaging over $10^4$ measurements.

### 6.1.1 . Fourier Analysis of the Generating Function

The knowledge of the GF at all times gives access to the values of the eigenenergies and the probabilities $|\langle\phi_k|\psi\rangle|^2 = |c_k|^2$. By performing a Fourier transform on the GF, denoted as $\tilde{F}(\nu)$, we can obtain direct information about the Hamiltonian spectra:

$$\begin{aligned} \tilde{F}(\nu) &= \sum_k |c_k|^2 \int_{-\infty}^{+\infty} e^{-itE_k} e^{-2\pi\nu t} dt, \\ &= \sum_k |c_k|^2 \, \delta\left(\nu + E_k/2\pi\right). \end{aligned} \tag{6.3}$$

Fig. 6.2 illustrates the Fourier transform of the generating functions depicted in Fig. 6.1. It becomes apparent that the Fourier transform of the GF provides direct access to information on the Hamiltonians' spectra and the probabilities associated with the coefficients of the initial wave function in the Hamiltonian eigenbasis via the points $(-E_k/2\pi, |c_k|^2)$.

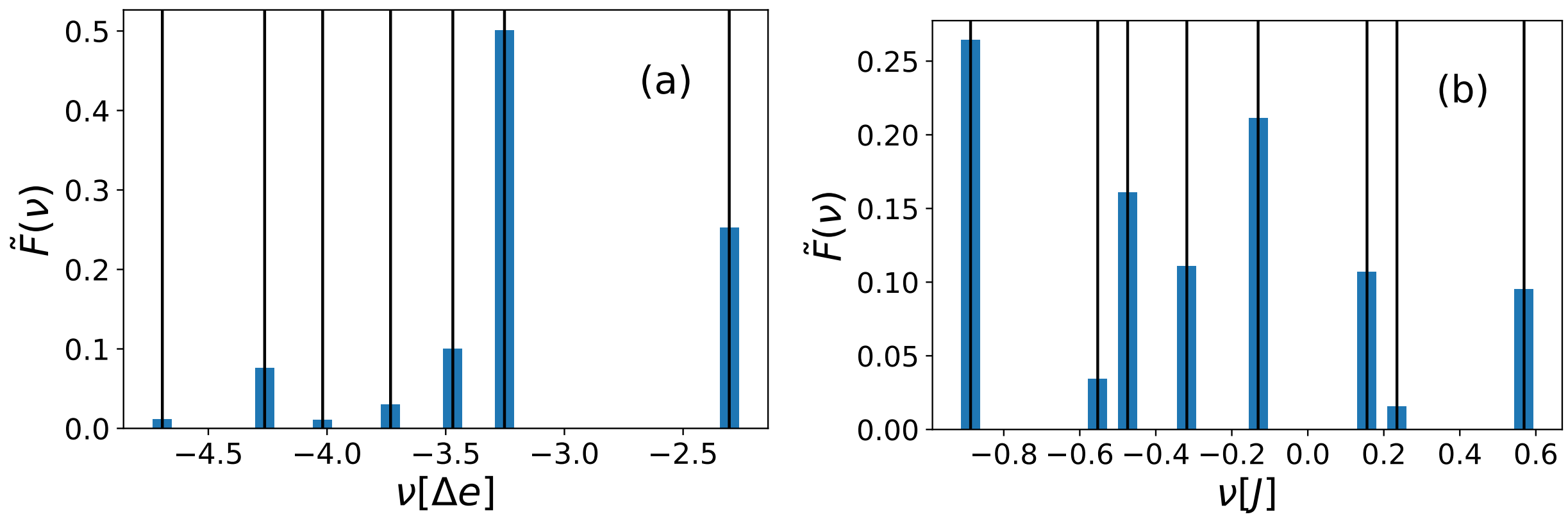

Figure 6.2: Fourier transform of the generating functions in Fig. 6.1 (blue bars). The black vertical lines represent the exact values $-E_k/2\pi$.

When obtaining the generating function on a quantum computer, the Trotter-Suzuki approximation imposes significant constraints on the maximum time evolution (denoted by $T/2$ below), as the circuit complexity can increase substantially for longer times. Additionally, due to the relation $\Delta E \Delta t \sim \hbar$, this approximation also limits the minimum energy resolution achievable when extracting energies. To observe this effect more clearly, let us consider the Fourier transform of the GF in the interval $[0, T/2]$. Given the property $F(-t) = F(t)^*$, this is equivalent to having the GF in the interval $[-T/2, T/2]$. Analytically, the GF in this interval can be regarded as the product of the total GF and a rectangular function $\Pi_{[-T/2,T/2]}$ defined as:

$$\Pi_{\left[-\frac{T}{2},\frac{T}{2}\right]} = \begin{cases} 0 & \text{if } t < -\frac{T}{2}, \\ 1 & \text{if } -\frac{T}{2} \leq t \leq \frac{T}{2}, \\ 0 & \text{if } t > -\frac{T}{2}. \end{cases} \tag{6.4}$$

Thus, we will have to find the Fourier transform of the function:

$$s(t) = F(t)\, \Pi_{\left[-\frac{T}{2},\frac{T}{2}\right]}. \tag{6.5}$$

Using that the Fourier transform of a rectangular function of amplitude $A$ (i.e., $A\Pi_{\left[-\frac{T}{2},\frac{T}{2}\right]}$) is equal to $\mathcal{F}\{A\Pi_{\left[-\frac{T}{2},\frac{T}{2}\right]}\} = AT\,\mathrm{sinc}\,(\nu T)$, where $\mathrm{sinc}\,(\nu) = \frac{\sin(\pi\nu)}{\pi\nu}$, we get:

$$\begin{aligned} \mathcal{F}\{s(t)\} \coloneqq \tilde{s}(\nu) &= \tilde{F}(\nu) * T\,\mathrm{sinc}\,(\nu T)\,, \\ &= T\sum_k |c_k|^2 \int_{-\infty}^{+\infty} \delta\left(\nu' + \frac{E_k}{2\pi}\right) \mathrm{sinc}\,[(\nu - \nu')\,T]\, d\nu', \\ &= T\sum_k |c_k|^2\, \mathrm{sinc}\left[\left(\nu + \frac{E_k}{2\pi}\right) T\right]. \end{aligned} \tag{6.6}$$

To illustrate the effect of the time interval $T$ on the Fourier transform, Fig. 6.3 presents $\tilde{s}(\nu)$ for the pairing and Hubbard Hamiltonians in Fig. 6.1 with different $T$ values. The various peaks can be clearly distinguished for sufficiently large values of $T$; however, for smaller values of $T$, the Fourier transform may not provide accurate information about either the eigenenergies $E_k$ or the amplitudes $|c_k|^2$. This will be further discussed in section 6.2.4.

As an alternative to the Fourier of the GF, we can extract the Hamiltonian moments $\langle H^k \rangle$ to perform spectra analysis using further post-processing methods. Here we mention that the eigenvalues $E_k$ and probabilities $|c_k|^2$, extracted from the Fourier method, provide access to the moments $\langle H^k \rangle$ through the relation:

$$\langle H^n \rangle = \sum_k |c_k|^2 E_k^n. \tag{6.7}$$

Given the complexity of implementing the Fourier transform, we examine an alternative for obtaining Hamiltonian moments using the $k^{th}$ derivative of the generating function, starting with its Taylor expansion:

$$\frac{d^k F(t)}{dt^k} = \sum_{n=k}^{\infty} \frac{(-i)^n}{(n-k)!} t^{n-k} \langle H^n \rangle. \tag{6.8}$$

Here, we realize that to get the $k^{th}$ moment $\langle H^k \rangle$ is enough to evaluate the $k^{th}$ derivative of the generating function at $t = 0$, i.e.,

$$\langle H^k \rangle = i^k \left. \frac{d^k F(t)}{dt^k} \right|_{t=0}. \tag{6.9}$$

Specific methods can extract the $k^{th}$ derivative from GF values; here, we employ the finite difference method, while the parameter-shift rule serves as another alternative [172, 173, 174]. The primary advantage of this approach over Fourier-based methods is that the time evolution required for computing the $k^{th}$ derivative is generally smaller, thereby reducing Trotter errors. In the following sections, we illustrate some methods that can be developed, provided that we can accurately obtain the $\langle H^k \rangle$ values. A critical analysis of the precision of these moments will be discussed later in this chapter.

## 6.2 . A Survey of Methods Based on Moments Estimates

During my thesis, I have delved into and formulated numerous methods that hinge on the comprehension of quantum systems' time evolution over brief intervals. The rationale behind exploring methods anchored in short times primarily stems from the significant restrictions current quantum devices face due to noise, as exemplified in Section 6.2.5. Some of these techniques are discussed below.

### 6.2.1 . t-Expansion Method

To estimate the ground state energy of a Hamiltonian, it is possible to use the imaginary time evolution method. This method starts by defining an imaginary evolution of an initial wave function $|\psi_0\rangle$, i.e.,

$$|\Psi(\tau)\rangle = \frac{e^{-\tau/2H}}{\sqrt{\langle \psi_0 | e^{-\tau H} | \psi_0 \rangle}} |\psi_0\rangle. \tag{6.10}$$

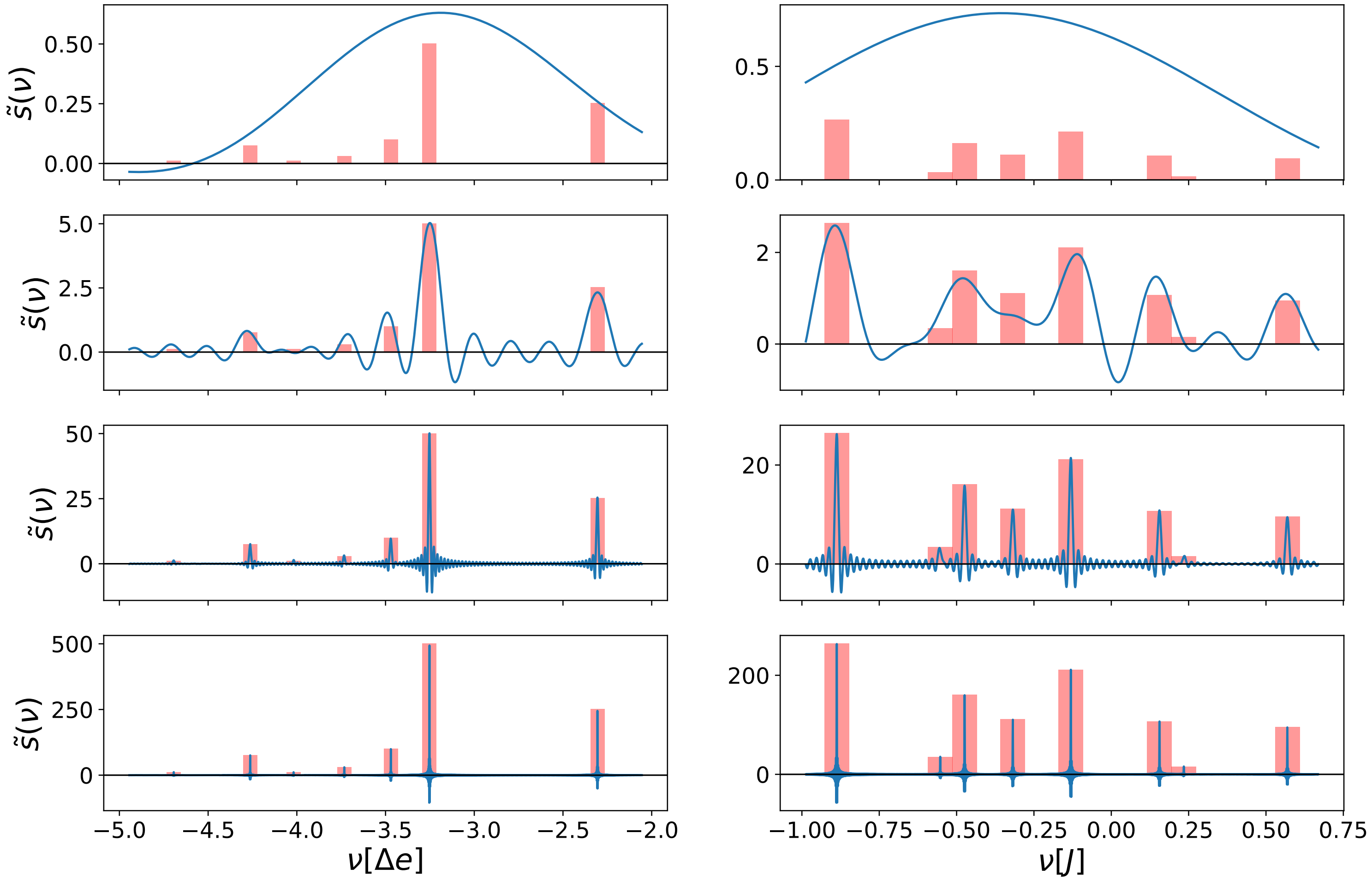

Figure 6.3: Effect of the time interval $T$ on the Fourier transform in Eq. 6.6. From top to bottom, the time intervals for the finite Fourier transform (solid blue line) considered were $1, 10, 10^2$, and $10^3$. The left (right) column corresponds to the results obtained using the pairing (Hubbard) Hamiltonian with the same parameters as in Fig. 6.1. The red bars are positioned at frequencies associated with the eigenvalues $\nu_k = -E_k/2\pi$ and have a height of $T\,|c_k|^2$. A black horizontal line at 0 is drawn for reference. We can observe that $E_k$ and $|c_k|^2$ become more distinguishable as $T$ increases.

If $\langle\psi_0|\Psi_{GS}\rangle \neq 0$, we know that $|\Psi(\tau)\rangle$ will converge to the ground state wave function when $\tau \to +\infty$, i.e.,:

$$E_{\rm GS} = \lim_{\tau\to+\infty} \langle\Psi(\tau)|H|\Psi(\tau)\rangle = \lim_{\tau\to+\infty} E(\tau). \tag{6.11}$$

Ref. [175] showed that $E(\tau)$ could be written in terms of the moments of the Hamiltonian estimated over the initial wave function $|\psi_0\rangle$:

$$E(\tau) = \frac{\langle He^{-\tau H}\rangle}{\langle e^{-\tau H}\rangle} = -\frac{d}{d\tau}\ln\langle e^{-\tau H}\rangle, \tag{6.12}$$

with

$$\ln\langle e^{-\tau H}\rangle = \sum_{k=0}^{+\infty} \frac{(-\tau)^k}{k!}\kappa_k, \tag{6.13}$$

where $\kappa_k$ is the cumulant of order $k$ of the Hamiltonian. These cumulants can be calculated from the moments of orders lower or equal to $k$ using the following recurrence relation:

$$\kappa_n = \langle H^n\rangle - \sum_{k=1}^{n-1} \binom{n-1}{k-1} \kappa_n \langle H^{n-k}\rangle, \tag{6.14}$$

that could be used iteratively with the condition $\kappa_1 = \langle H\rangle$. Eq. (6.13) indicates that the size of the interval $[0, \tau_{\rm max}]$, in which the function $E(\tau)$ is accurately approximated, increases with the number of moments or cumulants. In the following, I assume I can get accurate $\langle H^k\rangle$ values up to a required value of $k$. Following the procedure outlined in [175], we consider a Padé approximation of the derivative $dE(\tau)/d\tau$ instead of approximating the energy $E(\tau)$ directly, as it offers certain benefits. Specifically, we know that this derivative converges to 0 for $\tau \to +\infty$, and thus, we can ensure that the approximated $E(\tau)$ will consistently decrease as expected. Assuming we have access to only the lowest $M+2$ cumulants (or moments) of the Hamiltonian, $dE(\tau)/d\tau$ is approximated as

$$\frac{d}{d\tau}E(\tau) \simeq -\sum_{k=0}^{M} \frac{(-\tau)^k}{k!}\kappa_{k+2}. \tag{6.15}$$

We then replace this approximate form with a Padé approximation, denoted by Padé$[I,J](\tau)$ which has the form:

$$P[I,J](\tau) = \frac{\sum_{j=0}^{I} a_j\tau^j}{1+\sum_{k=1}^{J} b_k\tau^k}. \tag{6.16}$$

Since the integration of $P[I,J](\tau)$, should converge to a fixed energy, the decrease of $P[I,J](\tau)$ towards 0 must occur faster than $1/\tau$. This results in the additional constraint $J - I \geq 2$. Once a Padé approximation satisfying all the constraints has been found, $E(\tau)$ can be obtained by numerically integrating the derivative with respect to $\tau$. An illustration of this method for the pairing Hamiltonian of Fig. 6.1 panel (a) is shown in Fig. 6.4 and for the Hubbard Hamiltonian of Fig. 6.1 panel (b) in Fig. 6.5. Using the two models considered here, we can observe that this method is rather accurate in predicting the ground state energy. Note that the exact values of $\langle H^k\rangle$ have been used.

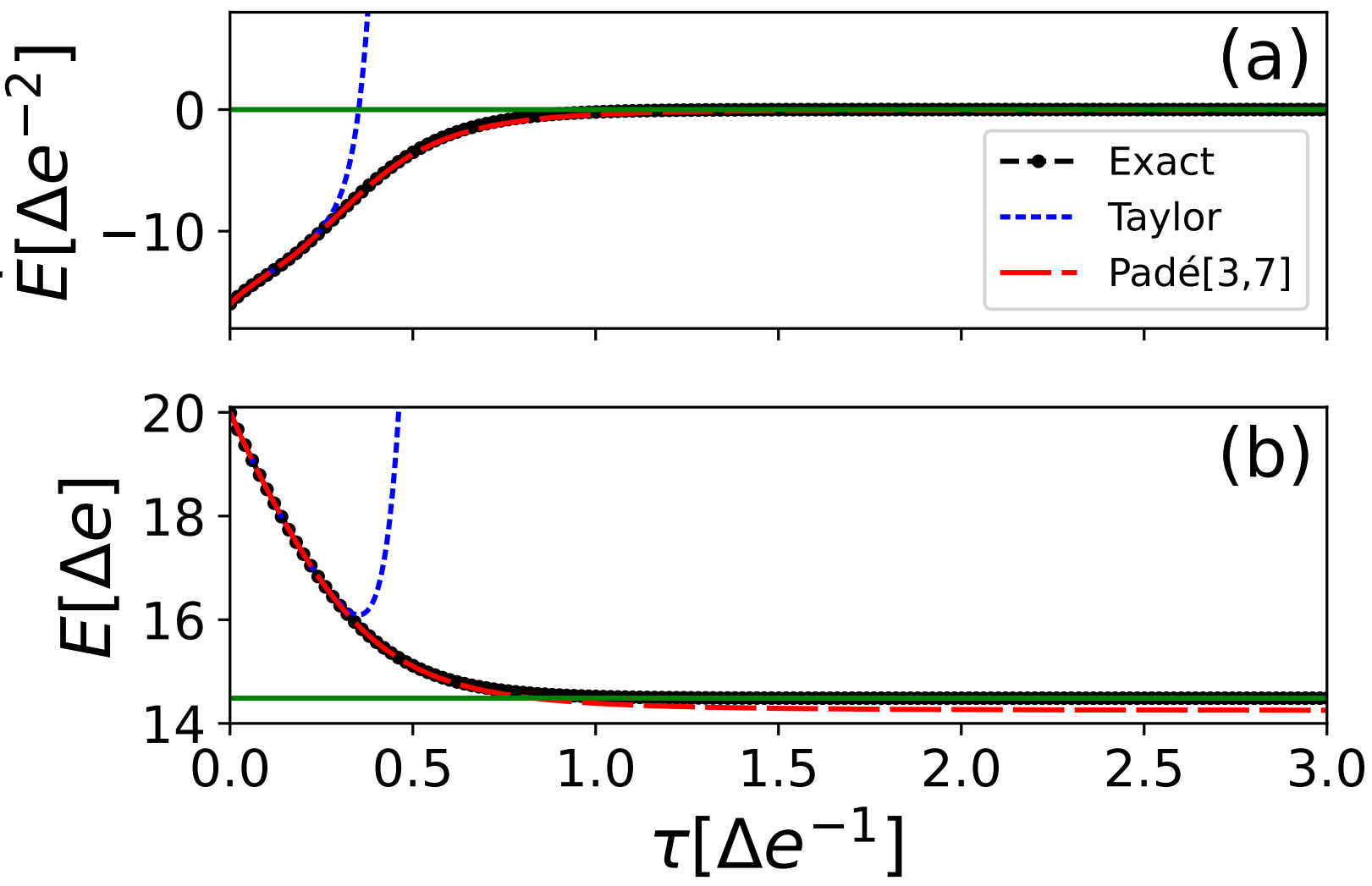

Figure 6.4: The $t$-expansion method is illustrated here as applied to the pairing model, using the same conditions as in panel (a) of Fig. 6.1. The function $E(\tau)$ and its derivative are respectively displayed in panels (b) and (a). In each panel, we present the exact solution given by Eq. (6.10) (black filled circles), the Taylor expansion from Eq. (6.15) with $M = 10$ (blue short-dashed line), and the Padé $[3, 7]$ approximation (red long-dashed line). It is important to note that $M = 10$ indicates that we used $M + 2 = 12$ cumulants as inputs. The exact imaginary-time solution (black-filled circle) was obtained by explicitly carrying out the imaginary-time evolution on a classical computer. In panel (b), the green horizontal line represents the exact ground state energy. From Ref. [71].

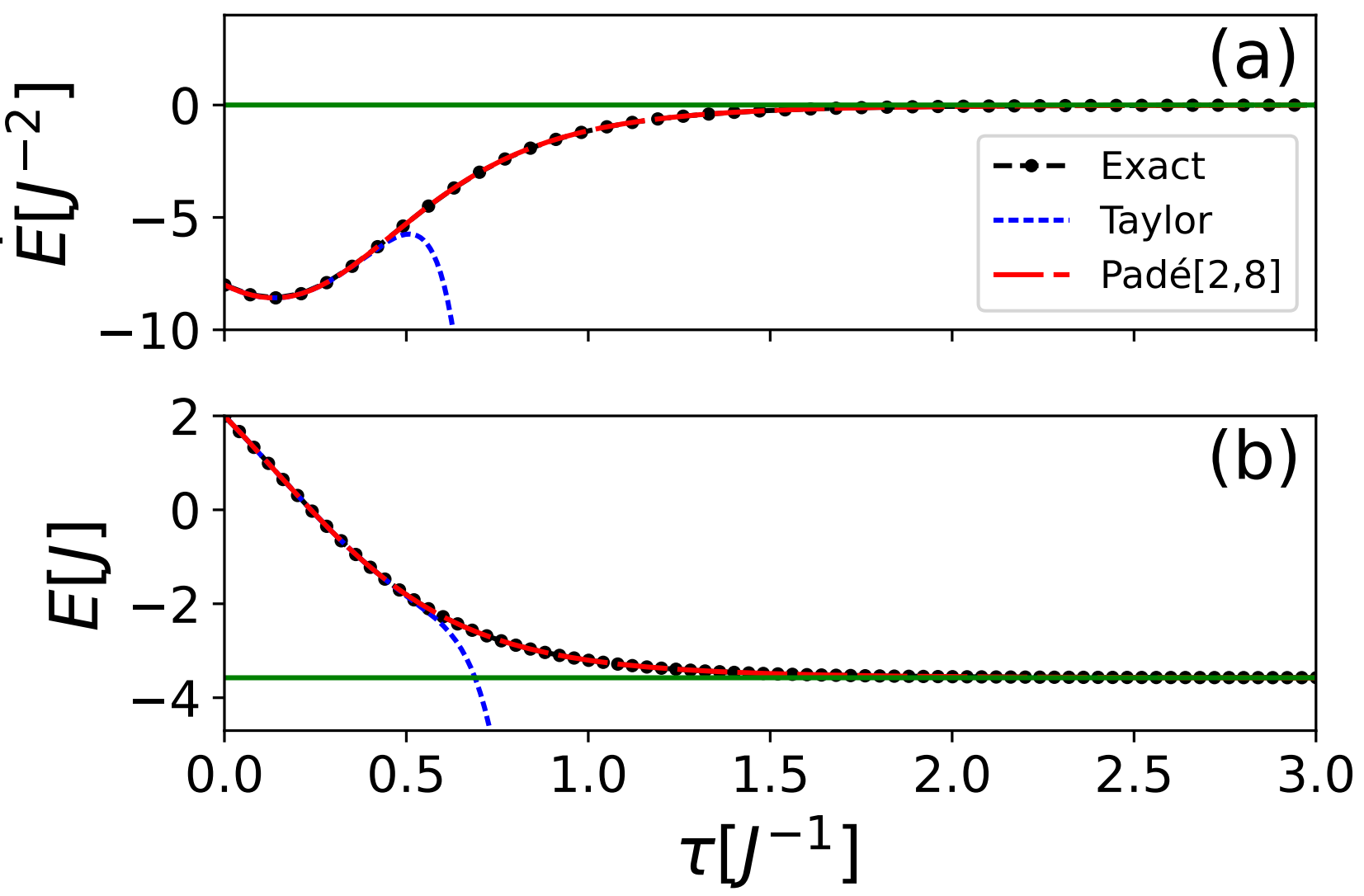

Figure 6.5: Similar to Fig. 6.4, this figure presents the results for the Fermi-Hubbard model using the parameters from panel (b) in Fig. 6.1. It is important to note that if only a single Slater determinant is employed as the initial condition instead of a mixture of six, the convergence towards the ground state energy will necessitate the inclusion of higher-order moments. From Ref. [71].

### 6.2.2 . Krylov Method

The Krylov method belongs to a broad family of quantum subspace expansion methods commonly used in classical computation to approximate the solution of an eigenvalue problem when a complete configuration interaction solution is prohibitive. It consists of iteratively constructing subspaces of the total Hilbert space with increasing complexity to then diagonalize the Hamiltonian defined in this reduced space to obtain approximations of the eigenenergies. Fig. 6.6 illustrates the principle of these types of methods. In the figure the subspace $\mathcal{H}_M$ is spanned by the states $\{|\Phi_0\rangle, \ldots, |\Phi_{M-1}\rangle\}$ which, themselves, are constructed iteratively, i.e., $|\Phi_k + 1\rangle$ is constructed from $|\Phi_k\rangle$. Famous examples of these kinds of methods, widely used in classical computation, are the Arnoldi and Lanczos methods [176].

The Krylov method generates different subspaces $\mathcal{H}_M$ using the powers of the Hamiltonian applied over the initial wave function, i.e., the subspace $\mathcal{H}_M$ is spanned by the non-orthogonal Krylov basis defined as $\{|\Phi_k\rangle = H^k|\Phi_0\rangle\} = \{|\Phi_0\rangle, H|\Phi_0\rangle, \ldots, H^{M-1}|\Phi_0\rangle\}$. Using this basis, we can define the overlap and Hamiltonian operators in the subspace $\mathcal{H}_M$:

$$\begin{cases} O_{kl} = \langle\Phi_l|\Phi_k\rangle = \langle H^{k+l}\rangle_0, \\ \\ H_{kl} = \langle\Phi_l|H|\Phi_k\rangle = \langle H^{k+l+1}\rangle_0, \end{cases} \tag{6.17}$$

where we have used the short-hand notation $\langle\Phi_0 |.| \Phi_0\rangle = \langle . \rangle_0$. We observe that all matrix elements of $O$ and $H$ in Eq. (6.17) can be expressed as moments of the initial state. To approximate the eigenvalues within this subspace, we first diagonalize the overlap matrix $O_{kl}$ in Eq. (6.17), resulting in a new set of orthonormal state vectors. We then diagonalize the Hamiltonian on this new basis. Detailed derivation and practical aspects of this process can be found in Appendix D.

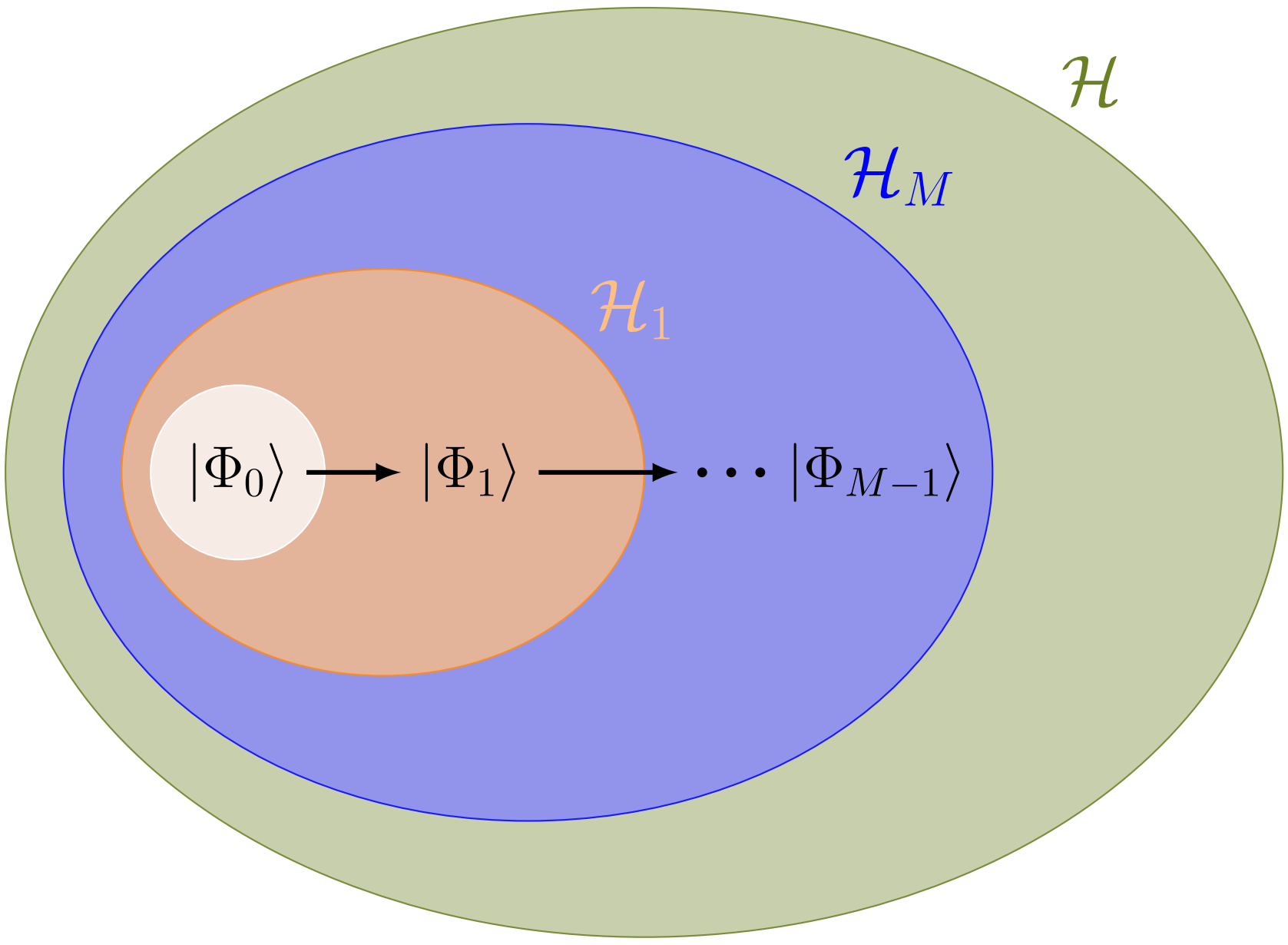

Figure 6.6: Schematics of the quantum subspace expansion technique's philosophy. Here, the eigenvalue problem is considered in subspaces with increasing dimensionality.

In Fig. 6.7, we present the evolution of the $E_\alpha^{(M)}$ values as a function of $M$ for the pairing Hamiltonian case in the strong coupling regime $g/\Delta e = 2.0$. This figure demonstrates that the energies acquired by diagonalizing the Hamiltonian $\tilde{H}$ in Eq. (D.5) with increasing $M$ converge towards some of the exact eigenvalues. The convergence is quicker for lower energies. A high accuracy for the ground state is observed at $M = 4$, corresponding to considering the first seven moments. Notably, better accuracy is achieved with fewer moments than those used in Fig. 6.4 for $g/\Delta e = 1$. Generally, the dimension of $\mathcal{H}_M$ is relatively small when compared with the total size of the Hilbert space, which, after accounting for particle number symmetry, was $70$ in the pairing model with four particles on eight levels with zero seniority. In both models, we systematically found that the diagonalization method offers access to excited states and appears to converge more rapidly to the ground state as the number of moments increases, enhancing its accuracy with $M$. Finally, it is worth noting that some excited states are not captured because they are orthogonal to the initial state. By exploring different initial states, one could anticipate obtaining eigenvalues not represented in Fig. 6.7.

#### 6.2.3 . Long-Time Evolution from Moments

The Krylov and $t$-expansion methods are designed to get an approximation of the energy spectrum. We show here that the Hamiltonian moments can also be used to obtain the approximate time evolution of the probability of survival. Using the $M$ states and eigenenergies $|E_\gamma^{(M)}\rangle$ and $\left\{E_\gamma^{(M)}\right\}$ obtained by diagonalizing the Hamiltonian in the $\mathcal{H}_M$ space, the evolution of the system

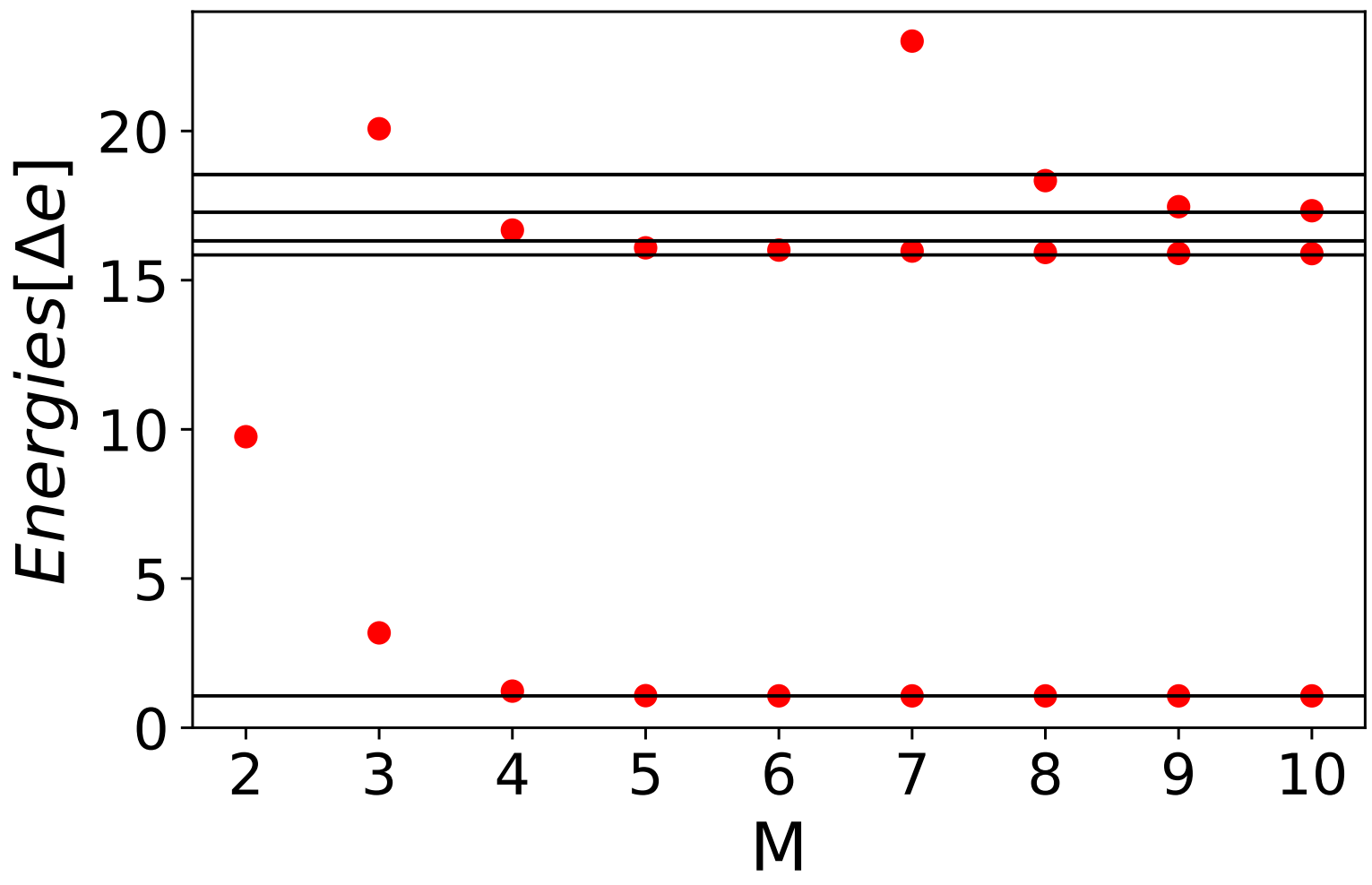

Figure 6.7: Illustration of the eigenvalue evolution (red-filled circles) obtained through the diagonalization of the Hamiltonian in the truncated Krylov basis for increasing values of $M$ in the pairing model. The conditions are the same as in panel (a) of Fig. 6.1, except that the pairing strength is set to $g/\Delta e = 2.0$. It is important to note that the number of associated moments used as inputs is given by $L = 2M - 1$, as we know that $\langle \Phi_0 | \Phi_0 \rangle = 1$. The horizontal black lines indicate the lowest exact eigenvalues of the pairing Hamiltonian.

can be approximated as:

$$|\Phi^{(M)}(t)\rangle = \sum_{\gamma=0}^{M-1} e^{-iE_\gamma^{(M)} t} |E_\gamma^{(M)}\rangle \langle E_\gamma^{(M)} | \Phi_0 \rangle. \tag{6.18}$$

The survival probability is then given by:

$$P_0^{(M)}(t) = |\langle \Phi_0 | \Phi^{(M)}(t) \rangle|^2 = \sum_{\gamma=0}^{M-1} e^{-iE_\gamma^{(M)} t} \left| \langle E_\gamma^{(M)} | \Phi_0 \rangle \right|^2 . \tag{6.19}$$

The computation of the overlaps $\langle E_\gamma^{(M)} | \Phi_0 \rangle$ is shown in Appendix D.1. Different evolutions of the survival probability obtained with various values of $M$ are illustrated in Fig. 6.8 and compared to the exact solution. In all cases, the approximate evolution matches the exact solution up to a certain time $t_{\max}(M)$; this time increases with $M$. We also observe that the evolution converges towards the exact solution as $M$ increases, even when the number of states included is significantly smaller than the size of the entire Hilbert space.

#### 6.2.4 . Difficulties in Obtaining the Hamiltonian Moments $\langle H^k \rangle$

To obtain the derivatives of the generating function at order $k$ in Eq. (6.9), we initially used the finite difference method (FDM). A comprehensive list of finite difference coefficients with varying levels of accuracy for estimating derivatives can be found in [177, 178]. While the FDM is suitable for obtaining accurate values for the first few moments, its precision diminishes as the order $k$ increases. It is worth noting that the previously discussed methods (Padé or Krylov-based) necessitate a rather precise determination of the moments. We have made substantial efforts to optimize both the time step and the number of points used in the finite difference. However, we concluded that even with noiseless quantum computers, the FDM could not achieve sufficient accuracy for computing $\langle H^k \rangle$ as $k$ increases. In the pairing model examined in Fig. 6.1, relatively good accuracy can be achieved with the FDM for moments up to $k = 10$ to $15$, depending on the interaction strength. We observed in practice that even a small error in the estimated moments could significantly impact the precision of post-processing. Furthermore, since the dimension of the Hilbert space increases with the number of degrees of freedom, we anticipate that the number of moments requiring high-precision calculations for a good convergence to the eigenenergies or for obtaining a good approximation of the evolution of survival probability will also increase compared to the cases presented in Fig. 6.7 and Fig. 6.8.

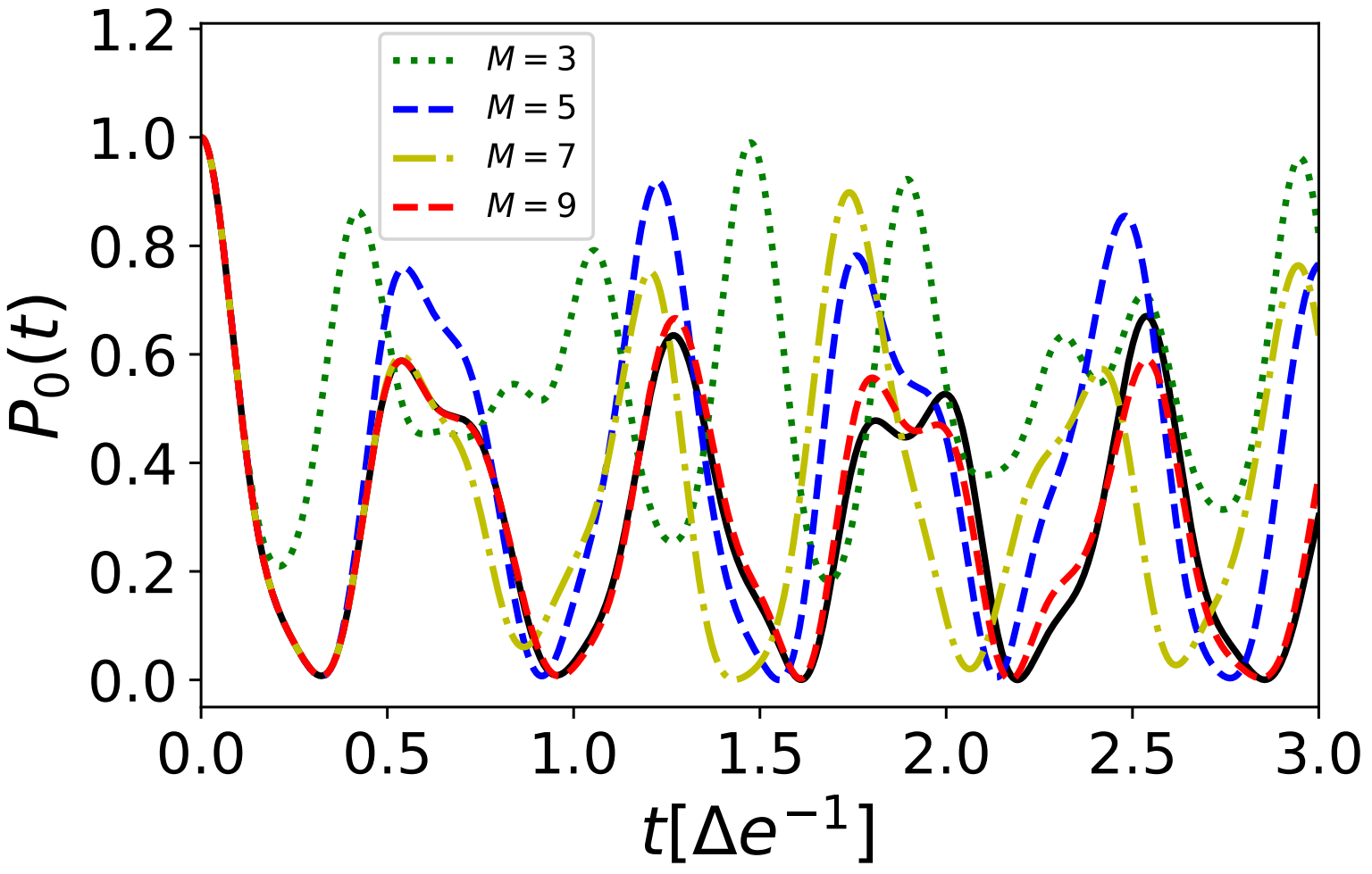

Figure 6.8: Evolution of the probability of survival in Eq. (6.19) as a function of time for increasing values of $M$ for the pairing model and $g/\Delta e = 2$. The black line shows the exact evolution.

In addition to the FDM approach, we extensively tested polynomial interpolation methods (both standard and Chebyshev) to obtain high precision on the moments. However, these polynomial methods can only achieve reasonable accuracy for the first few moments and are not accurate enough for higher $k$ values. Our conclusion is that the only approaches that attained global convergence of all moments with sufficient accuracy are those based on the Fourier transform of the generating function. Indeed, performing the Fourier transform provides an approximation of the initial state components on the eigenstates, denoted by $|c_\alpha|^2$, as well as a set of approximate eigenenergies $\widetilde{E}_\alpha$ (see Eq. (6.3)). From this information, one can easily approximate $\langle H^k \rangle$ using the formula $\sum_\alpha |c_\alpha|^2 \widetilde{E}^k_\alpha$. In the absence of noise on the Fourier transform, an accurate approximation of the moments to any order can be achieved, provided that the time-step used is sufficiently small to resolve the largest eigenenergy and the time interval $T$ is large enough to ensure good energy resolution. Using the Fourier transform usually requires computing the generating function for numerous time steps over a considerable $T$, significantly increasing the effort needed to compute the GF on the quantum computer. This aspect makes the approach less appealing, particularly

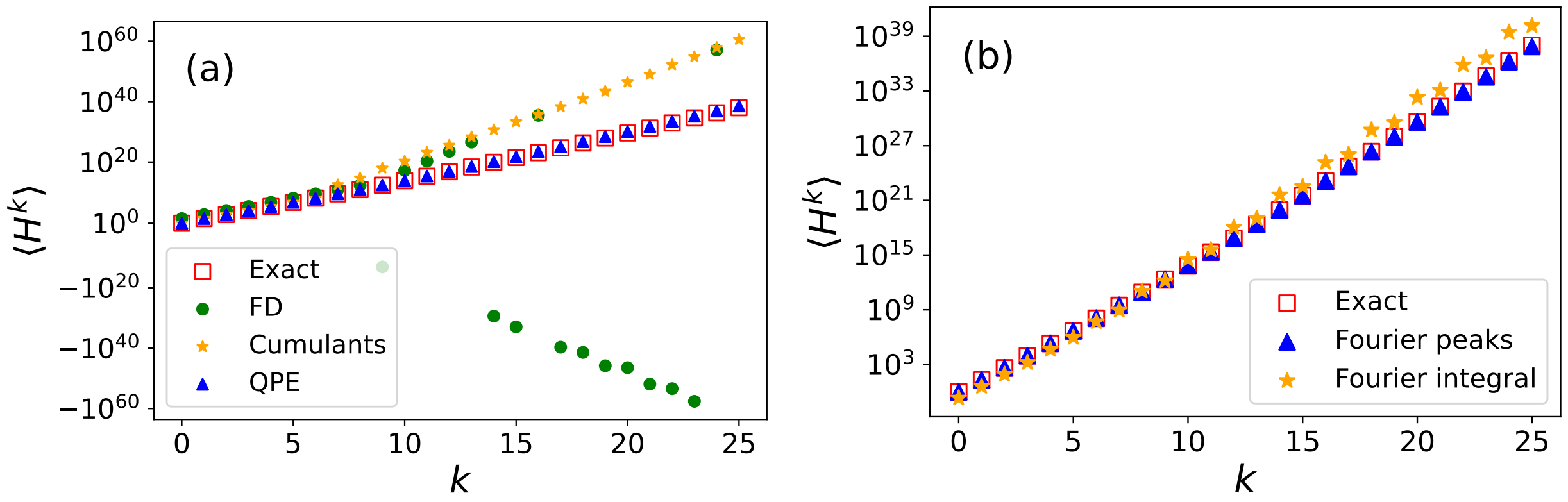

Figure 6.9: Some techniques used to obtain the moments of the generating function $\langle H^k \rangle$ for the pairing Hamiltonian with the parameters of the panel (a) in Fig. 6.1. (a) Red squares represent the exact moments computed via diagonalization. Green circles indicate the moments calculated using the FDM on Eq. (6.9). The moments denoted by orange stars were computed by producing the cumulants $\kappa_n$ from their generating function $Z(t) = \ln F(t)$ in Eq. (6.9), with $F(t)$ being the generating function of the moments, and then employing the recurrent relation in Eq. (6.14). The blue triangles were obtained using the Quantum Phase Estimation (QPE) algorithm with ten ancillary qubits and Eq. (6.7). (b) Once again, the red squares indicate the exact moments. In this case, the blue triangles mark the moments obtained using the peaks $\left(|\tilde{c}_k|^2, \tilde{E}_k\right)$ of the discrete Fourier transform with $2^{14}$ points and $T/2 = 802.03$. The probabilities $|\tilde{c}_k|^2$ were normalized to sum to 1. Finally, the orange stars were obtained using the $k^{th}$ derivative of the Fourier transform with Eq. (6.9), i.e., $\langle H^k \rangle = (-2\pi)^n \int_{-\infty}^{\infty} \tilde{s}(\nu)\, \nu^n d\nu$, in the discrete case.

compared to the QPE method, which is also based on the Fourier technique.

Fig. 6.9 illustrates some of the methods used to approximate the Hamiltonian moments for the pairing Hamiltonian of the panel (a) in Fig. 6.1. Further optimization of the temporal steps and the number of points used in the FDM could achieve increased accuracy. However, we observed that the bigger the $k$, the more points of the GF and the bigger the time step in the FDM needed to reach a good accuracy. One way to overcome the intrinsic error of the FDM approximation, which accumulates for higher-order derivatives, is to employ the parameter shift rule [173, 179, 180]. This approach promises to compute the gradients exactly, and for noiseless quantum computers, it would only be limited by shot noise. Although we have not explored this technique, it could be a promising area for future research.

#### 6.2.5 . Obtaining the Generating Function on an IBM Quantum Computer

Furthering our study of obtaining the Hamiltonian moments from a quantum computer via an FDM-based method, we have attempted to compute generating function values on some of the actual quantum processor units (QPU) available on the IBM quantum cloud. We focus here on the specific case of the *Santiago* QPU. Since the number of qubits is limited to 5 in this case, we considered the simple pairing case where a single pair could access two different single-particle levels with spacing $\Delta e = 1$. Such a case can be encoded on two qubits plus an extra ancillary qubit to perform the Hadamard test shown in Table 2.5. Raw results obtained with the *Santiago* QPU turned out to be strongly affected by noise. Therefore, we tried to implement some standard noise correction techniques. To test these error corrections, we used the *FakeSantiago* QPU that simulates the topology of the qubits and the noise of the real *Santiago* QPU using depolarizing, thermal relaxation, and read-out errors. We show in Fig. 6.10 the evolution of the real and imaginary parts of the generating function obtained with and without the noise. Results without noise correspond to the evolution obtained on a classical quantum computer emulator (i.e., qasm backend) of Qiskit.

In Fig. 6.10, we observe significant deviations in both the real and imaginary parts from the exact solution, even over short times without error correction, and eventually over longer times with error correction applied. These deviations stem from two sources: (i) the noise added in *FakeSantiago* to simulate the real device, and (ii) the discretization of time used in the Trotter-Suzuki method. Results obtained in Fig. 6.10 are calculated by simply assuming a single step in the Trotter-Suzuki technique; that is, for a given time $t$, the evolution time $\Delta t$ is directly equal to $t$. This was done to minimize the circuit depth and, thus, the effect of noise. While this approximation can be accurate for short times, a single Trotter step will induce deviations from the exact solution as $t$ increases. To illustrate this, we also show the result obtained with the QASM backend with no noise, the same Trotter-Suzuki time-step, and the same number of measurements in this figure. We see that, even in the absence of noise, some deviation from the exact solution occurs as $t$ increases. A simple solution to this problem is to increase the number of steps $n_t$ in the Trotter-Suzuki method, leading to $t/\Delta t = n_t$. However, the drawback is that the circuit depth significantly increases when $n_t$ is increased, even by a single unit. This induces a substantial increase in errors in the generating function that generally cannot be corrected by the methods discussed below. The results obtained for $n_t > 1$ with error corrections turned out to be worse than those of $n_t = 1$.

As an illustration of the effect of error correction, we present the results obtained after applying specific corrections in Fig. 6.10. We have used several correction methods to obtain the corrected results, including the read-out corrections of Ref. [181], supplemented by the post-selection correction of Ref. [182]. These two methods partially correct the noise observed in Fig. 6.10. We have also adapted the "reference correction" technique proposed in [182] to improve the results further. In this approach, we utilize the known value of the generating function at time $t = 0$ to construct a matrix $M$ that connects the noiseless measurements to the real measurements at this time. It is then assumed that the same matrix $M$ applies at all times.

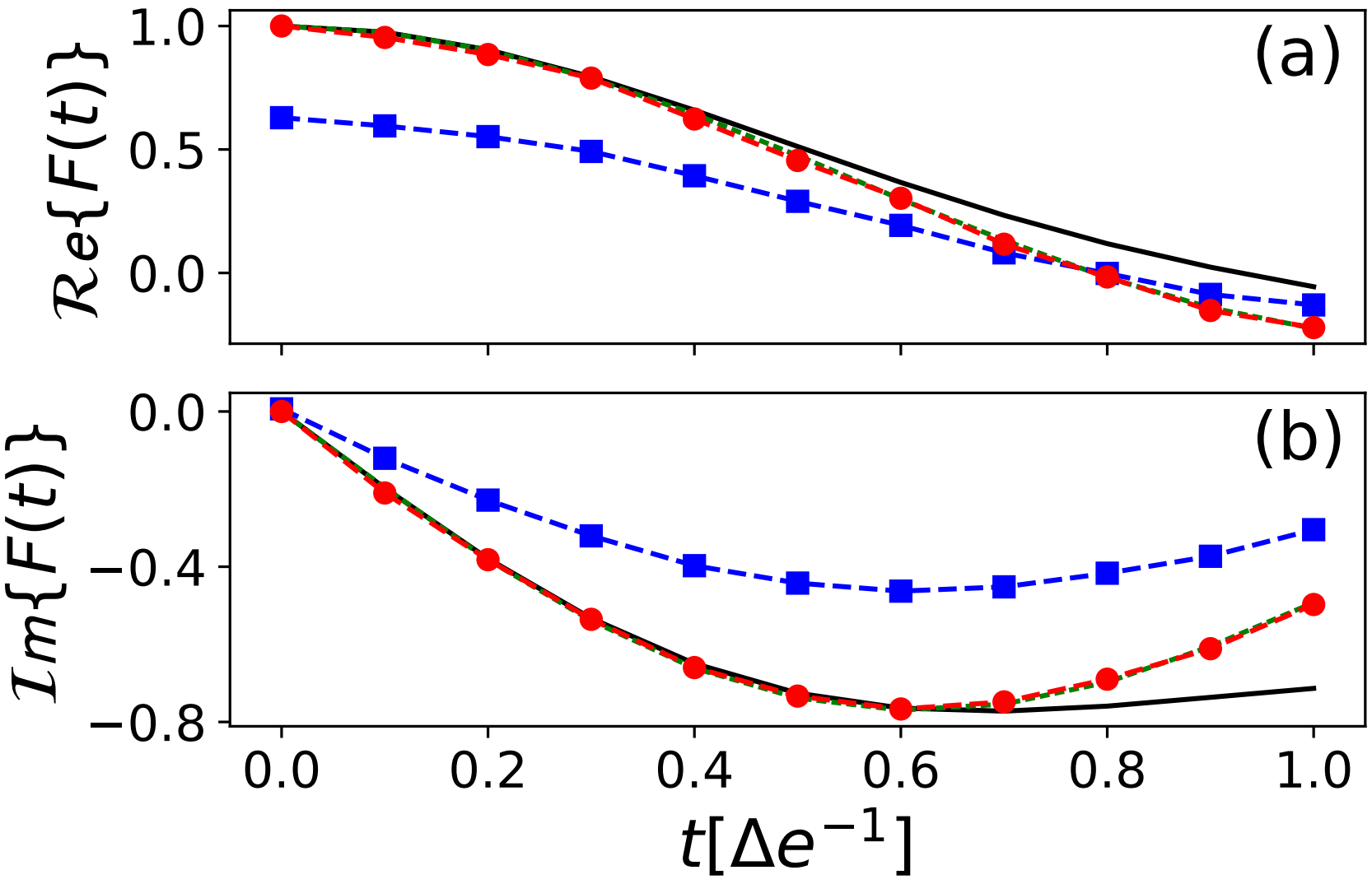

Figure 6.10: Real (a) and imaginary (b) parts of the generating function obtained for the case of a single pair on two levels with spacing $\Delta e = 1$ and coupling strength $g/\Delta e = 1$. The exact result obtained on a classical computer is shown as a solid black line. Results obtained with *FakeSantiago* backend without and with error corrections are shown respectively by blue squares and red circles. We also compare the results obtained with the QASM backend (no noise reference quantum calculation) as a green long-dashed line. In the quantum simulations, each point is obtained using $10^6$ measurements.

Results obtained using the combination of these three error correction techniques are shown with red circles in Fig. 6.10. We see that, with these methods, the error made in the *FakeSantiago* device can be rather accurately corrected. We finally mention that we also tried to apply the same protocol with the real *Santiago* device. Still, the results were noisier than on the fake device, and we could not obtain reasonable corrected results. It is worth noting that these results were obtained at the beginning of the thesis, and significant improvements may have been made to the performance of more recent quantum platforms. Concerning the computation of Hamiltonian moments, the accurate calculation of the generating function used in the FDM introduces another layer of complexity in addition to the challenges previously discussed in section 6.2.4. This extra layer of complication further diminishes the appeal of the FDM-based strategy. An alternative method, which bypasses the calculation of the Hamiltonian moments during Krylov computations, is presented in the following section.

### 6.3 . Quantum Krylov

To circumvent the difficulty of obtaining accurate high-order moments of the Hamiltonian, we explored the Quantum Krylov-based method [183, 184, 185] as an alternative. The starting point for this approach is to construct a reduced basis using the evolution operator of the Hamiltonian directly, which involves creating the following set of states:

$$\left\{|\Phi_0\rangle, e^{-i\tau_1 H}|\Phi_0\rangle, \ldots, e^{-i\tau_{M-1} H}|\Phi_0\rangle\right\}. \tag{6.20}$$

The overlap and Hamiltonian matrix elements in this new basis are constructed as follows:

$$\begin{cases} O_{kl} = \langle \Phi_l | \Phi_k \rangle = \langle e^{-i(\tau_k - \tau_l)H} \rangle_0, \\ H_{kl} = \langle \Phi_l | H | \Phi_k \rangle = \langle H e^{-i(\tau_k - \tau_l)H} \rangle_0. \end{cases} \tag{6.21}$$

We now apply the method used in Section 6.2.2 and detailed in Appendix D to approximate the eigenvalues of the Hamiltonian used in panel (a) of Fig. 6.1, but with $g/\Delta e = 0.5$. The results, using initial Hartree-Fock, Q-PAV, and Q-VAP states from Fig. 3.13, are shown in Fig. 6.11. As with the Krylov method discussed in the previous section, this approach converges to the eigenenergies of the Hamiltonian. Moreover, when combined with the Q-VAP initial state, the method demonstrates rapid convergence towards the ground state energy. However, it should be noted that this improvement comes at the cost of convergence towards the energies of the excited states, where a simple HF state offers better convergence. This behavior can be attributed to the fact that, as observed in Fig. 3.13, the Q-VAP state has almost no components in the excited eigenstates of the Hamiltonian.

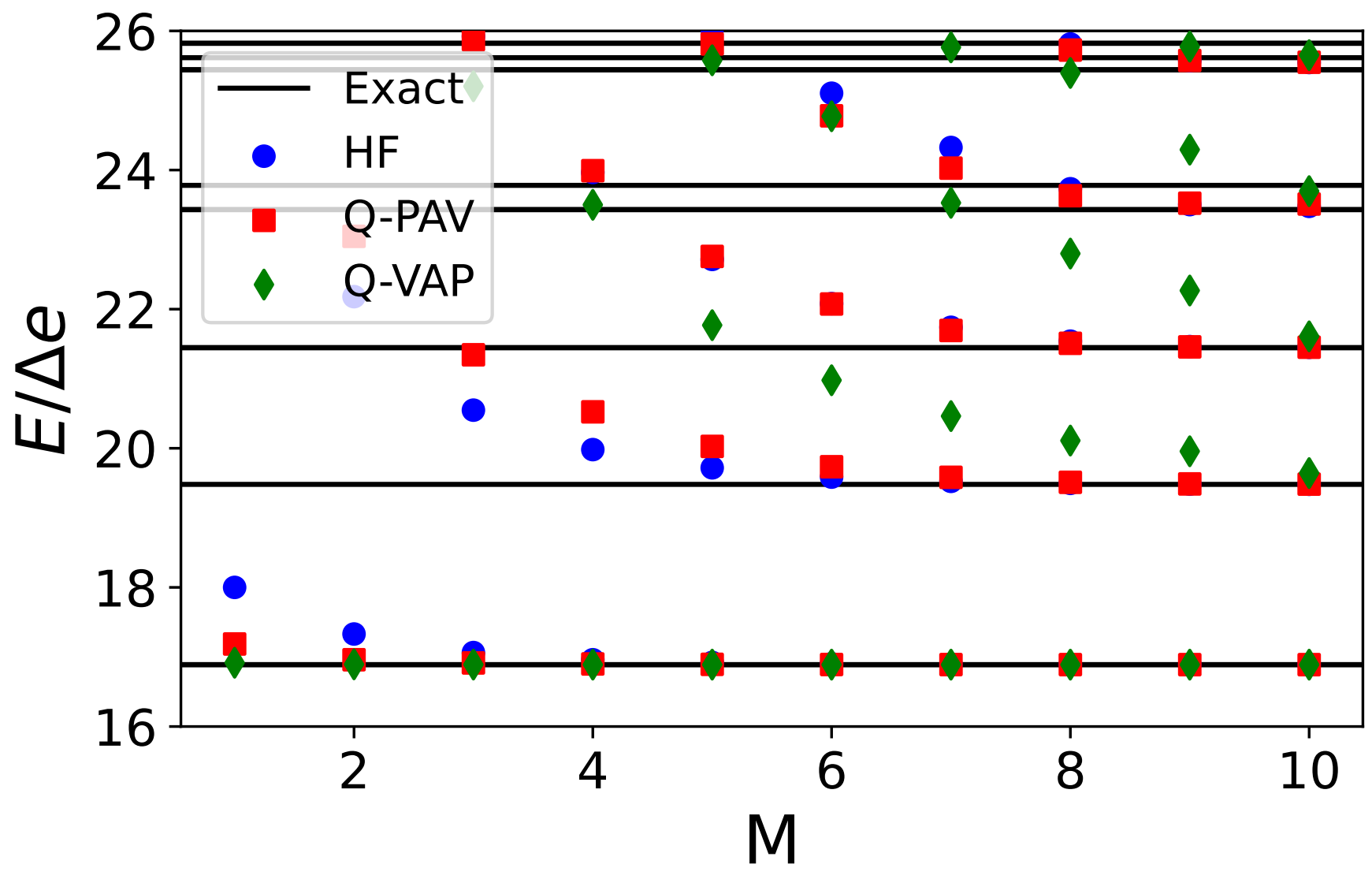

Figure 6.11: Illustration of the energies obtained using the Quantum Krylov method for the pairing problem with $4$ pairs on $8$ levels and $g/\Delta e = 0.5$. The results are derived using the set of times $\tau_i = i\Delta\tau$ for $i = 0, M-1$, and the initial states HF (blue circles), the Q-PAV (red squares), and Q-VAP (green diamonds) used in Fig. 3.13. The approximate energies are plotted as a function of the number of states $M$ on the basis. In this figure, we used $\Delta\tau \cdot \Delta e = 0.3$ and a threshold $\epsilon = 10^{-6}$ for state rejection when diagonalizing the overlap matrix. The black horizontal lines indicate the exact eigenenergies.

### 6.3.1 . Improving Convergence for Excited States

To access the first excited states' energies, one can utilize the knowledge of BCS theory. In the BCS framework, we can construct excited states using quasiparticle (QP) excitations over the BCS ground state in Eq. (3.22). In our case, since we are directly encoding pairs of particles in qubits, the excited states in BCS would correspond to pairs of quasiparticle excitations, i.e., 2QP, 4QP, etc. A $2k$QP quasiparticle excitation takes the form:

$$\begin{aligned} |\psi_{i_1,\dots,i_k}(\boldsymbol{\theta})\rangle = &\bigotimes_{m=1}^{k} \left[-\cos(\theta_{i_m})|0\rangle_{i_m} + \sin_{i_m}(\theta_{i_m})|1\rangle_{i_m}\right] \\ &\bigotimes_{p\neq(i_1,\dots,i_k)} \left[\sin(\theta_p)|0\rangle_p + \cos(\theta_p)|1\rangle_p\right]. \end{aligned} \tag{6.22}$$

These states have the associated energy:

$$E_{i_1,\dots,i_k} = \mathcal{E}_0 + 2\sum_{m=1}^{k} \mathcal{E}_{i_m}, \tag{6.23}$$

where, $\mathcal{E}_0$ is the BCS ground state energy and $\mathcal{E}_i$ is the quasiparticle energy, given by:

$$\mathcal{E}_i = \sqrt{(\varepsilon_i - \lambda)^2 + \Delta^2}, \tag{6.24}$$

with $\lambda$ the Fermi energy and $\Delta$ the pairing gap. All excited states produced by Eq. (6.22) are orthogonal to the BCS ground state in Eq. (3.22). We note that the lowest excited states are obtained by 2QP excitations on the level whose single-particle energy is closer to the Fermi one. To improve the convergence of the excited states using the quantum Krylov method in Fig. (6.11), we first construct the $2k$QP excited state from Eq. (6.22) using the angles obtained via the Q-VAP approach, and then use its projection as the initial state in the Krylov method. The orthogonality between the QP states and the BCS ground state is not maintained after projection. However, it is reasonable to expect that the projections of the QP excitations will exhibit a smaller overlap with the true ground state and a larger overlap with the actual excited states. The QPE algorithm serves as a valuable tool for confirming this hypothesis and for analyzing the projected QP excited states. In Fig. 6.12, we present the results of the QPE approach, starting from states with either 2QP or 4QP excitations. This figure demonstrates an increase in the excited-state components for the multiple QP excitations.

An illustration of the results achieved with the quantum Krylov approach applied to the projected 2QP state in Fig. 6.12(a) is depicted in Fig. 6.13. We see that the convergence towards the ground state deteriorates while the convergence towards the low-lying excited states' energies improves. We conducted systematic studies by varying the number and/or the levels of the QP excitations. In all cases, we observed an enhancement in the convergence of the excited states' energies compared to the case without QP excitations. However, it is important to note that predicting which states' energies will converge faster based on the type of excitations remains challenging.

### 6.3.2 . Comparison Between the QPE Algorithm and the Quantum Krylov Method

We can also compare the Quantum Krylov approach with the QPE method described in section 2.1.3 by examining the total time evolution required in each algorithm versus their respective approximations of the ground state energy. The results shown in Fig. 6.14 indicate that, generally, the quantum Krylov method is able to approximate the ground state energy using far fewer resources than the QPE. In particular, if the initial state has been properly prepared beforehand, for example, by performing a Q-VAP method, the quantum Krylov approach can rapidly increase the ground state energy estimation accuracy. However, it is important to note that the number of eigenenergies approximated by the Krylov-based method increases linearly with the subspace size. In contrast, with the QPE algorithm, we can access, in principle, all the eigenenergies, provided that enough ancilla qubits have been used.

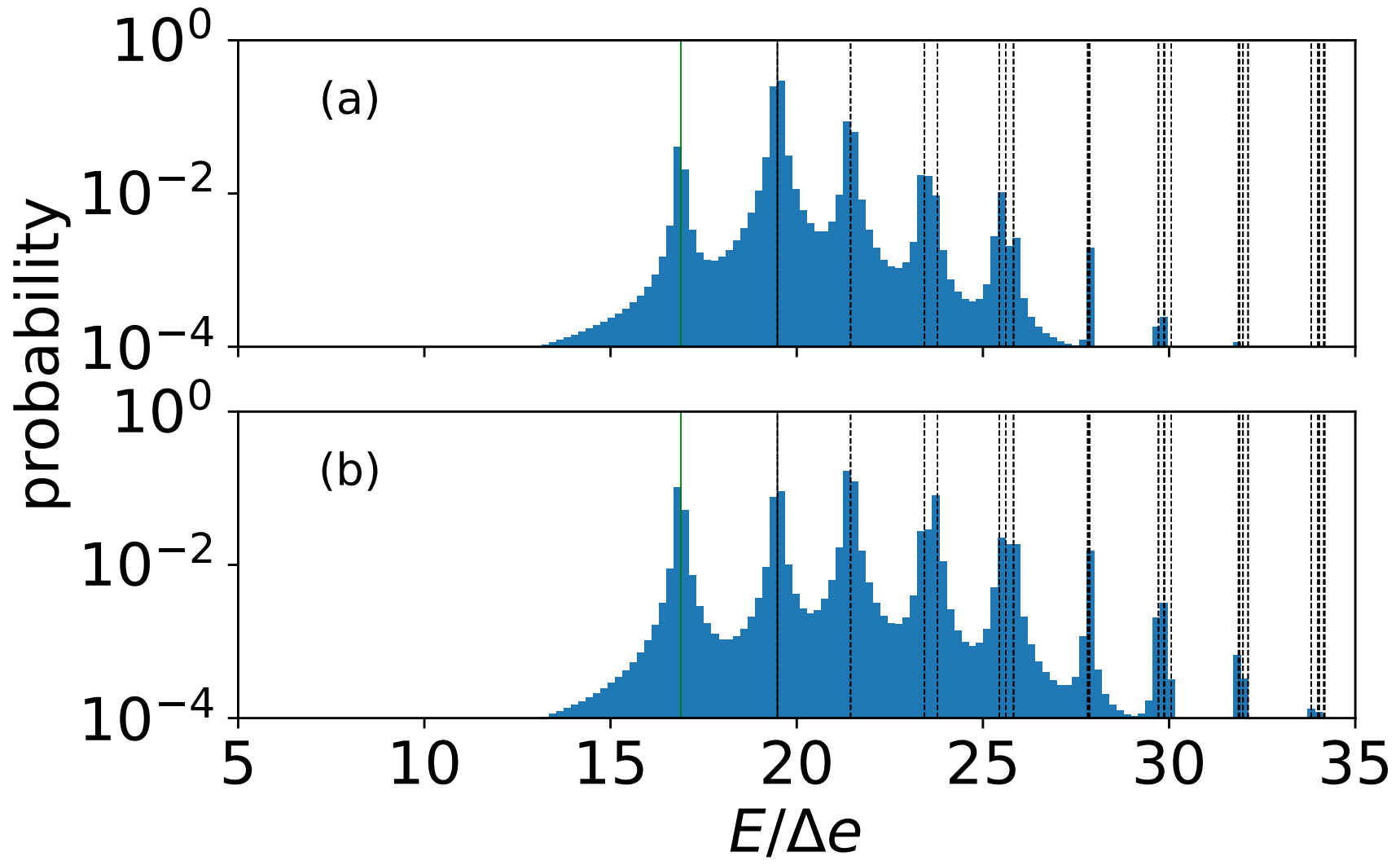

Figure 6.12: QPE algorithm with 8 ancillary qubits applied to the QP states in Eq. (6.22) using the pairing Hamiltonian with the same parameters as in Fig. (6.11). Panels (a) and (b) display results for 2QP and 4QP states, respectively. These figures can be compared to Fig. 3.13, where the QPE results for the BCS ground state are shown. The 2QP state corresponds to a QP excitation in the 4th single-particle level (i.e., $i_1 = 3$), while the 4QP state involves excitations in the 4th and 5th single-particle levels (i.e., $i_1, i_2 = 3, 4$). The green lines represent the ground state energy, the solid black line indicates the first excited state energy, and the black dotted lines show the remaining energies in the spectrum.

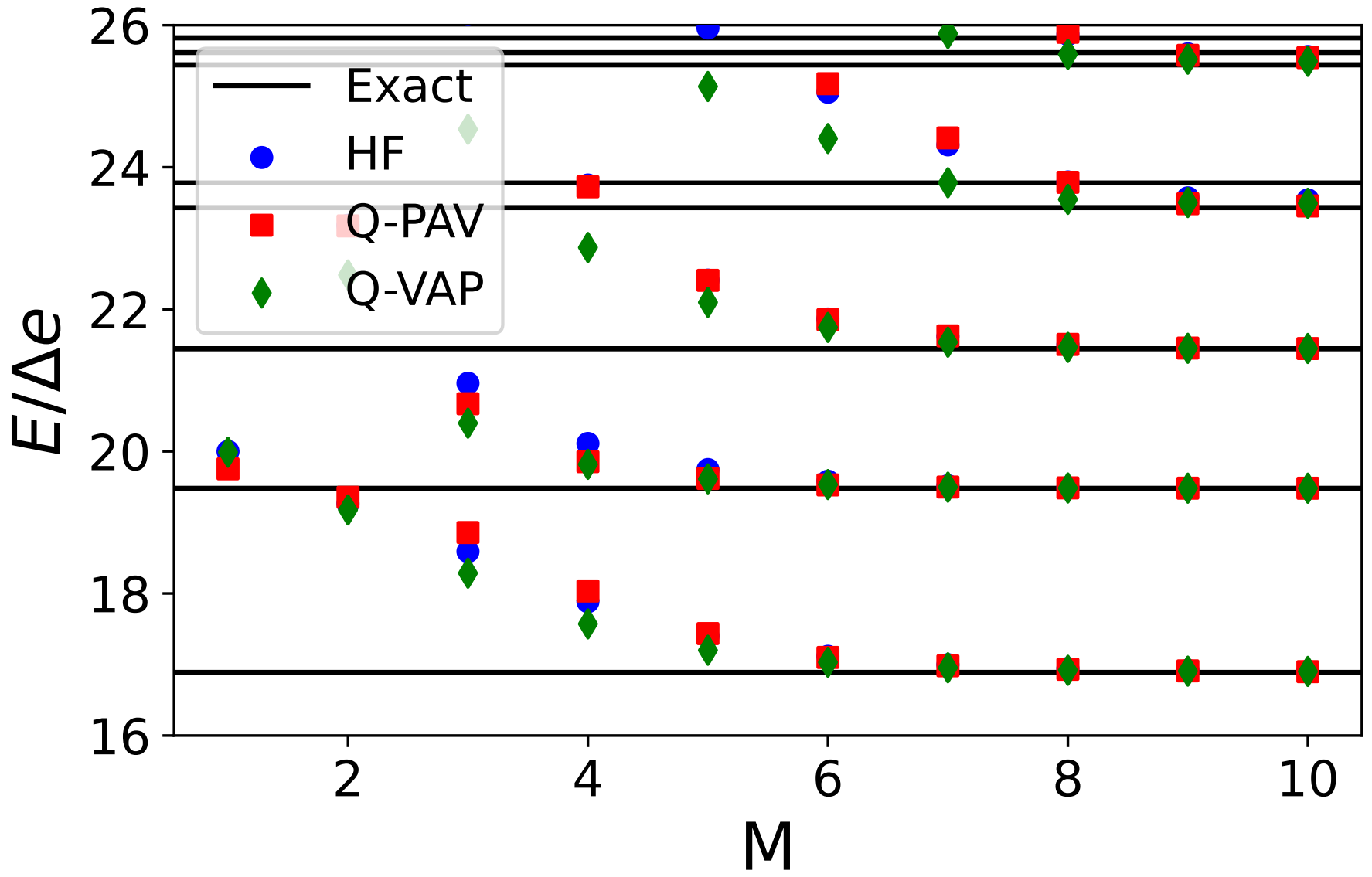

Figure 6.13: Same as in Fig. 6.11 but using as initial state the 2QP excited state used in Fig. 6.12(a).

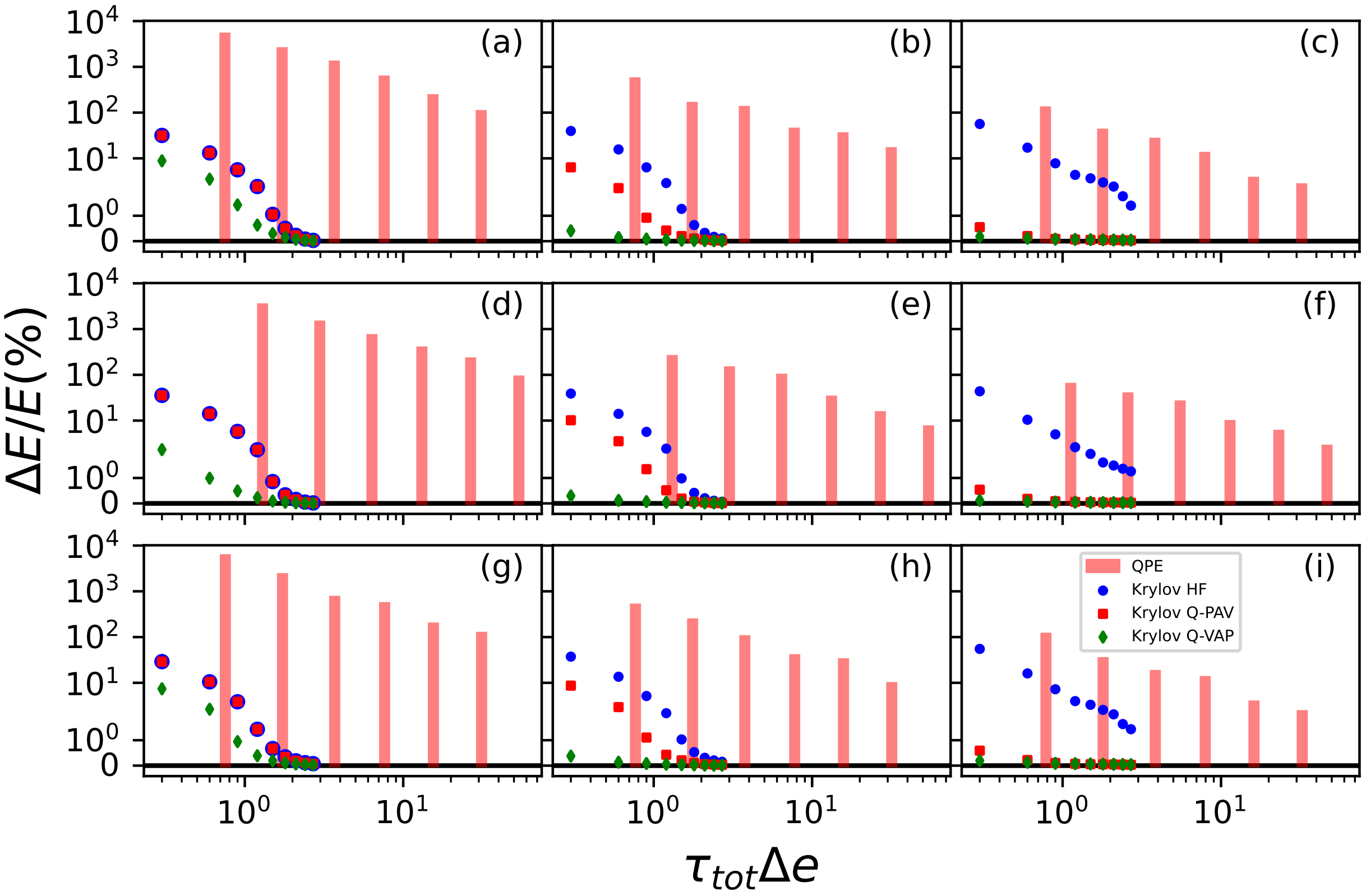

Figure 6.14: Illustration of the total time evolution $\tau_{tot}$ required to achieve a specific precision for the ground-state energy $\Delta E/E\,(\%)$ (Eq. (3.25)) using the quantum Krylov method with HF (blue circle), Q-PAV (red square), and Q-VAP (green diamond) initial states. These times are compared to the time necessary for the QPE method with $n_{QPE} = 3, 4, \ldots, 8$. The red boxes shown for the QPE correspond to the uncertainty in the ground-state energy deduced from the bin size of the main peak (see, for instance, Fig. 3.13). The left, middle, and right columns are obtained using different interaction strengths, $g/\Delta e = 0.2, 0.5$, and $1$, respectively. When $g < 0.3$, the system has not yet undergone a quantum phase transition; therefore, the HF and Q-PAV results are identical. Panels (a)–(c) are obtained for four pairs on eight doubly degenerated equidistant levels. Panels (d)–(f) are for two pairs on eight equidistant levels, and panels (g)–(i) are the same as panels (a)–(c) but with non-equidistant levels. In the latter case, each single-particle energy is shifted by a small random number.

## 6.4 . Conclusion

In this chapter, we presented and tested a variety of classical post-processing methods to extract information about the low-lying spectra of a Hamiltonian. Initially, we introduced the generating function of Hamiltonian moments with the goal of using short-time evolution on a quantum computer to access the aforementioned moments $\langle H^k \rangle$ and subsequently perform different post-processing methods. We showcased the process of obtaining the generating function on a real IBM quantum computer. At the time of using the quantum computing platform, it was challenging to derive useful results, so we illustrated some error mitigation techniques on a noisy simulator. Assuming accurate access to $\langle H^k \rangle$, we demonstrated how to enhance the ground state's accuracy using the $t$-expansion method, gather information about low-lying excited states' energies with the Krylov method, and approximate the long-time evolution of the survival probability.

Later, we discussed the challenges faced when attempting to obtain the moments $\langle H^k \rangle$ using the finite difference method on the generating function. In this section, we highlighted some methods employed to overcome these challenges and concluded that only those based on the Fourier transform could accurately approximate $\langle H^k \rangle$, rendering the computation of moments via the generating function less appealing. We then presented a quantum Krylov-based method that circumvents the computation of moments and improves the approximation of excited states' energies when combined with the BCS theory. Finally, we compared the performance of this method to the QPE algorithm in terms of the total time evolution required to achieve a certain accuracy.

# 7 - Conclusion and Outlook

This thesis primarily centers on incorporating the symmetry-breaking (SB)/symmetry-restoring (SR) technique into the quantum computing framework with the objective of describing precisely many-body interacting systems. This inclusion was achieved by formulating various projection methods, drawing upon the principles of the Quantum Phase Estimation (QPE) algorithm, the notion of a quantum Oracle, and the Classical Shadows method.

Before delving into the SB/SR method, the first chapter presents an overarching introduction to the quantum computing framework, which forms the foundation for the rest of the thesis. This chapter substantiates the use of quantum heuristics such as the Variational Quantum Eigensolver (VQE) algorithm by exploring the complexity class of the ground state estimation of the $k$-local Hamiltonian (defined in section 2.2.1). Within this context, our work on integrating the SB/SR method is positioned. Furthermore, as we conclude this chapter, we outline the treatment of the pairing and Hubbard Hamiltonians on a quantum computer. This treatment employs the Jordan-Wigner transformation to encode the physical problem and convert the fermionic operators into qubit equivalents. For a comprehensive picture, we also introduce other encoding types–namely, the parity and Brayvi-Kitaev ones.

The second chapter provides an in-depth exploration of the SB/SR method's implementation. It discusses the method's physical interpretation and its proficiency in capturing correlations more economically than a conventional symmetry-conserving framework. In the context of quantum computing, the SB/SR technique is realized through a fusion of the standard VQE algorithm with a projective operation. Building on this foundation, we delineate two strategies for ground state estimation, contingent on whether the parameters of the ansatz are altered before or after the projection. The first approach, wherein the projection follows the variation, we termed Quantum Projection After Variation (Q-PAV). Conversely, the second scenario, where the variation is subsequent to the projection, is denoted as Quantum Variation After Projection (Q-VAP). Three QPE-based projection methods with varying computational costs are introduced, utilizing the QPE, the Iterative QPE, and the Rodeo algorithms. These methods harness the potential of quantum circuits to filter a quantum state into the eigenspace associated with a specific eigenvalue. By contrast, "Oracle"-based methods, discussed in the third chapter, leverage an Oracle operator— a black-box operator that can distinguish the desired projection subspace from the rest of the Hilbert space. This is accomplished by applying a constant phase to the states within that subspace. Using this principle, we implemented four strategies for symmetry restoration: Oracle+Hadamard, Implicit projection, Grover/Hoyer, and an additional method hinging on the Linear Combination of Unitaries (LCU) algorithm. Considering the significant resources demanded by a general Oracle operator's implementation, our study finds that QPE-based methods, particularly the Iterative QPE approach, offer better efficiency in terms of quantum resource utilization compared to methods reliant on an Oracle.

A noteworthy observation in the third chapter is that the probability of implementing the Oracle using the LCU algorithm scales inversely with the number of qubits, whereas the probability of implementation of the projector through the same algorithm directly corresponds to the probability of the wavefunction being measured in the "Good" state. This insight may be advantageous in situations where the initial state has a significantly small component in the "Good" space.

Chapter four introduces a distinct approach to symmetry-restoration, crafted in response to the substantial constraints currently posed by Noisy Intermediate Scale Quantum (NISQ) platforms. It elucidates how, within the "Classical Shadows" paradigm, we can devise a projection technique for observables, such as the energy, limited solely to the post-processing or classical phase. Notably, this method significantly diminishes the utilization of quantum resources by increasing the use of classical ones. Moreover, we demonstrate how to optimize these computations by leveraging the orthogonality of the Pauli matrices and expanding the established derandomization procedure.

Lastly, the fifth chapter delves into various post-processing techniques that build on the ground state approximations yielded by the Q-PAV and Q-VAP methods. These techniques aim to either enhance the precision of the ground state energy or acquire information about the low-lying spectrum of a Hamiltonian. Certain strategies, such as the $t$-expansion and Krylov method, rely on the precise extraction of the Hamiltonian moments $\langle H^k \rangle$ via a finite difference method on the derivatives of the generating function computed on a quantum computer. Recognizing the challenges associated with accurately estimating these moments for larger orders of $k$, we pivoted our approach to introduce the Quantum Krylov method. This technique, akin to the traditional Krylov method, allows us to glean information about a Hamiltonian's low-lying spectrum directly from the values of the generating function, circumventing the need for the Hamiltonian moments' computation.

This thesis primarily concentrates on a specific topic—the symmetry-breaking problem in complex many-body systems in the context of using quantum computers. Prior to the research undertaken for this thesis, this area remained largely uncharted. Over the past few years, we have demonstrated the utility of a range of algorithmic tools in restoring disrupted symmetries during quantum simulations. While the strategies deployed by these methods to reach the same objective vary substantially, most stem from standard concepts developed throughout the history of quantum computing, such as the QPE algorithm, the Oracle concept, and quantum state tomography. These developments reflect the availability of a broad array of tools for accomplishing symmetry restoration within the SB/SR technique. In the realm of quantum computing, these tools might give access to results difficult for classical computers to reach. Furthermore, the integration of the SB/SR method into classical procedures, like many-body perturbation theory and coupled clusters, is an active area of research. This Ph.D. work could potentially be seen as one of the pioneering efforts toward realizing the possibility of implementing these classical methods within an SB/SR framework, all within a hybrid classical-quantum computing paradigm.

Owing to its adaptability, the SB/SR technique can be employed to estimate the ground state of any Hamiltonian that respects symmetries. This versatility makes it a powerful addition to the VQE algorithm, enabling it to achieve higher precision in ground-state approximation while reducing computational resources. Moreover, symmetries could also be disrupted by noise. Although not within the scope of this thesis, investigating the potential usefulness of the tools discussed herein for error mitigation could prove worthwhile. Even though certain methods may require additional resources, their collective diversity could form a comprehensive toolbox for error mitigation. This broad array of options allows for more customized solutions, optimizing error mitigation strategies to better align with specific algorithmic frameworks.

Another area of potential interest for the hybrid SB/SR framework developed in this thesis is quantum machine learning, which encompasses machine learning algorithms developed to harness the unique capabilities of quantum computing. These algorithms could potentially benefit from the SB/SR method, for instance, by improving the training of machine learning-related ansätze when they have symmetries to be respected. In addition, the classical shadows procedure, combined with the post-processing SR step, might allow a reduction in the computational resources required for extracting relevant data during the tomography process of a useful quantum state for quantum chemistry calculations. This approach would allow the extraction of observables only from a relevant part of the Hilbert space. Fundamentally, the SR step can be seen as a means to extract information about which states fulfill a particular condition and as a way to extract observables using only these states. This could have useful applications in Oracle-related quantum algorithms, such as search algorithms and others built upon these.

# Appendices

# A - Specificities of the Bravyi-Kitaev Transformation (BKT)

To describe in detail how to perform the BKT, we can start by constructing the transformation matrix from the Fock space to the qubit space shown in Fig. A.1.

$$\beta_{2^{x+1}} = \left(\begin{array}{c|c} \beta_{2^x} & 0 \\ \hline 0 & \\ \leftarrow 1 \rightarrow & \beta_{2^x} \end{array}\right) \qquad \beta^{-1}_{2^{x+1}} = \left(\begin{array}{c|c} \beta^{-1}_{2^x} & 0 \\ \hline 0 & \\ \leftarrow 0 \quad 1 & \beta^{-1}_{2^x} \end{array}\right)$$

Figure A.1: The matrix $\beta_n$ transforms fermionic states, represented as vectors $(b_0, b_1, \ldots, b_{N-1})^T$, into vectors in the Brayvi-Kitaev basis. To obtain $\beta_n$, we take the $(n \times n)$ matrix from the top-left corner of matrix $\beta_{2^{x+1}}$ for $2^x < n \leq 2^{x+1}$. The initial value is $\beta_1 = [1]$, and a row of ones in the bottom left quadrant is indicated by $\leftarrow 1 \rightarrow$. We also show the recursion pattern to construct the inverse transform $\beta^{-1}_{2^{x+1}}$. All the entries in the left-bottom quadrant of the inverse transform are zeros except for the rightmost bottom entry, which is 1.

Now, it is useful to define several subsets of the qubits, which contain the information needed to apply fermionic operators to the state. These sets are defined below:

1. The update set, $U(i)$, is the set of qubits that must be updated when the occupation of some orbital $i$ is changed. Specifically, $U(i)$ includes the qubits that store partial sums that depend on the occupation of orbital $i$.

2. The parity set, $P(j)$, is the set of qubits necessary to determine the parity of the set of orbitals with index $< i$.

3. The flip set, $F(i)$, is the set of qubits that determine whether qubit $i$ and orbital $i$ have equal or opposite occupation. Specifically, the qubits in $F(i)$ hold the parity of the occupation of orbitals with indices less than $i$ that contribute to the partial sum stored in qubit $i$.

4. Since the flip set is a subset of the parity set, it is convenient to define the complement of the flip set in the parity set, the remainder set $R(i) \equiv P(i) \setminus F(i)$.

It is noteworthy that all sets linked to qubit $i$ are defined so as to ensure that this qubit never constitutes a member of any of these sets. In the JWT, we have $U(i) = F(i) = \emptyset$ and $P(i) = \{j | j < i\}$, on the other hand, for the parity encoding we have $U(i) = \{j | j > i\}$ and $F(i) = P = (i) = \{i - 1\}$. For the BKT, the sets are filled as follows:

1. $U$: To determine the qubits that belong to the update set for a given orbital $j$, we can examine the column of the transformation matrix $\beta_n$ corresponding to $j$. Since $\beta_n$ is a lower diagonal matrix, only qubits with indices greater than $j$ will be contained in $U(j)$. Thus, the update set will be composed of the non-zero entries of the column $j$ such that $i > j$, i.e., $U(j) = \left\{ i | i > j \text{ and } [\beta_n]_{ij} \neq 0 \right\}$. Notice that we use $j$ to refer to the fermionic orbitals and $i$ to refer to the qubits, following the standard convention in the matrix $\beta_n$. In the BKT, the qubits with even indices represent the occupation state of orbital $j$, while the qubits with odd indices store partial sums. Therefore, the update set for orbital $j$ will only contain odd qubit indices.

2. $P$: To determine the parity set of a given orbital, we need to examine the rows of the matrix $\pi_n \beta_n^{-1}$, where $\pi_n = \Pi_n - I^{\otimes n}$ and $\Pi$ is given in Eq. (2.28). Here $[\pi_n]_{ij}$ is the transformation matrix that maps the fermionic space to the parity space spanned by $\{|p_i\rangle\}^{\otimes n}$, where each $p_i$ encodes the parity of the set of orbitals $\{j\}$ such that $j < i$. Thus, the rows of $\left[\pi_n \beta_n^{-1}\right]_i j$ indicate which set of qubits are involved in storing the parity $p_i$. Since the matrix $\pi_n \beta_n^{-1}$ is lower-diagonal, we can define the parity set as $P(i) = \left\{ j | j < i \text{ and } \left[\pi_n \beta_n^{-1}\right]_{ij} \neq 0 \right\}$.

3. $F$: Finally, to determine the flip set of a given orbital, we need to examine the rows of the matrix $\left[\beta_n^{-1}\right]_{ij}$. The non-zero entries of row $i$ indicate the indices of the qubits that encode the occupation of orbital $i$. However, by definition, the flip set excludes the qubit associated with the orbital itself (i.e., $i = j$). Thus, by using that $\beta_n^{-1}$ is a lower diagonal matrix, we can define the flip set as $F(i) = \left\{ j | j < i \text{ and } \left[\beta_n^{-1}\right]_{ij} \neq 0 \right\}$. As qubits with even indices $j$ hold the occupation of orbital $i = j$, the flip set is empty for all even $j$.

Fig. A.2 shows the matrices involved in defining the elements of the sets for $n = 8$. It is worth noting that [116] gives a way to recursively define the update, parity, and flip sets without explicitly constructing the matrices.

$$\beta_8 = \begin{pmatrix} 1&0&0&0&0&0&0&0\\ 1&1&0&0&0&0&0&0\\ 0&0&1&0&0&0&0&0\\ 1&1&1&1&0&0&0&0\\ 0&0&0&0&1&0&0&0\\ 0&0&0&0&1&1&0&0\\ 0&0&0&0&0&0&1&0\\ 1&1&1&1&1&1&1&1 \end{pmatrix} \quad \pi_8\beta_8^{-1} = \begin{pmatrix} 0&0&0&0&0&0&0&0\\ 1&0&0&0&0&0&0&0\\ 0&1&0&0&0&0&0&0\\ 0&1&1&0&0&0&0&0\\ 0&0&0&1&0&0&0&0\\ 0&0&0&1&1&0&0&0\\ 0&0&0&1&0&1&0&0\\ 0&0&0&1&0&1&1&0 \end{pmatrix} \quad \beta_8^{-1} = \begin{pmatrix} 1&0&0&0&0&0&0&0\\ 1&1&0&0&0&0&0&0\\ 0&0&1&0&0&0&0&0\\ 0&1&1&1&0&0&0&0\\ 0&0&0&0&1&0&0&0\\ 0&0&0&0&1&1&0&0\\ 0&0&0&0&0&0&1&0\\ 0&0&0&1&0&1&1&1 \end{pmatrix}$$

$U(5)$ $P(5)$ $F(5)$

Figure A.2: Transformation matrices used to get the different sets for the Brayvi-Kitaev transformation for $n = 8$, i.e., $x = 2$. The update, parity, and flip set are obtained by using $\beta_8$, $\pi_8\beta_8^{-1}$ and $\beta_8^{-1}$ respectively. All sums in the matrix multiplications are modulo 2. As an example, we highlight in red the matrix elements that play a role in determining the elements of the different sets for the 5$^{\text{th}}$ orbital. The complete listing of elements of each set for $n = 8$ is provided in Table A.1.

**Brayvi-Kitayev operators** With the update, parity, and flip sets of the Brayvi-Kitaev transformation, we can define the transformation of the fermionic creation and annihilation operators. For

| Set | Qubit index | | | | | | | |
|---|---|---|---|---|---|---|---|---|
| | 0 | 1 | 2 | 3 | 4 | 5 | 6 | 7 |
| $U(i)$ | $\{1,3,7\}$ | $\{3,7\}$ | $\{3,7\}$ | $\{7\}$ | $\{5,7\}$ | $\{7\}$ | $\{7\}$ | $\emptyset$ |
| $P(i)$ | $\emptyset$ | $\{0\}$ | $\{1\}$ | $\{1,2\}$ | $\{3\}$ | $\{3,4\}$ | $\{3,5\}$ | $\{3,5,6\}$ |
| $F(i)$ | $\emptyset$ | $\{0\}$ | $\emptyset$ | $\{1,2\}$ | $\emptyset$ | $\{4\}$ | $\emptyset$ | $\{3,5,6\}$ |

Table A.1: Elements of the update set $U(i)$, parity set $P(i)$ and flip set $F(i)$ for the $n = 8$ Brayvi-Kiatev transformation.

even-indexed qubits, this is relatively simple, as they only store the occupation of their corresponding orbital, i.e., $q_j = b_j$. Thus, we only need to apply $X$ gates to the qubits in the update set and $Z$ gates to the ones in the parity set:

$$
\begin{aligned}
c_\alpha^\dagger &= X_{U(\alpha)} \otimes Q_\alpha^+ \otimes Z_{P(\alpha)}, \\
c_\alpha &= X_{U(\alpha)} \otimes Q_\alpha^- \otimes Z_{P(\alpha)}.
\end{aligned}
\tag{A.1}
$$

with $Q^\pm$ the qubit creation/annihilation operators as defined for Eq. (2.27). Given that the flip set is not empty for odd indices, the occupation of the qubit might be flipped from that of the electronic state. Taking that into account, the transformation of the operators, in this case, is given by:

$$
\begin{aligned}
c_\alpha^\dagger &= X_{U(\alpha)} \otimes \Theta_\alpha^+ \otimes Z_{R(\alpha)}, \\
c_\alpha &= X_{U(\alpha)} \otimes \Theta_\alpha^- \otimes Z_{R(\alpha)}.
\end{aligned}
\tag{A.2}
$$

where $\Theta_\alpha^\pm = Q^\pm \otimes E_{F(\alpha)} - Q^\mp \otimes O_{F(\alpha)}$ and $E_S$(resp. $O_S$) is the projector over the even (resp. odd) states of a set $S$. These projectors are defined as:

$$
E_S = \frac{1}{2}\left(I_S + Z_S\right), \qquad O_S = \frac{1}{2}\left(I_S - Z_S\right),
$$

where $I_S$ (resp. $Z_S$) corresponds to the tensor product of the matrix $I$ (resp. $Z$) over the qubits that belong to $S$. By defining the set $\rho(i)$ as:

$$
\rho(i) = \begin{cases} P(i) & \text{if } i \text{ even}, \\ R(i) & \text{if } i \text{ odd}, \end{cases}
$$

we can summarize the transformation relations of the fermionic operators as:

$$
\begin{aligned}
c_\alpha^\dagger &= \frac{1}{2}\left(X_{U(\alpha)} \otimes X_\alpha \otimes Z_{P(\alpha)} - iX_{U(\alpha)} \otimes Y_\alpha \otimes Z_{\rho(\alpha)}\right), \\
c_\alpha &= \frac{1}{2}\left(X_{U(\alpha)} \otimes X_\alpha \otimes Z_{P(\alpha)} + iX_{U(\alpha)} \otimes Y_\alpha \otimes Z_{\rho(\alpha)}\right).
\end{aligned}
\tag{A.3}
$$

Table A.2 presents the transformation of common operators in the many-body Hamiltonian to the three different encodings.

| Operator | Transformation | | |
|---|---|---|---|
| | Jordan-Wigner | Parity | Brayvi-Kitaev |
| $c_\alpha^\dagger c_\alpha$ | $\frac{1}{2}\left(I_\alpha - Z_\alpha\right)$ | $\frac{1}{2}\left(I_\alpha I_{\alpha-1} - Z_\alpha Z_{\alpha-1}\right)$ | $\frac{1}{2}\left(I_{F(\alpha)} - Z_{\underline{F(\alpha)}}\right)$ |
| $c_\alpha^\dagger c_\beta^\dagger c_\beta c_\alpha$ | $\frac{1}{4}\left(I_\alpha - Z_\alpha\right)\left(I_\beta - Z_\beta\right)$ | $\frac{1}{4}\left(I_\alpha I_{\alpha-1} - Z_\alpha Z_{\alpha-1}\right)\left(I_\beta I_{\beta-1} - Z_\beta Z_{\beta-1}\right)$ | $\frac{1}{4}\left(I - Z_{\underline{F(\alpha)}} - Z_{\underline{F(\beta)}} + Z_{\underline{F_{\alpha\beta}}}\right)$ |
| $c_\alpha^\dagger c_\beta + c_\beta^\dagger c_\alpha$ | $X_\alpha \bigotimes_{k=\beta+1}^{\alpha-1} Z_k \otimes \left(-X_\beta\right)$ | $\left(Z_\alpha Y_{\alpha-1}\right) \bigotimes_{k=\beta+1}^{\alpha-2} X_k \otimes \left(X_\beta Z_{\beta-1}\right)$ | $\frac{1}{2} X_{U_{\alpha\beta} \setminus \delta_{\alpha\beta}} Y_{\delta_{\alpha\beta}} Z_{P^0_{\alpha\beta \setminus \delta_{\alpha\beta}}} \left(Y_\beta X_\alpha - X_\alpha Y_\beta\right)$ |

Table A.2: In descending order, the transformation of the number, exchange, and excitation operators using the Jordan-Wigner, parity, and Brayvi-Kitaev transformation. For the Jordan-Wigner (resp. parity) calculations, we have assumed $\alpha > \beta$ (resp. $\alpha > \beta + 1$). In the Brayvi-Kitaev transformation, the new sets are defined as $\underline{F(i)} \equiv F(i) \cup \{i\}$, $\underline{F_{ij}} \equiv \underline{F(i)} \bigtriangleup \underline{F(j)} = \left(\underline{F(i)} \cup \underline{F(j)}\right) \setminus \left(\underline{F(i)} \cap \underline{F(j)}\right)$, $U_{ij} \equiv U(i) \bigtriangleup U(j)$, $\delta_{ij} \equiv U(i) \cap P(j)$ and $P_{ij}^0 \equiv P(i) \bigtriangleup P(j)$. $U(i)$, $P(i)$, and $F(i)$ are the sets used for the Brayvi-Kitaev transformation (see section 2.2.2).

# B - Derivation of the General Rotation $R\left(\lambda\right)$

The conditions so that the $G$ matrix in Eq (4.5) can be written in the form of the $F$ matrix Eq. (4.6) are:

$$G = F\begin{pmatrix} -\left[\left(1-e^{i\phi}\right)g+e^{i\phi}\right] & \left(1-e^{i\phi}\right)\sqrt{g(1-g)}e^{i\varphi} \\ \left(1-e^{i\phi}\right)\sqrt{g(1-g)} & \left[\left(1-e^{i\phi}\right)g-1\right]e^{i\varphi} \end{pmatrix} \tag{B.1}$$

$$= \begin{pmatrix} \cos(\lambda)e^{iv} & -\sin(\lambda)e^{i(v-u)} \\ \sin(\lambda)e^{i(v+u)} & \cos(\lambda)e^{iv} \end{pmatrix}. \tag{B.2}$$

We note that the terms in the diagonal of the $F$ matrix are equal, imposing this condition on $G$ (left side of Eq (B.2)) gives:

$$-\left[\left(1-e^{i\phi}\right)g+e^{i\phi}\right] = \left[\left(1-e^{i\phi}\right)g-1\right]e^{i\varphi},$$

$$\left(e^{i\phi}-1\right)g\left(e^{i\varphi}+1\right) = e^{i\phi}-e^{i\varphi}.$$

If we divide by $2ie^{i(\phi+\varphi)/2}$ on both sides we get:

$$2g\sin\left(\frac{\phi}{2}\right)\cos\left(\frac{\varphi}{2}\right) = \sin\left(\frac{\phi-\varphi}{2}\right).$$

Using the identity $\sin\left(x-y\right) = \sin\left(x\right)\cos\left(y\right) - \sin\left(y\right)\cos\left(x\right)$ we get:

$$\sin\left(\frac{\varphi}{2}\right)\cos\left(\frac{\phi}{2}\right) = \sin\left(\frac{\phi}{2}\right)\cos\left(\frac{\varphi}{2}\right)\left(1-2g\right),$$

or, equivalently:

$$\tan\left(\frac{\varphi}{2}\right) = \tan\left(\frac{\phi}{2}\right)\left(1-2g\right). \tag{B.3}$$

Now, if we equal the lower left block of both sides in Eq (B.2):

$$\left(1-e^{i\phi}\right)\sqrt{g(1-g)} = \sin(\lambda)e^{i(v-u)}.$$

If we take the module on both sides, we get the following:

$$|1-e^{i\phi}| = \frac{|\sin\left(\lambda\right)|}{\sqrt{g(1-g)}}.$$

$\lambda$ can be seen as the angle of the general rotation we would like to implement. As the angle between $|\Psi\rangle$ and $|\psi_g\rangle$ always lies in the interval $[0, \frac{\pi}{2}]$, we would like to implement rotations in this interval. Thus we have $|\sin\left(\lambda\right)| = \sin\left(\lambda\right)$. The previous equation becomes:

$$\frac{\sin\left(\lambda\right)}{\sqrt{g(1-g)}} = \sqrt{\left(1-e^{i\phi}\right)\left(1-e^{-i\phi}\right)},$$

$$2\left(1-\cos\left(\phi\right)\right) = \frac{\sin^2\left(\lambda\right)}{g(1-g)},$$

$$\cos\left(\phi\right) = 1 - \frac{\sin^2\left(\lambda\right)}{2g(1-g)}. \tag{B.4}$$

We should notice that Eq (B.4) imposes a condition on the choice of $\lambda$. Indeed, given that $-1 \leq \cos(\phi) \leq 1$, we have:

$$-1 \leq 1 - \frac{\sin^2(\lambda)}{2g(1-g)} \leq 1,$$
$$0 \leq \sin^2(\lambda) \leq 4g(1-g).$$

As $g = \sin^2(\theta)$ we get:

$$0 \leq \sin(\lambda) \leq \sin(2\theta).$$

As $\arcsin(x)$ is increasing in the interval $[0, \pi/2]$, we get that:

$$0 \leq \lambda \leq 2\theta. \tag{B.5}$$

Dividing the equation for the upper-right block's equality by the equation for the lower-left one in Eq. (B.2), we obtain the following equation:

$$e^{i\varphi} = -e^{2iu},$$
$$e^{i\varphi} = e^{i(\pi-2u)}.$$

This leads to the relation:

$$u = \frac{\pi - \varphi}{2}. \tag{B.6}$$

To summarize, if we want to implement a rotation of angle $\lambda$ given that we already know $g$, we use Eq (B.4) to extract $\phi$, then Eq (B.3) to get $\varphi$, after Eq (B.6) to get $u$, and finally, we can use the equation of the upper left entry in Eq (B.2) to get $v$:

$$e^{iv} = -\frac{1}{\cos(\lambda)} \left[ \left(1 - e^{i\phi}\right) g + e^{i\phi} \right],$$
$$v = Arg\left( -\frac{1}{\cos(\lambda)} \left[ \left(1 - e^{i\phi}\right) g + e^{i\phi} \right] \right), \tag{B.7}$$

where $Arg$ denotes the argument function. Applying all the presented conditions, the rotation matrix of angle $\lambda$ can then be written as:

$$R(\lambda) = \begin{pmatrix} \cos(\lambda) & -\sin(\lambda) \\ \sin(\lambda) & \cos(\lambda) \end{pmatrix} = e^{-iv} \begin{pmatrix} 1 & 0 \\ 0 & e^{-iu} \end{pmatrix} G(\phi, \varphi) \begin{pmatrix} 1 & 0 \\ 0 & e^{iu} \end{pmatrix}.$$

for an angle $\lambda \leq 2\theta$.

# C - Success Probability Using the LCU Method

We define $p_S$ as the success probability corresponding to the probability of measuring only 0s in the ancillary qubits. We would like to maximize this probability while minimizing the quantum resources required to implement the LCU. As discussed in Section 4.4.2 and noted in the literature [155, 158], there is some flexibility in the choice of the operators $B$ and $E$ depicted in Fig. 4.4. In this context, we consider three possible choices for these operators, referred to as LCU-B, LCU-H, and LCU-L (we assume again that the operator to be applied to the state is $A = \sum_{k=0}^{m-1} \alpha_k V_k$):

1. **LCU-B:** In this case, the operator $B$ initializes the LCU register as:

$$B|0\rangle^{\otimes n_{LCU}} = \frac{1}{\mathcal{N}_B} \sum_{k=0}^{m-1} \sqrt{\alpha_k}|k\rangle, \tag{C.1}$$

   with $\mathcal{N}_B = \sqrt{\sum_{k=0}^{m-1} \alpha_k}$. Here, we used the condition $\alpha_k \geq 0$ and $\alpha_k \in \mathbb{R}$ for all $k$. This prescription is used together with $E = B^\dagger$ and leads to the success probability:

$$p_S^B = \frac{\|A|\Psi\rangle\|^2}{\mathcal{N}_B^4}. \tag{C.2}$$

2. **LCU-H:** This case is the one presented in Ref. [158] where:

$$B|0\rangle^{\otimes n_{LCU}} = \frac{1}{\mathcal{N}_H} \sum_{k=0}^{m-1} \alpha_k|k\rangle, \tag{C.3}$$

   with $\mathcal{N}_H = \sqrt{\sum_{k=0}^{m-1} |\alpha_k|^2}$. A first possible choice for $E$ is $\bigotimes H^{n_{LCU}}$. The corresponding success probability is given by:

$$p_S^H = \frac{\|A|\Psi\rangle\|^2}{2^{n_{LCU}} \mathcal{N}_H^2}, \tag{C.4}$$

   with the constraint $n_{LCU} = \lceil \log_2 m \rceil$, where $\lceil \log_2 m \rceil$ denotes the lowest integer such that $n_{LCU} \geq \log_2 m$. This approach has the advantage of reducing the quantum resources compared to the LCU-B case.

3. **LCU-L:** The last case is the one proposed in [123] where $B$ is still given by Eq. (C.3), while $E = L^\dagger$ is deduced from Eq. (4.27). The success probability is then given by

$$p_S^L = \frac{\|A|\Psi\rangle\|^2}{m\mathcal{N}_H^2}. \tag{C.5}$$

## C.1 . Comparison of $p_S$ for the Projector and Oracle

In section 4.4.2, we discussed the possibility of using the LCU approach to implement either the projector associated with the oracle or the oracle itself. Similar conclusions can be drawn for any other symmetry. In this section, we estimate the success probability of performing these two operations in the context of particle number restoration. Employing the notation from section 3.2.1, the coefficients and operators in the projector take the following form:

$$\alpha_k = \frac{e^{-i\phi_k p'}}{n+1}, \qquad V_k = e^{i\phi_k N},$$

with $\phi_k = 2\pi k/(n+1)$, $p'$ the number of particles we wish to project onto, and $n$ is the number of qubits in the system register. Consequently, it follows that:

$$\mathcal{N}_B = 1, \; \mathcal{N}_H = \frac{1}{\sqrt{n+1}}.$$

We then deduce that:

$$p_S^B = p_S^L = p_G, \tag{C.6}$$

$$p_S^H = \frac{p_G\,(n+1)}{2^{\lceil \log_2(n+1)\rceil}} \leq p_G, \tag{C.7}$$

where $p_G = |\langle \Psi_G|\Psi\rangle|^2$ and $|\Psi_G\rangle$ is the projected (good) state. We now consider the implementation of the oracle itself. In this case, we have:

$$\alpha_k = \delta_{k0} e^{i\mu} + \left(e^{i\varphi} - e^{i\mu}\right) e^{-i\phi_k p'}/(n+1),$$

while the $V_k$ are the same as for the projector. A lengthy but straightforward calculation gives the different success probabilities:

$$p_S^B = \frac{1}{1 + (n^2-1)\,x + 2n\sqrt{x\,(1-x)}},$$

with $x = \frac{2n}{(n+1)^2}\left[1 - \cos\left(\varphi - \mu\right)\right]$, together with:

$$p_S^H = \frac{1}{2^{\lceil \log_2(n+1)\rceil}\left[1 + x\,(n-1)\right]},$$

and

$$p_S^L = \frac{1}{(n+1)\left[1 + x\,(n-1)\right]}.$$

In Fig C.1, we display the evolutions of the three probabilities $p_S^B$, $p_S^H$, and $p_S^L$ as a function of $n$ for the case $(\varphi, \mu) = (\pi, 0)$, which corresponds to the standard oracle used in Grover search (see Section 2.1.3). In this case, we have $x = 4n/(n+1)^2$, and for the large $n$ limit, we deduce that both $p_S^B$ and $p_S^L$ scale as $n^{-1}$, while $p_S^H$ scales as $1/2^{\lceil \log_2(n+1)\rceil}$. We observe that the LCU-B has a marginally higher success probability; however, this advantage may be offset in practice by the fact that both LCU-H and LCU-L necessitate fewer operations.

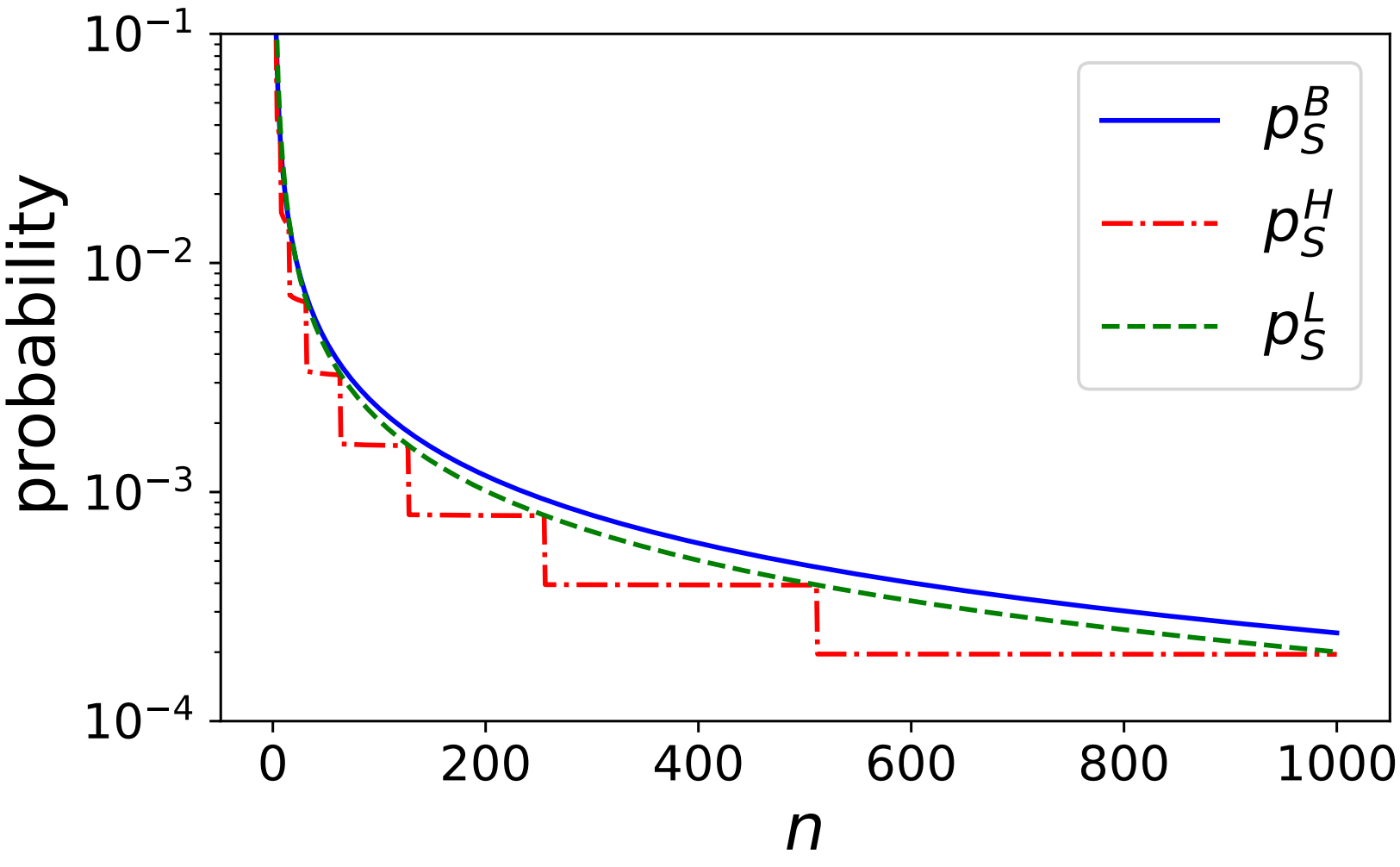

Figure C.1: Evolution of the 3 success probabilities $p_S^B$, $p_S^H$, and $p_S^L$ as function on $n$ for the set of angles $(\varphi, \mu)$ equal to $(\pi, 0)$. It is important to note that the steps in $p_S^H$ result from the factor $\lceil \log_2 (n+1) \rceil$.

# D - Eigenvalues and Eigenstates in the Subspace $\mathcal{H}_M$

Let us consider that we have the $2M-1$ moments needed to construct the overlap and Hamiltonian matrices in the subspace $\mathcal{H}_M$ as shown in Eq. (6.17). As the matrix elements of both $O$ and $H$ become large very rapidly, we should normalize the states in the Krylov basis as well as the two matrices $O_{kl}$ and $H_{kl}$, i.e.:

$$|\Phi_k\rangle \to \frac{|\Phi_k\rangle}{\sqrt{\langle\Phi_k|\Phi_k\rangle}}, \qquad O_{kl} \to \frac{O_{kl}}{\sqrt{O_{kk}O_{ll}}}, \qquad H_{kl} \to \frac{H_{kl}}{\sqrt{O_{kk}O_{ll}}}. \tag{D.1}$$

Then, we obtain the matrices $U$ and $D$ in $O = UDU^\dagger$ via classical diagonalization. The matrix $D$ is diagonal with positive eigenvalues $\{d_k\}_{k=0,\dots,M-1}$ associated to the eigenstates $\{|\xi_k\rangle\}_{k=0,\dots,M-1}$. If some of the eigenvalues $d_k$ are 0, we can dispense some of the states $|\xi_\alpha\rangle$ as they are nonlinear independent states. To obtain the eigenvectors $|\xi_\alpha\rangle$ in terms of the states of the Krylov basis, we can observe the projector in the $\mathcal{H}_M$ subspace for both basis, i.e.:

$$P_{\mathcal{H}_M} = \sum_{\alpha=0}^{M-1} |\xi_\alpha\rangle\langle\xi_\alpha| = \sum_{i,j=0}^{M-1} |\Phi_i\rangle O^{-1}_{ij}\langle\Phi_j|. \tag{D.2}$$

Using the matrices $U, D$ in the decomposition of $O$, $O^{-1}_{ij} = \sum_\alpha U_{i\alpha}\frac{1}{d_\alpha}U^*_{j\alpha}$, the previous equation becomes:

$$\begin{aligned}\sum_{\alpha=0}^{M-1} |\xi_\alpha\rangle\langle\xi_\alpha| &= \sum_{i,j=0}^{M-1} |\Phi_i\rangle \left(\sum_\alpha U_{i\alpha}\frac{1}{d_\alpha}U^*_{j\alpha}\right)\langle\Phi_j|,\\ |\xi_\alpha\rangle\langle\xi_\alpha| &= \left(\sum_{i=0}^{M-1}|\Phi_i\rangle\frac{U_{i\alpha}}{\sqrt{d_\alpha}}\right)\left(\sum_{j=0}^{M-1}\frac{U^*_{j\alpha}}{\sqrt{d_\alpha}}\langle\Phi_i|\right).\end{aligned} \tag{D.3}$$

Defining the matrix $X_{ij} = \frac{U_{ij}}{\sqrt{d_j}}$, the vectors $|\xi_\alpha\rangle$ become:

$$|\xi_\alpha\rangle = \sum_{i=0}^{M-1}|\Phi_i\rangle X_{i\alpha}. \tag{D.4}$$

Now, we can express the Hamiltonian in the new orthonormal basis $\{|\xi_\alpha\rangle\}$:

$$\tilde{H}_{\alpha\beta} = \langle\xi_\alpha|H|\xi_\beta\rangle = \sum_{ij} X^*_{i\alpha}X_{j\alpha}\langle\Phi_i|H|\Phi_j\rangle. \tag{D.5}$$

By diagonalizing $\tilde{H} = TET^\dagger$, we get the approximation of the spectra in the subspace $\mathcal{H}_M$, $\{E_i\}$, and the associated eigenstates $|E_j\rangle = T_{ij}$.

## D.1 . Approximation Probability of Survival

To approximate the long-time evolution of the probability of survival by using the approximated energies $E_\gamma$, we need to compute the overlaps $\langle E_\gamma|\Phi_0\rangle$ in Eq. (6.18). To do this, we first establish the relationship between the Krylov basis $|\Phi_k\rangle$ and the eigenbasis of the Hamiltonian in the $\mathcal{H}_M$ space $|E_\gamma\rangle$. This can be done via the closure relations over the basis $|\xi_\alpha\rangle$ and $|E_\lambda\rangle$:

$$|\Phi_k\rangle = \sum_{\alpha=0}^{M-1} |\xi_\alpha\rangle\langle\xi_\alpha|\Phi_k\rangle = \sum_{\alpha,\lambda=0}^{M-1} |E_\lambda\rangle\langle E_\lambda|\xi_\alpha\rangle\langle\xi_\alpha|\Phi_k\rangle. \tag{D.6}$$

The elements of the $T$ matrix found in the diagonalization of $\tilde{H}_{\alpha\beta}$ give us the overlaps $\langle E_\lambda|\xi_\alpha\rangle$, i.e., $T^*_{\alpha\lambda} = \langle E_\lambda|\xi_\alpha\rangle$. To obtain the overlaps $\langle\xi_\alpha|\Phi_k\rangle$, we can use the fact that the matrix $U$ in Eq. (D.3) is unitary, i.e., $\sum_{\alpha=0}^{M-1} U^*_{k\alpha}U_{l\alpha} = \delta_{k,l}$. Starting from Eq. (D.4) we have:

$$\sum_{\alpha=0}^{M-1} U^*_{k\alpha}\sqrt{d_\alpha}|\xi_\alpha\rangle = \sum_{\alpha=0}^{M-1} U^*_{k\alpha} \sum_{l=0}^{M-1} |\Phi_l\rangle U_{l\alpha} = \sum_{l=0}^{M-1} |\Phi_l\rangle \underbrace{\sum_{\alpha=0}^{M-1} U^*_{k\alpha}U_{l\alpha}}_{\delta_{kl}} = |\Phi_k\rangle. \tag{D.7}$$

Given that $|\Phi_k\rangle = \sum_{\alpha=0}^{M-1} |\xi_\alpha\rangle\langle\xi_\alpha|\Phi_k\rangle$, we get:

$$\langle\xi_\alpha|\Phi_k\rangle = U^*_{k\alpha}\sqrt{d_\alpha} = d_\alpha X^*_{k\alpha}. \tag{D.8}$$

Replacing the previous results in Eq. (D.6) we obtain:

$$|\Phi_k\rangle = \sum_{\alpha,\lambda=0}^{M-1} |E_\lambda\rangle T^*_{\alpha\lambda} d_\alpha X^*_{k\alpha}. \tag{D.9}$$

The overlaps $\langle E_\gamma|\Phi_k\rangle$ are simply computed as:

$$\langle E_\gamma|\Phi_k\rangle = \sum_{\alpha=0}^{M-1} T^*_{\alpha\gamma} d_\alpha X^*_{k\alpha}. \tag{D.10}$$

## References

[1] R. P. Feynman, *Simulating physics with computers*, Int J Theor Phys **21** (1982) 467. doi:10.1007/BF02650179.

[2] J. Preskill, *Quantum Computing in the NISQ era and beyond*, Quantum **2** (2018) 79. doi:10.22331/q-2018-08-06-79.

[3] P. W. Shor, *Algorithms for quantum computation: discrete logarithms and factoring*, Proceedings 35th Annual Symposium on Foundations of Computer Science (1994) 124. doi:10.1109/SFCS.1994.365700.

[4] B. Zeng, X. Chen, D. Zhou, and X. Wen, *Quantum Information Meets Quantum Matter*, Springer (2019). doi:10.1007/978-1-4939-9084-9.

[5] D. Zhang et al., *Selected topics of quantum computing for nuclear physics*, Chinese Phys. **B 30** (2021) 020306. doi:10.1088/1674-1056/abd761.

[6] I. Stetcu, A, Baroni, and J. Carlson, *Variational approaches to constructing the many-body nuclear ground state for quantum computing*, Phys. Rev. **C 105** (2022) 064308. doi:10.1103/PhysRevC.105.064308.

[7] C. Yudong et al., *Quantum Chemistry in the Age of Quantum Computing*, Chemical Reviews **119** (2019) 10856. doi:10.1021/acs.chemrev.8b00803.

[8] A. J. McCaskey et al. *Quantum chemistry as a benchmark for near-term quantum computers*, npj Quantum Inf **5** (2019) 99. doi:10.1038/s41534-019-0209-0.

[9] B. Bauer, S. Bravyi, M. Motta, and G. K. Chan, *Quantum Algorithms for Quantum Chemistry and Quantum Materials Science* Chemical Reviews **120** (2020) 12685. doi:10.1021/acs.chemrev.9b00829.

[10] "Binario cropped.png" by MdeVicente, licensed under CC0 1.0 Universal (CC 1.0) Public Domain Dedication https://creativecommons.org/publicdomain/zero/1.0/deed.en, available from https://commons.wikimedia.org/wiki/File:Binario_cropped.png.

[11] "Energy levels.svg" by Rehua (SVG), Rozzychan (Original), licensed under Attribution-ShareAlike 3.0 Unported (CC BY-SA 3.0) https://creativecommons.org/licenses/by-sa/3.0/deed.en, available from https://commons.wikimedia.org/wiki/File:Energy_levels.svg.

[12] "Sucrose molecule.svg" by Medium69 and William Crochot, licensed under Attribution-ShareAlike 4.0 International (CC BY-SA 4.0) https://creativecommons.org/licenses/by-sa/4.0/deed.en, available from https://commons.wikimedia.org/wiki/File:Sucrose_molecule.svg.

[13] "U.S. Department of Energy - Science - 304 015 003 (16334175965).jpg" by U.S. Department of Energy's Argonne National Laboratory from the United States, Public Domain, available from https://commons.wikimedia.org/wiki/File:U.S._Department_of_Energy_-_Science_-_304_015_003_(16334175965).jpg. Text added by Edgar Andres Ruiz Guzman.

[14] "Bose Einstein condensate.png" by NIST/JILA/CU-Boulder, Public Domain, available from https://commons.wikimedia.org/wiki/File:Bose_Einstein_condensate.png. Text added by Edgar Andres Ruiz Guzman.

[15] "YL1M0sph.png" by Pickwick, licensed under Attribution-ShareAlike 3.0 Unported (CC BY-SA 3.0) https://creativecommons.org/licenses/by-sa/3.0/deed.en, available from https://commons.wikimedia.org/wiki/File:YL1M0sph.png.

[16] "YL2M0sph.png" by Pickwick, licensed under Attribution-ShareAlike 3.0 Unported (CC BY-SA 3.0) https://creativecommons.org/licenses/by-sa/3.0/deed.en, available from https://commons.wikimedia.org/wiki/File:YL2M0sph.png.

[17] "YL3M2sph.png" by Pickwick, licensed under Attribution-ShareAlike 3.0 Unported (CC BY-SA 3.0) https://creativecommons.org/licenses/by-sa/3.0/deed.en, available from https://commons.wikimedia.org/wiki/File:YL3M2sph.png.

[18] "YL2M1sph.png" by Pickwick, licensed under Attribution-ShareAlike 3.0 Unported (CC BY-SA 3.0) https://creativecommons.org/licenses/by-sa/3.0/deed.en, available from https://commons.wikimedia.org/wiki/File:YL2M1sph.png.

[19] T. Ayral, P. Besserve, D. Lacroix, and E.A. Ruiz Guzman, *Quantum computing with and for many-body physics*, arXiv:2303.04850 (2023). doi:10.48550/arXiv.2303.04850.

[20] J.L. Friar, G.L. Payne, V.G.J. Stoks, and J.J. de Swart, *Triton calculations with the new Nijmegen potentials*, Physics Letters **B 311** (1993) 4. doi:10.1016/0370-2693(93)90523-K.

[21] B. S. Pudliner et al., *Quantum Monte Carlo calculations of nuclei with* $A <\sim 7$, Phys. Rev. **C 56** (1997) 1720. doi:10.1103/PhysRevC.56.1720.

[22] S. Quaglioni and P. Navrátil, *Ab initio no-core shell model and microscopic reactions: Recent achievements*, Few-Body Syst **44** (2008) 337. doi:10.1007/s00601-008-0322-7.

[23] I. Shavitt and R. J. Bartlett, *Many-body Methods in Chemistry and Physics: MBPT and Coupled-Cluster Theory*, Cambridge University Press, Cambridge, UK, (2009). doi:10.1017/CBO9780511596834.

[24] R. J. Bartlett and M. Musiał, *Coupled-cluster theory in quantum chemistry*, Rev. Mod. Phys. **79** (2007) 291. doi:10.1103/RevModPhys.79.291.

[25] H. Hergert et al., *The In-Medium Similarity Renormalization Group: A novel ab initio method for nuclei*, Physics Reports **621** (2016) 165. doi:10.1016/j.physrep.2015.12.007.

[26] A. F. Lisetskiy, *Ab-initio shell model with a core*, Phys. Rev. **C 48** (2008) 044302. doi:10.1103/PhysRevC.78.044302.

[27] S. R. Stroberg, H. Hergert, S. K. Bogner, and J. D. Holt, *Nonempirical Interactions for the Nuclear Shell Model: An Update*, Annual Review of Nuclear and Particle Science **69** (2019) 307. doi:10.1146/annurev-nucl-101917-021120.

[28] P. Ring, P. Schuck, and M. R. Strayer, *The Nuclear Many-Body Problem*, Physics Today **36** (1983) 70. doi:10.1063/1.2915762.

[29] J. P. Blaizot et al., *Quantum Theory of Finite Systems and Quantum Many-Particle Systems*, Physics Today **41** (1988) 106. doi:10.1063/1.2811565.

[30] J. A. Sheikh, *Symmetry restoration in mean-field approaches*, J. Phys. G: Nucl. Part. Phys. **48** (2021) 123001. doi:10.1088/1361-6471/ac288a.

[31] HY. Huang, R. Kueng, and J. Preskill, *Predicting many properties of a quantum system from very few measurements*, Nat. Phys. **16** (2020) 1050. doi:10.1038/s41567-020-0932-7

[32] A. Smith, M. S. Kim, F. Pollmann and J. Knolle, *Simulating quantum many-body dynamics on a current digital quantum computer*, npj Quantum Inf **5** (2019) 106. doi:10.1038/s41534-019-0217-0.

[33] D. Lacroix, E. A. Ruiz Guzman, and P. Siwach, *Symmetry breaking/symmetry preserving circuits and symmetry restoration on quantum computers* Eur. Phys. J. **A 59** (2023) 3. doi:10.1140/epja/s10050-022-00911-7.

[34] J. Fraxanet, T. Salamon, and M. Lewenstein, *The Coming Decades of Quantum Simulation*, arXiv:2204.08905v2 (2022). doi:10.48550/arXiv.2204.08905.

[35] A. J. Daley, *Practical quantum advantage in quantum simulation*, Nature **607** (2022) 667. doi:10.1038/s41586-022-04940-6.

[36] O. Ezratty, *Understanding Quantum Technologies 2022*, arXiv:2111.15352 (2022). 10.48550/arXiv.2111.15352.

[37] K. N. Schymik et al., *In situ equalization of single-atom loading in large-scale optical tweezer arrays*, Phys. Rev. **A 106** (2022) 022611. 10.1103/PhysRevA.106.022611.

[38] Y. Chew et al., *Ultrafast energy exchange between two single Rydberg atoms on a nanosecond timescale*, Nat. Photon. **16** (2022) 724. 10.1038/s41566-022-01047-2.

[39] C. D. Bruzewicz, J. Chiaverini, R. McConnell, and M. S. Jeremy, *Trapped-ion quantum computing: Progress and challenges*, Applied Physics Reviews **6** (2019) 021314. 10.1063/1.5088164.

[40] K. R. Brown, J. Chiaverini, J. M. Sage, and H. H́affner, *Materials challenges for trapped-ion quantum computers*. Nat Rev Mater **6** (2021) 892. 10.1038/s41578-021-00292-1.

[41] Infineon, *Trapped ion quantum computing* (2020) https://www.infineon.com/cms/en/product/promopages/trapped-ions/

[42] A. Laucht et al., *Roadmap on quantum nanotechnologies*, Nanotechnology **32** (2021) 162003. doi:10.1088/1361-6528/abb333.

[43] M. Ruf, *Quantum networks based on color centers in diamond*, Journal of Applied Physics **130** (2021) 070901. doi:10.1063/5.0056534.

[44] "NIST Physicists Coax Six Atoms into Quantum 'Cat' State (5884514942).jpg" by National Institute of Standards and Technology, Public Domain, available from https://commons.wikimedia.org/wiki/File:NIST_Physicists_Coax_Six_Atoms_into_Quantum_%27Cat%27_State_(5884514942).jpg.

[45] "4 Qubit, 4 Bus, 4 Resonator IBM Device (Jay M. Gambetta, Jerry M. Chow, and Matthias Steffen, 2017).png" by Jay M. Gambetta, Jerry M. Chow & Matthias Steffen, licensed under Attribution 4.0 International (CC BY 4.0) https://creativecommons.org/licenses/by/4.0/deed.en, available from https://commons.wikimedia.org/wiki/File:4_Qubit,_4_Bus,_4_Resonator_IBM_Device_(Jay_M._Gambetta,_Jerry_M._Chow,_and_Matthias_Steffen,_2017).png.

[46] "OAM qubit.svg" by JozumBjada, licensed under Attribution-ShareAlike 4.0 International (CC BY-SA 4.0) https://creativecommons.org/licenses/by-sa/4.0/deed.en, available from https://commons.wikimedia.org/wiki/File:OAM_qubit.svg. The image was cropped by Edgar Andres Ruiz Guzman.

[47] "Nitrogen-vacancy center.png" by National Institute of Standards and Technology, Public Domain, available from https://commons.wikimedia.org/wiki/File:Nitrogen-vacancy_center.png.

[48] I. Bloch, J. Dalibard, and W. Zwerger, *Many-body physics with ultracold gases*, Rev. Mod. Phys. **80** (2008) 885. doi:10.1103/RevModPhys.80.885.

[49] D. Loss and D.P. DiVincenzo, *Quantum computation with quantum dots*, Phys. Rev. **A 57** (1998) 120. doi:10.1103/PhysRevA.57.120.

[50] A. Browaeys and T. Lahaye, *Many-body physics with individually controlled Rydberg atoms*, Nat. Phys. **16** (2020) 132. doi:10.1038/s41567-019-0733-z.

[51] L. Henriet et al., *Quantum computing with neutral atoms*, Quantum **4** (2020) 327. 10.22331/q-2020-09-21-327.

[52] Qiskit Development Team, *Qiskit: An Open-source Framework for Quantum Computing*, (2021). doi:10.5281/zenodo.2573505.

[53] J. Koch et al., *Charge-insensitive qubit design derived from the Cooper pair box*, Phys. Rev. **A 76** (2007) 042319. doi:10.1103/PhysRevA.76.042319.

[54] J. A. Schreier et al., *Suppressing charge noise decoherence in superconducting charge qubits*, Phys. Rev. **B 77** (2008) 180502. doi:10.1103/PhysRevB.77.180502.

[55] J. Benhelm, G. Kirchmair, C. Roos, and R. Blatt, *Towards fault-tolerant quantum computing with trapped ions*, Nat. Phys. **4** (2008) 463. doi:10.1038/nphys961.

[56] T. P. Harty et al.,*High-fidelity preparation, gates, memory, and readout of a trapped-ion quantum bit*, Phys. Rev. Lett. **113** (2014) 220501. doi:10.1103/PhysRevLett.113.220501.

[57] C. J. Ballance et al., *High-Fidelity Quantum Logic Gates Using Trapped-Ion Hyperfine Qubits*, Phys. Rev. Lett. **117** (2016) 060504. doi:10.1103/PhysRevLett.117.060504.

[58] K. Maller et al., *Rydberg-blockade controlled-NOT gate and entanglement in a two-dimensional array of neutral-atom qubits*, Phys. Rev. **A 92** (2015) 022336. doi:10.1103/PhysRevA.92.022336.

[59] Y.-Y. Jau, et al., *Entangling atomic spins with a Rydberg-dressed spin-flip blockade*, Nat. Phys. **12** (2016) 71. doi:10.1038/nphys3487.

[60] M. Saffman, *Quantum computing with atomic qubits and Rydberg interactions: progress and challenges*, J. Phys. B: At. Mol. Opt. Phys. **49** (2016) 202001. doi:10.1088/0953-4075/49/20/202001.

[61] T. Ladd et al., *Quantum computers*, Nature **464** (2010) 45. doi:10.1038/nature08812.

[62] M. A. Nielsen and I. L. Chuang, *Quantum information and quantum computation.*, Cambridge University Press (2010). doi:10.1017/CBO9780511976667.

[63] M. Motta and J. Rice, *Emerging quantum computing algorithms for quantum chemistry*, WIREs Comput Mol Sci. **12** (2022) e1580. doi:10.1002/wcms.1580.

[64] K. Seki and S. Yunoki, *Spatial, spin, and charge symmetry projections for a Fermi-Hubbard model on a quantum computer*, Phys. Rev. **A 105** (2022) 032419. doi:10.1103/PhysRevA.105.032419.

[65] A. Y. Kitaev, *Quantum computations: algorithms and error correction*, Russian Mathematical Surveys **52** (1997) 1191. doi:10.1070/RM1997v052n06ABEH002155.

[66] M. Born, *Zur Quantenmechanik der Stoßvorgänge*, Z. Physik **37** (1926) 863. doi:10.1007/BF01397477.

[67] F. de Lima Marquezino, R. Portugal, and C. Lavor, *A primer on quantum computing*, Springer (2019). doi:10.1007/978-3-030-19066-8.

[68] R. Cleve, A. Ekert, C. Macchiavello, and M.Mosca, *Quantum algorithms revisited*, Proc. R. Soc. Lond. **A 454** (1998) 339. doi:10.1098/rspa.1998.0164.

[69] G. Benenti, G. Casati, and G. Strini, *Principles of Quantum Computation and Information Vol I*, WORLD SCIENTIFIC (2004). doi:10.1142/5528.

[70] D. Lacroix, *Symmetry-Assisted Preparation of Entangled Many-Body States on a Quantum Computer*, Phys. Rev. Lett. **125** (2020) 230502. doi:10.1103/PhysRevLett.125.230502.

[71] E. A. Ruiz Guzman and D. Lacroix, *Accessing ground-state and excited-state energies in a many-body system after symmetry restoration using quantum computers*, Phys. Rev. C **105** (2022) 024324. doi:10.1103/PhysRevC.105.024324.

[72] D. J. Bernstein and T. Lange, *Post-quantum cryptography*, Nature **549** (2017) 188. doi:10.1038/nature23461.

[73] D. Beckman, A. N. Chari, S. Devabhaktuni, and J. Preskill, *Efficient networks for quantum factoring*, Phys. Rev. **A 54** (1996) 1034. doi:10.1103/PhysRevA.54.1034.

[74] E. W. Weisstein *Number Field Sieve*, from MathWorld–A Wolfram Web Resource (2023). `https://mathworld.wolfram.com/NumberFieldSieve.html`

[75] F. Boudot et al., *The State of the Art in Integer Factoring and Breaking Public-Key Cryptography*, IEEE Security & Privacy **20** (2022) 80. doi:10.1109/MSEC.2022.3141918.

[76] H.F. Trotter, *On the Product of Semi-Groups of Operators*, Proc. Am. Math. Soc. **10** (1959) 545. doi:10.2307/2033649.

[77] M. Suzuki, *Generalized Trotter's formula and systematic approximants of exponential operators and inner derivations with applications to many-body problems*, Commun.Math. Phys. **51** (1976) 183. doi:10.1007/BF01609348.

[78] D.W. Berry, G. Ahokas, R. Cleve, and B. C. Sanders, *Efficient Quantum Algorithms for Simulating Sparse Hamiltonians*, Commun. Math. Phys. **270** (2007) 359. doi:10.1007/s00220-006-0150-x

[79] N. Hatano and M. Suzuki, *Finding Exponential Product Formulas of Higher Orders*, Springer (2005). doi:10.1007/11526216_2.

[80] S. McArdle et al., *Quantum computational chemistry*, Rev. Mod. Phys. **92** (2020) 015003. doi:10.1103/RevModPhys.92.015003.

[81] G. Brassard and P. Hoyer, *An exact quantum polynomial-time algorithm for Simon's problem*, Proceedings of the Fifth Israeli Symposium on Theory of Computing and Systems (1997) 12. doi:10.1109/ISTCS.1997.595153.

[82] L. K. Grover, *Quantum Computers Can Search Rapidly by Using Almost Any Transformation*, Phys. Rev. Lett. **80** (1998) 4329. doi:10.1103/PhysRevLett.80.4329.

[83] L. K. Grover, *Quantum Mechanics Helps in Searching for a Needle in a Haystack*, Phys. Rev. Lett. **79** (1997) 325. doi:10.1103/PhysRevLett.79.325.

[84] P. Hoyer, *On Arbitrary phases in quantum amplitude amplification*, Phys. Rev. **A 62**, (2000) 052304. doi:10.1103/PhysRevA.62.052304.

[85] A. Roggero et al., *Quantum computing for neutrino-nucleus scattering*, Phys. Rev. **D 101** (2020) 074038. doi:10.1103/PhysRevD.101.074038.

[86] B. Hall, A. Roggero, A. Baroni, and J. Carlson, *Simulation of collective neutrino oscillations on a quantum computer*, Phys. Rev. **D 104** (2021) 063009. doi:10.1103/PhysRevD.104.063009.

[87] V. Amitrano, *Trapped-ion quantum simulation of collective neutrino oscillations*, Phys. Rev. **D 107** (2023) 023007. doi:10.1103/PhysRevD.107.023007.

[88] S. Baeurle, *Grand canonical auxiliary field Monte Carlo: a new technique for simulating open systems at high density*, Computer Physics Communications **157** (2004) 201. doi:10.1016/j.comphy.2003.11.001.

[89] D. Gottesman, *Theory of fault-tolerant quantum computation*, Phys. Rev. **A 57** (1998) 127. doi:10.1103/PhysRevA.57.127.

[90] D. Gottesman, *The Heisenberg representation of quantum computers*, Group 22: Proceedings of the XXII International Colloquium on Group Theoretical Methods in Physics (1998) 32. doi:10.48550/arXiv.quant-ph/9807006.

[91] S. Aaronson and D. Gottesman, *Improved simulation of stabilizer circuits*, Phys. Rev. **A 70** (2004) 052328. doi:10.1103/PhysRevA.70.052328.

[92] C. P. Williams, *Explorations in Quantum Computing*, Springer (2010). doi:10.1007/978-1-84628-887-6.

[93] E. Bernstein and U. Vazirani, *Quantum complexity theory*, SIAM J. Comput. **26** (1997) 1411.doi:10.1137/S0097539796300921.

[94] A. Y. Kitaev, A. H. Shen, and M. N. Vyalyi, *Classical and Quantum Computation*, American Mathematical Society (2002). doi:10.5555/863284.

[95] J. Watrous, *Quantum Computational Complexity*, Springer (2012) 2361. doi:10.1007/978-1-4614-1800-$9_1$47.

[96] T. Hogg, *Quantum search heuristics*, Phys. Rev. **A 61** (2000) 052311. doi:10.1103/PhysRevA.61.052311.

[97] E. Farhi, J. Goldstone, S. Gutmann, and L. Zhou, *A Quantum Approximate Optimization Algorithm* (2014). doi:10.48550/arXiv.1411.4028.

[98] S. Mandrà, G. G. Guerreschi, and A. Aspuru-Guzik, *Faster than classical quantum algorithm for dense formulas of exact satisfiability and occupation problems*, New J. Phys. **18** (2016) 073003. doi:10.1088/1367-2630/18/7/073003.

[99] J. Kempe, A. Kitaev, and O. Regev, *The Complexity of the Local Hamiltonian Problem*, Springer (2005) 372. doi:10.1007/978-3-540-30538-$5_3$1.

[100] D. Gosset, J. C. Mehta, and T. Vidick, *QCMA hardness of ground space connectivity for commuting Hamiltonians*, Quantum **1** (2017) 16. doi:10.22331/q-2017-07-14-16.

[101] S. Lee et al., *Is there evidence for exponential quantum advantage in quantum chemistry?* (2022). doi:10.48550/arXiv.2208.02199.

[102] N. Schuch, I. Cirac, and F. Verstraete, *Computational Difficulty of Finding Matrix Product Ground States*, Phys. Rev. Lett. **100** (2008) 250501. doi:10.1103/PhysRevLett.100.250501.

[103] N. Bansal, S. Bravyi, and B. M. Terhal, *Classical Approximation Schemes for the Ground-State Energy of Quantum and Classical Ising Spin Hamiltonians on Planar Graphs*, Quantum Info. Comput. **9** (2009) 701. doi:10.48550/arXiv.0705.1115.

[104] F Barahona, *On the computational complexity of Ising spin glass models*, J. Phys. A: Math. Gen. **15** (1982) 3241. doi:10.1088/0305-4470/15/10/028.

[105] S. Bravyi, *Monte Carlo Simulation of Stochastic Hamiltonians*, Quantum Info. Comput. **15** (2015) 1122. doi:10.48550/arXiv.1402.2295.

[106] A. M. Childs, D. Gosset, and Z. Webb, *The Bose-Hubbard Model is QMA-complete*, Springer (2014) 308. doi:10.1007/978-3-662-43948-$7_2$6.

[107] N. Schuch and F. Verstraete, *Computational complexity of interacting electrons and fundamental limitations of density functional theory*, Nature Phys **5** (2009) 732. doi:10.1038/nphys1370.

[108] B. O'Gorman, S.Irani, J. Whitfield, and B. Fefferman, *Intractability of Electronic Structure in a Fixed Basis*, PRX Quantum **3** (2022) 020322. doi:10.1103/PRXQuantum.3.020322.

[109] L. Bittel and M. Kliesch, *Training Variational Quantum Algorithms Is NP-Hard*, Phys. Rev. Lett. **127** (2021) 120502. doi:10.1103/PhysRevLett.127.120502.

[110] P. Jordan and E. Wigner, *Über das Paulische Äquivalenzverbot*, Z. Physik **47** (1928) 631. doi:10.1007/BF01331938.

[111] J.T. Seeley, M.J. Richard, and P.J. Love, *The Bravyi-Kitaev transformation for quantum computation of electronic structure*, J. Chem. Phys. **137** (2012) 224109. doi:10.1063/1.4768229.

[112] S.B. Bravyi and A.Y. Kitaev, *Fermionic Quantum Computation*, Ann. Phys. **298** (2002) 210. doi:10.1006/aphy.2002.6254.

[113] M. Steudtner and S. Wehner, *Fermion-to-qubit mappings with varying resource requirements for quantum simulation*, New J. Phys. **20** (2018) 063010. doi:10.1088/1367-2630/aac54f.

[114] N. Moll, A. Fuhrer, P. Staar, and I. Tavernelli, *Optimizing qubit resources for quantum chemistry simulations in second quantization on a quantum computer*, J. Phys. A: Math. Theor. **49** (2016) 295301. doi:10.1088/1751-8113/49/29/295301.

[115] S. Bravyi, J. M. Gambetta, A. Mezzacapo, and K. Temme, *Tapering off qubits to simulate fermionic Hamiltonians*, arXiv:1701.08213v1 (2017). doi:10.48550/arXiv.1701.08213.

[116] A. Tranter et al., *The Bravyi–Kitaev transformation: Properties and applications*, Int. J. Quantum Chem. **115** (2015) 1431. doi:10.1002/qua.24969.

[117] J. von Delft and D.C. Ralf, *Spectroscopy of discrete energy levels in ultrasmall metallic grains*, Phys. Rep. **345** (2001) 61. doi:10.1016/S0370-1573(00)00099-5.

[118] V. Zelevinsky and A. Volya, *Nuclear pairing: New perspectives*, Phys. Atom. Nuclei **66** (2003) 1781. doi:10.1134/1.1619492.

[119] J. Dukelsky, S. Pittel, and G. Sierra, *Colloquium: Exactly solvable Richardson-Gaudin models for many-body quantum systems*, Rev. Mod. Phys. **76** (2004) 643. doi:10.1103/RevModPhys.76.643

[120] D.M. Brink and R.A. Broglia, *Nuclear Superfluidity: Pairing in Finite Systems*, Cambridge University Press (2005). doi:10.1017/CBO9780511534911.

[121] A. Khamoshi, F.A. Evangelista, and G. E. Scuseria, *Correlating AGP on a quantum computer*, Quantum Sci. Technol. **6** (2021) 014004. doi:10.1088/2058-9565/abc1bb.

[122] E. A. Ruiz Guzman and D. Lacroix, *Calculation of generating function in many-body systems with quantum computers: technical challenges and use in hybrid quantum-classical methods*, arXiv (2021). doi:10.48550/arXiv.2104.08181.

[123] E. A. Ruiz Guzman and D. Lacroix, *Restoring broken symmetries using quantum search "oracles"*, Phys. Rev. **C 107** (2023) 034310. doi:10.1103/PhysRevC.107.034310.

[124] D. Jaksch et al., *Cold Bosonic Atoms in Optical Lattices*, Phys. Rev. Lett. **81** (1998) 3108. doi:10.1103/PhysRevLett.81.3108.

[125] M. Greiner et al., *Quantum phase transition from a superfluid to a Mott insulator in a gas of ultracold atoms*, Nature **415** (2002) 39. doi:10.1038/415039a.

[126] A. Peruzzo et al., *A variational eigenvalue solver on a photonic quantum processor*, Nat Commun **5** (2014) 4213. doi:10.1038/ncomms5213.

[127] D. J. Gross, *The role of symmetry infundamentalphysics*, Proc. Natl. Acad. Sci. **93** (1996) 14256. doi:10.1073/pnas.93.25.14256.

[128] A. Messiah, *Mecanique Quantique, Volume II*, American Journal of Physics **28** (1960) 580. doi:10.1119/1.1935901.

[129] I. G. Kaplan and D. M. Bishop, *Symmetry of Many-Electron Systems*, Physics Today **28** (1975) 55. doi:10.1063/1.2998922.

[130] M. Hammermesh and C. Flammer, *Group Theory and Its Application to Physical Problems*, Physics Today **16** (1963) 62. doi:10.1063/1.3050758.

[131] D. Lacroix and D. Gambacurta, *Projected quasiparticle perturbation theory*, Phys. Rev. **C 86** (2012) 014306. doi:10.1103/PhysRevC.86.014306.

[132] D. Lacroix, *Review of mean-field theory*, Ecole Joliot-Curie (30 years) "Physics at the femtometer scale" (2011) url:in2p3-00624365.

[133] A. J. Beekman, L. Rademaker, and J. V. Wezel. *An introduction to spontaneous symmetry breaking*, SciPost Phys. Lect. Notes (2019). doi:10.21468/SciPostPhysLectNotes.11.

[134] O. F. Syljuåsen, *Entanglement and spontaneous symmetry breaking in quantum spin models*, Phys. Rev. **A 68** (2003) 060301. doi:10.1103/PhysRevA.68.060301.

[135] M. Bender, P. H. Heenen, and P. G. Reinhard, *Self-consistent mean-field models for nuclear structure*, Rev. Mod. Phys. **75** (2003) 121. doi:10.1103/RevModPhys.75.121.

[136] L. M. Robledo, T. R. Rodríguez and R. R. Rodríguez-Guzmán, *Mean field and beyond description of nuclear structure with the Gogny force: a review*, J. Phys. G: Nucl. Part. Phys. **46** (2018) 013001. doi:10.1088/1361-6471/aadebd.

[137] M. C. Tran, Y. Su, D. Carney, and J. M. Taylor, *Faster Digital Quantum Simulation by Symmetry Protection*, PRX Quantum **2** (2021) 010323. doi:10.1103/PRXQuantum.2.010323.

[138] X. Bonet-Monroig, R. Sagastizabal, M. Singh, and T. E. O'Brien, *Low-cost error mitigation by symmetry verification*, Phys. Rev. **A 98** (2018) 062339. doi:10.1103/PhysRevA.98.062339.

[139] W. J. Huggins, *Virtual Distillation for Quantum Error Mitigation*, Phys. Rev. **X 11** (2021) 041036. doi:10.1103/PhysRevX.11.041036.

[140] R. Sagastizabal, *Experimental error mitigation via symmetry verification in a variational quantum eigensolver* Phys. Rev. **A 100** (2019) 010302. doi:10.1103/PhysRevA.100.010302.

[141] S. McArdle, Sam, X. Yuan, and S. Benjamin, *Error-Mitigated Digital Quantum Simulation*, Phys. Rev. Lett. **122** (2019) 180501. doi:10.1103/PhysRevLett.122.180501.

[142] P. A. M. Dirac, *The Principles of Quantum Mechanics*, Nature **136** (1935) 411. doi:10.1038/136411a0.

[143] P. Siwach and D. Lacroix, *Filtering states with total spin on a quantum computer*, Phys. Rev. **A 104** (2021) 062435. doi:10.1103/PhysRevA.104.062435.

[144] A. Marzuoli, M. Rasetti, *Spin network quantum simulator*, Physics Letters **A 306** (2002) 79. doi:10.1016/S0375-9601(02)01600-6.

[145] A. Marzuoli and M. Rasetti, *Computing spin networks*, Ann. Phys. **318** (2005) 345. doi:10.1016/j.aop.2005.01.005.

[146] S. P. Jordan, *Permutational quantum computing*, Quantum Inf. Comput. **10** (2010) 470. doi:10.5555/2011362.2011369.

[147] J. Tilly et al., *The Variational Quantum Eigensolver: a review of methods and best practices*, Phys. Rep. **986** (2022) 1. doi:10.1016/j.physrep.2022.08.003.

[148] Elegant themes `https://www.elegantthemes.com/blog/freebie-of-the-week/beautiful-flat-icons-for-free`.

[149] Flaticon flaticon.es/iconos-gratis/computacion-cuantica

[150] B. Atalay, D. M. Brink, and A. Mann, *A product form for projection operators*, Physics Letters **B 46** (1973) 145. doi:10.1016/0370-2693(73)90666-7.

[151] K. Choi et al., *Rodeo Algorithm for Quantum Computing*, Phys. Rev. Lett. **127** (2021) 040505. doi:10.1103/PhysRevLett.127.040505.

[152] Z. Qian et al., *Demonstration of the Rodeo Algorithm on a Quantum Computer*, arXiv:2110.07747v1 (2021). doi:arXiv.2110.07747.

[153] M. Bee-Lindgren et al., *Rodeo Algorithm with Controlled Reversal Gates*, arXiv:2208.13557v1 (2022), doi:10.48550/arXiv.2208.13557.

[154] G. L. Long, *General quantum interference principle and duality computer*, Common. Theor. Phys. **45** (2006) 825. doi:10.1088/0253-6102/45/5/013.

[155] A. M. Childs and N. Wiebe, *Hamiltonian simulation using linear combinations of unitary operations*, Quantum Inf. Compt. **12** (2012) 901. doi:10.5555/2481569.2481570.

[156] D. W. Berry et al., *Exponential improvement in precision for simulating sparse Hamiltonians*, Proceedings of the 46th ACM Symposium on Theory of Computing (2014) 283. doi:10.1145/2591796.2591854.

[157] D. Berry, A. Childs, and R. Kothari, *Hamiltonian Simulation with Nearly Optimal Dependence on all Parameters* Proceedings of the 56th IEEE Symposium on Foundations of Computer Science (2015) 792. doi:10.1109/FOCS.2015.54.

[158] S. Wei, H. Li, and G. Long, *A Full Quantum Eigensolver for Quantum Chemistry Simulations*, Research **2020** (2020). doi:10.34133/2020/1486935.

[159] A. J. da Silva, and D. K. Park, *Linear-depth quantum circuits for multi-qubit controlled gates*, Phys. Rev. **A 106** (2022) 042602. doi:10.1103/PhysRevA.106.042602.

[160] V. V. Shende, S. S. Bullock, and I. L. Markov, *Synthesis of Quantum Logic Circuits*, IEEE Trans. on Computer-Aided Design **25** (2006) 1000. doi:10.1109/TCAD.2005.855930.

[161] S.S. Bullock and I.L. Markov, *Asymptotically optimal circuits for arbitrary n-qubit diagonal comutations*, Quant. Inf. and Comp., **4** (2004) 27. doi:10.5555/2011572.2011575.

[162] J. Welch, D. Greenbaum, S. Mostame, and A. Aspuru-Guzik, *Efficient quantum circuits for diagonal unitaries without ancillas*, New J. Phys. **16** (2014) 033040. doi:10.1088/1367-2630/16/3/033040.

[163] G.M. D'Ariano, M. G. A. Paris, M. F. Sacchi, *Quantum Tomography*, Adv. Imaging Electron Phys. **128** (2003) 205. doi:10.48550/arXiv.quant-ph/0302028

[164] G. Torlai et al. *Quantum process tomography with unsupervised learning and tensor networks*. Nat Commun **14** (2023) 2858. doi:10.1038/s41467-023-38332-9

[165] M. Cramer et al., *Efficient quantum state tomography*, Nat Commun **1** (2010) 149. doi:10.1038/ncomms1147

[166] G. Torlai et al., *Neural-network quantum state tomography*, Nature Phys **14** (2018) 447. doi:10.1038/s41567-018-0048-5

[167] J. Carrasquilla et al., *Reconstructing quantum states with generative models*, Nat Mach Intell **1** (2019) 155. doi:10.1038/s42256-019-0028-1

[168] A. Scott, *Shadow Tomography of Quantum States*, Proceedings of the 50th Annual ACM SIGACT Symposium on Theory of Computing (2018) 325. doi:10.1145/3188745.3188802

[169] HY. Huang, R. Kueng, and J. Preskill, *Efficient Estimation of Pauli Observables by Derandomization*, Phys. Rev. Lett. **127** (2021) 030503. doi:10.1103/PhysRevLett.127.030503

[170] T. Tsuchimochi, Y. Mori, and S. L. Ten-no, *Spin-projection for quantum computation: A low-depth approach to strong correlation*, Phys. Rev. Res. **2** (2020) 043142. 10.1103/PhysRevResearch.2.043142

[171] R. D. Somma, *Quantum eigenvalue estimation via time series analysis*, New J. Phys. **21** (2019) 123025. doi:10.1088/1367-2630/ab5c60.

[172] K. Mitarai, M. Negoro, M. Kitagawa, and K. Fujii, *Quantum circuit learning*, Phys. Rev. **A 98** (2018) 032309. doi:10.1103/PhysRevA.98.032309.

[173] M. Schuld et al., *Evaluating analytic gradients on quantum hardware*, Phys. Rev. **A 99** (2019) 032331. doi:10.1103/PhysRevA.99.032331.

[174] A. F. Izmaylov, R. A. Lang, and T. Yen, *Analytic gradients in variational quantum algorithms: Algebraic extensions of the parameter-shift rule to general unitary transformations*, Phys. Rev. **A 104** (2021) 062443. doi:10.1103/PhysRevA.104.062443.

[175] D. Horn, and M. Weinstein, *The $t$ expansion: A nonperturbative analytic tool for Hamiltonian system*, Phys. Rev. **D 30** (1984) 1256. doi:10.1103/PhysRevD.30.1256.

[176] Y. Saad, *Numerical Methods for Large Eigenvalue Problems*. Revised Edition, SIAM, Philadelphia. (2011) doi:10.1137/1.9781611970739.

[177] B. Fornberg, *Generation of Finite Difference Formulas on Arbitrarily Spaced Grids*, Math. Comp. **51** (1988) 699 doi:10.1090/S0025-5718-1988-0935077-0.

[178] Wiki page for finite difference – `https://en.wikipedia.org/wiki/Finite_difference_coefficient`

[179] A. Mari, T. R. Bromley, and N. Killoran, *Estimating the gradient and higher-order derivatives on quantum hardware* Phys. Rev. **A 103** (2021) 012405. doi:10.1103/PhysRevA.103.012405.

[180] D. Wierichs, J. Izaac, C. Wang, and C. Y. Lin, *General parameter-shift rules for quantum gradients*, Quantum **6** (2022) 677. doi:10.22331/q-2022-03-30-677.

[181] A. Blais et al., *Cavity quantum electrodynamics for superconducting electrical circuits: An architecture for quantum computation*, Phys. Rev. **A 69** (2004) 062320. doi:10.1103/PhysRevA.69.062320

[182] S. Sun et al., *Quantum Computation of Finite-Temperature Static and Dynamical Properties of Spin Systems Using Quantum Imaginary Time Evolution*, PRX Quantum **2** (2021) 010317. 10.1103/PRXQuantum.2.010317.

[183] C. L. Cortes and S. K. Gray, *Quantum Krylov subspace algorithms for ground- and excited-state energy estimation*, Phys. Rev. **A 105** (2022) 022417. doi:10.1103/PhysRevA.105.022417.

[184] N. H. Stair, R. Huang, and F. A. Evangelista, *A Multireference Quantum Krylov Algorithm for Strongly Correlated Electrons*, J. Chem. Theory Comput. **16** (2020) 2236. doi:10.1021/acs.jctc.9b01125.

[185] R. M. Parrish and P. L. McMahon, *Quantum Filter Diagonalization: Quantum Eigendecomposition without Full Quantum Phase Estimation* arXiv (2019). doi:10.48550/arXiv.1909.08925.

[186] F. Bloch, *Nuclear Induction*, Phys. Rev. **70** (1946) 460. doi:10.1103/PhysRev.70.460.

[187] B. Bauer et al., *Hybrid Quantum-Classical Approach to Correlated Materials*, Phys. Rev. **X 6** (2016) 031045. doi:10.1103/PhysRevX.6.031045.

[188] G. Fano and S.M. Blinder, *Mathematical Physics in Theoretical Chemistry*, Elsevier, (2019). https://www.elsevier.com/books/mathematical-physics-in-theoretical-chemistry/blinder/978-0-12-813651-5.

[189] Vojtech Havlicek, Sergii Strelchuk, and Kristan Temme, *Classical algorithm for quantum SU(2) Schur sampling* Phys. Rev. **A 99** (2019) 062336. doi:10.1103/PhysRevA.99.062336.

[190] J. Ripoche, T. Duguet, J. -P. Ebran, and D. Lacroix, *Combining symmetry breaking and restoration with configuration interaction: Extension to $z$-signature symmetry in the case of the Lipkin model*, Phys. Rev. **C 97** (2018) 064316. doi:10.1103/PhysRevC.97.064316.

[191] M. Plesch and Č, Brukner, *Quantum-state preparation with universal gate decompositions*, Phys. Rev. **A 83** (2011) 032302. doi:10.1103/PhysRevA.83.032302.

universite PARIS-SACLAY

ÉCOLE DOCTORALE
Particules, hadrons, énergie et noyau : instrumentation, imagerie, cosmos et simulation (PHENIICS)

**Title:** Symmetry breaking and restoration for many-body problems treated on quantum computers

**Abstract:** The Symmetry-Breaking/Symmetry-Restoration methodology is a well-established tool in many-body physics, which serves to enhance the approximation of a Hamiltonian's ground state energy within a variational framework. At its core, this technique involves intentionally breaking in the wave function ansatz the symmetries respected by the Hamiltonian during the search for the ground state. This process allows us to capture complex correlations between particles at a lower numerical cost compared to a symmetry-preserving framework. Given that the true ground state must be consistent with the problem symmetries, a symmetry-restoration step is a priori needed to project the symmetry-breaking ansatz back into the appropriate subspace of the total Hilbert space.
This thesis mainly investigates how to implement the symmetry-breaking/symmetry-restoration scheme using quantum computers. The Variational Quantum Eigensolver (VQE) is the quantum algorithm used for the variational component. Its goal is to prepare an approximation of the ground state using a parametric quantum state ansatz by minimizing the energy associated with the Hamiltonian. The Bardeen-Cooper-Schrieffer (BCS) ansatz was employed as a symmetry-breaking ansatz, and the applications were demonstrated using the pairing and Hubbard Hamiltonians. We also used the Jordan-Wigner transformation to encode the operators based on the Hamiltonians. With these elements, we can identify two ways of using the symmetry-breaking/symmetry-restoration process in conjunction with the VQE method: we can either vary the parameters of a symmetry-breaking ansatz before or after the symmetry restoration. The first approach is identified as Quantum Projection After Variation, while the latter is named Quantum Variation After Projection.
One of the achievements of this thesis is the development of diverse techniques for symmetry restoration. Some of them are based on the Quantum Phase Estimation (QPE) algorithm, including the Iterative QPE and Rodeo methods. We also investigated other techniques based on the quantum "Oracle" concept, such as the Oracle+Hadamard and Grover-Hoyer methods. The Linear Combination of Unitaries algorithm was used to either directly implement the projection or the oracle in the "Oracle"-based methods. We also adopt the Classical Shadow formalism to restore symmetries, intending to optimize the quantum resources required for the symmetry-restoration step. The different methods for symmetry restoration are illustrated through the thesis for the particle number and total spin symmetries.
In the final part, we present hybrid quantum-classical techniques that can help extract valuable information about the low-lying spectrum of a Hamiltonian. Assuming we can accurately extract the Hamiltonian moments from their generating function using a quantum computer, we introduce two methods for spectra analysis: the t-expansion method, which combines imaginary time evolution and the Padé approximation, and the Krylov method, a quantum subspace expansion method that can provide information about the low-lying eigenvalues of the Hamiltonian and the evolution of the survival probability. Furthermore, we also present the Quantum Krylov method. This technique gives similar information to the Krylov method but without the need to estimate the Hamiltonian moments, a task that can be difficult on near-term quantum computers.

**Titre:** Brisure et restauration de symétrie pour les problèmes à plusieurs corps traités sur ordinateurs quantiques

**Résumé:** La méthode de Brisure de Symétrie/Restauration de Symétrie est un outil largement utilisé pour traiter un ensemble de particules en interaction. Cette méthode permet souvent d'améliorer l'approximation de l'énergie de l'état fondamental d'un Hamiltonien dans un cadre variationnel. Dans cette technique, certaines symétries du système sont brisées intentionnellement dans la fonction d'onde utilisée lors de la recherche de l'état fondamental. Ce processus permet de capturer des corrélations complexes entre les particules à un coût numérique plus faible par rapport à un cadre qui préserve ces symétries. Étant donné que le véritable état fondamental doit respecter les symétries du problème, une étape de restauration de symétrie est a priori nécessaire pour revenir dans le sous-espace approprié de l'espace de Hilbert total.

Cette thèse examine comment mettre en œuvre le schéma de Brisure de Symétrie/Restauration de Symétrie à l'aide d'ordinateurs quantiques. L'algorithme quantique que nous utilisons pour la composante variationnelle est la méthode de "Variational Quantum Eigensolver (VQE)". Son but est de préparer une approximation de l'état fondamental en utilisant une fonction d'onde paramétrisée afin de minimiser l'énergie associée au Hamiltonien. L'ansatz quantique de Bardeen-Cooper-Schrieffer (BCS) a été utilisé comme ansatz brisant les symétries, et les techniques développées ont été testées en utilisant les Hamiltoniens d'appariement et le modèle de Hubbard. Nous avons également utilisé la transformation de Jordan-Wigner pour encoder les opérateurs présents dans le Hamiltonien. Avec ces outils, à l'instar des calculs sur ordinateurs classiques, deux manières d'utiliser le processus de Brisure de Symétrie/Restauration de Symétrie en conjonction avec la méthode VQE sont développés : nous pouvons varier les paramètres d'un ansatz de Brisure de Symétrie avant ou après la restauration de symétrie. La première approche est identifiée comme Projection Quantique Après Variation, tandis que la dernière est nommée Variation Quantique Après Projection.

L'un des aboutissements de cette thèse est le développement de diverses techniques pour la restauration de symétrie. Certaines d'entre elles sont basées sur l'algorithme d'Estimation de Phase Quantique (QPE), ce qui comprend les méthodes de QPE itérative et de Rodéo. Nous avons également étudié d'autres techniques basées sur le concept "d'Oracles quantiques", telles que les méthodes Oracle+Hadamard et Grover-Hoyer. L'algorithme de Combinaison Linéaire d'opérateurs Unitaires a été utilisé pour implémenter directement soit la projection, soit l'oracle. Nous avons également adapté le formalisme de "Classical Shadows" pour la restauration de symétries dans le but d'optimiser les ressources quantiques nécessaires. Les différentes méthodes de restauration de symétrie sont illustrées à travers la thèse pour les symétries associées au nombre de particules et au spin total.

Dans la dernière partie de la thèse, des techniques dites hybrides quantiques-classiques sont présentées, qui peuvent aider à extraire des informations précieuses sur le spectre de basse énergie du Hamiltonien. En supposant qu'il est possible d'extraire avec précision les moments du Hamiltonien à partir de leur fonction génératrice à l'aide d'un ordinateur quantique, nous introduisons deux méthodes d'analyse des spectres : la méthode de t-expansion, qui combine l'évolution en temps imaginaire et l'approximation de Padé, et la méthode de Krylov, une méthode d'expansion du sous-espace quantique qui peut fournir des informations sur les valeurs propres du Hamiltonien et sur l'évolution de la probabilité de survie. De plus, nous présentons également des résultats de la méthode Krylov Quantique. Cette technique donne des informations similaires à la méthode de Krylov, mais sans avoir besoin d'estimer les moments du Hamiltonien, une tâche qui reste difficile dans les ordinateurs quantiques actuels.